\documentclass[12pt,a4paper]{article}

\usepackage[british]{babel}

\usepackage[a4paper,top=2cm,bottom=2cm,left=2.5cm,right=2.5cm,marginparwidth=1.75cm]{geometry}

\usepackage[numbers]{natbib}
\usepackage{amsmath}
\usepackage{graphicx}
\usepackage[colorlinks=true, allcolors=blue]{hyperref}
\usepackage{hyperref}
\usepackage[title]{appendix}
\usepackage{mathrsfs}
\usepackage{amsfonts}
\usepackage{booktabs} % For \toprule, \midrule, \botrule
\usepackage{caption}  % For \caption
\usepackage{threeparttable} % For table footnotes
\usepackage{algorithm}
\usepackage{algorithmicx}
\usepackage{algpseudocode}
\usepackage{listings}
\usepackage{enumitem}
\usepackage{chngcntr}
\usepackage{booktabs}
\usepackage{lipsum}
\usepackage{subcaption}
\usepackage{authblk}
\usepackage[T1]{fontenc}    % Font encoding
\usepackage{csquotes}       % Include csquotes
\usepackage{diagbox}
\usepackage{subcaption}
\usepackage{graphicx}
\usepackage{float}
\usepackage{caption}
\usepackage{subcaption}
\usepackage{tikz}
\usepackage{amsmath}
\usepackage{color}

\usepackage[symbol]{footmisc}

\usepackage{setspace}
\usepackage{titlesec}
\titleformat{\section} % Redefine section numbering format
  {\normalfont\Large\bfseries}{\thesection.}{1em}{}
  
\usepackage{lineno} % Add the lineno package

\rightlinenumbers % Right-align line numbers

\usepackage{float}   % for customizing caption position
\usepackage{caption} % for customizing caption format


\title{Hybrid machine learning models based on physical patterns to accelerate CFD simulations: a short guide on autoregressive models}

\author[1,*]{A. Sengupta}
\author[1]{R. Abadía-Heredia}
\author[1]{A. Hetherington}
\author[1]{J. Miguel Pérez}
\author[1]{S. Le Clainche}
\affil[1]{\small ETSI Aeronautica y del Espacio, Universidad Politecnica de Madrid, Plaza Cardenal
Cisneros, 3, Madrid, 28040, Spain}
\affil[*]{Corresponding author: \texttt{a.sengupta@upm.es}(Arindam Sengupta) }

\date{}  % Remove date

\begin{document}
\maketitle

\begin{abstract}

Accurate modeling of the complex dynamics of fluid flows is a fundamental challenge in computational physics and engineering. This study presents an innovative integration of High-Order Singular Value Decomposition (HOSVD) with Long Short-Term Memory (LSTM) architectures to address the complexities of reduced-order modeling (ROM) in fluid dynamics. HOSVD improves the dimensionality reduction process by preserving multidimensional structures, surpassing the limitations of Singular Value Decomposition (SVD). The methodology is tested across numerical and experimental data sets, including two- and three-dimensional (2D and 3D) cylinder wake flows, spanning both laminar and turbulent regimes. 

The emphasis is also on exploring how the depth and complexity of LSTM architectures contribute to improving predictive performance. Simpler architectures with a single dense layer effectively capture the periodic dynamics, demonstrating the network’s ability to model non-linearities and chaotic dynamics. The addition of extra layers provides higher accuracy at minimal computational cost. These additional layers enable the network to expand its representational capacity, improving the prediction accuracy and reliability. 

The results demonstrate that HOSVD outperforms SVD in all tested scenarios, as evidenced by using different error metrics. Efficient mode truncation by HOSVD-based models enables the capture of complex temporal patterns, offering reliable predictions even in challenging, noise-influenced data sets. The findings underscore the adaptability and robustness of HOSVD-LSTM architectures, offering a scalable framework for modeling fluid dynamics.

\end{abstract}


%-------------------------------------------
% Paper Body
%-------------------------------------------
%--- Section ---%
\section{Introduction}

Modeling the intricacies of complex fluid flows is essential for applications in numerous natural and industrial processes. Advancements in numerical and computational techniques have transformed the study of fluid flows, enabling precise simulations of phenomena ranging from laminar to turbulent regimes. However, achieving high-fidelity fluid flow simulations using numerical methods, such as Computational Fluid Dynamics (CFD), often incurs a substantial computational cost. Conventional CFD methodologies, such as direct numerical simulation (DNS) and large-eddy simulation (LES), involve solving complex problems with high-dimensional systems but are computationally expensive \cite{abadia2022predictive, lopez2021model, le2019prediction}. These methods often require substantial computational resources and extensive processing times to generate high-dimensional data, making them a difficult option to employ for many real-world applications.

To address these challenges, researchers have increasingly turned to machine learning (ML) techniques and reduced order models (ROMs), which have shown remarkable success in learning complex patterns and temporal dynamics from data. ML algorithms, particularly deep learning (DL) models with artificial neural networks (ANN), can process vast amounts of data to identify underlying structures and predict future states. Convolutional Neural Networks (CNNs) and Long Short-Term Memory (LSTM) networks are two prominent ML optimization architectures that have been adeptly employed in various applications, including image recognition, time-series forecasting, and, more recently, fluid dynamics. Each offers unique advantages that address the specific challenges inherent in fluid flow prediction. LSTMs are effective in capturing long-term dependencies in sequential data \cite{siami2018comparison}. In fluid dynamics, LSTMs are utilized to capture and predict the temporal evolution of flow patterns and dynamic behaviors over time. Their ability to process sequential data makes them particularly effective for handling the temporal complexities of fluid flows, such as those seen in turbulent wake dynamics or oscillatory jet flows. CNNs are particularly effective for classification and image recognition tasks \cite{jacobsen2017multiscale}. The strengths of CNNs in feature extraction and pattern recognition make them highly suitable for applications in the field of fluid mechanics \cite{guo2016convolutional}. Using their capacity to identify local patterns, such as vortices or boundary layer structures, CNNs provide an essential tool for analyzing the spatial characteristics of fluid flows. For example, Guastoni et al. \cite{guastoni2020prediction} used CNNs to predict wall-bounded turbulence, achieving substantial improvements in predictive accuracy over traditional methods. Drikakis and Sofos \cite{drikakis2023can} explored how deep neural networks can model turbulent flows, showcasing the ability to capture intricate flow patterns that traditional methods may miss. Similarly, Vinuesa et al. \cite{vinuesa2023transformative} discussed in their paper how ML techniques have been used to improve the modeling of turbulence and the overall performance of computational fluid dynamics (CFD) simulations. They highlighted that ML approaches can efficiently handle the complex, high-dimensional data typical in fluid mechanics, providing significant advances in predictive accuracy and computational efficiency.

Another important aspect of a DL model is its architecture. A well-designed architecture is crucial to effectively capture the temporal evolution and nonlinear dynamics of fluid flow phenomena. The proper development of these architectures improves the ability of the model to predict complex fluid flows with greater accuracy and efficiency, providing a robust framework for solving computationally intensive problems in fluid mechanics. LSTMs have been integrated with ROMs, autoencoders, and hybrid frameworks to enhance their predictive capabilities and improve computational efficiency. An application of this is the work by Wiewel et al. \cite{wiewel2019latent}, who introduced an LSTM-based latent space physics framework where raw high-dimensional data were first reduced using autoencoders into a latent space. In this latent space, the LSTMs modeled the temporal evolution of the flow dynamics. The architecture relied on the ability of autoencoders to compress complex spatial data into manageable representations, while the LSTMs provided temporal forecasting capabilities. This approach was particularly effective in cylinder wake problems, where the architecture captured nonlinear temporal interactions in a computationally efficient manner. Building on this, Wiewel et al. \cite{wiewel2020latent} extended their work with a latent-space subdivision approach. This architecture used multiple LSTM models operating in subdivided latent spaces, with each LSTM responsible for a specific subset of the dynamics. This refinement improved the stability of the architecture over long prediction horizons and its adaptability to complex scenarios like jet flows. By handling each component of the latent space separately, the architecture was able to better account for different temporal dynamics and ensure accurate forecasting over extended timescales. LSTM and CNNs can also be combined to create hybrid architectures. Han et al. \cite{han2021hybrid} demonstrated this by proposing a hybrid deep neural network architecture designed to model unsteady flow fields with moving boundaries, such as oscillating cylinders. In this architecture, CNN layers were employed to extract spatial features from flow-field snapshots, capturing localized patterns such as vortices and wake structures. These spatial features were then passed to the LSTM layers, which are well-suited for modeling temporal dependencies, enabling accurate predictions of the flow's future states. This hybrid design effectively integrates spatial and temporal dynamics, making it particularly effective in predicting unstable wake behavior.

However, there is no fixed methodology tailored to fluid mechanics problems, as the optimal architecture configuration often depends on the specific characteristics of the flow under study. This challenge has led to the development of case-specific architectures, as shown in the studies above. Factors such as the number of layers, the selection of neurons per layer, and other critical parameters must be investigated to determine the optimal architecture for fluid mechanics problems. The study by Cao et al. \cite{cao2025automatic} further reinforces this notion. They proposed a Pareto-based optimization method to automatically discover the best composite architectures. Instead of defining a single "best" architecture, they construct a Pareto front, optimizing for multiple objectives such as accuracy, training time, and model complexity. This study reflects the need for tailored deep learning models, where architecture selection is guided by the unique dynamics of the system under consideration.

On the other hand, ROMs have surfaced as a promising approach to achieving both efficiency and accuracy for cleaning and data extraction from CFD simulations, experimental, and real-world datasets. ROMs significantly reduce the computational cost by projecting high-dimensional data onto a lower-dimensional subspace, capturing the most essential features of the system \cite{pant2021deep}. Several studies have demonstrated the successful application of ROMs in fluid dynamics. Rowley and Dawson \cite{rowley2017model} provided a comprehensive review of model reduction techniques, including Proper Orthogonal Decomposition (POD) \cite{schmid2010dynamic}, Dynamic Mode Decomposition (DMD) \cite{sirovich1987turbulence}, and other methods for flow analysis and control, showcasing the efficiency of POD and DMD in reducing computational costs. Schmid \cite{schmid2010dynamic} and Kutz et al. \cite{kutz2016dynamic} explored the utility of dynamic mode decomposition to capture the dynamics of complex systems with experimental and numerical data, demonstrating its role in improving traditional ROMs. Similarly, Taira et al. \cite{taira2017modal} elucidated the merits of Singular Value Decomposition (SVD) \cite{golub1971singular} and DMD in fluid dynamics, emphasizing its ability to capture essential fluid characteristics with minimal computational effort. SVD serves as one of the primary techniques for calculating POD modes, as highlighted by Le Clainche \cite{le2020introduction}. By decomposing the flow field into orthogonal spatial modes and their corresponding temporal coefficients, SVD enables efficient dimensionality reduction while preserving the most significant features of the flow dynamics. This dual capability makes SVD a cornerstone of reduced-order modeling in fluid dynamics, offering both computational efficiency and physical interpretability. Begiashvili et al. \cite{begiashvili2023data} provides a comprehensive review of the most widely used modal decomposition techniques, including DMD, SVD, spectral POD \cite{sieber2016spectral}, higher-order DMD \cite{vega2020higher}, and resolvent analysis \cite{jovanovic2005componentwise}, emphasizing their strengths and limitations in fluid dynamics applications. The study highlights how each method performs in developing ROMs and extracting dominant patterns from complex flow fields, offering critical insights into their suitability for various fluid dynamic problems. Both POD and DMD are powerful and complementary techniques for identifying flow structures in fluid dynamics. POD modes are orthogonal in space, meaning that each POD mode captures multiple frequencies. Conversely, DMD modes are orthogonal in time, with each mode associated with a single frequency, making DMD particularly well-suited for identifying flow instabilities. In linear flows, the DMD modes correspond to the linear stability modes, while in periodic flows, they represent Fourier modes. In saturated flows, the DMD modes capture groups of flow structures that effectively model the flow dynamics \cite{gomez2012four}. This distinction allows POD to excel in capturing dominant spatial structures and DMD to provide detailed temporal dynamics.

The fusion of ROMs with deep learning architectures has attracted a lot of attention in recent times. These models, termed hybrid ROMs, can capture the underlying physics while using deep learning architectures to construct highly robust predictive models \cite{le2022data, abadia2022predictive}. Abadía-Heredia et al. \cite{abadia2022predictive} introduced a hybrid ROM that integrates POD with CNN and LSTM networks to predict fluid flow in complex scenarios. Their model significantly reduced computational time from tens of hours to a few minutes while maintaining accuracy and demonstrated the potential of combining POD with deep learning architectures to obtain efficient and accurate fluid dynamics predictions. This hybrid approach leverages the strengths of both methodologies to model complex, non-linear interactions within the flow field. In addition, Abadía-Heredia et al. \cite{abadia2024exploring} presented hybrid ROMs that integrate SVD with neural networks and compared these with purely deep learning-based predictive ROMs, which use autoencoders (AEs) for dimensionality reduction. The study demonstrated that hybrid ROMs are more robust and require fewer snapshots for training while maintaining accuracy across different cases, including turbulent flows. Furthermore, hybrid ROMs offer greater generalizability by allowing neural network architectures to remain consistent as they reduce the data to a few dominant POD modes. Le Clainche et al. \cite{le2022data} demonstrated that the combination of machine learning with DMD effectively captures the main flow instabilities and energy-producing mechanisms in turbulent flows, resulting in accurate long-term predictions of wall-shear stress. San and Maulik \cite{san2018machine} applied artificial neural networks (ANN) in conjunction with POD for real-time prediction of flow fields, showcasing the potential of hybrid ROMs in fluid dynamics. In another study, Xu et al. \cite{xu2022data} proposed a hybrid model that integrates POD with advanced deep learning models, including LSTM and CNN networks, for fluid flow prediction. The model used POD to extract dominant spatial modes from high-dimensional flow data, significantly reducing computational complexity. The reduced temporal coefficients were then fed into the LSTM and CNN models for sequential forecasting. All of these studies clearly highlight the advantages of integrating modal decomposition techniques with DL models for forecasting in the domain of fluid mechanics.

In this work, we present a novel hybrid ROM model that integrates High-Order Singular Value Decomposition (HOSVD) \cite{tucker1966some} with different LSTM architectures to predict flow field solutions. HOSVD significantly improves the traditional SVD by extending its capabilities to multi-way data arrays (tensors). The main advantage of using SVD/HOSVD is that data is reorganized into matrices, allowing it to work with a reduced quantity of data. This provides a comprehensive framework for dimensionality reduction that can capture the flow physics and patterns within the data. This work builds on the foundational work of Abadía-Heredia et al. \cite{abadia2022predictive}, which integrated POD with deep learning models. The methodology involves applying HOSVD to the temporal data to extract the dominant modes, which are then used as input to neural networks for temporal evolution prediction. This approach enhances the dimensionality reduction process, leading to more accurate and efficient temporal predictions. Another objective is to explore the design of LSTM-based deep learning architecture. By constructing and testing various LSTM architectures with different levels of complexity, the objective is to understand the relationship between architectural depth and the predictive performance required for various cases with different levels of complexity. This work seeks to optimize LSTM architectures, analyzing factors such as layers, number of neurons, learning rate, and other key parameters to better understand their impact on modeling the intricate temporal dynamics of fluid flows. This analysis is crucial for identifying how deep these architectures need to be to effectively model the intricate spatio-temporal dynamics of fluid flows. The performance of the enhanced ROM has been evaluated in complex fluid dynamics problems involving both numerical and experimental data. These test cases include a range of fluid flow scenarios that encompass turbulent and laminar flows.

In addition to introducing this novel methodology, this article also serves as a guide for building autoregressive models using DL architectures. This guide is designed to assist beginners and intermediate users interested in developing hybrid ROMs for temporal forecasting. The article compiles essential information on data preprocessing, dimensionality reduction, network design, debugging, and training strategies, offering step-by-step insights. Furthermore, it tries to consolidate best practices from experienced developers. To the authors’ knowledge, this is the first practical guide dedicated to developing physics-based hybrid ROMs aimed at accelerating CFD simulations while maintaining high predictive accuracy.

\begin{figure}[h!]
    \centering
    \includegraphics[width=1\textwidth,height=0.65\textwidth,keepaspectratio]{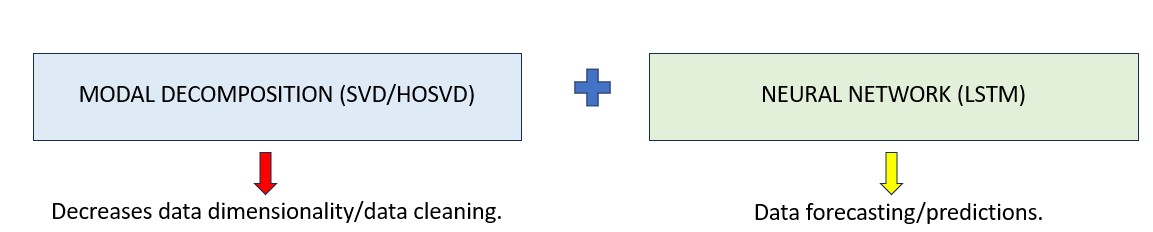}
    \caption{Combination of modal decomposition and neural network for Hybrid ROM.}
    \label{fig:hybrid_rom}
\end{figure}

The article has been divided into four sections. Section 2 includes the developed methodology combining SVD and HOSVD with deep learning architectures. Section 3 introduces the cases and discusses the results obtained. Conclusions are presented in Section 4.

%--- Section ---%
\section{Methodology}\label{sec2}

This section outlines the development of a hybrid ROM that integrates SVD and HOSVD with LSTM architectures to improve predictive accuracy and computational efficiency. Both decomposition techniques have been implemented separately and compared. Determining the appropriate network depth is essential, as deeper networks can model complex patterns but risk overfitting when data is limited, whereas shallower networks may fail to capture intricate fluid flow dynamics. Choosing optimal hyperparameters, such as learning rate, batch size, and sequence length, is another hurdle, as these parameters greatly impact model performance and often require extensive experimentation to fine-tune. The selection of the number of neurons per layer is equally important as too few neurons may lead to underfitting, while an excess can result in overfitting and increased computational costs. Additionally, managing computational resources is vital, as complex architectures will incur additional costs \cite{balderas2024optimizing}. The complexity of the model must be carefully balanced with the available resources to ensure practical training and deployment. A broad overview of these foundational aspects, not specific to any case but general principles and strategies that were followed for the development of the optimization model, are presented in the Appendix.

This integrated approach addresses the challenges of processing high-dimensional spatio-temporal data by effectively capturing the underlying dynamics of fluid flows, which are crucial for simulations in engineering and research applications. This section is organized as follows: (i) data structure, (ii) modal decomposition techniques, (iii) deep learning model, (iv) data preprocessing, and (v) autoregression. 

\begin{figure}[h!]
    \centering
    \includegraphics[width=1\textwidth, height=1\textwidth,keepaspectratio]{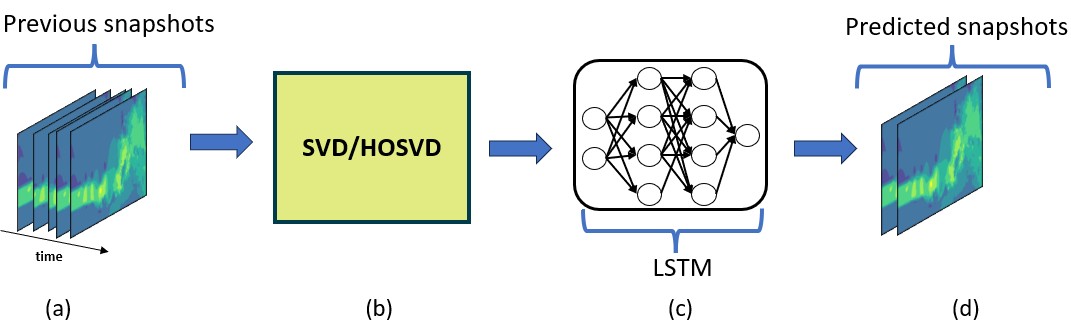}
    \caption{Illustration of the hybrid ROM methodology.}
    \label{fig:meth}
\end{figure}

The proposed hybrid ROM methodology is summarized in Figure~\ref{fig:meth}. The approach consists of the following key steps:

\begin{enumerate}
    \item[(a)] The model is provided with $N$ previous snapshots of the flow field, capturing its evolution over time.
    \item[(b)] The dimensionality of the snapshot data is reduced using SVD or HOSVD, resulting in a set of dominant spatial modes and corresponding temporal coefficients. 
    \item[(c)] The temporal coefficients obtained from SVD/HOSVD are used as input to train the DL model with an LSTM architecture, which learns their temporal evolution. The trained LSTM model predicts the temporal evolution of the coefficients over a specified future time horizon.
    \item[(d)] The predicted coefficients are then used to reconstruct the solution and predict the evolution of the flow field over time. Autoregression has been implemented to predict over multiple timesteps.
\end{enumerate}

\subsection{Data structure}

In this study, data are organized in the form of a multidimensional array (\textit{snapshot tensor}) to capture the temporal dynamics of fluid flow fields. This data is arranged in the snapshot matrix as \cite{abadia2022predictive}:

\begin{equation}
 	\boldsymbol{X} = \boldsymbol{{V}}_1^K = [\boldsymbol{V}_1, \boldsymbol{V}_2, \ldots, \boldsymbol{V}_k, \boldsymbol{V}_{k+1}, \ldots, \boldsymbol{V}_{K-1}, \boldsymbol{V}_K], 
 	\label{eq:snapshot_tensor}
\end{equation}

where $\boldsymbol{V}_k$ represents the variable of the flow field at time instant $\boldsymbol{t_k}$, and $\boldsymbol{K}$ is the total number of snapshots. The snapshot tensor is structured to include both spatial and temporal information, ensuring a comprehensive representation of the flow field dynamics. The data dimensions are given by $\boldsymbol{J} \times \boldsymbol{K}$, where $\boldsymbol{J}$ denotes the total number of spatial grid points along the streamwise, spanwise, and normal directions, respectively. 

The data sets used in this study include velocity components for two and three-dimensional flow fields. For two-dimensional flow fields, the data sets include the streamwise and normal velocity components (\(\boldsymbol{v}_x\) and \(\boldsymbol{v}_y\)), while for three-dimensional flow fields, the datasets include the streamwise, normal, and spanwise velocity components (\(\boldsymbol{v}_x\), \(\boldsymbol{v}_y\), and \(\boldsymbol{v}_z\)).

The generalized snapshot tensor for three-dimensional flow data with \(\boldsymbol{T}_{\mathrm{var}} = 3\) physical variables (velocity components) can be represented as follows \cite{hetherington2024modelflows}:
\begin{equation}
\begin{aligned}
    \boldsymbol{V}_{1j_2j_3j_4k} &= \boldsymbol{v}_x(x_{j_2}, y_{j_3}, z_{j_4}, t_k), \\
    \boldsymbol{V}_{2j_2j_3j_4k} &= \boldsymbol{v}_y(x_{j_2}, y_{j_3}, z_{j_4}, t_k), \\
    \boldsymbol{V}_{3j_2j_3j_4k} &= \boldsymbol{v}_z(x_{j_2}, y_{j_3}, z_{j_4}, t_k),
\end{aligned}
\label{eq:3d_tensor_specific}
\end{equation}
where:
\begin{itemize}
    \item \(j_1 = 1, 2, 3\) indexes the three velocity components (\(\boldsymbol{v}_x\), \(\boldsymbol{v}_y\), \(\boldsymbol{v}_z\)),
    \item \(j_2, j_3, j_4\) index the discrete spatial grid points in the \(x\), \(y\), and \(z\) directions, respectively,
    \item \(k\) indexes the temporal snapshots.
\end{itemize}

For two-dimensional flow fields, the datasets are represented by a fourth-order tensor since the spanwise (\(z\)) direction is not included. This organization allows for the application of SVD/HOSVD, which extracts dominant temporal modes. These are then used to train the LSTM network for temporal predictions. Depending on the specific case, the shape of the tensor can vary significantly, ranging from tensors that incorporate only one velocity component to those that account for all three velocity components (streamwise, spanwise, and vertical) or even some including several variables (i.e., pressure, several species in reactive flows, several pollutants in air pollution modeling, etc.). Predictions are generated for both two-dimensional and three-dimensional flow fields, ensuring that the methodology is comprehensive across distinct spatial configurations. The tensors are structured to incorporate all relevant physical variables and spatial-temporal dependencies, providing a robust representation of the flow dynamics.

\subsection{Modal Decomposition Techniques}
Singular Value Decomposition (SVD) is a fundamental matrix factorization technique extensively used in numerical analysis, machine learning, and data-driven reduced-order modeling. SVD decomposes a given snapshot matrix \( \boldsymbol{V}_1^K \) (representing spatio-temporal data) into three components: spatial modes \( \boldsymbol{U} \), singular values \( \boldsymbol{\Sigma} \), and temporal modes \( \boldsymbol{T}^\top \). The decomposition is expressed as \cite{abadia2022predictive}:
\begin{equation}
\boldsymbol{V}_1^K = \boldsymbol{U} \, \boldsymbol{\Sigma} \, \boldsymbol{T}^\top,
\label{eq:SVD}
\end{equation}
Where:
\begin{itemize}
    \item \( \boldsymbol{V}_1^K \) is the \( J \times K \) snapshot matrix, where \( J = T_{\text{var}} \cdot N_x \cdot N_y \cdot N_z \) is the total number of spatial degrees of freedom, and \( K \) is the number of temporal snapshots,
    \item \( \boldsymbol{U} \) is the \( J \times J \) orthogonal matrix of spatial modes,
    \item \( \boldsymbol{\Sigma} \) is the \( J \times K \) diagonal matrix of singular values, arranged in decreasing order, representing the energy associated with each mode,
    \item \( \boldsymbol{T}^\top \) is the \( K \times K \) orthogonal matrix of temporal modes, where \( \boldsymbol{T}^\top \) denotes the transpose of the temporal modes matrix.
\end{itemize}

The singular values \( \sigma_i \) in \( \boldsymbol{\Sigma} \) represent the importance of each corresponding mode in describing the energy or variance of the data. Larger singular values correspond to modes that capture more significant features of the dataset. These singular values are arranged in descending order within the diagonal matrix \( \boldsymbol{\Sigma} \). To select the most dominant modes while filtering out less significant ones, a subset of \( N \) dominant modes is chosen such that \cite{abadia2022predictive}:
\begin{equation}
\frac{\sigma_{N+1}}{\sigma_1} \leq \varepsilon,
\label{eq:tolerance}
\end{equation}
Where \( \sigma_1 \) is the largest singular value, \( \sigma_{N+1} \) is the first excluded singular value, and \( \varepsilon \) is a user-defined tolerance controlling the energy threshold or significance of the retained modes.

Using the dominant singular values and the corresponding spatial and temporal modes, the snapshot matrix \( \boldsymbol{V}_1^K \) can be approximated as \cite{abadia2022predictive}:
\begin{equation}
\boldsymbol{V}_1^K \simeq \boldsymbol{U}_N \, \boldsymbol{\Sigma}_N \, \boldsymbol{T}_N^\top,
\label{eq:reduced_SVD}
\end{equation}
where \( \boldsymbol{U}_N \) contains the first \( N \) spatial modes, \( \boldsymbol{\Sigma}_N \) contains the most significant singular values, and \( \boldsymbol{T}_N^\top \) contains the corresponding temporal modes.

This framework effectively combines dimensionality reduction with temporal evolution, enabling efficient and accurate reconstruction and prediction of flow dynamics at reduced computational cost. By truncating the smaller singular values, SVD enables dimensionality reduction and noise suppression, making it a powerful tool for applications such as data cleaning, image compression, and Principal Component Analysis (PCA) \cite{compton2020singular, wall2003singular}.

HOSVD extends these principles of SVD to multidimensional arrays. HOSVD preserves the inherent multifaceted nature of data, making it particularly suitable for applications where spatial, temporal, and other complex dependencies need to be maintained \cite{Tuck1963a, de2000multilinear, vega2020higher}. This approach allows for a more accurate and efficient representation of high-dimensional data, capturing the significant features along each mode \cite{kolda2009tensor, lu2011survey}. As a result, HOSVD is particularly advantageous for applications in fields such as fluid dynamics, image processing, and data compression, where preserving the multidimensional relationships within the data is crucial for accurate analysis and prediction.

HOSVD decomposes databases organized in tensor form, where SVD is applied to each one of the fibers of the tensor. For instance, the HOSVD of the fourth-order tensor \( \boldsymbol{\mathcal{X}} \) is presented as:

\begin{equation}
\mathcal{X}_{j_1j_2j_3k} \simeq \sum_{p_1=1}^{P_1} \sum_{p_2=1}^{P_2} \sum_{p_3=1}^{P_3} \sum_{n=1}^{N} \boldsymbol{S}_{p_1p_2p_3n}
   \boldsymbol{U}^{(1)}_{j_1p_1} \boldsymbol{U}^{(2)}_{j_2p_2} \boldsymbol{U}^{(3)}_{j_3p_3} \boldsymbol{T}_{kn},    
   \label{eq:HOSVD_4th_order}
\end{equation}

where \( \boldsymbol{S}_{p_1p_2p_3n} \) is the \emph{core tensor}, another fourth-order tensor, and the columns of the matrices
\( \boldsymbol{U}^{(1)} \), \( \boldsymbol{U}^{(2)} \), \( \boldsymbol{U}^{(3)} \), and \( \boldsymbol{T} \) are known as the \textit{modes} of the decomposition.

The first set of modes (i.e., the columns of the matrices \( \boldsymbol{U}^{(l)} \) for \( l = 1,2,3 \)) correspond to the number of components of the database and the spatial variables, so they are known as the \textit{spatial HOSVD modes}, while the columns of the matrix \( \boldsymbol{T} \) correspond to the time variable and are referred to as the \textit{temporal HOSVD modes}.

The decomposition involves singular values corresponding to each mode, signifying the importance of each mode. Similarly to SVD, dimensionality reduction is achieved by retaining significant modes based on a tolerance \( \varepsilon \) applied to each set \cite{hetherington2024modelflows}:

\begin{equation}
\frac{\sigma^{(1)}_{P_1+1}}{\sigma^{(1)}_1} \leq \varepsilon_1, \quad 
\frac{\sigma^{(2)}_{P_2+1}}{\sigma^{(2)}_1} \leq \varepsilon_2, \quad 
\frac{\sigma^{(3)}_{P_3+1}}{\sigma^{(3)}_1} \leq \varepsilon_3, \quad 
\frac{\sigma^{t}_{N+1}}{\sigma^{t}_1} \leq \varepsilon_4.
\label{eq:HOSVD_tolerances_horizontal_numbered}
\end{equation}

After dimensionality reduction, the tensor is approximated as \cite{hetherington2024modelflows}:
\begin{equation}
\boldsymbol{\mathcal{X}} \simeq \sum_{n=1}^{N} \boldsymbol{W}_{j_1j_2j_3n} \, \hat{\boldsymbol{T}}_{kn},
\label{eq:HOSVD_reduced}
\end{equation}
Where:
\begin{itemize}
    \item \( \boldsymbol{W}_{j_1j_2j_3n} \) are the spatial modes.
    \item \( \hat{\boldsymbol{T}}_{kn} = \sigma^t_n \, \boldsymbol{T}_{kn} \) are the temporal modes.
\end{itemize}

This reduced representation ensures that the essential dynamics of the data are preserved while significantly reducing computational complexity, enabling applications in predictive modeling and real-time analysis.

\subsection{Deep Learning Model}
Long Short-Term Memory Networks are a specialized type of recurrent neural network (RNN) designed to address the challenges of learning long-term dependencies in sequential data \cite{hochreiter1997long}. As highlighted by Yu et al. \cite{yu2019review}, LSTM networks have demonstrated unparalleled success in various applications, ranging from speech recognition to time-series forecasting, making them a cornerstone of modern deep learning research. Standard RNNs struggle with problems such as vanishing and exploding gradients, making them inefficient at handling temporal dependencies when the gap between relevant inputs is large. LSTM networks overcome this limitation by introducing gating mechanisms that control the flow of information through the network, allowing them to effectively "remember" and "forget" data as needed \cite{Goodfellow-et-al-2016}.

 Multiple architectures have been implemented to forecast the temporal evolution of reduced-order flow modes derived through dimensionality reduction techniques. The model begins with an input layer, followed by an LSTM layer, and ends with dense layers, forming a structured architecture. This combination enables the model to effectively generalize across diverse flow scenarios, providing accurate and reliable forecasts. Simpler architectures have fewer trainable parameters, which reduces computational costs. Hence, the approach starts by implementing the simplest architectures and progressively increasing complexity, allowing for a systematic evaluation of how network depth impacts performance. The architectures are structured as follows:

\begin{itemize}
    \item \textbf{LSTM 1 Dense:}  
    The simplest architecture consists of an input layer connected directly to an LSTM layer, followed by a single dense layer as output. The LSTM layer captures temporal dependencies, and the dense layer maps these representations to the output space using a linear activation function. This architecture focuses on leveraging the ability of LSTM to model sequential patterns without additional transformations, providing a baseline to evaluate the impact of deeper configurations.

    \begin{table}[h!]
    \centering
    \resizebox{0.9\textwidth}{!}{
    \begin{tabular}{|c|l|c|l|l|}
    \hline
    Layer & Layer Details &  Neurons & Activation Function & Dimension \\ \hline
    0        & Input         & modes          & -        & (seq\_len, modes) \\ \hline
    1        & LSTM          & 128        & -            & 128 \\ \hline
    2        & Dense         & modes & Linear            & modes \\ \hline
    \end{tabular}}
    \caption{Layer details for LSTM 1 Dense architecture.}
    \label{tab:lstm_arch1}
    \end{table}

    \item \textbf{LSTM 2 Dense:}  
    This architecture extends the first LSTM 1 dense architecture by introducing an intermediate dense layer with a non-linear activation function, such as LeakyReLU, before the final output dense layer. The additional dense layer refines the temporal features learned by the LSTM, allowing the network to capture more intricate relationships within the data. This architecture balances simplicity and additional feature transformation, potentially improving performance on moderately complex data sets.

    \begin{table}[h!]
    \centering
    \resizebox{0.9\textwidth}{!}{
    \begin{tabular}{|c|l|c|l|l|}
    \hline
    Layer & Layer Details &  Neurons & Activation Function & Dimension \\ \hline
    0        & Input         & modes         & -            & (seq\_len, modes) \\ \hline
    1        & LSTM          & 128        & -               & 128 \\ \hline
    2        & Dense         & 64         & Leaky ReLU      & 64 \\ \hline
    3        & Dense         & modes & Linear               & modes \\ \hline
    \end{tabular}}
    \caption{Layer details for LSTM 2 Dense architecture.}
    \label{tab:lstm_arch2}
    \end{table}

    \item \textbf{LSTM Time-Distributed:}  
    The final architecture features a single LSTM layer followed by a time-distributed layer. The time-distributed layer applies dense to each of the horizon timesteps individually. Unlike the original implementation by Abadía-Heredia et al. \cite{abadia2022predictive}, which uses a forecast horizon of 6, this work adopts a horizon of 2. Abadía-Heredia et al. developed a predictive model, which has been adapted into an autoregressive framework to align with the overall design of this study. This will help compare the results between single-step and multi-step temporal predictions. In addition, to maintain similarity, the number of neurons was changed from 100 to 128.

    \begin{table}[h!]
    \centering
    \resizebox{0.9\textwidth}{!}{
    \begin{tabular}{|c|l|c|l|l|}
    \hline
    Layer & Layer Details           &  Neurons & Activation Function & Dimension \\ \hline
    0        & Input                  & modes          & -     & (seq\_len, modes) \\ \hline
    1        & LSTM                   & 128        & -                   & 128 \\ \hline
    2        & Dense                  & 64         & Leaky ReLU          & 64 \\ \hline
    3       & Time-Distributed (Dense)  & 32         & Leaky ReLU          & 32 \\ \hline
    4       & Dense                   & modes       & Linear        & modes \\ \hline
    \end{tabular}}
    \caption{Layer details for LSTM Time-Distributed architecture.}
    \label{tab:lstm_arch4}
    \end{table}

\end{itemize}

The varying depth of dense layers in these architectures aims to explore the optimal configuration for modeling temporal data. Shallower architectures rely on the inherent ability of LSTM to capture temporal dependencies, while deeper architectures leverage additional dense layers for complex feature transformations. By systematically testing these configurations, this study seeks to determine the ideal balance between model complexity and generalization, ensuring robust predictions across diverse fluid dynamics scenarios.

The choice of activation function often depends on the task and the network architecture. Leaky ReLU was chosen as the activation function for this study due to its ability to address the limitations of standard ReLU in handling negative input values effectively. This feature makes it particularly advantageous for modeling both laminar and turbulent flow regimes. 
\begin{equation}
   \text{Leaky ReLU}(\boldsymbol{z}) = 
   \begin{cases} 
    \boldsymbol{z}, & \text{if } \boldsymbol{z} \geq 0 \\ 
    0.01\boldsymbol{z}, & \text{if } \boldsymbol{z} < 0 
   \end{cases}
   \label{eq:leaky_relu}
\end{equation}

In laminar flows characterized by smooth and well-defined dynamics, Leaky ReLU preserves small negative gradients, ensuring the network's learning capacity in regions with minimal activation. In turbulent flows, where complex variations dominate the flow field, Leaky ReLU mitigates the issue of "dead neurons" by allowing non-zero gradients for negative inputs, thus enabling the network to capture intricate flow patterns. Hyperbolic tangent is the default activation function for the LSTM layer \cite{kerasLSTM}. 

The architecture is optimized using the Mean Absolute Error (MAE) as the loss function and the ADAM optimizer \cite{kingma2014adam}, ensuring robust convergence and effective handling of the diverse flow regimes considered in this study.
\begin{equation}
    \text{MAE} = \frac{1}{n} \sum_{i=1}^{n} \left| \mathbf{y}_i - \mathbf{\hat{y}}_i \right|
    \label{eq:mae}
\end{equation}

where \( \mathbf{y}_i \) represents the true values, \( \mathbf{\hat{y}}_i \) denotes the predicted values, and \( n \) is the total number of observations.

The Adam optimizer combines the strengths of momentum and RMSprop for efficient training \cite{kingma2014adam}. Momentum computes an exponentially weighted average of past gradients, which smooths updates and accelerates convergence in the most relevant directions. Simultaneously, Adam employs adaptive learning rates for each parameter, adjusting them based on recent gradient magnitudes. This dynamic adjustment ensures stable updates, even in the presence of noisy gradients or non-stationary objectives.

\subsubsection{Hyperparameter Tuning}

Tuning crucial parameters is another important step in optimizing deep learning models, as it directly impacts model performance and convergence. Hyperparameters are parameters established before training begins, such as the learning rate, batch size, number of layers, and number of units per layer. Their values significantly influence the effectiveness with which the model learns from the data. 

Each architecture has been hyper-tuned for the parameters of batch size, learning rate, and sequence length to ensure optimal performance across each specific test case. Sequence length determines how much historical data (time steps) the model sees as input. The batch size determines the number of samples processed before updating the model, with larger batches often allowing higher learning rates. The learning rate controls the step size during optimization and should be tuned to ensure stable and efficient training.

Selecting these hyperparameters is a problem-specific task, as there is no universal configuration that guarantees optimal performance. Bayesian optimization, coarse-to-fine adjustment, grid search, and random search are some of the most common techniques implemented for hyperparameter tuning. Choosing the ranges of these parameters will strictly depend on the dataset used, refining them through experimentation, periodicity, and domain knowledge. Monitoring training performance and using validation data are key to finding the optimal values. 

The number of neurons in the LSTM layer was tuned and set to 128, which was found to be optimal across all test cases. This choice ensures a balance between model complexity and predictive accuracy without excessive overfitting. For the dense layers, 64 and 32 neurons were selected for progressive feature compression, where the network gradually reduces the dimensionality of the extracted temporal features before producing the final output. This design helps avoid abrupt dimensionality reductions, which could lead to the loss of critical flow dynamics. 
 
Bayesian optimization was implemented to find the optimal values for the various LSTM architectures. The ranges for the hyperparameter tuning have been presented in Table \ref{tab:hyperparam_ranges}.
\begin{table}[h!]
\centering
\begin{tabular}{|c|c|c|c|}
\hline
\textbf{Parameter}     & \textbf{Laminar data sets}       & \textbf{Turbulent data sets} \\ \hline
\textbf{Batch Size}    & 4 to 32                          & 4 to 32                  \\ \hline
\textbf{Learning Rate} & \(10^{-3} \) to \(10^{-4}\)     & \(10^{-4} \) to \(10^{-5}\) \\ \hline
\textbf{Sequence Length} & 5 to 20                         & 20 to 50                \\ \hline
\end{tabular}
\caption{Hyperparameter ranges for batch size, learning rate, and sequence length for the data sets used in this study.}
\label{tab:hyperparam_ranges}
\end{table}

A well-designed deep learning model involves thoughtful considerations about data preparation, computational efficiency, and model architecture to optimize performance and generalization. An overview of these aspects for model development is presented in Appendix A3.

\subsection{Data Preprocessing}

Effective data management is critical for training reliable and robust deep learning models. One of the foundational principles is the division of data into distinct sets: training, development (dev), and test sets. This division enables an unbiased evaluation and systematic optimization of the model. The data is typically divided into training, development, and test sets in ratios such as 80-10-10 or 70-20-10, etc. \cite{kumar2020data, muraina2022ideal}. However, the proper division of the data should depend on the size of the data set. For smaller data sets, the dev and test sets might use larger proportions of 20\% or 30\%, while larger data sets often adopt splits such as 98-1-1, as even small percentages yield sufficient examples for evaluation. In some cases, data sets are split only into training and development sets, particularly when the data set size is small or when cross-validation is used for evaluation \cite{coursera_deeplearning}. By excluding a separate test set, all available data can be utilized for model training and iterative optimization. 
In this study, the data set is divided into training and development/test sets in a ratio of 80:20, where the development and test sets are identical. Specifically, the columns of the temporal modes matrix \( \hat{\boldsymbol{T}} \) are partitioned into two smaller matrices used for training and development/testing of the predictive model. The dimensions of these matrices are \( N \times T_{\text{train}} \) (training) and \( N \times T_{\text{test}} \) (test/development), where the total number of temporal snapshots \( T \) satisfies:

\begin{equation}
T_{\text{train}} + T_{\text{test}} = T.
\end{equation}

\begin{figure}[ht!]
    \centering
    \includegraphics[width=1\textwidth, height=1\textwidth,keepaspectratio]{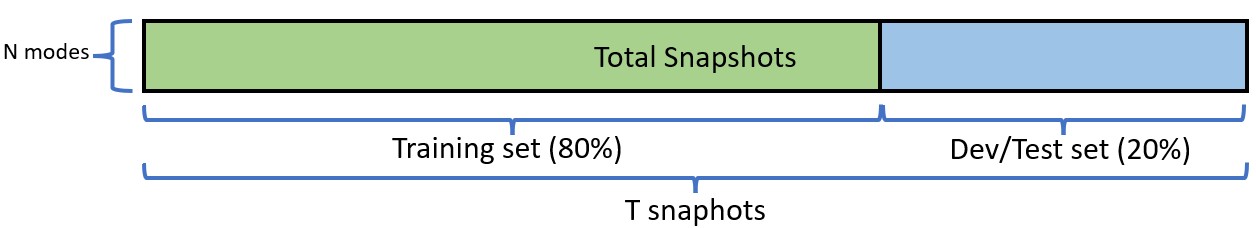}
    \caption{Snapshots used for training and development in the predictive models.}
    \label{fig:train}
\end{figure}

A key step to ensure consistent and unbiased learning during the training and development phases is to normalize the data to optimize neural network performance. It ensures that input features are standardized, which can significantly improve the efficiency and stability of the training process. One of the primary reasons for normalization is to address issues arising from the differing scales of the input features. Without normalization, features with large value ranges can dominate those with smaller ranges, resulting in elongated cost function contours \cite{coursera_deeplearning}. 

Among the various normalization techniques, two of the most commonly used methods are Min-Max normalization and Z-score normalization \cite{patro2015normalization}. In this work, Z-score normalization has been employed for preprocessing. The input tensor is normalized by subtracting its mean and dividing by its standard deviation:
\begin{equation}
    \boldsymbol{\mathcal{X}}_{\text{norm}} = \frac{\boldsymbol{\mathcal{X}} - \boldsymbol{\mu}}{\boldsymbol{\sigma}}
    \label{eq:zscore_normalization}
\end{equation}
where \(\boldsymbol{\mathcal{X}}\) represents the input tensor, \(\boldsymbol{\mu}\), and \(\boldsymbol{\sigma}\) are the mean and the standard deviation, respectively.

This method is particularly effective for handling features with outliers or when features are on different scales. Maintaining standardized input ranges enhances model generalization on unseen data, ultimately balancing feature scales and improving the training stability and convergence speed of the neural network. More on normalization techniques can be found in the papers by Hetherington et al. \cite{hetherington2024modelflows} and Corrochano et al. \cite{corrochano2024hierarchical}. 

\subsubsection{Rolling Window Approach}

The rolling window method is a widely used technique in time series analysis and forecasting, particularly effective for capturing temporal dependencies and trends in sequential data \cite{abadia2022predictive,amor2016recursive, li2014rolling}. The implemented sequence generator utilizes a rolling window approach to prepare input-output pairs for time-series forecasting tasks. This method is particularly suited for recurrent neural networks (RNNs), LSTM, or similar sequential models, where the temporal dependency between consecutive data points plays a crucial role in prediction accuracy.

The rolling window mechanism is designed to slide over the time-series data, extracting fixed-length sequences (\textit{seq\_len}) as inputs and the subsequent time steps (\textit{horizon}) as outputs. This approach ensures that the model can learn temporal patterns effectively while maintaining a consistent input-output structure. 

Specifically, for each batch of data:
\begin{itemize}
    \item A window of size \textit{seq\_len} is used to extract the input sequence, denoted as \(\mathbf{X} \in \mathbb{R}\), where \(\mathbf{X}\) represents the set of input sequences consisting of \textit{seq\_len} consecutive time steps from the data set.
    \item The output target, \(\mathbf{y} \in \mathbb{R}\), is constructed by taking the next \textit{horizon} time steps immediately following the input sequence. Here, \(\mathbf{y}\) represents the set of corresponding output targets that the model aims to predict.
    
\end{itemize}

The rolling window approach allows the generator to traverse the data set in overlapping segments, ensuring that each time step contributes to both input and output sequences. This overlap is critical for extracting meaningful temporal correlations, especially in scenarios with limited data. 

\begin{figure}[h!]
    \centering
    \includegraphics[width=0.65\textwidth, height=0.65\textwidth, keepaspectratio]{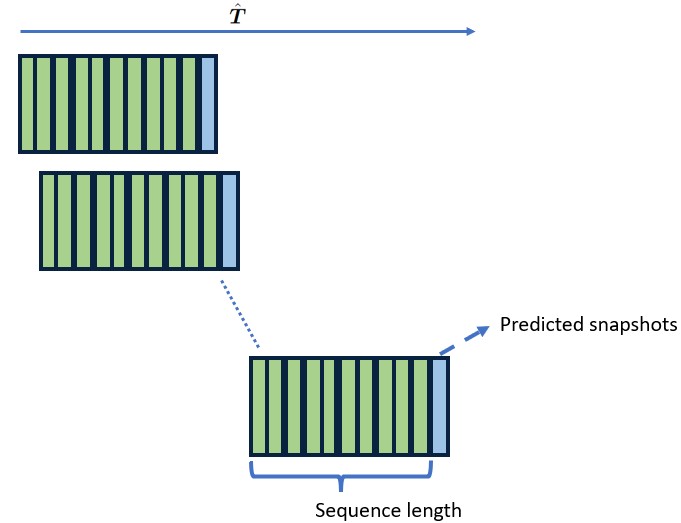}
    \caption{Illustration of the rolling window mechanism \cite{abadia2022predictive}.}
    \label{fig:rolling_window}
\end{figure}

\subsection{Autoregression}

 Autoregression relies on the regression of a variable against one or more past values of itself \cite{article21}. This approach is particularly effective for sequential data, as it builds predictions step-by-step by feeding prior predictions as inputs to predict subsequent values.

\begin{figure}[ht!]
    \centering
    \includegraphics[width=1\textwidth, height=0.75\textheight, keepaspectratio]{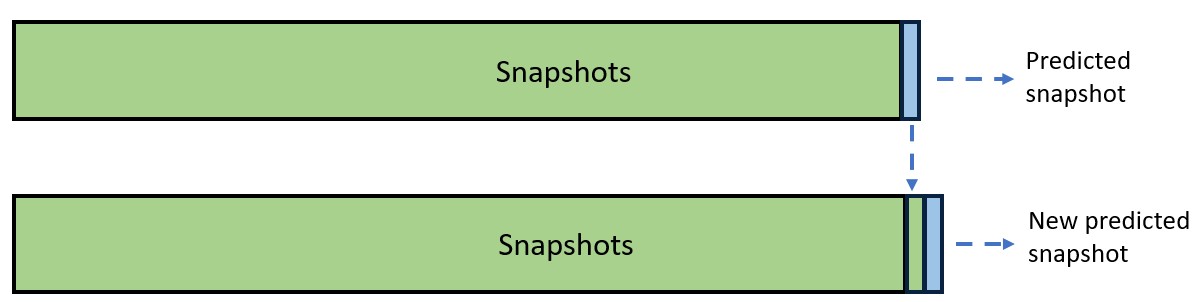}
    \caption{Illustration of the autoregressive process for a horizon of one.}
    \label{fig:autoregression}
\end{figure}

Figure~\ref{fig:autoregression} illustrates the autoregressive process for a horizon of one. In this method, the predicted snapshot is fed back into the model to generate the next predicted snapshot, effectively building a sequence iteratively. This technique ensures that the temporal dependencies are preserved and learned efficiently by the model.

The autoregressive process for temporal prediction can be described in the following steps:

\begin{itemize}
    \item \textbf{Initialization:} The input data, consisting of \( T \) snapshots, is represented as:
    \begin{equation}
    \boldsymbol{X}_0 = \{\boldsymbol{x}_1, \boldsymbol{x}_2, \ldots, \boldsymbol{x}_T\},
    \label{eq:input_snapshots}
    \end{equation}
    where \( \boldsymbol{x}_i \) represents the \( i \)-th snapshot.

    \item \textbf{Future Prediction:} The model predicts the next snapshot \( \boldsymbol{\hat{x}}_{T+1} \) using the past snapshots \( \boldsymbol{X}_0 \) as input:
    \begin{equation}
    \boldsymbol{\hat{x}}_{T+1} = f(\boldsymbol{x}_T, \boldsymbol{x}_{T-1}, \ldots, \boldsymbol{x}_{T-q+1}; \theta),
    \label{eq:one_step_prediction}
    \end{equation}
    Where \( f \) is the predictive model parameterized by \( \theta \), and \( q \) is the window size indicating the number of past snapshots used for prediction. The same follows for multi-step predictions.

    \item \textbf{Feedback Mechanism:} The predicted snapshot \( \boldsymbol{\hat{x}}_{T+1} \) is appended to the existing set of snapshots to form the new input for subsequent predictions:
    \begin{equation}
    \boldsymbol{X}_1 = \{\boldsymbol{x}_2, \boldsymbol{x}_3, \ldots, \boldsymbol{x}_T, \boldsymbol{\hat{x}}_{T+1}\}.
    \label{eq:feedback_mechanism}
    \end{equation}

    \item \textbf{Recursive Prediction:} The process is repeated to predict \( H \) future snapshots:
    \begin{equation}
    \boldsymbol{\hat{x}}_{T+h} = f(\boldsymbol{\hat{x}}_{T+h-1}, \boldsymbol{\hat{x}}_{T+h-2}, \ldots, \boldsymbol{\hat{x}}_{T+h-q+1}; \theta), \quad h = 2, 3, \ldots, H.
    \label{eq:recursive_prediction}
    \end{equation}
\end{itemize}

These equations generalize the autoregressive framework, where the predictive model \( f \) learns the temporal dependencies between snapshots and iteratively generates future predictions. This framework ensures that the sequential nature of the data is effectively preserved, and the iterative nature of this method enables the model to propagate predictions over longer horizons, albeit with potential accumulation of errors.

\subsection{Metrics for Comparison}

Evaluating the performance of models in fluid dynamics requires reliable metrics that not only assess the accuracy of predictions but also provide insights into potential errors and variations. For such assessments, uncertainty quantification (UQ) has been implemented along with metrics such as the average Relative Root Mean Squared Error (RRMSE) to evaluate prediction reliability and measure the accuracy and robustness of the models.

\subsubsection{Relative Root Mean Squared Error (RRMSE)}

RRMSE is a normalized version of Root Mean Squared Error (RMSE), providing a relative measure of error by comparing the model prediction deviations with the range or mean of the observed data. It is defined as:
\begin{equation}
\textbf{RRMSE} = \frac{\sqrt{\frac{1}{\boldsymbol{N}} \sum_{i=1}^{\boldsymbol{N}} (\boldsymbol{y}_i - \hat{\boldsymbol{y}}_i)^2}}{\bar{\boldsymbol{y}}}
\label{eq:rrmse}
\end{equation}
where \(\boldsymbol{y}_i\) represents the observed values, \(\hat{\boldsymbol{y}}_i\) the predicted values, \(\boldsymbol{N}\) the total number of samples, and \(\bar{\boldsymbol{y}}\) the mean of the observed values. The average of which has been presented in the results section.

RRMSE provides a relative error metric that is independent of the scale of the data, making it suitable for comparing models across different data sets or scales. Lower RRMSE values indicate higher prediction accuracy.

\subsubsection{Uncertainty Quantification}
Accurate predictions in fluid dynamics are crucial, but it is equally important to quantify the uncertainties associated with these predictions to understand their reliability and robustness. Uncertainty quantification (UQ) allows the assessment of potential errors and variations in predictions, which is essential to make informed decisions in various fluid dynamics applications. 

In this analysis, UQ is performed by calculating the normalized error between the original tensor data and its reconstructed counterpart. The errors across each velocity component are quantified using probability density functions (PDFs), offering a statistical depiction of the uncertainties. This approach allows us to visualize and interpret error distributions effectively, highlighting the performance differences between the HOSVD and SVD methods. By analyzing these error distributions, areas can be identified where the models perform well and where improvements may be necessary, ultimately enhancing the predictive accuracy of our models.

The normalized error for each velocity component is given by \cite{hetherington2023low}:

\begin{equation}
\boldsymbol{\epsilon}_u = \frac{\boldsymbol{U}_{\text{true}} - \boldsymbol{U}_{\text{pred}}}{\left| \boldsymbol{U}_{\text{true}} - \boldsymbol{U}_{\text{pred}} \right|_{\text{max}}}
\end{equation}
\begin{equation}
\boldsymbol{\epsilon}_v = \frac{\boldsymbol{V}_{\text{true}} - \boldsymbol{V}_{\text{pred}}}{\left| \boldsymbol{V}_{\text{true}} - \boldsymbol{V}_{\text{pred}} \right|_{\text{max}}}
\end{equation}
\begin{equation}
\boldsymbol{\epsilon}_w = \frac{\boldsymbol{W}_{\text{true}} - \boldsymbol{W}_{\text{pred}}}{\left| \boldsymbol{W}_{\text{true}} - \boldsymbol{W}_{\text{pred}} \right|_{\text{max}}}
\end{equation}

here, \( \boldsymbol{U}_{\text{true}} \), \( \boldsymbol{V}_{\text{true}} \), and \( \boldsymbol{W}_{\text{true}} \) represent the true values of the velocity components, while \( \boldsymbol{U}_{\text{pred}} \), \( \boldsymbol{V}_{\text{pred}} \), and \( \boldsymbol{W}_{\text{pred}} \) are the predicted values. The errors are normalized by the maximum absolute error to facilitate comparison between different velocity components.

The methodology section has detailed the integration of SVD and HOSVD with LSTM networks, along with the implementation of the rolling-window method to effectively process temporal data. All of the LSTM architectures have been hyper-tuned for optimal predictive accuracy, the results of which are presented in the next section. This comprehensive approach is designed to improve the predictive capabilities and computational efficiency of the model.

\section{Test Cases}
In the investigation of dynamic fluid behavior, the adopted approach has been subjected to various scenarios of fluid flows to establish its efficacy and adaptability. The test cases encompass a diverse range of data types and flow regimes, including laminar and turbulent flows, as well as numerical and experimental data sets.
The proposed method, which integrates SVD/HOSVD with deep learning architectures, is tested on three fluid dynamics problems: 

\begin{itemize}
    \item \textbf{Laminar Flow Past a Circular Cylinder (Numerical):} 

The flow around a circular cylinder is a well-known benchmark in fluid dynamics used to validate numerical methods and experimental techniques. The dynamics of the flow are governed by the Reynolds number (Re), which is defined based on the cylinder diameter \(D\). At low Reynolds numbers, the flow is steady; however, as Re increases, significant flow transitions occur.

The motion of an incompressible Newtonian fluid is governed by the Navier-Stokes equations, expressed as:

\begin{equation}
    \frac{\partial \mathbf{u}}{\partial t} + (\mathbf{u} \cdot \nabla) \mathbf{u} = -\frac{1}{\boldsymbol{\rho}} \nabla \mathbf{p} + \boldsymbol{\nu} \nabla^2 \mathbf{u}
\end{equation}

\begin{equation}
    \nabla \cdot \mathbf{u} = 0
\end{equation}

where \( \mathbf{u} = (u_x, u_y, u_z) \) is the velocity field, \( \mathbf{p} \) is the pressure, \( \boldsymbol{\rho} \) is the fluid density, and \( \boldsymbol{\nu} \) is the kinematic viscosity. The first equation represents momentum conservation, while the second enforces the incompressibility condition.

 In the 2D case, the simulation is performed at Re = 130. At this Reynolds number, the flow is characterized by unsteady behavior due to the formation of a von Kármán vortex street in the wake of the cylinder. The computational domain extends \(15D\) upstream and \(50D\) downstream in the streamwise direction, with the cylinder diameter \(D\) normalized to \(D = 1\). The dataset has been obtained from Ref \cite{modelflows_cylinder}, where the spatial dimensions are  \(n_x = 100\) points in the streamwise direction, \(n_y = 100\) points in the vertical direction, and \(n_t = 500\) the temporal steps, with a time step \(\Delta t = 0.2\).

 \begin{figure}[h!]
    \centering
      \begin{minipage}{0.48\textwidth}
        \centering
        \begin{tikzpicture}
            \node[anchor=south west, inner sep=0] (image) at (0,0) {\includegraphics[width=0.85\textwidth]{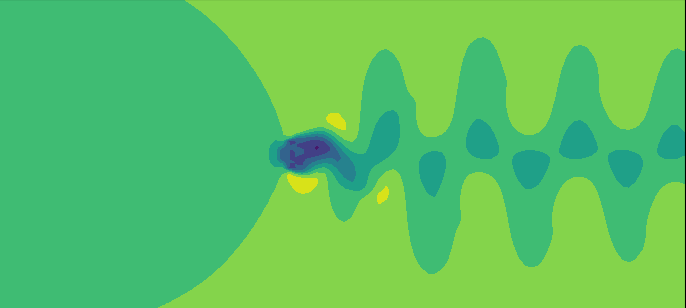}};
            \node[anchor=north, rotate=90, font=\scriptsize] at (-0.6,1.5) {Y};
            \node[anchor=north, font=\scriptsize] at (3.4,-0.2) {X};
        \end{tikzpicture}
    \end{minipage}
    \hfill
    \begin{minipage}{0.48\textwidth}
        \centering
        \begin{tikzpicture}
            \node[anchor=south west, inner sep=0] (image) at (0,0) {\includegraphics[width=0.85\textwidth]{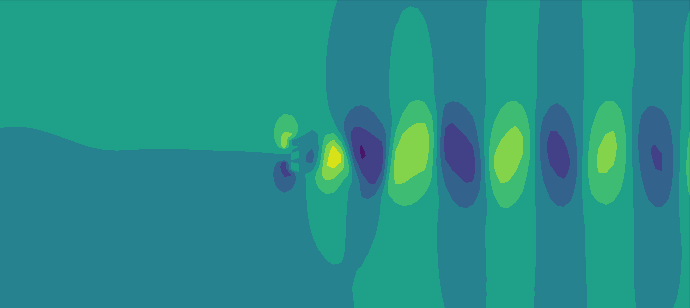}};
            \node[anchor=north, font=\scriptsize] at (3.4,-0.2) {X};
        \end{tikzpicture}
    \end{minipage}
    
    \caption{Ground truth snapshots for the 2D cylinder case. From left to right: streamwise velocity and normal velocity components.}
    \label{fig:svd_snapshots_2d}
\end{figure}
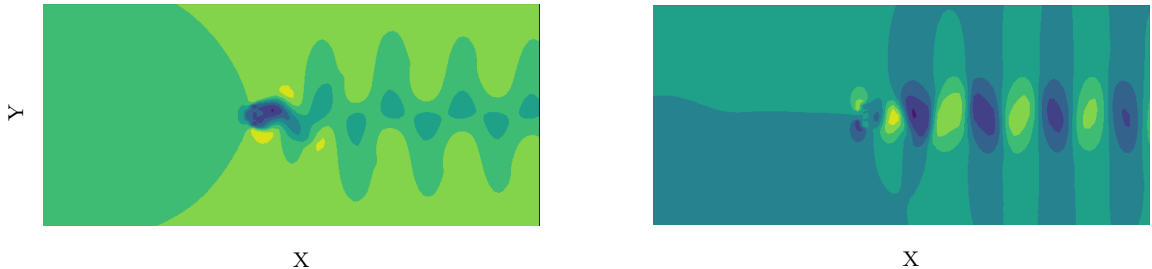

 Extending the problem to three dimensions introduces spanwise variations and requires capturing all three velocity components. The flow becomes unsteady at Re = 46, where a Hopf bifurcation initiates flow with two-dimensional oscillations. Beyond Re = 189, a second bifurcation causes three-dimensionality for specific spanwise wavelengths. The boundary conditions are consistent with the 2D case but account for the spanwise velocity component. The domain has a spatial dimension of \(n_x = 100\), \(n_y = 40\), \(n_z = 64\) in the spanwise direction, and \(n_t = 599\) temporal snapshots, with a time step \(\Delta t = 1\). The data set presented by Le Clainche et al. \cite{le2018spatio} has been utilized for the 3D cylinder case. The 2D simulation considers \(n_{var} = 2\) variables, representing the velocity components (\(u, v\)) in the streamwise and vertical directions, while the 3D simulation expands to \(n_{var} = 3\) variables to include the spanwise velocity component (\(w\)).

 \begin{figure}[h!]
    \centering
      \begin{minipage}{0.48\textwidth}
        \centering
        \begin{tikzpicture}
            \node[anchor=south west, inner sep=0] (image) at (0,0) {\includegraphics[width=0.85\textwidth]{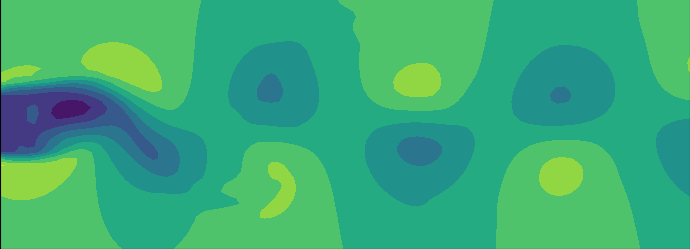}};
            \node[anchor=north, rotate=90, font=\scriptsize] at (-0.6,1.3) {Y};
            \node[anchor=north, font=\scriptsize] at (3.4,-0.2) {X};
        \end{tikzpicture}
    \end{minipage}
    \hfill
    \begin{minipage}{0.48\textwidth}
        \centering
        \begin{tikzpicture}
            \node[anchor=south west, inner sep=0] (image) at (0,0) {\includegraphics[width=0.85\textwidth]{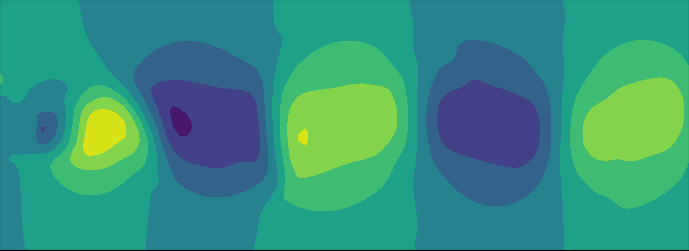}};
            \node[anchor=north, font=\scriptsize] at (3.4,-0.2) {X};
        \end{tikzpicture}
    \end{minipage}
     \hfill
    \begin{minipage}{0.48\textwidth}
        \centering
        \begin{tikzpicture}
            \node[anchor=south west, inner sep=0] (image) at (0,0) {\includegraphics[width=0.85\textwidth]{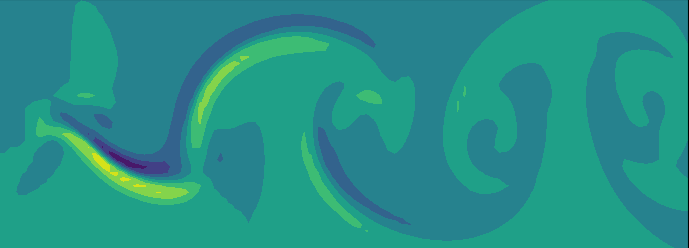}};
            \node[anchor=north, rotate=90, font=\scriptsize] at (-0.6,1.3) {Y};
            \node[anchor=north, font=\scriptsize] at (3.4,-0.2) {X};
        \end{tikzpicture}
    \end{minipage}
    \caption{Ground truth snapshots for the 3D cylinder case. Top (left to right): streamwise velocity and normal velocity components. Bottom: spanwise velocity components.}
    \label{fig:svd_snapshots_3D}
\end{figure}
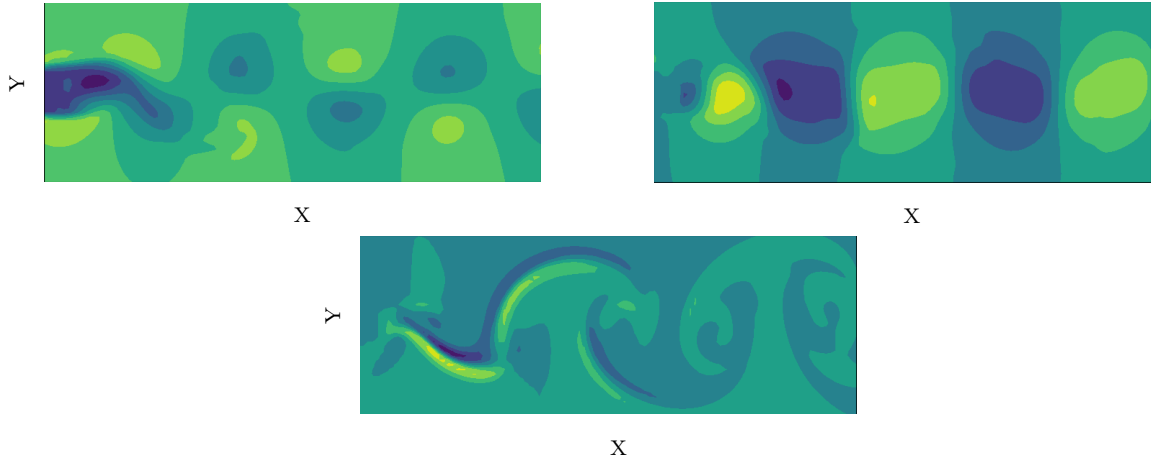

The three-dimensional cylinder flow data set is particularly challenging due to the strong coupling of dynamics across the three spatial dimensions. The flow is transitory, with the spanwise component only beginning to develop near the 300 snapshot mark. This data set serves as an excellent benchmark for testing and validating the developed hybrid model, providing insights into its robustness. Both cases are simulated using the open-source spectral element solver Nek5000, which solves the incompressible Navier-Stokes equations.

\begin{figure}[h!]
    \centering
    \includegraphics[width=0.8\textwidth]{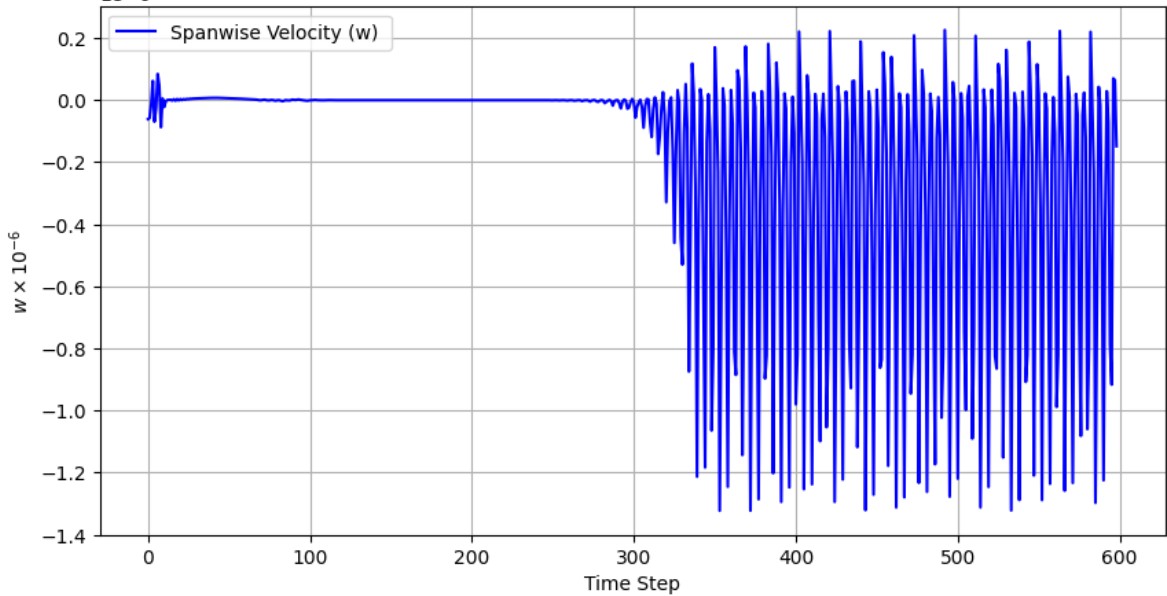} % Replace with actual image filename
    \caption{Temporal evolution of the spanwise velocity component (\( w \)) for the 3D cylinder with the data collected at a representative point in the wake of a cylinder.}
    \label{fig:spanwise_velocity}
\end{figure}

   \item \textbf{Turbulent Flow Past a Circular Cylinder (Experimental):} 

The third test case involves an experimental study of the turbulent wake flow behind a circular cylinder, as presented by Mendez et al. \cite{mendez2020multiscale}. The experiment utilized Particle Image Velocimetry (PIV) data obtained in the low-speed wind tunnel at the Von Kármán Institute, offering a realistic dataset with inherent noise due to measurement uncertainties. 

The cylinder used in the experiment has a diameter of \(d = 5 \, \text{mm}\) and a length of \(L = 200 \, \text{mm}\). The wind tunnel has an exposed measurement area of approximately \(70 \, \text{mm} \times 26 \, \text{mm}\). The dataset focuses on steady-state one, where the Reynolds number is around 2600. The experimental domain is represented with \(n_x = 111\) points in the streamwise direction and \(n_y = 301\) points in the normal direction. The data set captures two velocity components (\(n_{var} = 2\)): the streamwise velocity (\(u\)) and the normal velocity (\(v\)). Measurements were collected over \(n_t = 4000\) snapshots, with \(\Delta t = 0.33 \, \text{s}\).

This dataset provides an additional layer of complexity for the validation of numerical models, as it incorporates the challenges of experimental noise and turbulent flow characteristics.

\begin{figure}[h!]
    \centering
      \begin{minipage}{0.48\textwidth}
        \centering
        \begin{tikzpicture}[yscale=0.01]
            \node[anchor=south west, inner sep=0] (image) at (0,0) 
            {\includegraphics[width=0.85\textwidth]{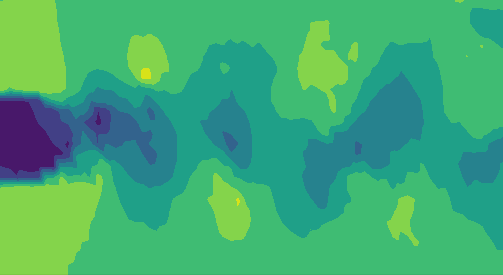}};
            \node[anchor=north, rotate=90, font=\scriptsize] at (-0.6,160.2) {Y}; 
            \node[anchor=north, font=\scriptsize] at (3.4,-0.2) {X};
        \end{tikzpicture}
    \end{minipage}
    \hfill
    \begin{minipage}{0.48\textwidth}
        \centering
        \begin{tikzpicture}[yscale=0.01] 
            \node[anchor=south west, inner sep=0] (image) at (0,0) 
            {\includegraphics[width=0.85\textwidth]{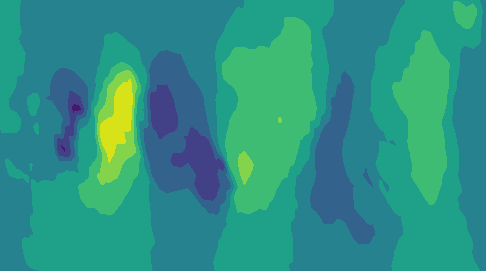}};
            \node[anchor=north, font=\scriptsize] at (3.4,-0.2) {X};
        \end{tikzpicture}
    \end{minipage}
    
    \caption{Ground truth snapshots for the experimental cylinder case. From left to right: streamwise velocity and normal velocity components.}
    \label{fig:svd_snapshots_VKI}
\end{figure}
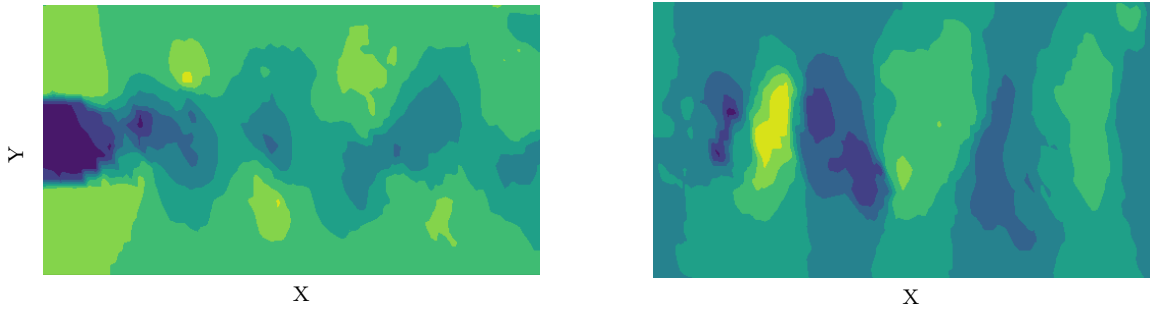

\end{itemize}

To further illustrate the characteristics of each test case, Table 1 summarizes the type of data, nature of flow, and Reynolds number for each scenario. The shape of the dataset is presented in the format \((n_{var}, n_x, n_y, n_z, n_t)\), where for the 2D case the spatial resolution is defined by \(n_x\) and \(n_y\), and for the 3D case \(n_z\) represents the resolution in the spanwise direction along with \(n_x\), \(n_y\), and \(n_t\) representing temporal snapshots.

\begin{table}[h!]
\centering
\resizebox{\textwidth}{!}{%
\begin{tabular}{|c|c|c|c|c|}
\hline
\textbf{Test Case} & \textbf{Data Type} & \textbf{Nature} & \textbf{Reynolds Number} & \textbf{Tensor Shape} \\ \hline
2D Cylinder & Numerical & Laminar & 130 & (2, 100, 100, 500) \\ \hline
3D Cylinder & Numerical & Laminar & 280 & (3, 100, 40, 64, 599) \\ \hline
Experimental Cylinder & Experimental & Turbulent & 2600 & (2, 111, 301, 4000) \\ \hline
\end{tabular}%
}
\caption{Summary of fluid dynamics test cases including data type, nature of flow, Reynolds number, and tensor shape.}
\label{tab:test_cases}
\end{table}

\subsection{Selection of Modes}

\begin{figure}[H]
    \centering
    % First subplot (2D Cylinder at the top)
    \begin{subfigure}[b]{0.45\textwidth}
        \centering
        \includegraphics[width=\textwidth]{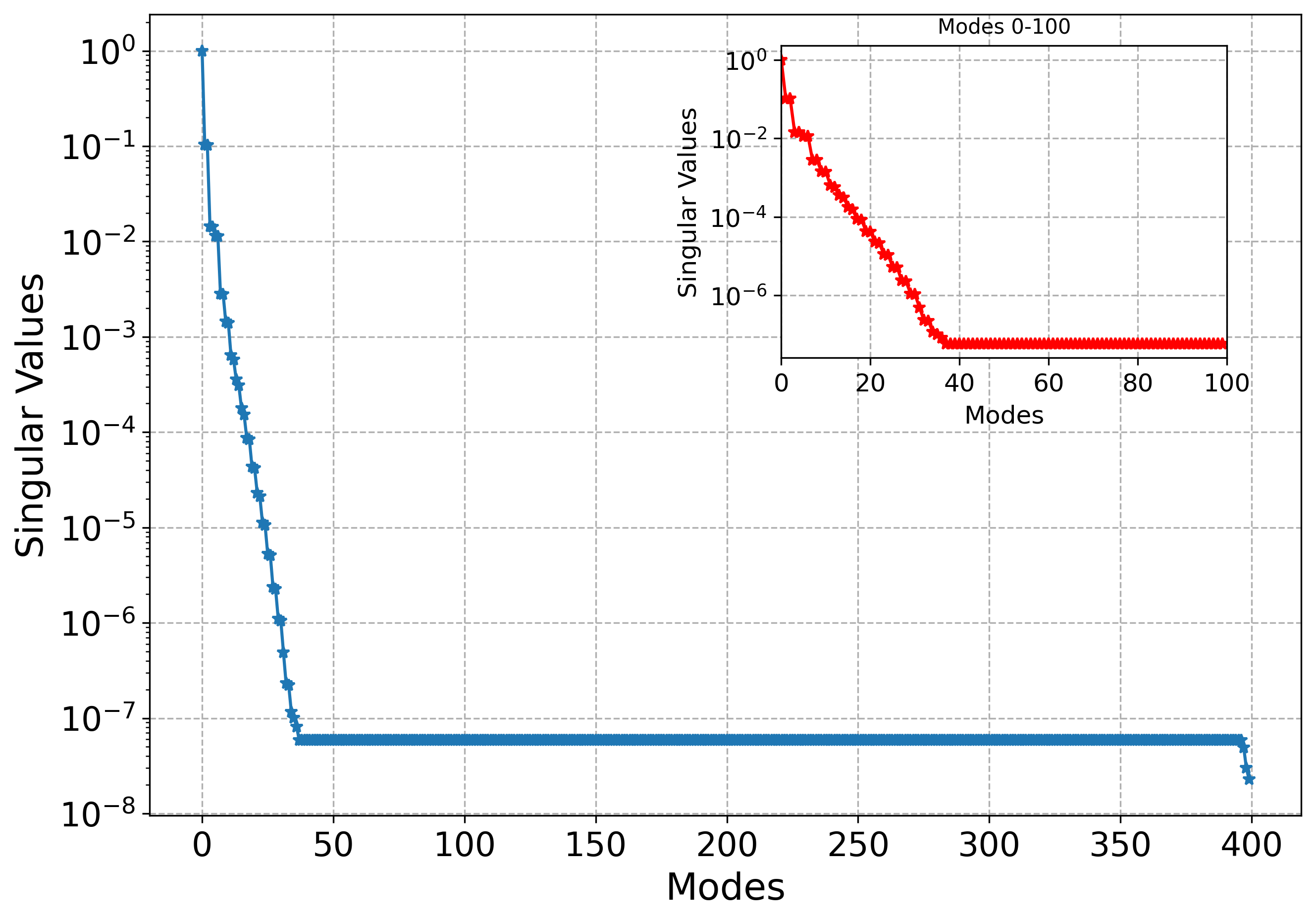}
        \caption{2D Cylinder}
        \label{fig:sv1}
    \end{subfigure}
    \\ % Line break for separating the top image from the bottom row
    % Second subplot (3D Cylinder)
    \begin{subfigure}[b]{0.45\textwidth}
        \centering
        \includegraphics[width=\textwidth]{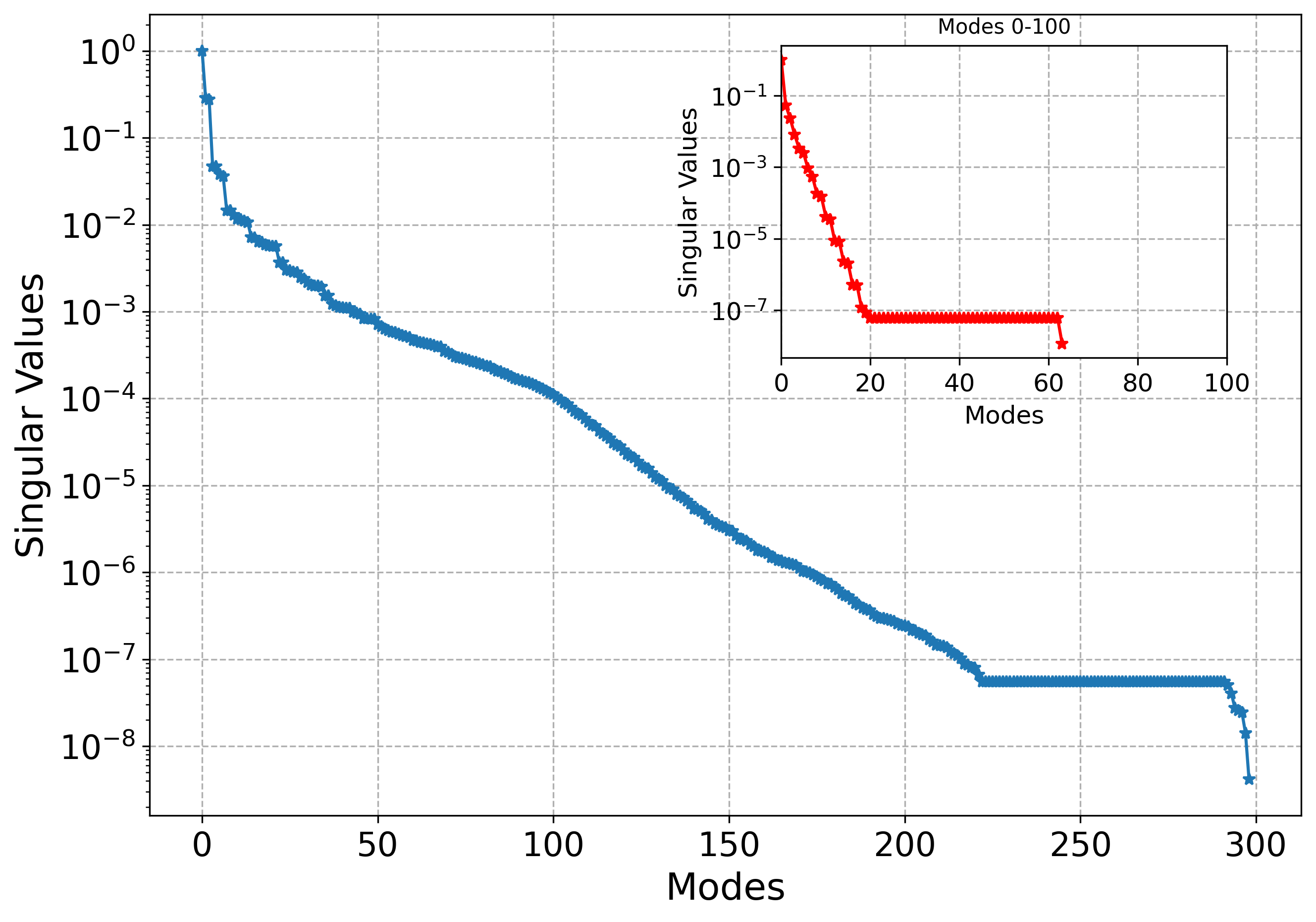}
        \caption{3D Cylinder}
        \label{fig:sv2}
    \end{subfigure}
    % Third subplot (Experimental Cylinder)
    \begin{subfigure}[b]{0.45\textwidth}
        \centering
        \includegraphics[width=\textwidth]{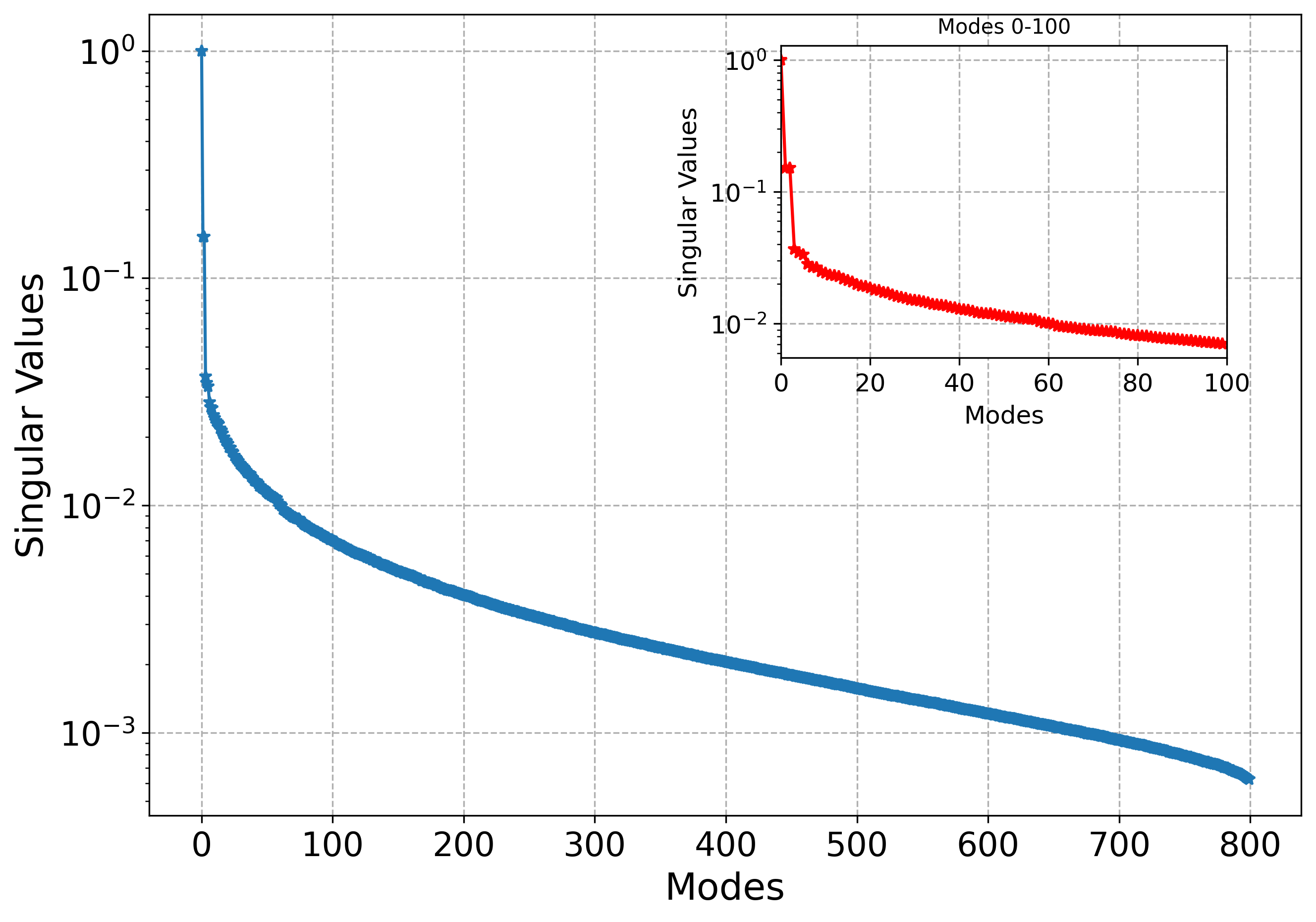}
        \caption{Experimental Cylinder}
        \label{fig:sv3}
    \end{subfigure}
    \caption{Singular values versus modes for the 2D cylinder, 3D cylinder, and experimental cylinder. Top: (a) 2D cylinder. Bottom (from left to right): (b) 3D cylinder and (c) experimental cylinder.}
    \label{fig:singular_values}
\end{figure}

Choosing the correct number of modes is crucial as they capture the dominant flow structures while discarding irrelevant noise. The selection is based on the decay of singular values, ensuring that only the most relevant modes are retained for reconstruction. For laminar datasets, selecting 8-12 modes provided an optimal balance, with 10 modes chosen for the predictions. In the case of the turbulent dataset, 4-6 modes worked best, and 5 modes were used. Figure~\ref{fig:singular_values} presents the singular values versus modes curves for the three cases.

In the 2D cylinder case, all significant flow dynamics can be represented by approximately 30-40 modes. Beyond this range, the singular values drop to \( \mathcal(10^{-8}) \), which corresponds to machine errors. These low-amplitude modes primarily capture spatial redundancies or numerical noise originating from the CFD simulations. Thus, selecting 8-12 modes is justified, as the singular values within this range are still large enough to retain the dominant coherent structures of the flow. This selection ensures that the primary flow dynamics are well-reconstructed with minimal error. The same can be observed for the 3D cylinder case. It is important to note that this dataset represents a transient flow solution, meaning that some of the modes correspond to decaying transient behavior. If too many modes are included, the reconstruction may incorporate errors related to these transient decaying modes. Therefore, selecting a moderate number of modes is essential to balance accuracy while avoiding the inclusion of unnecessary transient noise.

For the case of the experimental turbulent cylinder, selecting 4-6 modes is sufficient to capture the dominant flow structures. The magnitude of the singular values for these modes is \( \geq 10^{-2} \), meaning that reconstruction errors in the range of 5-20\% are expected. Additionally, experimental measurement uncertainties are at least 5\%, indicating that including modes with smaller singular value amplitudes would introduce unnecessary measurement noise into the reconstruction. The evolution of the first ten modes for this dataset has been presented in Appendix A2.

\subsection{Comparison of Temporal Predictions with Ground Truth}

In this subsection, the results of the predictive models have been presented, comparing the predicted snapshots against the ground truth data. The ensuing figures demonstrate the proficiency of the integrated approach, employing HOSVD alongside robust deep learning frameworks, in capturing the intricate dynamics of fluid flows. Comparisons have been made with SVD-based models developed using the same methodology and hyperparameters. The temporal evolution of the predictions was compared with the ground truth to assess how well the models captured dynamic behavior over time. 

 For UQ, the error probability distribution has been plotted for all velocity components, with the streamwise component (\(u\)) marked in red, the normal component (\(v\)) in green, and the spanwise component (\(w\)) in blue. These comparisons provide a comprehensive analysis of the predictive capabilities of the architectures and the dimensionality reduction methods across all test cases.

It will also be particularly interesting to observe how the same architecture behaves in different test cases. To investigate this, each of the three LSTM architectures will be tested with similar neurons and activation functions in all three cases. This approach allows for a consistent evaluation of the architectures' adaptability and effectiveness in handling varying flow scenarios. The following tables summarize the tuned hyperparameters used for the LSTM architectures across different test cases.
\paragraph{}

\begin{table}[h!]
\centering
\begin{tabular}{|c|c|c|c|}
\hline
\multicolumn{4}{|c|}{\textbf{2D Cylinder}} \\ \hline
\textbf{Architecture} & \textbf{Learning Rate} & \textbf{Batch Size} & \textbf{Sequence Length} \\ \hline
LSTM 1 Dense          & 0.001                  & 12                  & 20                       \\ \hline
LSTM 2 Dense          & 0.001                 & 32                  & 5                      \\ \hline
LSTM Time-Distributed          & 0.001                 & 20                  & 5                      \\ \hline

\end{tabular}
\caption{Tuned hyperparameters for 2D cylinder flow.}
\label{tab:2d_cylinder}
\end{table}

\begin{table}[h!]
\centering
\begin{tabular}{|c|c|c|c|}
\hline
\multicolumn{4}{|c|}{\textbf{3D Cylinder}} \\ \hline
\textbf{Architecture} & \textbf{Learning Rate} & \textbf{Batch Size} & \textbf{Sequence Length} \\ \hline
LSTM 1 Dense          & 0.001                  & 8                  & 20                       \\ \hline
LSTM 2 Dense          & 0.001                 & 16                  & 5                      \\ \hline
LSTM Time-Distributed          & 0.001                  & 20                 & 15                      \\ \hline
\end{tabular}
\caption{Tuned hyperparameters for 3D cylinder flow.}
\label{tab:3d_cylinder}
\end{table}

\begin{table}[h!]
\centering
\begin{tabular}{|c|c|c|c|}
\hline
\multicolumn{4}{|c|}{\textbf{Experimental Cylinder}} \\ \hline
\textbf{Architecture} & \textbf{Learning Rate} & \textbf{Batch Size} & \textbf{Sequence Length} \\ \hline
LSTM 1 Dense          & 0.0001                & 20                  & 25                       \\ \hline
LSTM 2 Dense          & 0.0001                  & 32                  & 35                       \\ \hline
LSTM Time-Distributed          & 0.0001                 & 24                  & 20                       \\ \hline
\end{tabular}
\caption{Tuned hyperparameters for experimental cylinder flow.}
\label{tab:exp_wake_flow}
\end{table}

\subsubsection{Case: Laminar Flow Past a Circular Cylinder (2D) }

\begin{figure}[h!]
    \centering
    \textbf{(a) Ground Truth} \\[5pt]
    \begin{minipage}{0.5\textwidth}
        \centering
        \includegraphics[width=\textwidth]{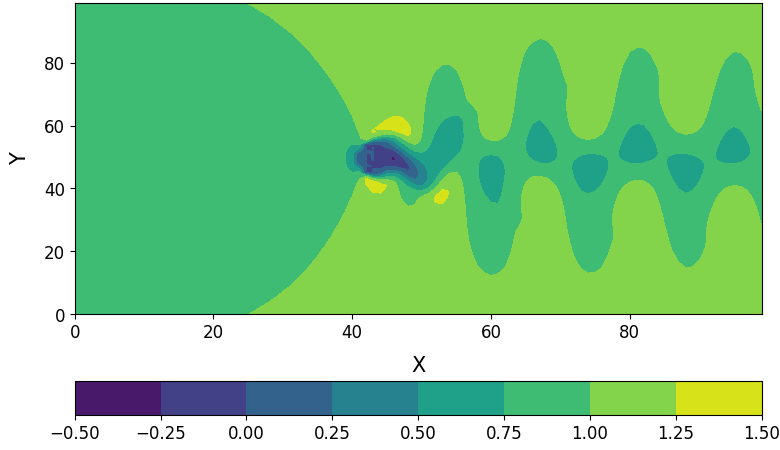}
    \end{minipage}
    \\[15pt]
   % Centered Label for (b) HOSVD and (b) SVD
    \makebox[\textwidth]{\textbf{(b) HOSVD} \hspace{5cm} \textbf{(b) SVD}} \\[5pt]
    
    % Row 1
    \begin{minipage}{0.48\textwidth}
        \centering
        \begin{tikzpicture}
            \node[anchor=south west, inner sep=0] (image) at (0,0) {\includegraphics[width=0.85\textwidth]{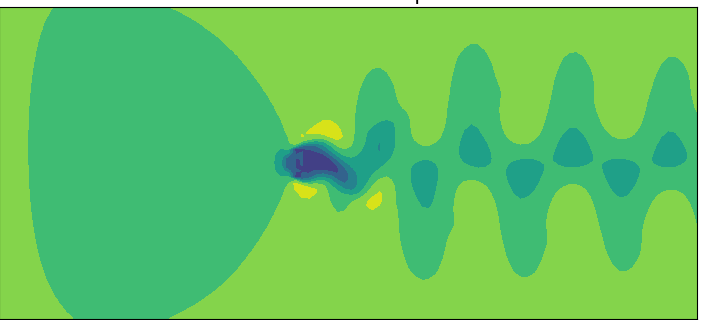}};
            \node[anchor=north, rotate=90, font=\scriptsize] at (-0.6,1.5) {Y};
            \node[anchor=north, font=\scriptsize] at (3.4,-0.2) {X};
        \end{tikzpicture}
    \end{minipage}
    \hfill
    \begin{minipage}{0.48\textwidth}
        \centering
        \begin{tikzpicture}
            \node[anchor=south west, inner sep=0] (image) at (0,0) {\includegraphics[width=0.85\textwidth]{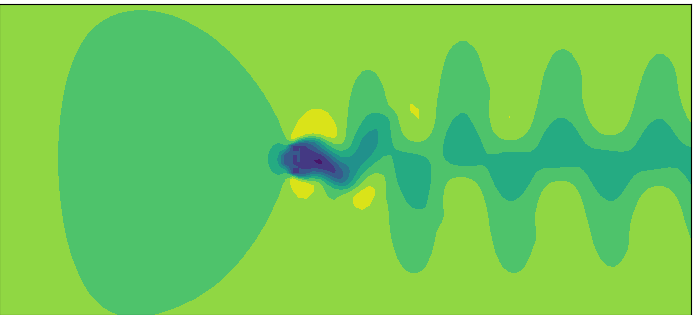}};
            \node[anchor=north, font=\scriptsize] at (3.4,-0.2) {X};
        \end{tikzpicture}
    \end{minipage}
    \\[10pt]
    
    % Row 2
    \begin{minipage}{0.48\textwidth}
        \centering
        \begin{tikzpicture}
            \node[anchor=south west, inner sep=0] (image) at (0,0) {\includegraphics[width=0.85\textwidth]{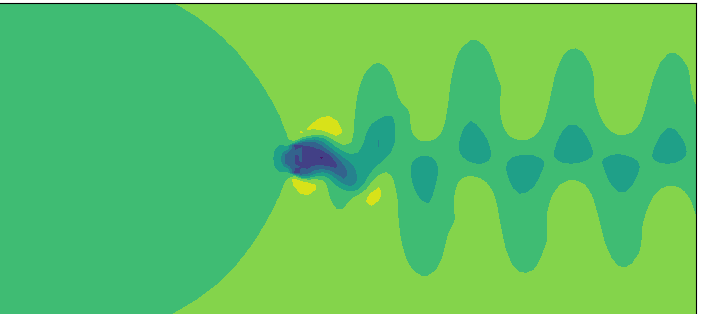}};
            \node[anchor=north, rotate=90, font=\scriptsize] at (-0.6,1.5) {Y};
            \node[anchor=north, font=\scriptsize] at (3.4,-0.2) {X};
        \end{tikzpicture}
    \end{minipage}
    \hfill
    \begin{minipage}{0.48\textwidth}
        \centering
        \begin{tikzpicture}
            \node[anchor=south west, inner sep=0] (image) at (0,0) {\includegraphics[width=0.85\textwidth]{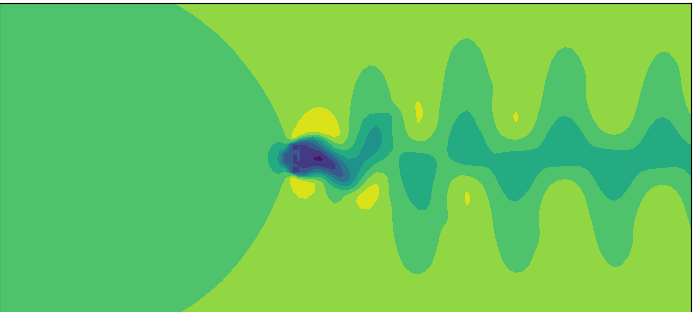}};
            \node[anchor=north, font=\scriptsize] at (3.4,-0.2) {X};
        \end{tikzpicture}
    \end{minipage}
    \\[10pt]
    
    % Row 3
    \begin{minipage}{0.48\textwidth}
        \centering
        \begin{tikzpicture}
            \node[anchor=south west, inner sep=0] (image) at (0,0) {\includegraphics[width=0.85\textwidth]{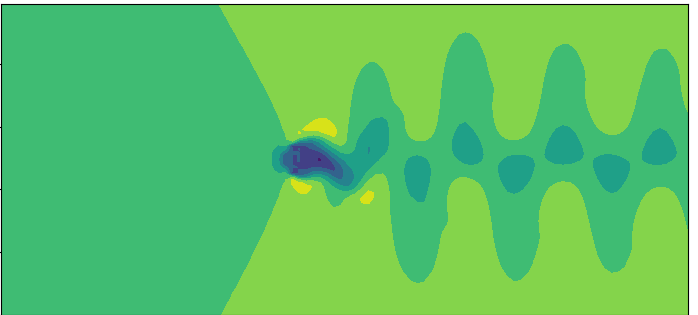}};
            \node[anchor=north, rotate=90, font=\scriptsize] at (-0.6,1.5) {Y};
            \node[anchor=north, font=\scriptsize] at (3.4,-0.2) {X};
        \end{tikzpicture}
    \end{minipage}
    \hfill
    \begin{minipage}{0.48\textwidth}
        \centering
        \begin{tikzpicture}
            \node[anchor=south west, inner sep=0] (image) at (0,0) {\includegraphics[width=0.85\textwidth]{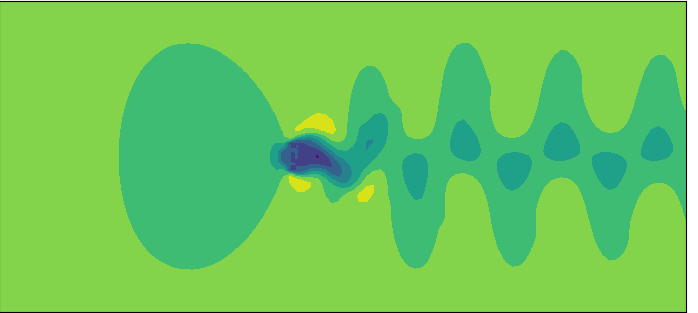}};
            \node[anchor=north, font=\scriptsize] at (3.4,-0.2) {X};
        \end{tikzpicture}
    \end{minipage}

    \caption{(a) Ground truth streamwise velocity components at \(t = 428\). 
    (b) Comparison of the predicted streamwise velocity components for HOSVD (left column) and SVD (right column) across different LSTM architectures. From top to bottom: predictions for LSTM with 1 Dense, 2 Dense, and Time-Distributed architectures.}
    \label{fig:combined_figure}
\end{figure}
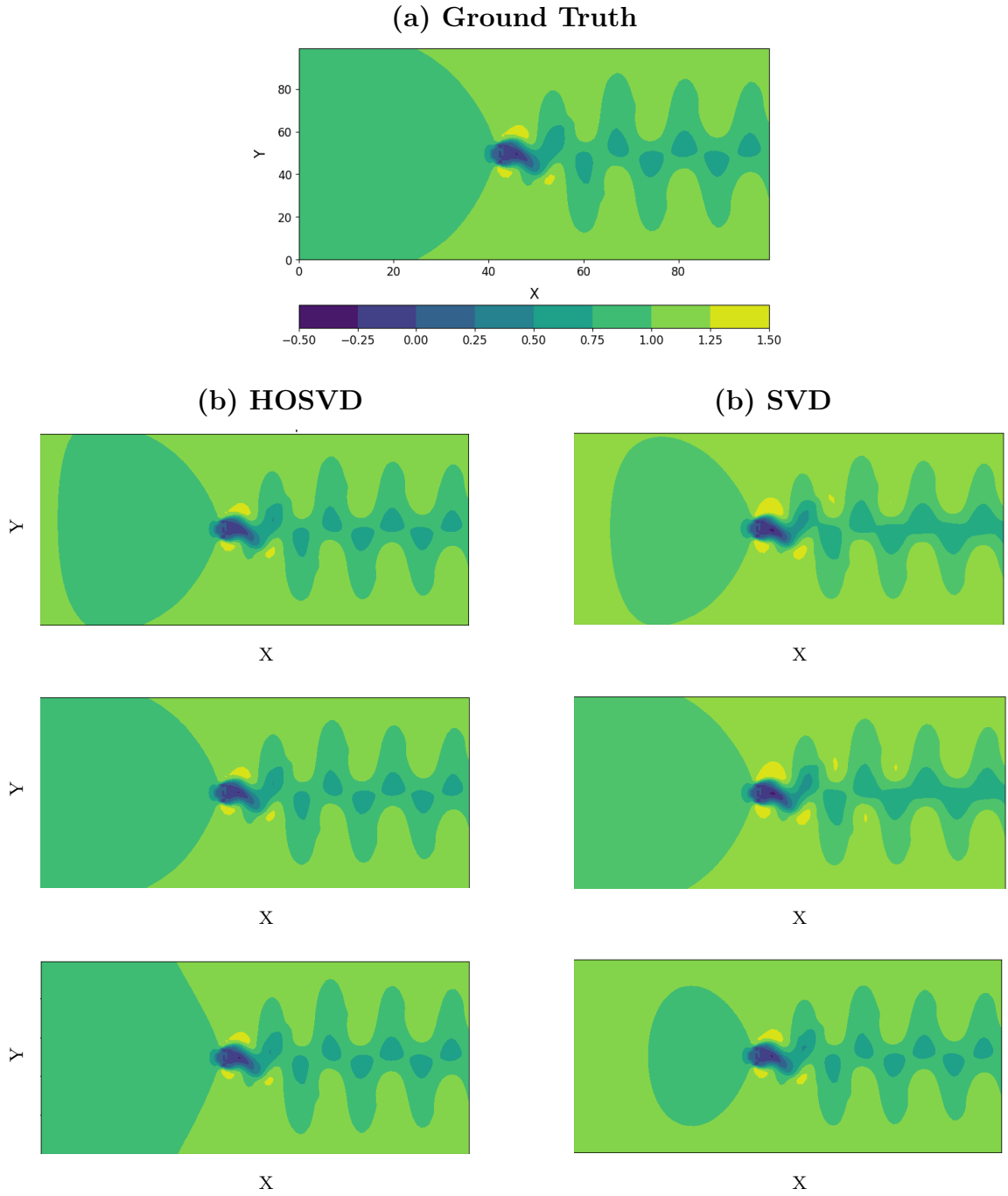

\begin{figure}[h!]
    \centering
    \textbf{(a) Ground Truth} \\[5pt]
    \begin{minipage}{0.5\textwidth}
        \centering
        \includegraphics[width=\textwidth]{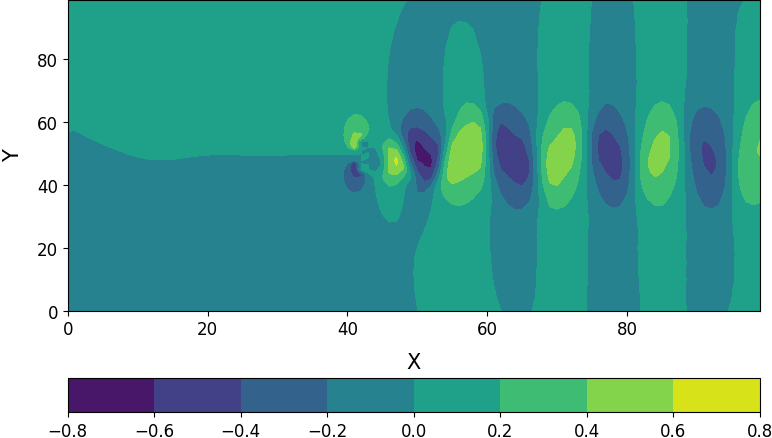}
    \end{minipage}
    \\[15pt]
   % Centered Label for (b) HOSVD and (b) SVD
    \makebox[\textwidth]{\textbf{(b) HOSVD} \hspace{5cm} \textbf{(b) SVD}} \\[5pt]
    
    % Row 1
    \begin{minipage}{0.48\textwidth}
        \centering
        \begin{tikzpicture}
            \node[anchor=south west, inner sep=0] (image) at (0,0) {\includegraphics[width=0.85\textwidth]{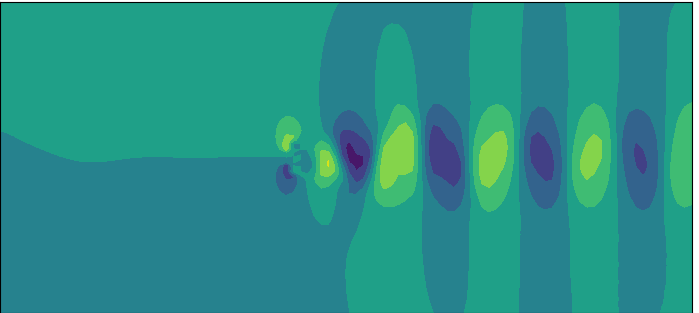}};
            \node[anchor=north, rotate=90, font=\scriptsize] at (-0.6,1.5) {Y};
            \node[anchor=north, font=\scriptsize] at (3.4,-0.2) {X};
        \end{tikzpicture}
    \end{minipage}
    \hfill
    \begin{minipage}{0.48\textwidth}
        \centering
        \begin{tikzpicture}
            \node[anchor=south west, inner sep=0] (image) at (0,0) {\includegraphics[width=0.85\textwidth]{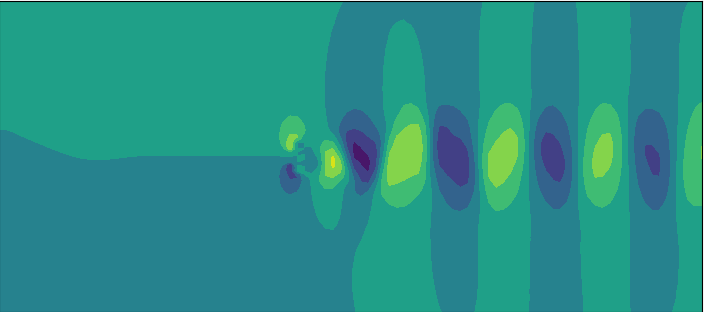}};
            \node[anchor=north, font=\scriptsize] at (3.4,-0.2) {X};
        \end{tikzpicture}
    \end{minipage}
    \\[10pt]
    
    % Row 2
    \begin{minipage}{0.48\textwidth}
        \centering
        \begin{tikzpicture}
            \node[anchor=south west, inner sep=0] (image) at (0,0) {\includegraphics[width=0.85\textwidth]{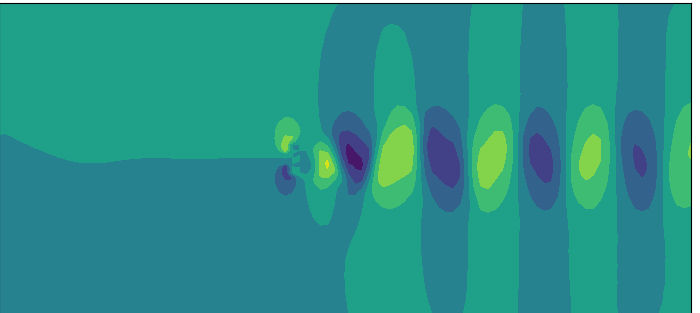}};
            \node[anchor=north, rotate=90, font=\scriptsize] at (-0.6,1.5) {Y};
            \node[anchor=north, font=\scriptsize] at (3.4,-0.2) {X};
        \end{tikzpicture}
    \end{minipage}
    \hfill
    \begin{minipage}{0.48\textwidth}
        \centering
        \begin{tikzpicture}
            \node[anchor=south west, inner sep=0] (image) at (0,0) {\includegraphics[width=0.85\textwidth]{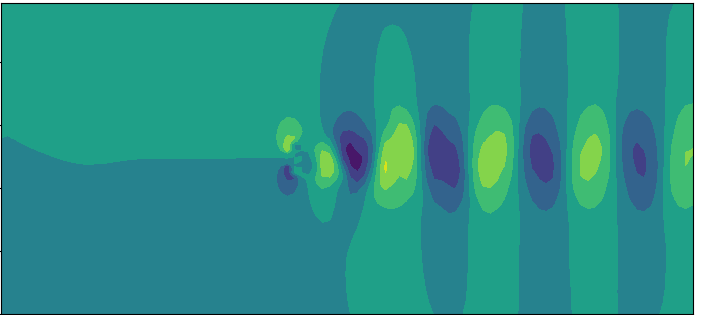}};
            \node[anchor=north, font=\scriptsize] at (3.4,-0.2) {X};
        \end{tikzpicture}
    \end{minipage}
    \\[10pt]
    
    % Row 3 
    \begin{minipage}{0.48\textwidth}
        \centering
        \begin{tikzpicture}
            \node[anchor=south west, inner sep=0] (image) at (0,0) {\includegraphics[width=0.85\textwidth]{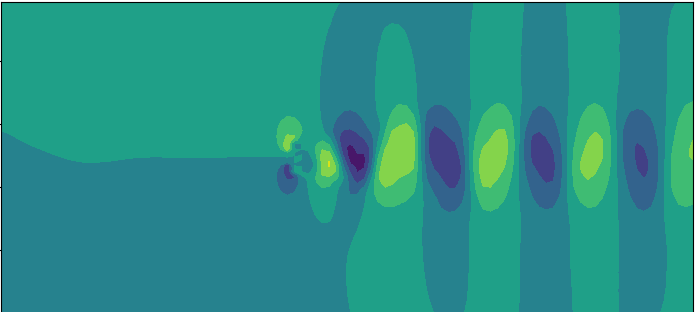}};
            \node[anchor=north, rotate=90, font=\scriptsize] at (-0.6,1.5) {Y};
            \node[anchor=north, font=\scriptsize] at (3.4,-0.2) {X};
        \end{tikzpicture}
    \end{minipage}
    \hfill
    \begin{minipage}{0.48\textwidth}
        \centering
        \begin{tikzpicture}
            \node[anchor=south west, inner sep=0] (image) at (0,0) {\includegraphics[width=0.85\textwidth]{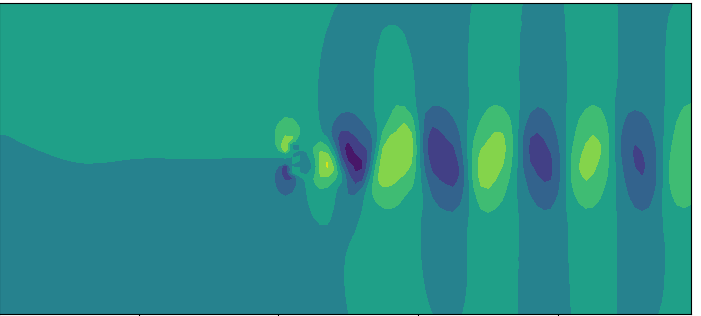}};
            \node[anchor=north, font=\scriptsize] at (3.4,-0.2) {X};
        \end{tikzpicture}
    \end{minipage}

    \caption{(a) Ground truth normal velocity components at \(t = 428\). 
    (b) Comparison of the predicted normal velocity components for HOSVD (left column) and SVD (right column) across different LSTM architectures. From top to bottom: predictions for LSTM with 1 Dense, 2 Dense, and Time-Distributed architectures.}
    \label{fig:lstm_normal_predictions}
\end{figure}

This test case focuses on the classic flow past a two-dimensional cylinder at Re = 130, a regime in which laminar vortices form in the wake of the cylinder. The original data set consists of a tensor that contains 500 snapshots. A total of 100 snapshots have been predicted to compare with the last 100 ground truth snapshots. Figure~\ref{fig:combined_figure} and ~\ref{fig:lstm_normal_predictions} illustrate the ground truth and the predictions for the streamwise and normal velocity components, respectively, at the time step \(t = 428\), providing a reference for evaluating the model's predictive performance. 
The figures below present the predicted streamwise and normal velocity components for both the HOSVD and SVD approaches using LSTM architectures with 1 dense, 2 dense, and time-distributed layers. The first column represents predictions from the HOSVD-based method, while the second column corresponds to predictions from the SVD-based method. Each row sequentially depicts the results for different architectures: the first row for LSTM with 1 dense, the second row for LSTM with 2 dense, and the third row for LSTM with time-distributed. After the snapshots, the average RRMSE was presented, followed by the temporal evolution of the flow at a representative point in the wake and the UQ results. This provides a deeper insight into the accuracy of the predictions over time.

 The most notable observation is that HOSVD provides smoother and more consistent predictions compared to SVD. This can be attributed to its ability to handle the multi-dimensional nature of the tensor data more effectively, preserving spatial and temporal correlations during dimensionality reduction. For the streamwise velocity components, HOSVD handles regions with sharp transitions and high gradients more effectively, ensuring better alignment with the ground truth. In contrast, for certain snapshots, like at t = 428, SVD predicts an averaged flow in some areas, demonstrating its limitations in preserving temporal consistency. However, the combination of SVD with the time-distributed architecture performs reasonably well, generating snapshots visually comparable to those obtained from the HOSVD models. The normal velocity predictions from the HOSVD and SVD models are quite similar to the ground truth and are robust, capturing subtle variations and maintaining consistency across the entire flow field.

\begin{table}[h!]
\centering
\begin{tabular}{|c|c|c|}
\hline
\textbf{Architecture} & \textbf{HOSVD (\%)} & \textbf{SVD (\%)} \\ \hline
LSTM 1 Dense          & 0.4                  & 0.6               \\ \hline
LSTM 2 Dense          & 0.2                  & 0.5               \\ \hline
LSTM Time-Distributed  & 0.4                  & 0.5               \\ \hline
\end{tabular}
\caption{RRMSE values for HOSVD and SVD across LSTM architectures for the streamwise velocity component.}
\label{tab:rrmse_streamwise}
\end{table}

\begin{table}[h!]
\centering
\begin{tabular}{|c|c|c|}
\hline
\textbf{Architecture} & \textbf{HOSVD (\%)} & \textbf{SVD (\%)} \\ \hline
LSTM 1 Dense          & 5.1                 & 7.8               \\ \hline
LSTM 2 Dense          & 3.8                 & 5             \\ \hline
LSTM Time-Distributed  & 4.2                & 5.4             \\ \hline
\end{tabular}
\caption{RRMSE values for HOSVD and SVD across LSTM architectures for the normal velocity component.}
\label{tab:rrmse_normal}
\end{table}

For quantitative analysis, the RRMSE values for the streamwise and normal velocity components are summarized in Tables \ref{tab:rrmse_streamwise} and \ref{tab:rrmse_normal}. In the streamwise direction, HOSVD demonstrates improved predictive accuracy over SVD in the 1 dense, 2 dense, and time-distributed architectures. For the normal component, error values are generally higher due to the presence of large zero regions in the flow field. Since the model predicts these values with machine error, it increases the RRMSE. But for both the decomposition techniques and across all the architectures, the RRMSE values are quite similar, with HOSVD maintaining a slight advantage over SVD.

\begin{figure}[h!]
    \centering
    
    % Centered Label for (b) HOSVD and (b) SVD
    \makebox[\textwidth]{\textbf{(a) HOSVD} \hspace{5cm} \textbf{(b) SVD}} \\[5pt]
    % Row 1
    \begin{minipage}{0.48\textwidth}
        \centering
        \begin{tikzpicture}
            \node[anchor=south west, inner sep=0] (image) at (0,0) {\includegraphics[width=1.9\textwidth, height=0.6\textwidth, keepaspectratio]{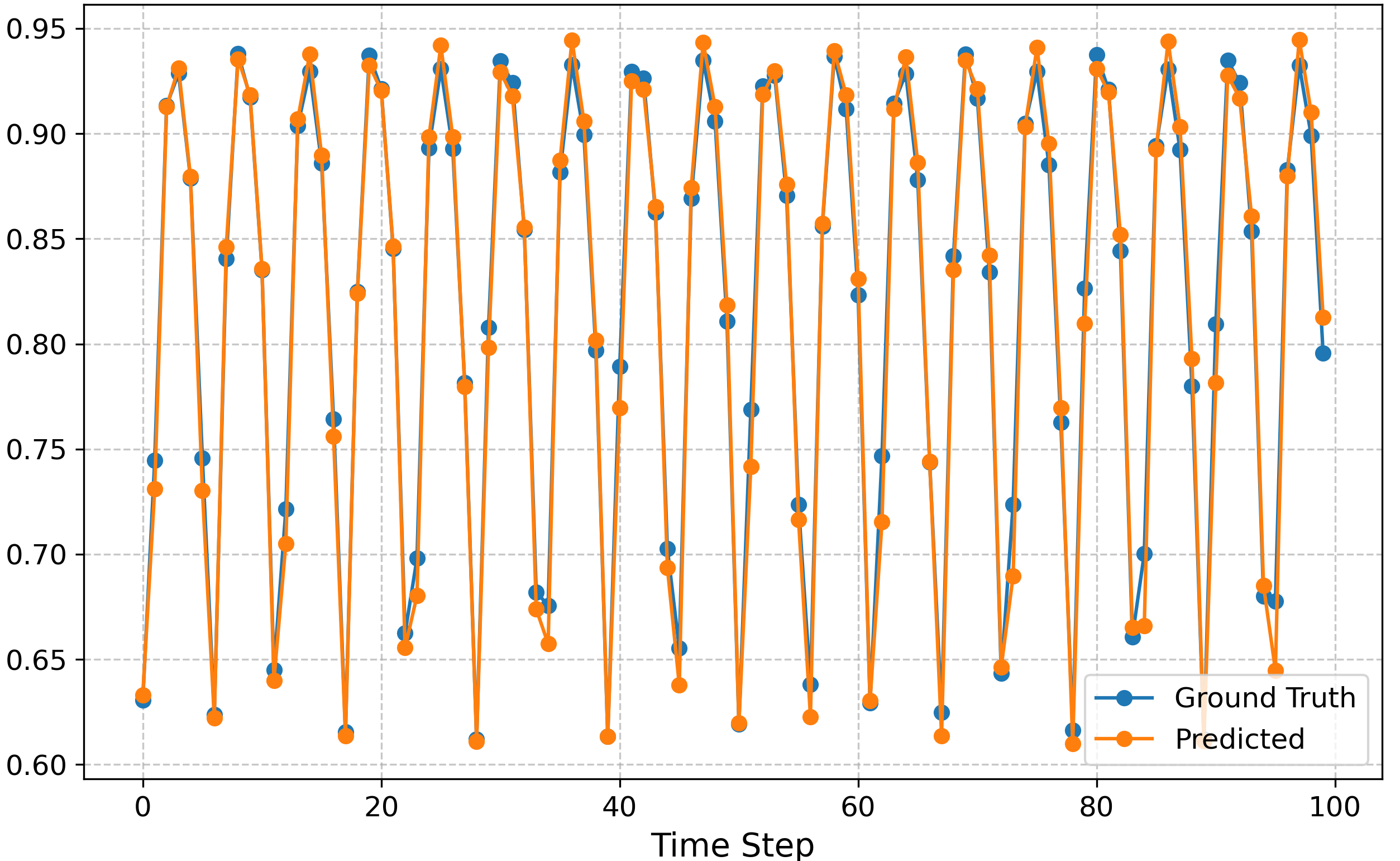}};
            \node[anchor=north, rotate=90, font=\footnotesize\bfseries] at (-0.45,2.5) {u};
        \end{tikzpicture}
    \end{minipage}
    \hfill
    \begin{minipage}{0.48\textwidth}
        \centering
        \includegraphics[width=1.9\textwidth, height=0.6\textwidth, keepaspectratio]{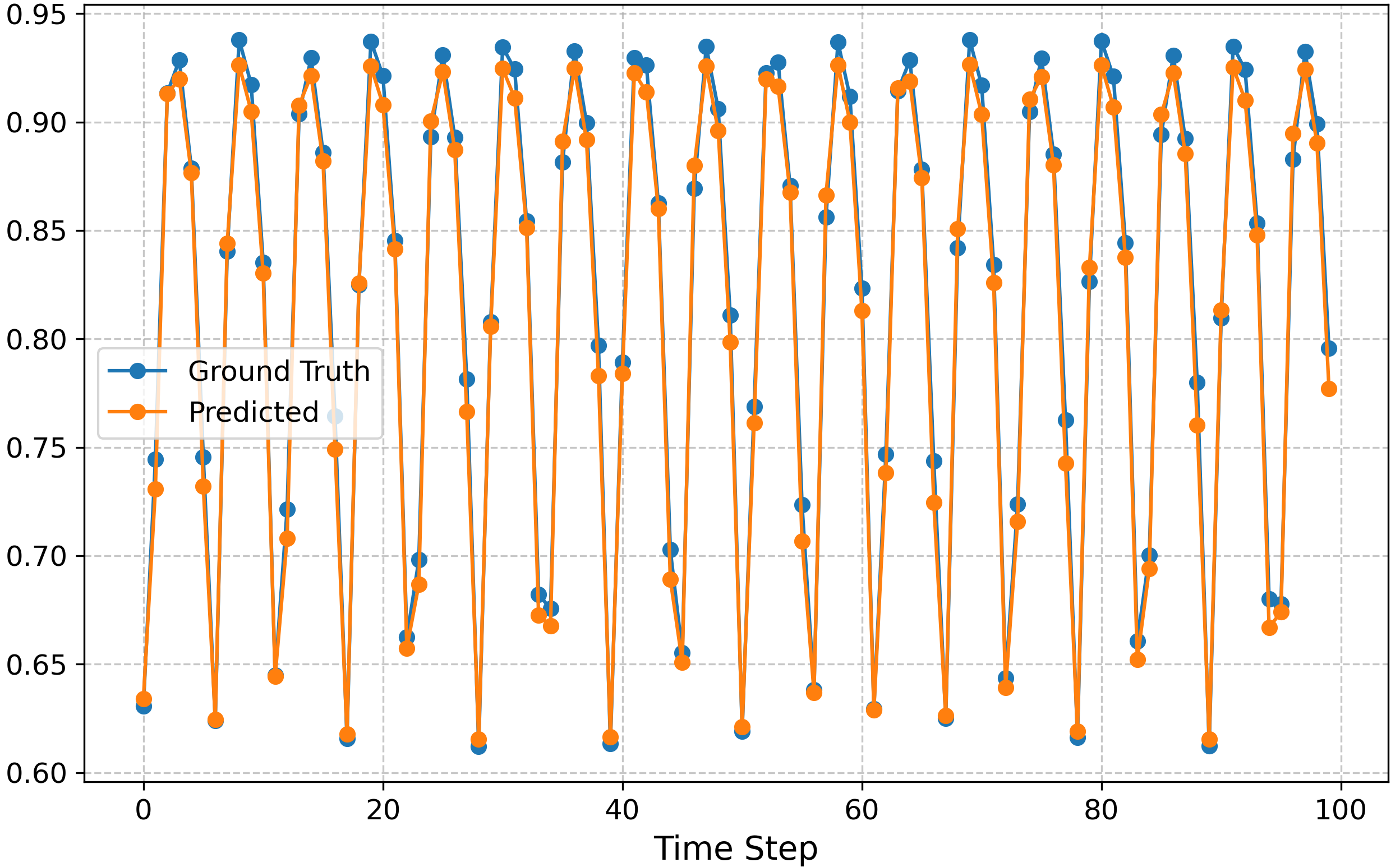}
    \end{minipage}
    \\[15pt]
    % Row 2
    \begin{minipage}{0.48\textwidth}
        \centering
        \begin{tikzpicture}
            \node[anchor=south west, inner sep=0] (image) at (0,0) {\includegraphics[width=1.9\textwidth, height=0.6\textwidth, keepaspectratio]{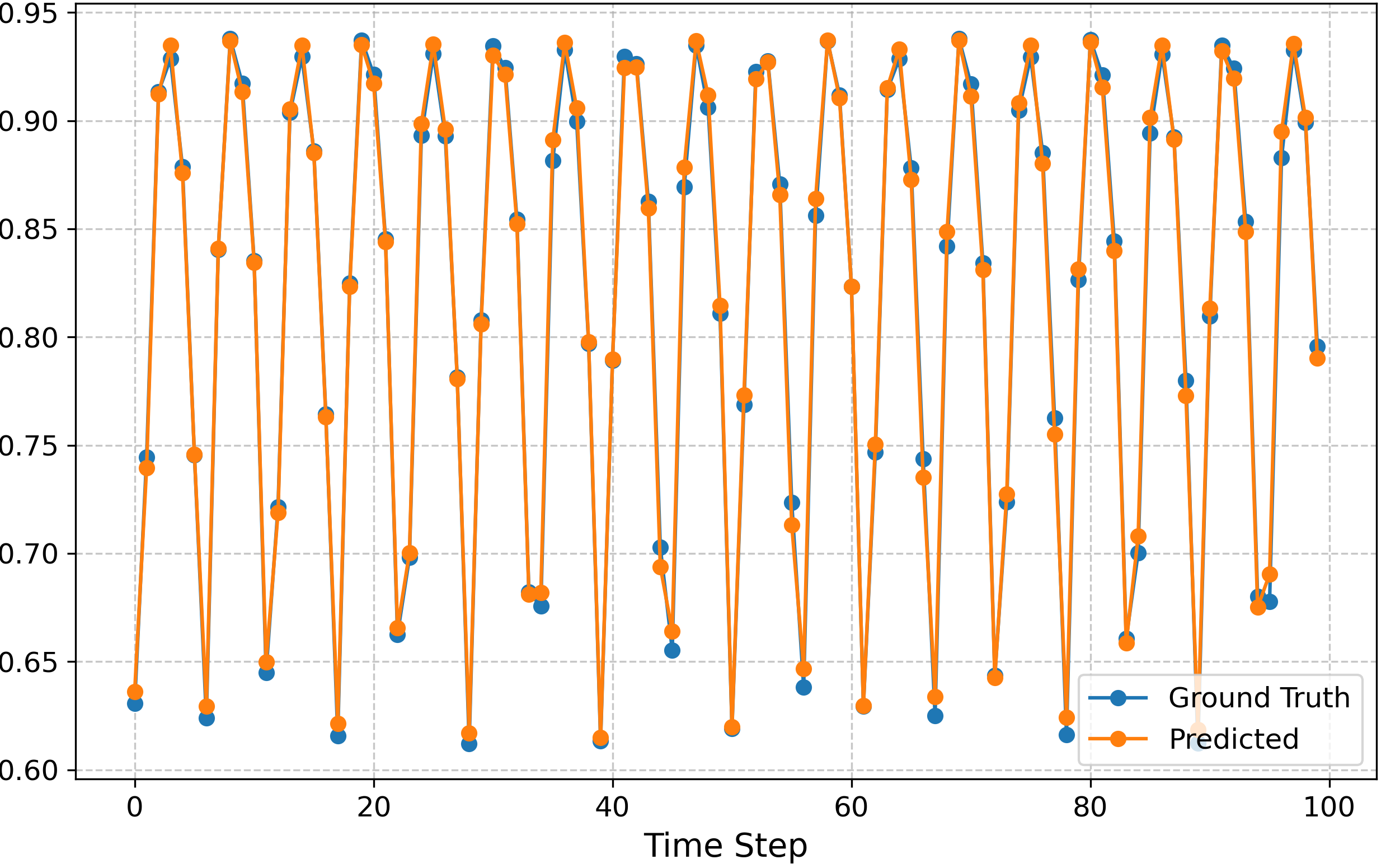}};
            \node[anchor=north, rotate=90, font=\footnotesize\bfseries] at (-0.45,2.5) {u};
        \end{tikzpicture}
    \end{minipage}
    \hfill
    \begin{minipage}{0.48\textwidth}
        \centering
        \includegraphics[width=1.9\textwidth, height=0.6\textwidth, keepaspectratio]{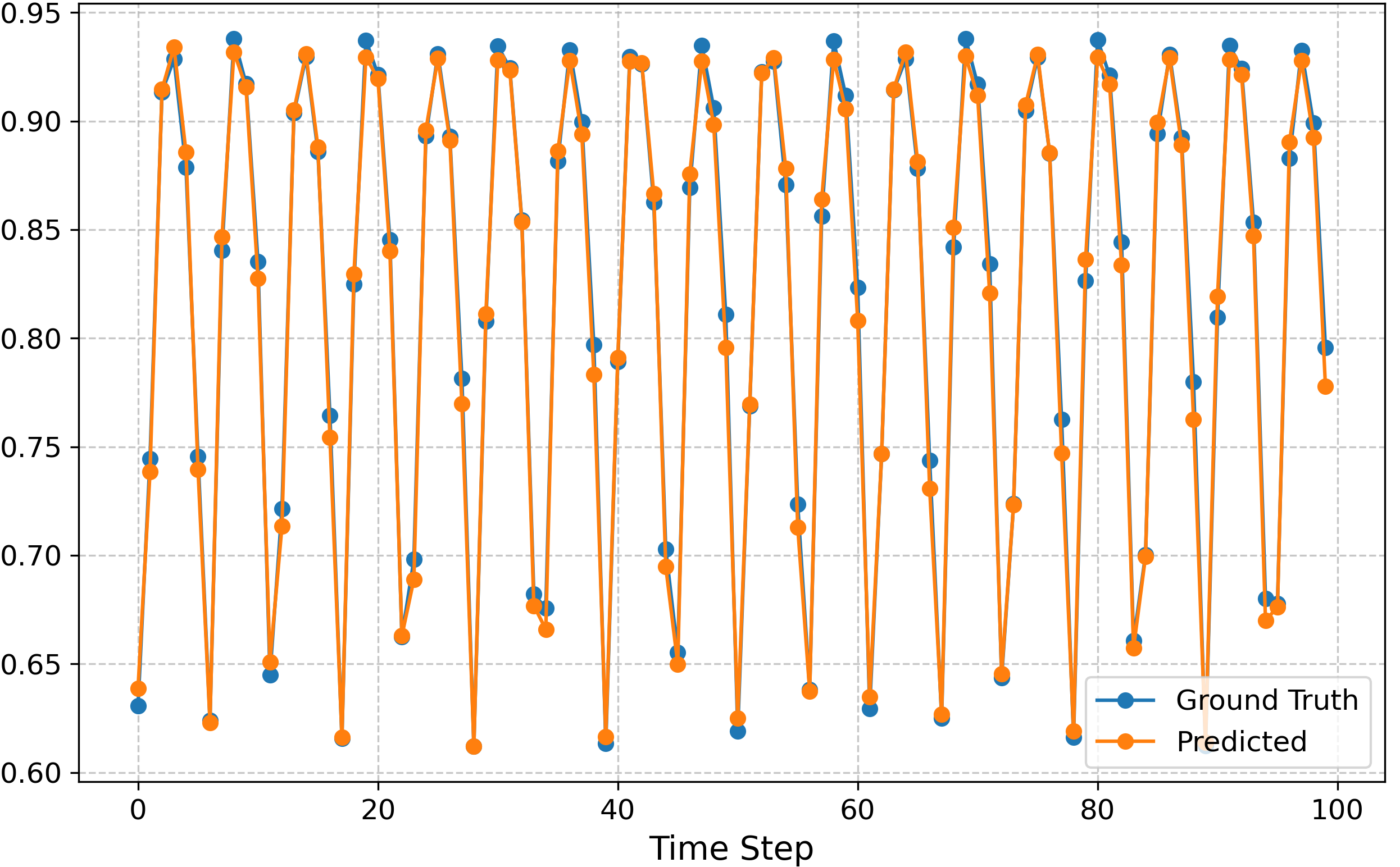}
    \end{minipage}
    \\[15pt]
    % Row 3
    \begin{minipage}{0.48\textwidth}
        \centering
        \begin{tikzpicture}
            \node[anchor=south west, inner sep=0] (image) at (0,0) {\includegraphics[width=2.3\textwidth, height=0.6\textwidth, keepaspectratio]{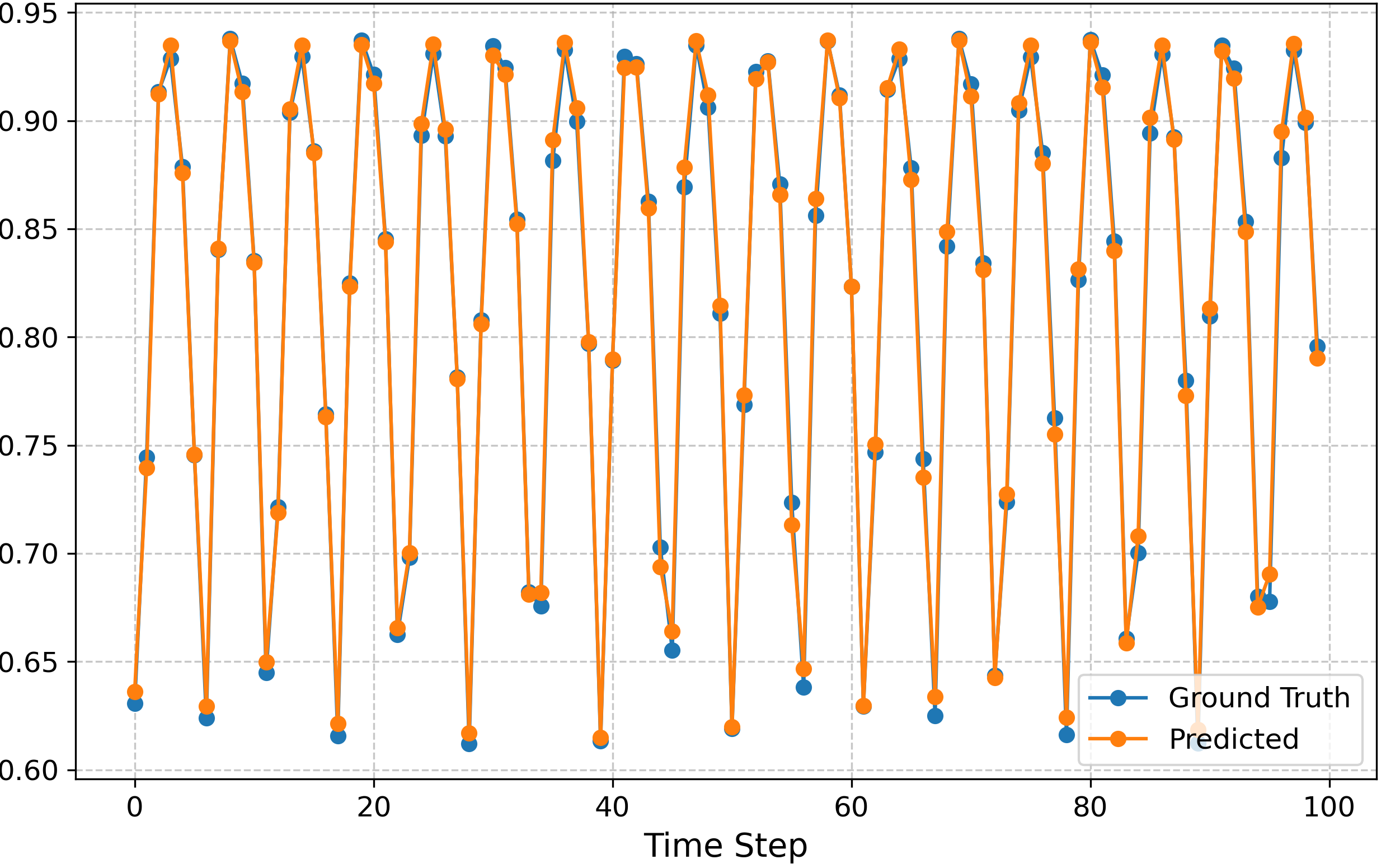}};
            \node[anchor=north, rotate=90, font=\footnotesize\bfseries] at (-0.45,2.5) {u};
        \end{tikzpicture}
    \end{minipage}
    \hfill
    \begin{minipage}{0.48\textwidth}
        \centering
        \includegraphics[width=2.3\textwidth, height=0.6\textwidth, keepaspectratio]{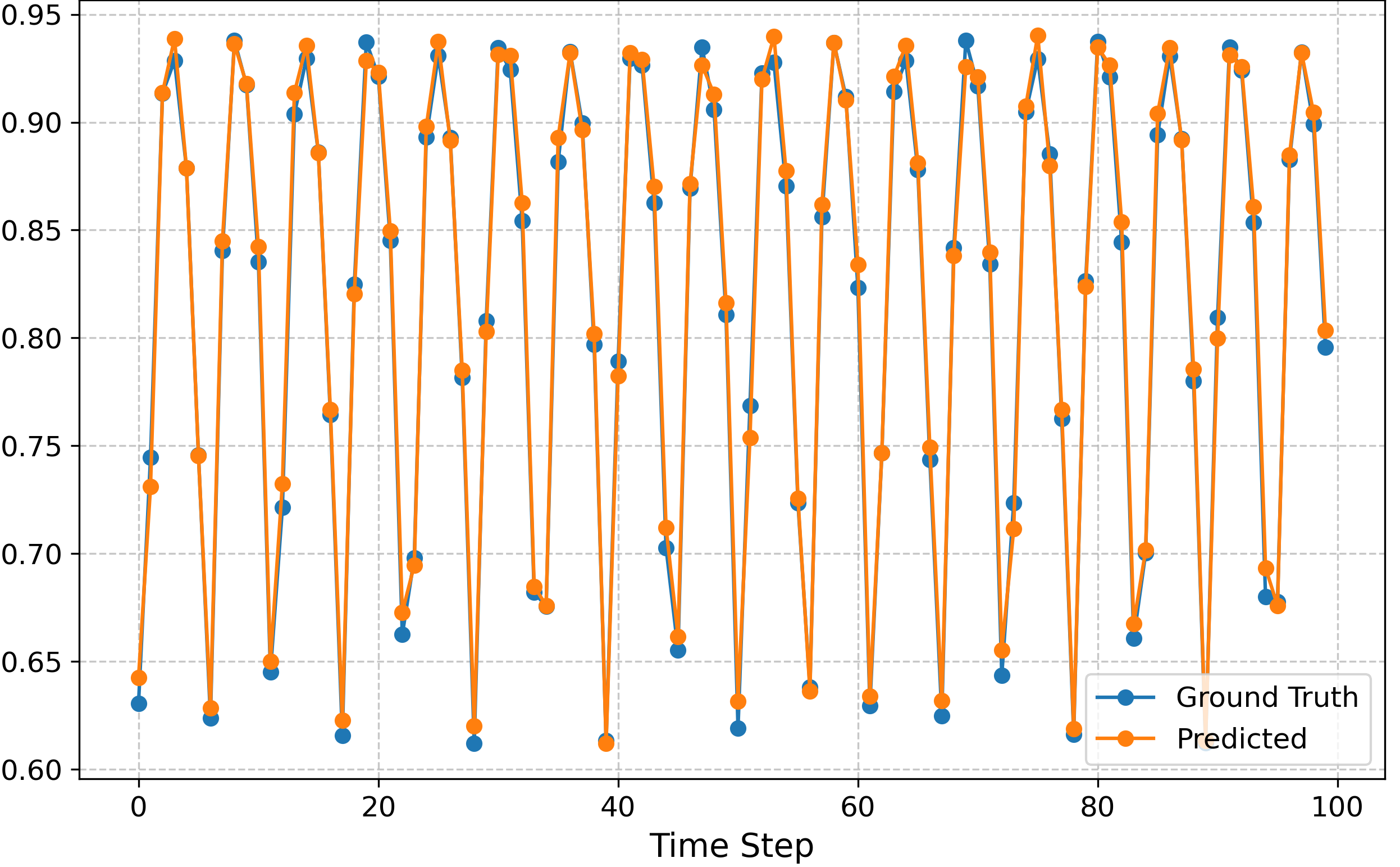}
    \end{minipage}
    \caption{Comparison of the predicted temporal evolution of streamwise velocity components for (a) HOSVD (left column) and (b) SVD (right column) across different LSTM architectures. From top to bottom: LSTM 1 Dense, LSTM 2 dense, and LSTM Time-Distributed.}
    \label{fig:streamwise_evolution}
\end{figure}

\begin{figure}[h!]
    \centering
    
    % Centered Label for (b) HOSVD and (b) SVD
    \makebox[\textwidth]{\textbf{(a) HOSVD} \hspace{5cm} \textbf{(b) SVD}} \\[5pt]
    % Row 1
    \begin{minipage}{0.48\textwidth}
        \centering
        \begin{tikzpicture}
            \node[anchor=south west, inner sep=0] (image) at (0,0) {\includegraphics[width=1.9\textwidth, height=0.6\textwidth, keepaspectratio]{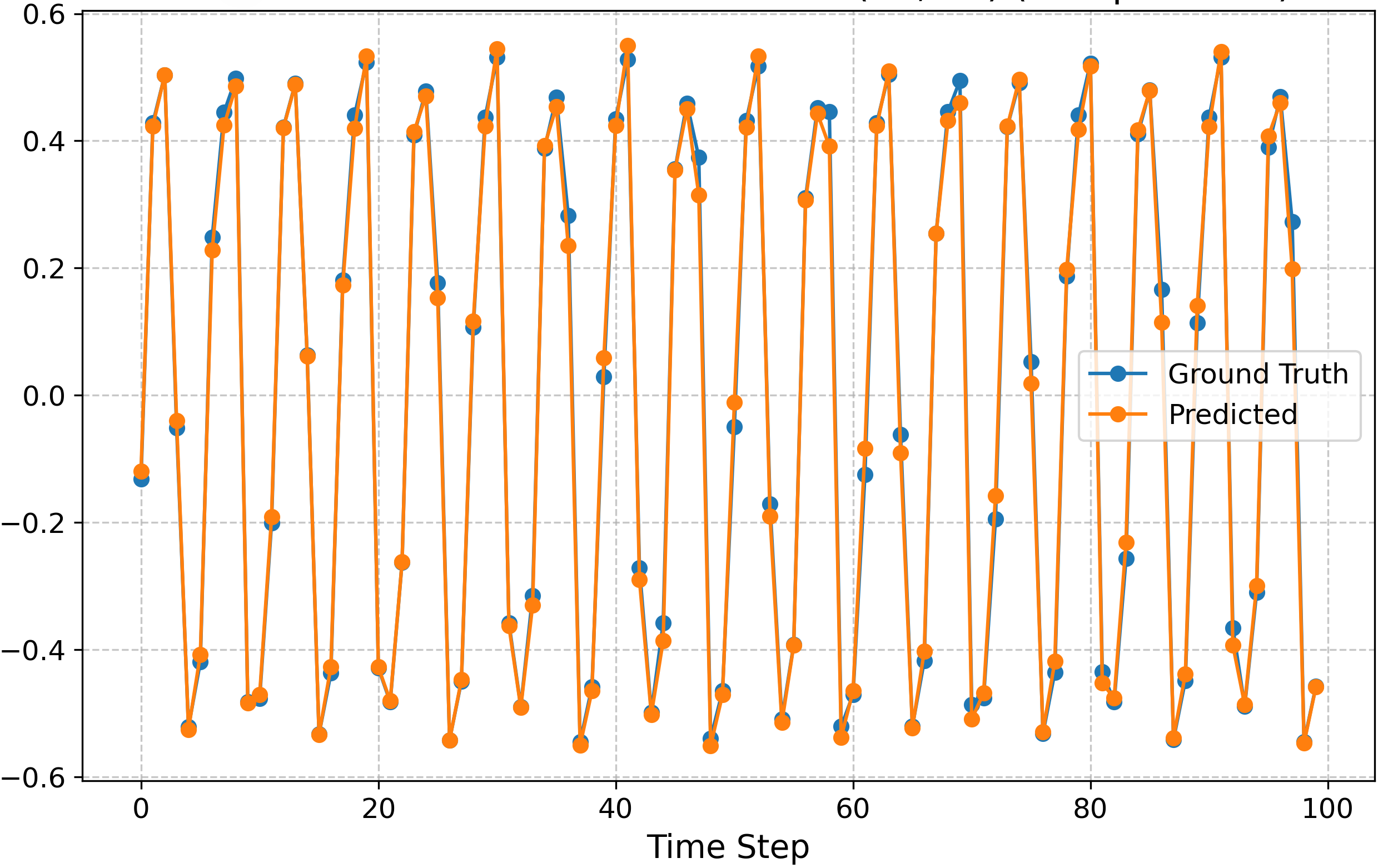}};
            \node[anchor=north, rotate=90, font=\footnotesize\bfseries] at (-0.4,2.5) {v};
        \end{tikzpicture}
    \end{minipage}
    \hfill
    \begin{minipage}{0.48\textwidth}
        \centering
        \includegraphics[width=1.9\textwidth, height=0.6\textwidth, keepaspectratio]{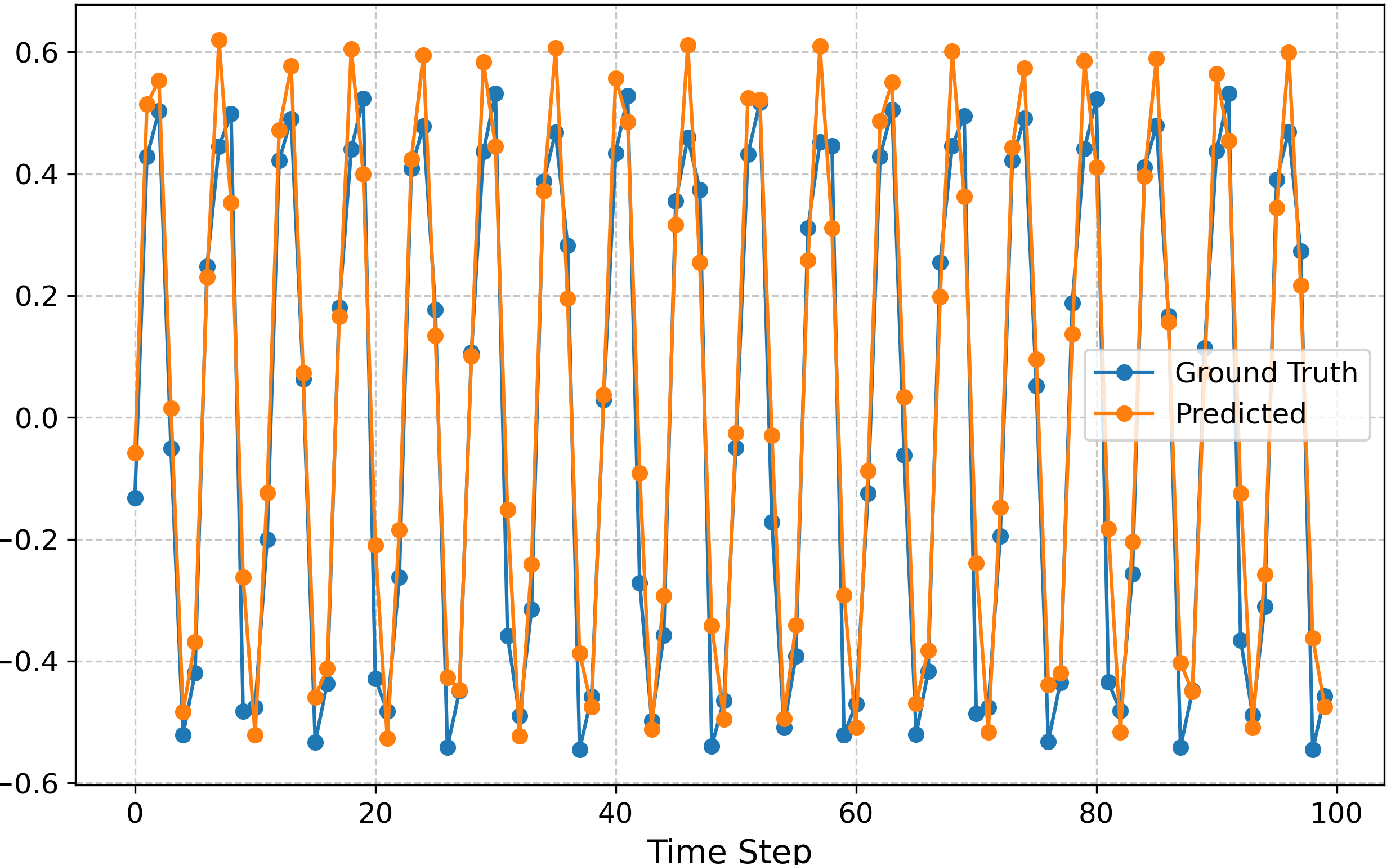}
    \end{minipage}
    \\[15pt]
    % Row 2
    \begin{minipage}{0.48\textwidth}
        \centering
        \begin{tikzpicture}
            \node[anchor=south west, inner sep=0] (image) at (0,0) {\includegraphics[width=1.9\textwidth, height=0.6\textwidth, keepaspectratio]{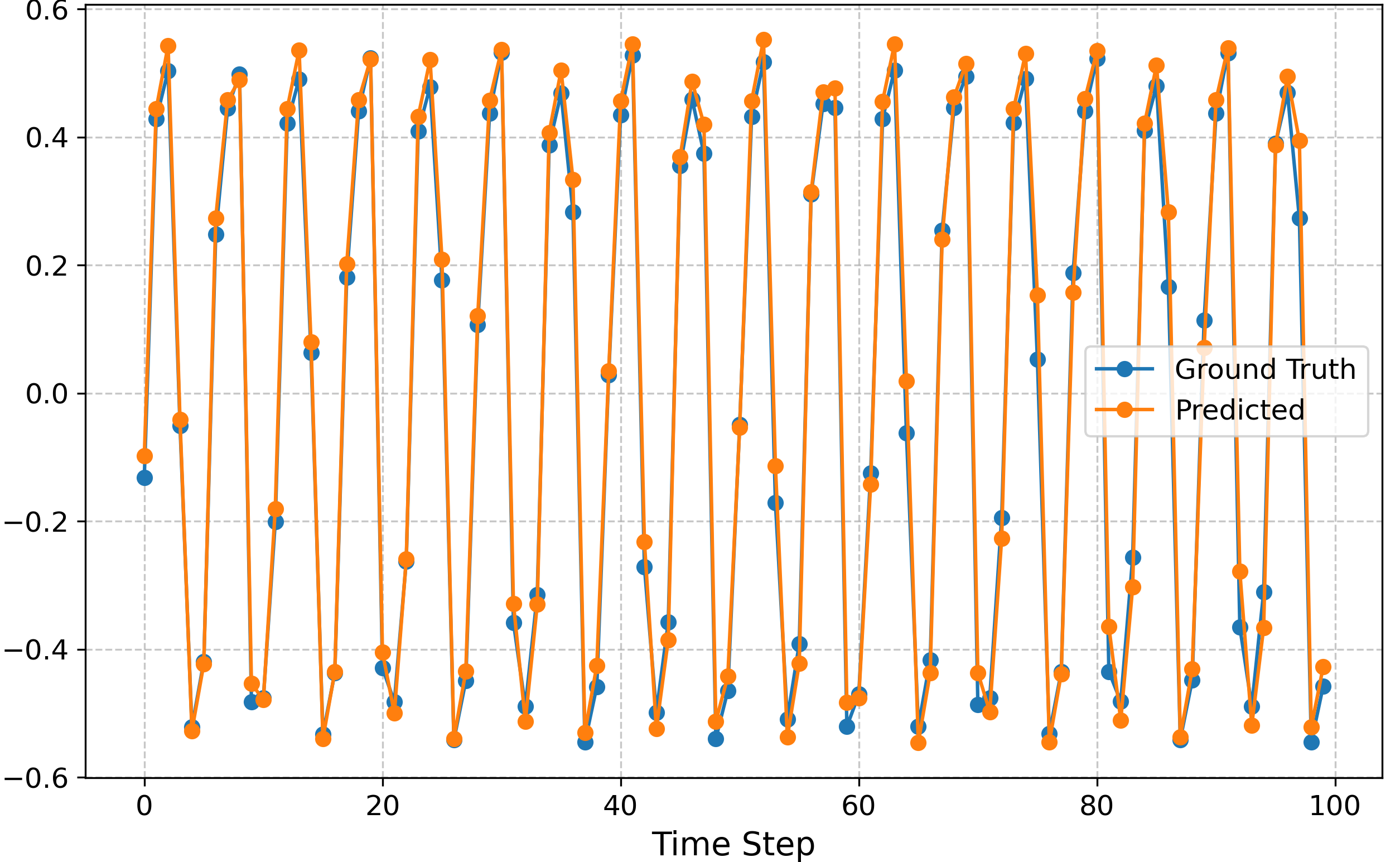}};
            \node[anchor=north, rotate=90, font=\footnotesize\bfseries] at (-0.4,2.5) {v};
        \end{tikzpicture}
    \end{minipage}
    \hfill
    \begin{minipage}{0.48\textwidth}
        \centering
        \includegraphics[width=1.9\textwidth, height=0.6\textwidth, keepaspectratio]{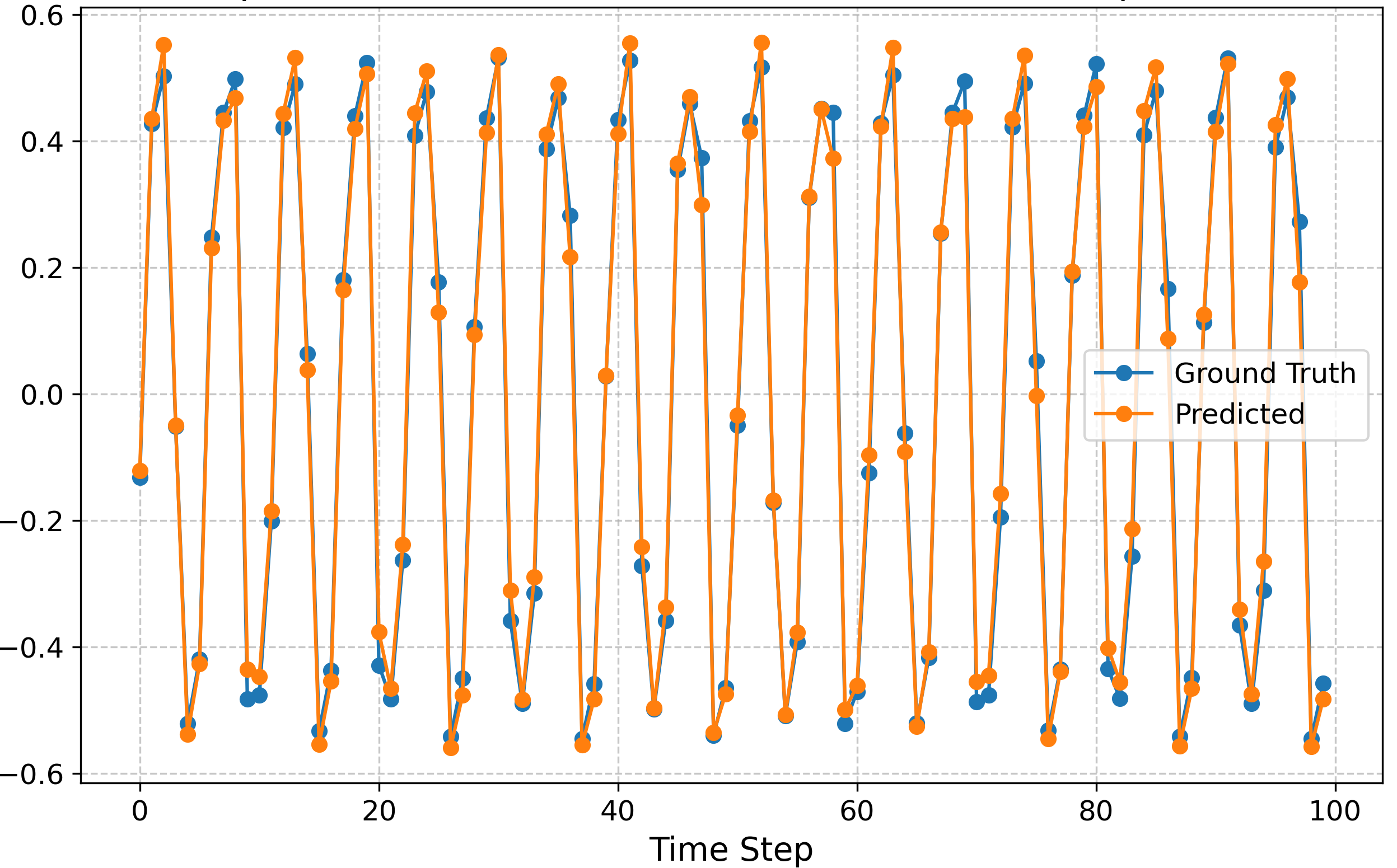}
    \end{minipage}
    \\[15pt]
    % Row 3
    \begin{minipage}{0.48\textwidth}
        \centering
        \begin{tikzpicture}
            \node[anchor=south west, inner sep=0] (image) at (0,0) {\includegraphics[width=1.9\textwidth, height=0.6\textwidth, keepaspectratio]{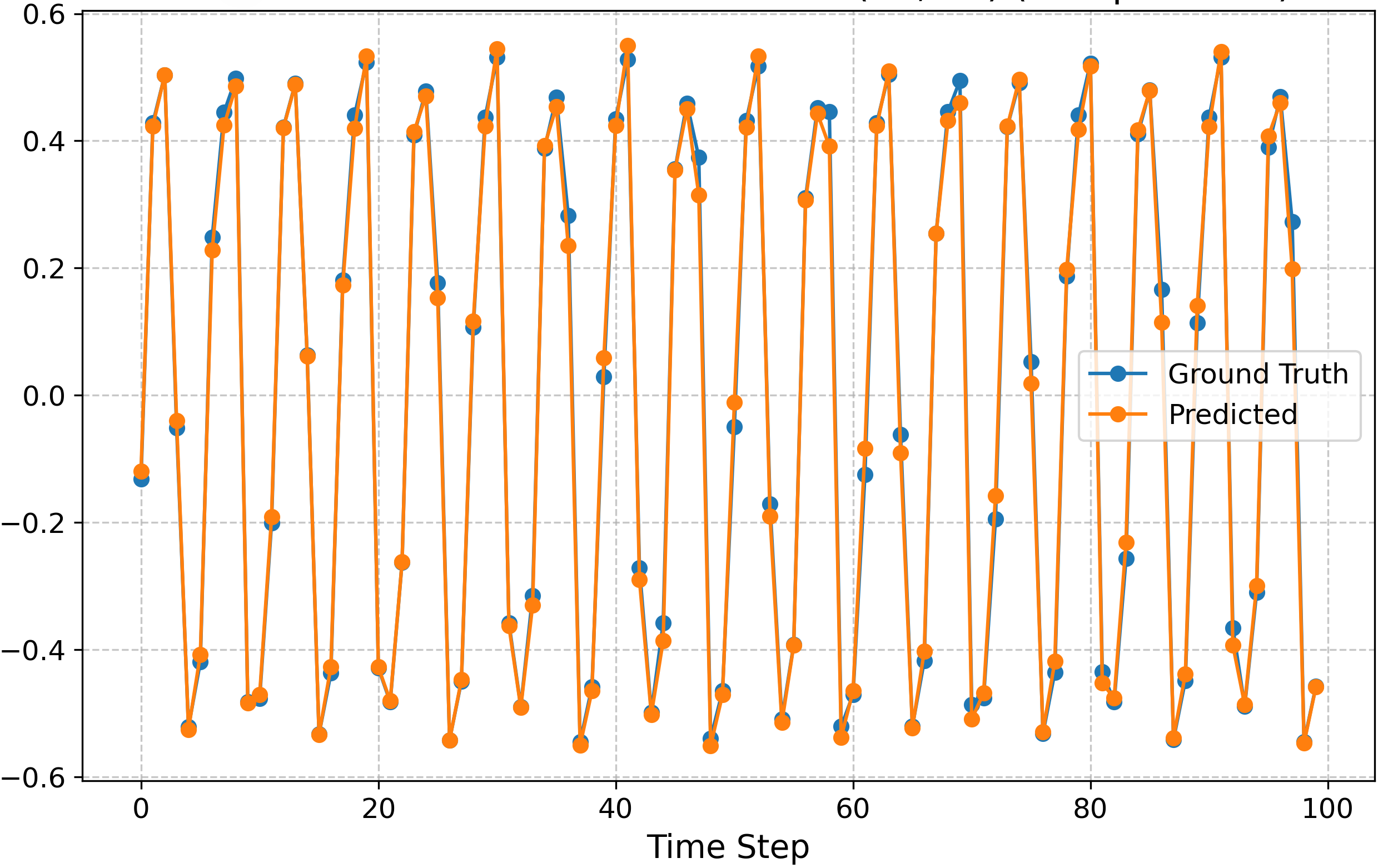}};
            \node[anchor=north, rotate=90, font=\footnotesize\bfseries] at (-0.4,2.5) {v};
        \end{tikzpicture}
    \end{minipage}
    \hfill
    \begin{minipage}{0.48\textwidth}
        \centering
        \includegraphics[width=1.9\textwidth, height=0.6\textwidth, keepaspectratio]{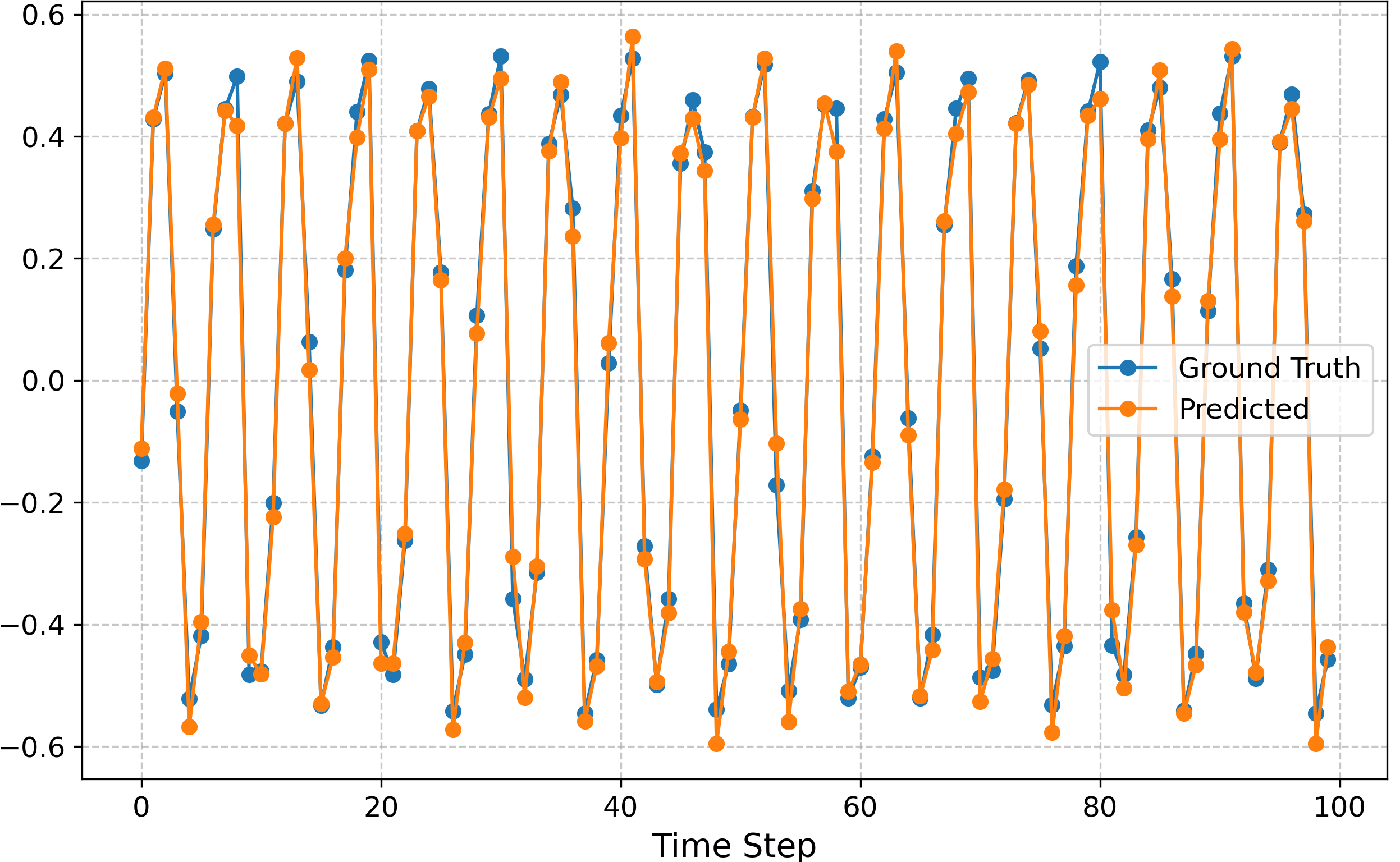}
    \end{minipage}
    \caption{ Comparison of the predicted temporal evolution of normal velocity components for (a) HOSVD (left column) and (b) SVD (right column) across different LSTM architectures. From top to bottom: LSTM 1 Dense, LSTM 2 dense, and LSTM Time-Distributed.}
    \label{fig:normal_velocity_evolution}
\end{figure}

The temporal evolution plots in Figures \ref{fig:streamwise_evolution} and \ref{fig:normal_velocity_evolution} present the comparison between the ground truth and the predicted streamwise and normal velocity components, respectively. The predicted temporal patterns for the streamwise velocity of both HOSVD and SVD closely follow the ground truth, demonstrating the ability of the models to capture the periodicity of the streamwise velocity component. Minor deviations are observed in the peaks and troughs of the predictions, with SVD in the vicinity of the peaks. Both the predicted streamwise and the predicted normal velocity patterns align well with the ground truth. The differences between HOSVD and SVD are minimal, with both methods achieving a high degree of temporal accuracy across all architectures.

In terms of the LSTM architectures themselves, all three configurations—1 dense, 2 dense, and time-distributed perform robustly, demonstrating their ability to accurately capture the flow dynamics when paired with robust dimensionality reduction techniques. The results emphasize that even relatively simple architectures, when optimized with appropriate hyperparameters such as learning rate, batch size, and sequence length, can achieve competitive results. This finding highlights the importance of proper parameter selection and demonstrates that increasing model complexity does not always yield significant improvements, especially when a simpler network can already provide sufficient predictive accuracy.

Further analysis is conducted on the performance of the predictive models by evaluating uncertainty quantification for the HOSVD and SVD approaches. UQ provides insights into the reliability of the predicted snapshots.
\begin{figure}[h!]
    \centering
    % Centered Label for (b) HOSVD and (b) SVD
    \makebox[\textwidth]{\textbf{(a) HOSVD} \hspace{5cm} \textbf{(b) SVD}} \\[5pt]
    \begin{minipage}{0.48\textwidth}
        \centering
        \includegraphics[width=1.9\textwidth, height=0.57\textwidth, keepaspectratio]{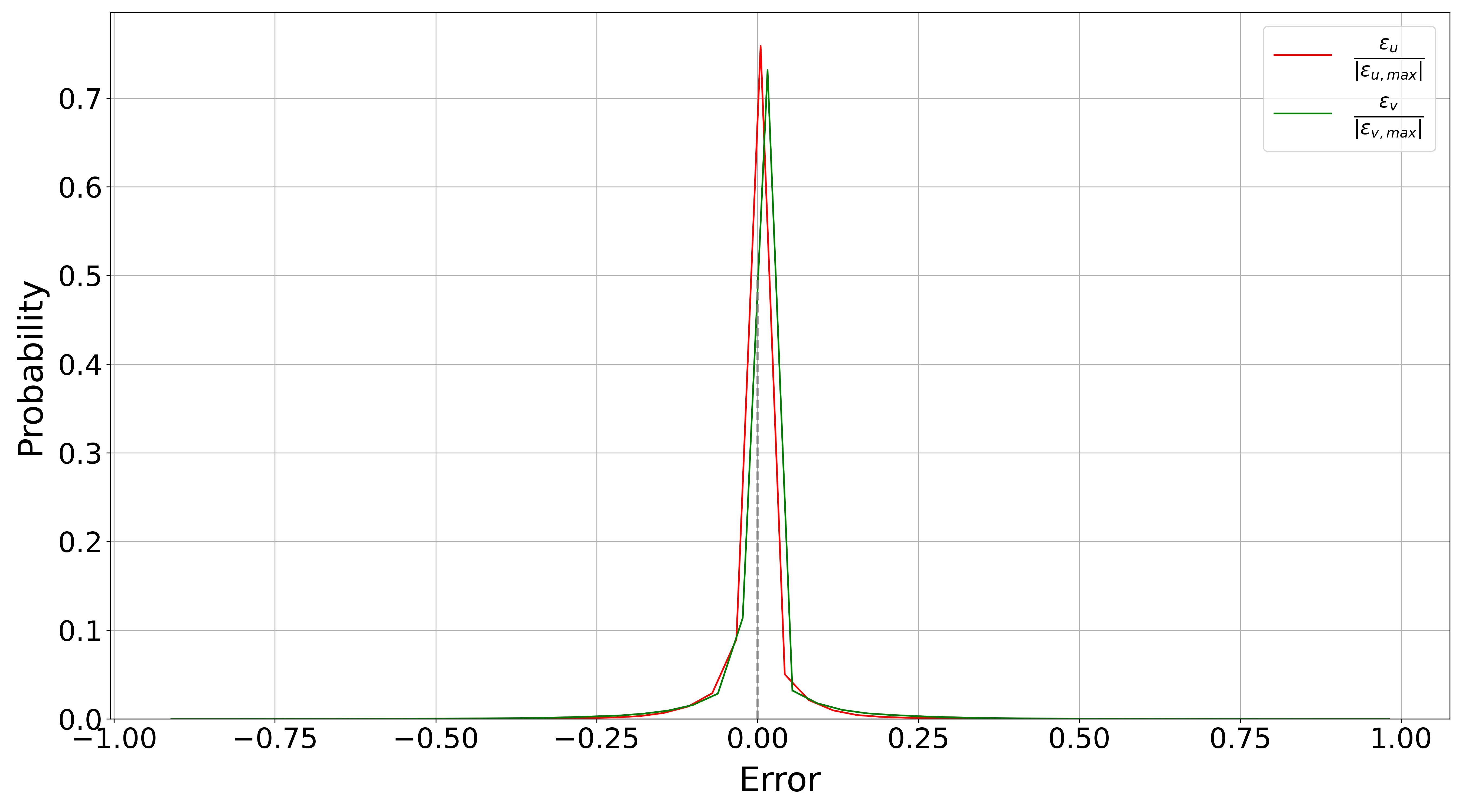}
    \end{minipage}
    \hfill
    \begin{minipage}{0.48\textwidth}
        \centering
        \includegraphics[width=1.9\textwidth, height=0.57\textwidth, keepaspectratio]{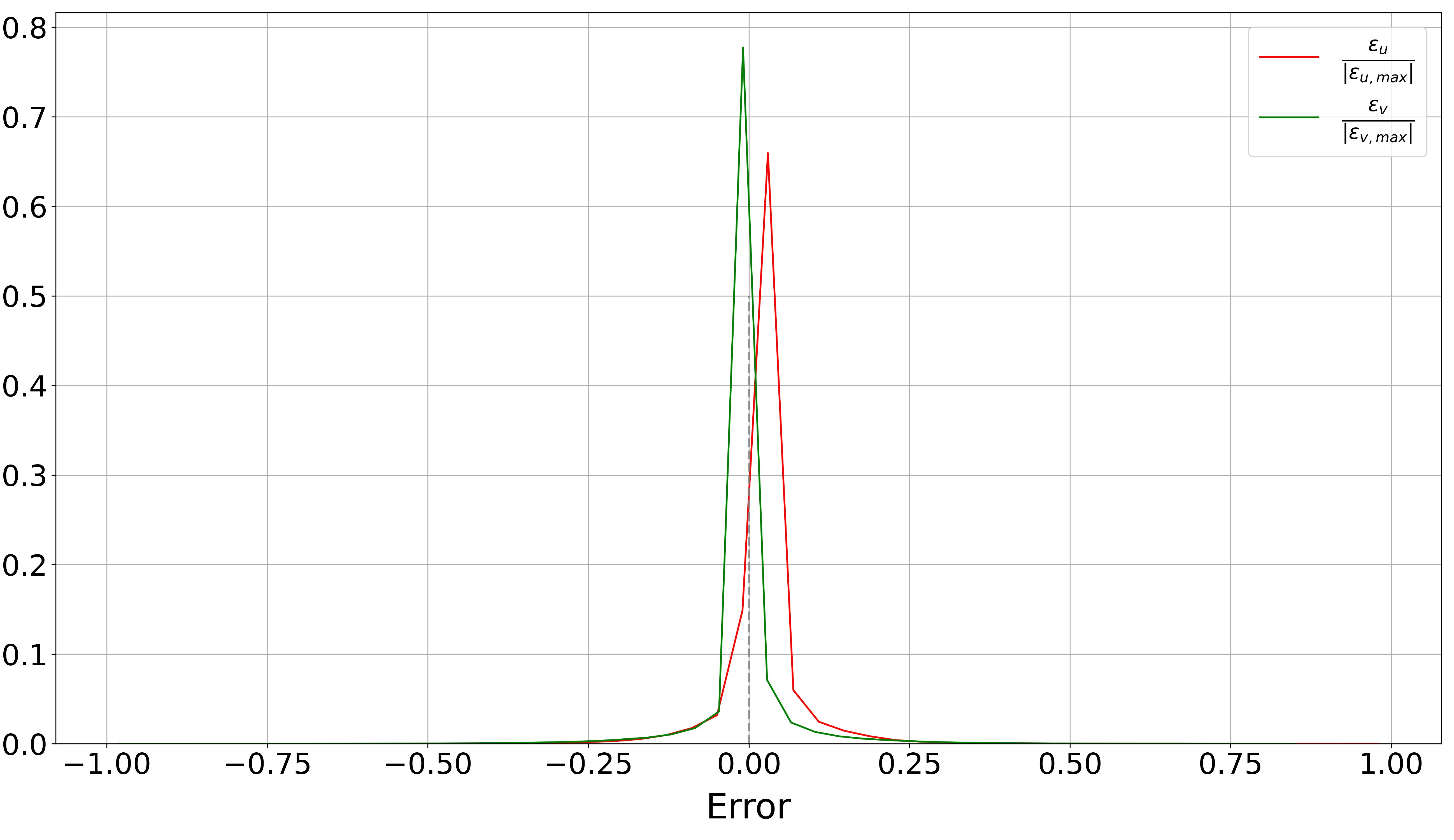}
    \end{minipage}
    \\[10pt]
    \begin{minipage}{0.48\textwidth}
        \centering
        \includegraphics[width=1.9\textwidth, height=0.57\textwidth, keepaspectratio]{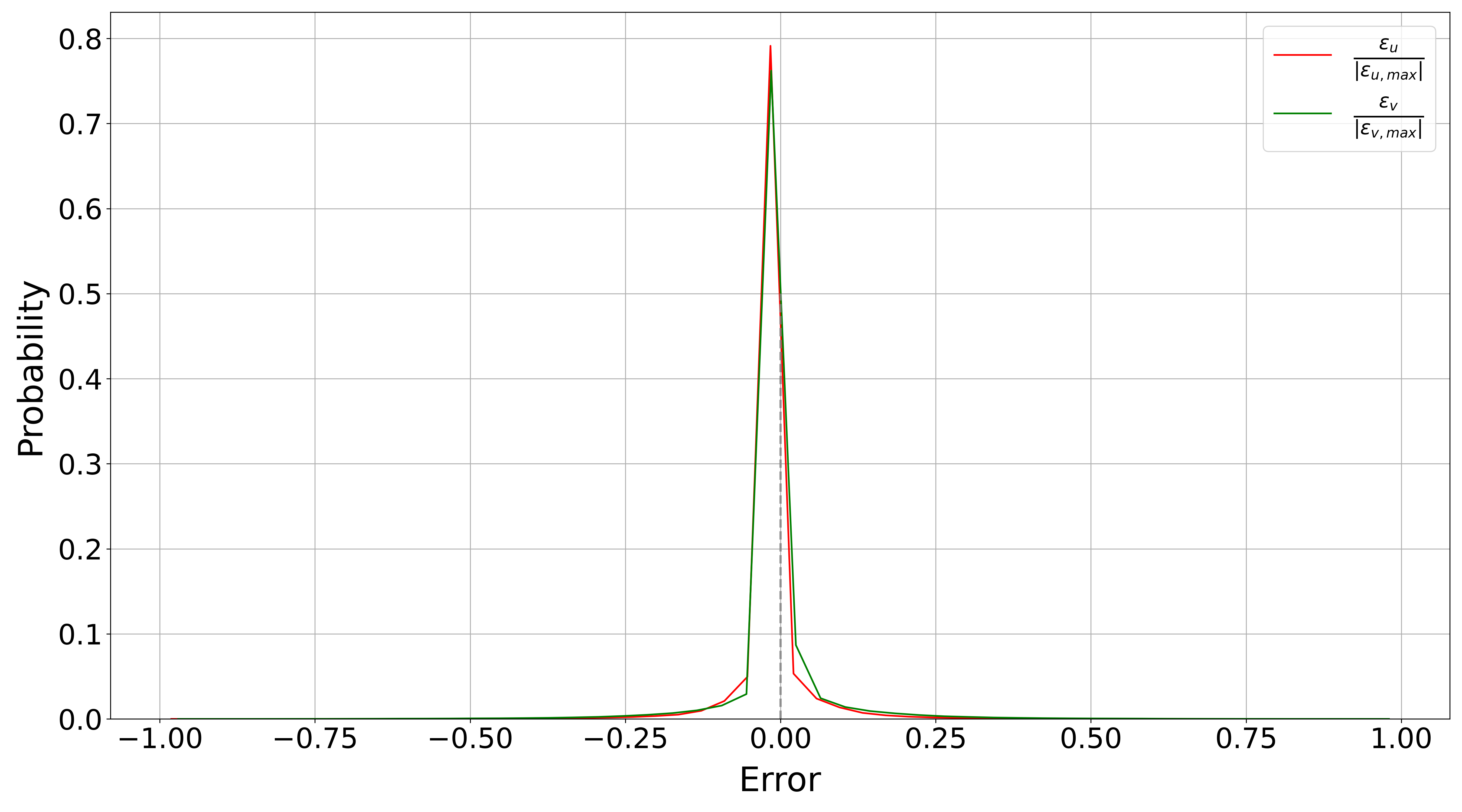}
    \end{minipage}
    \hfill
    \begin{minipage}{0.48\textwidth}
        \centering
        \includegraphics[width=1.9\textwidth, height=0.57\textwidth, keepaspectratio]{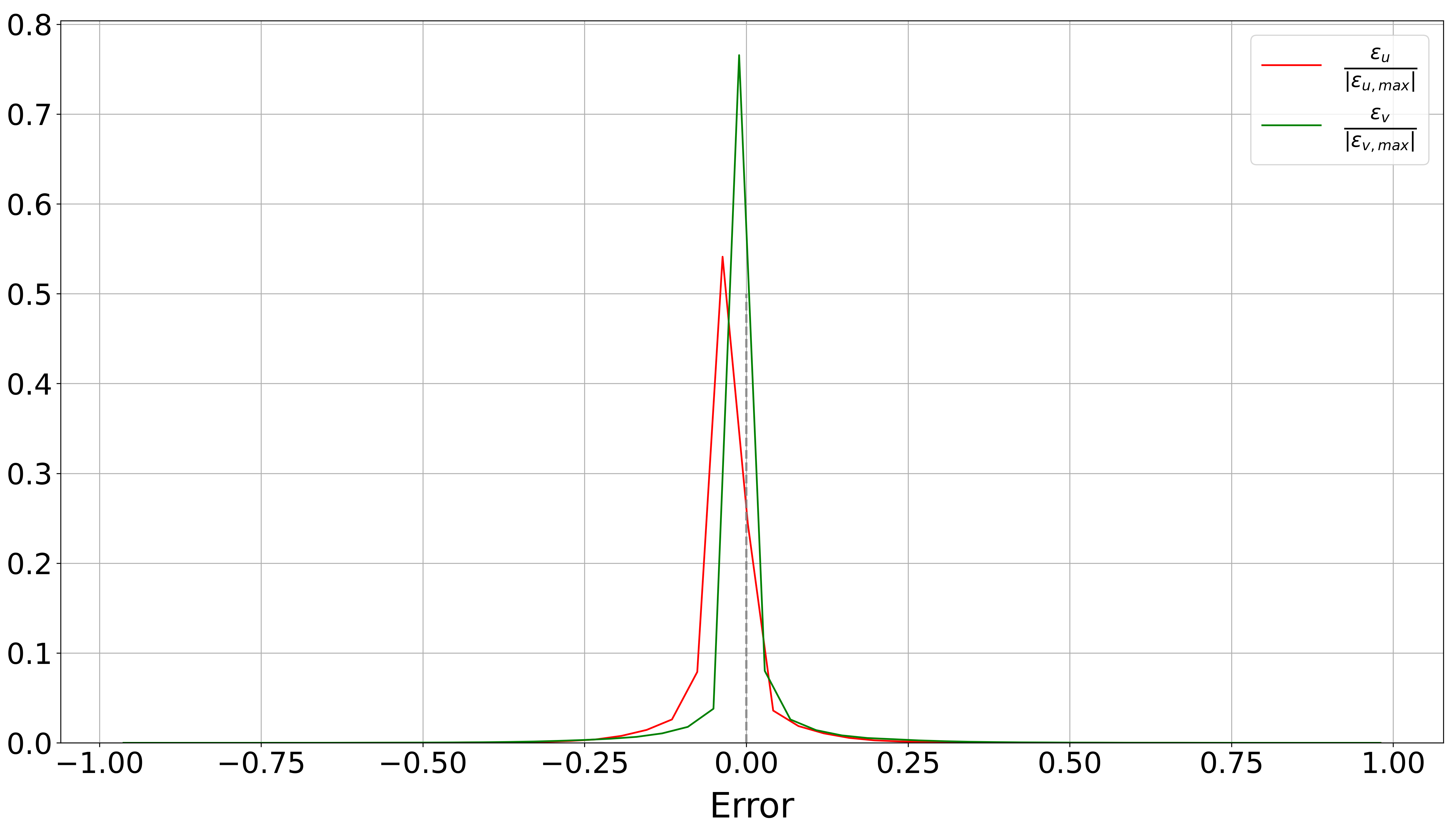}
    \end{minipage}
    \\[10pt]
    \begin{minipage}{0.48\textwidth}
        \centering
        \includegraphics[width=1.9\textwidth, height=0.57\textwidth, keepaspectratio]{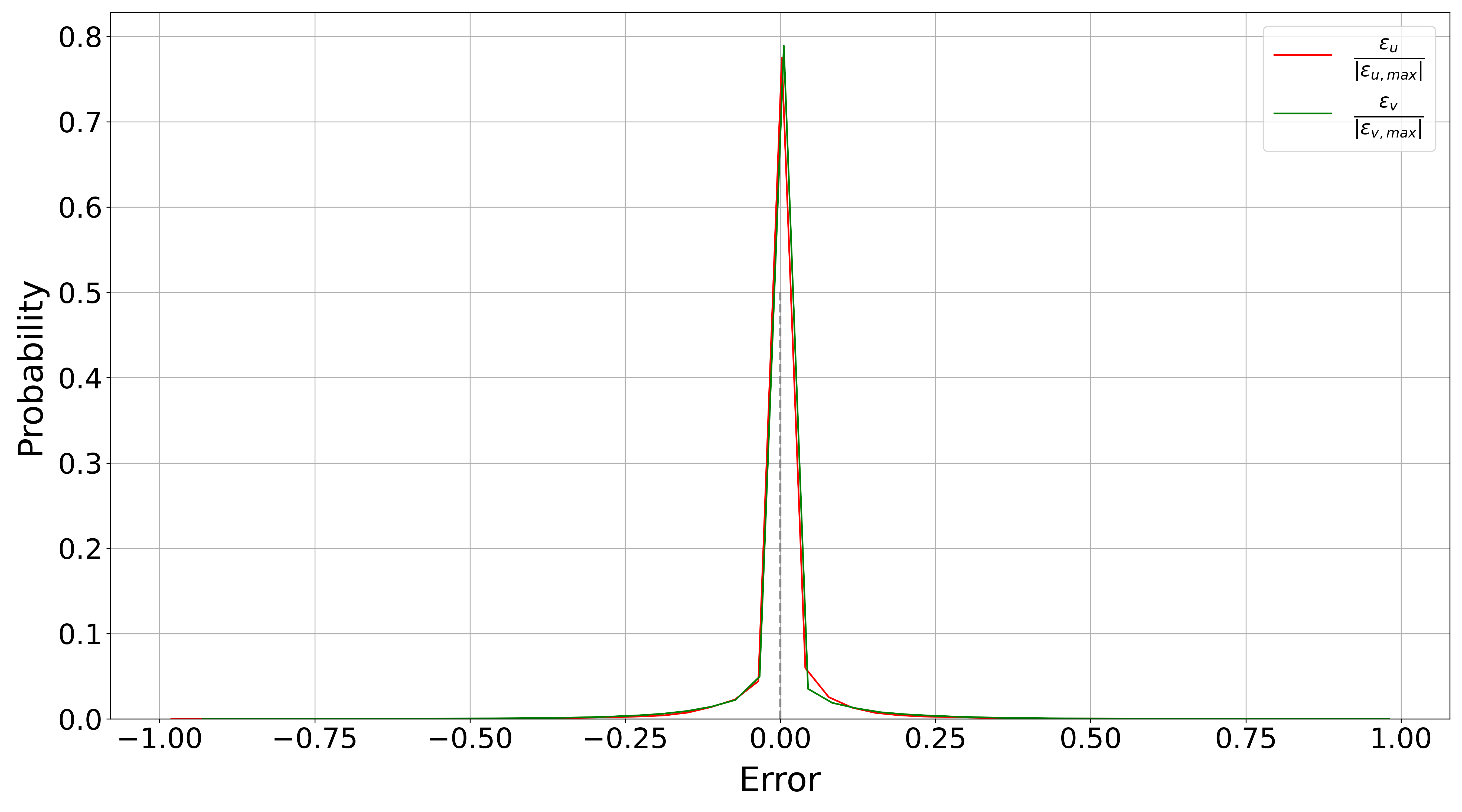}
    \end{minipage}
    \hfill
    \begin{minipage}{0.48\textwidth}
        \centering
        \includegraphics[width=1.9\textwidth, height=0.57\textwidth, keepaspectratio]{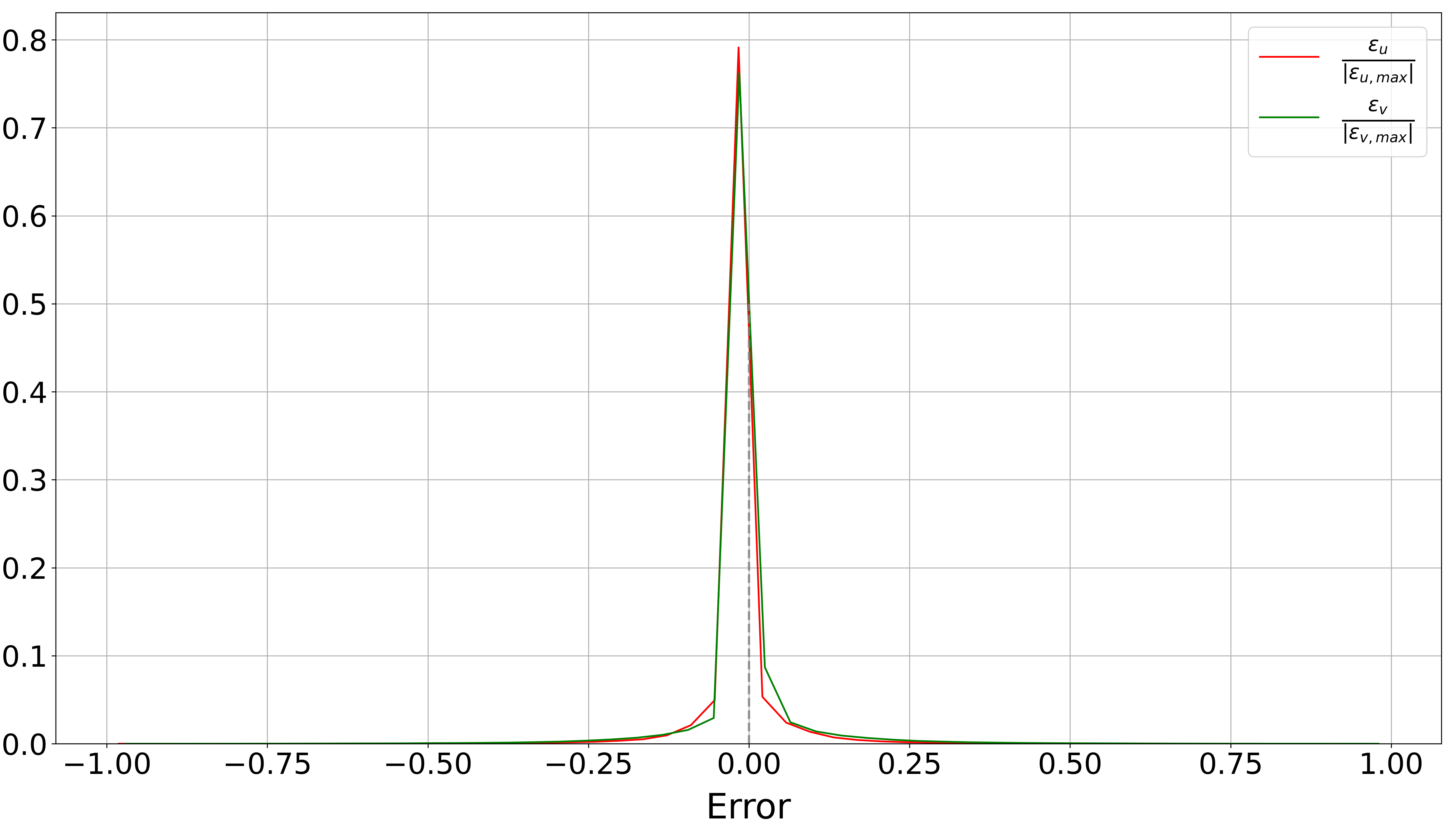}
    \end{minipage}
    \caption{Uncertainty quantification (UQ) results for (a) HOSVD (left column) and (b) SVD (right column) across LSTM architectures. From top to bottom: predictions for LSTM with 1 Dense, 2 Dense, and Time-Distributed architectures for the 2D cylinder flow. Histograms have been constructed using 50 bins.}
    \label{fig:uq_results}
\end{figure}

Figure \ref{fig:uq_results} presents the UQ results for the velocity components across the LSTM architectures. In general, HOSVD demonstrates more concentrated error distributions around zero compared to SVD. For instance, in the 1 dense and 2 dense architectures, the HOSVD error distributions exhibit narrower and more peaked profiles, indicating reduced variability and higher prediction reliability. The HOSVD-based LSTM 2 dense and time-distributed models show the best results with both the components following a normal distribution profile and a probability of about 80\% for 0 error. In contrast, SVD displays wider and skewed error distributions, suggesting slightly lower accuracy and consistency in capturing flow dynamics except for the time-distributed architecture, which shows comparable error distributions with HOSVD, reflecting the capacity of deeper architectures to compensate for some of the limitations of SVD.

\subsubsection{Case: Laminar Flow Past a Circular Cylinder (3D)}

The flow past a three-dimensional cylinder extends the dynamics observed in two-dimensional cases. This data set captures the flow behavior at Re = 280, where the flow transitions to a fully three-dimensional state, showcasing intricate patterns and interactions in the spanwise direction.

A total of 599 snapshots were collected during the simulation, but only the last 299 snapshots, representing the saturated flow regime, were used in this study. As mentioned before, the flow in the spanwise direction only begins to form between 200 to 300 snapshots. Hence, the 80-20 split was not followed. Instead, the last 99 snapshots were used for validation, and the preceding 200 for training. The data set encompasses the three velocity components \((u, v, w)\), representing the streamwise, normal, and spanwise flow directions, respectively.

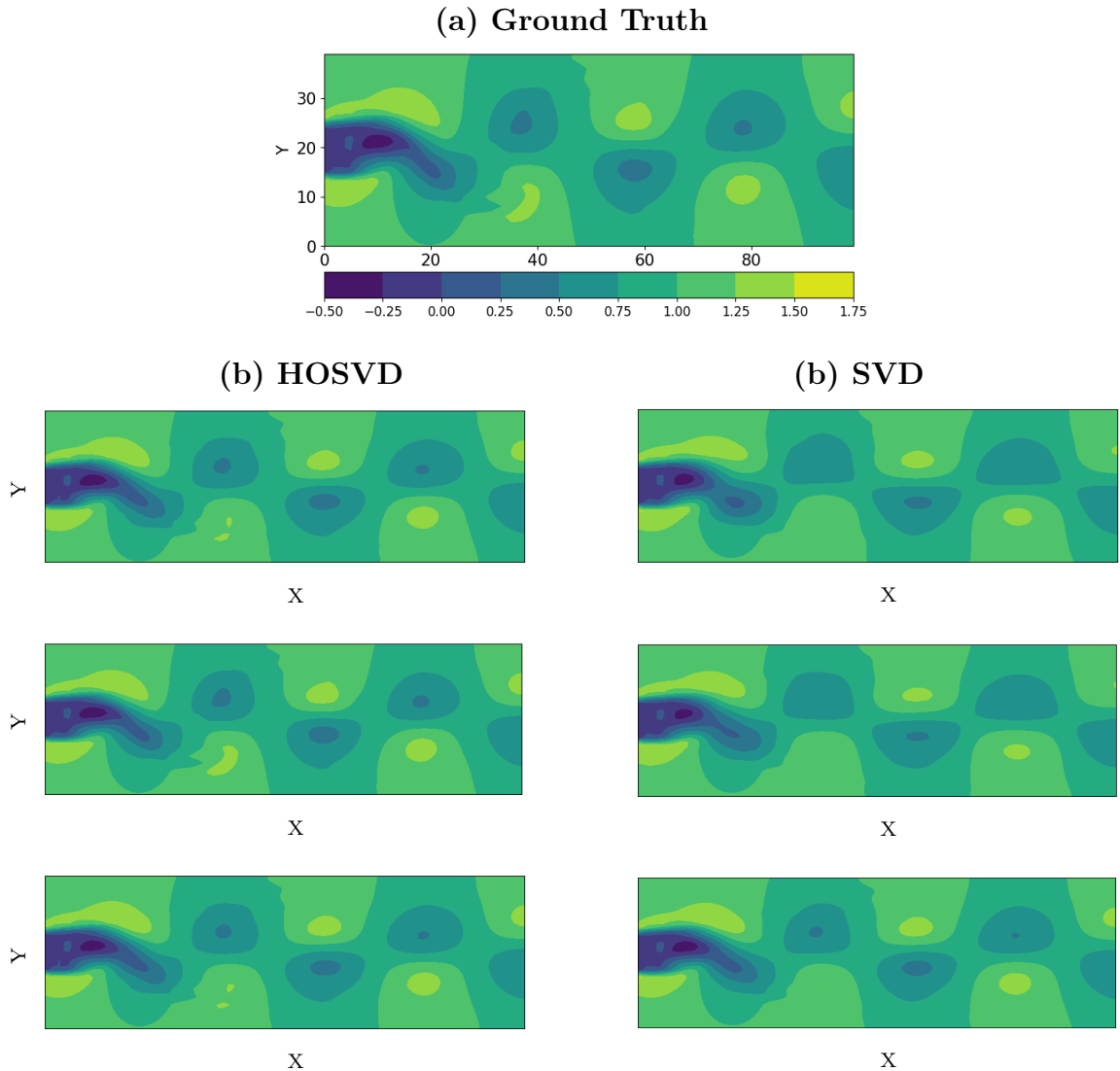
\begin{figure}[h!]
    \centering
    \textbf{(a) Ground Truth} \\[5pt]
    \begin{minipage}{0.5\textwidth}
        \centering
        \includegraphics[width=\textwidth]{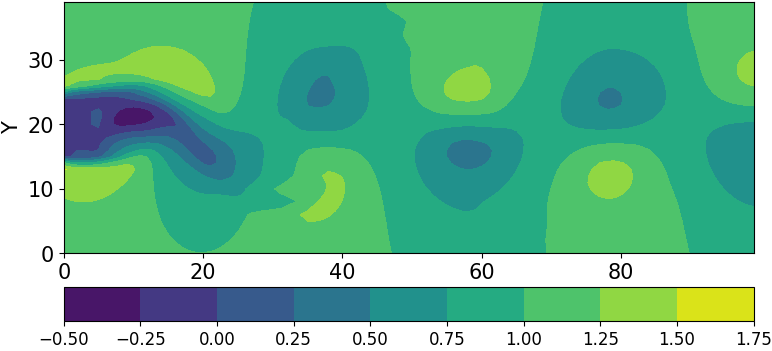}
    \end{minipage}
    \\[15pt]
   % Centered Label for (b) HOSVD and (b) SVD
    \makebox[\textwidth]{\textbf{(b) HOSVD} \hspace{5cm} \textbf{(b) SVD}} \\[5pt]
    
    % Row 1
    \begin{minipage}{0.48\textwidth}
        \centering
        \begin{tikzpicture}
            \node[anchor=south west, inner sep=0] (image) at (0,0) {\includegraphics[width=0.85\textwidth]{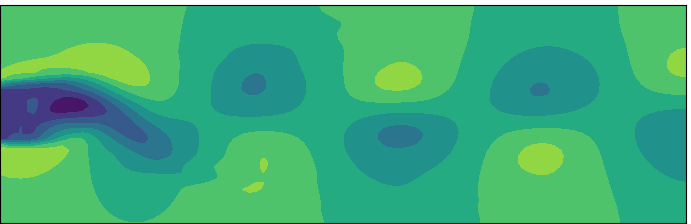}};
            \node[anchor=north, rotate=90, font=\scriptsize] at (-0.6,1) {Y};
            \node[anchor=north, font=\scriptsize] at (3.4,-0.2) {X};
        \end{tikzpicture}
    \end{minipage}
    \hfill
    \begin{minipage}{0.48\textwidth}
        \centering
        \begin{tikzpicture}
            \node[anchor=south west, inner sep=0] (image) at (0,0) {\includegraphics[width=0.85\textwidth]{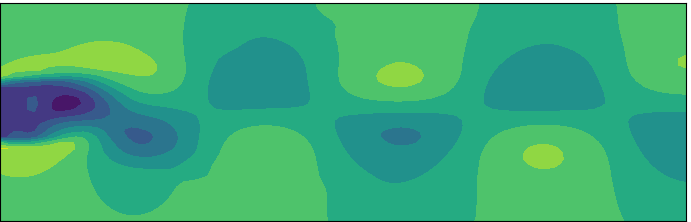}};
            \node[anchor=north, font=\scriptsize] at (3.4,-0.2) {X};
        \end{tikzpicture}
    \end{minipage}
    \\[10pt]
    
    % Row 2
    \begin{minipage}{0.48\textwidth}
        \centering
        \begin{tikzpicture}
            \node[anchor=south west, inner sep=0] (image) at (0,0) {\includegraphics[width=0.85\textwidth]{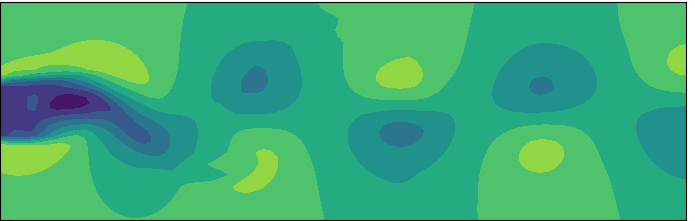}};
            \node[anchor=north, rotate=90, font=\scriptsize] at (-0.6,1) {Y};
            \node[anchor=north, font=\scriptsize] at (3.4,-0.2) {X};
        \end{tikzpicture}
    \end{minipage}
    \hfill
    \begin{minipage}{0.48\textwidth}
        \centering
        \begin{tikzpicture}
            \node[anchor=south west, inner sep=0] (image) at (0,0) {\includegraphics[width=0.85\textwidth]{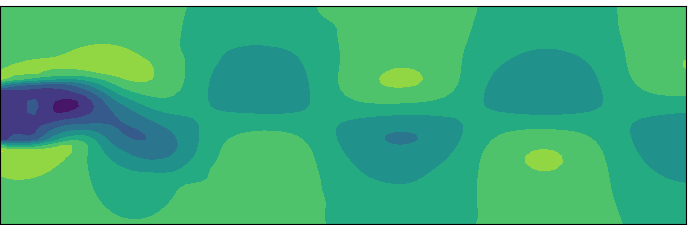}};
            \node[anchor=north, font=\scriptsize] at (3.4,-0.2) {X};
        \end{tikzpicture}
    \end{minipage}
    \\[10pt]
    
    % Row 3 
    \begin{minipage}{0.48\textwidth}
        \centering
        \begin{tikzpicture}
            \node[anchor=south west, inner sep=0] (image) at (0,0) {\includegraphics[width=0.85\textwidth]{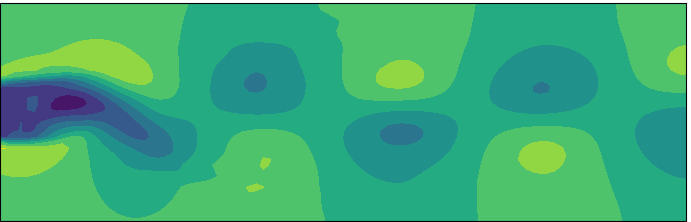}};
            \node[anchor=north, rotate=90, font=\scriptsize] at (-0.6,1) {Y};
            \node[anchor=north, font=\scriptsize] at (3.4,-0.2) {X};
        \end{tikzpicture}
    \end{minipage}
    \hfill
    \begin{minipage}{0.48\textwidth}
        \centering
        \begin{tikzpicture}
            \node[anchor=south west, inner sep=0] (image) at (0,0) {\includegraphics[width=0.85\textwidth, height=0.27\textwidth]{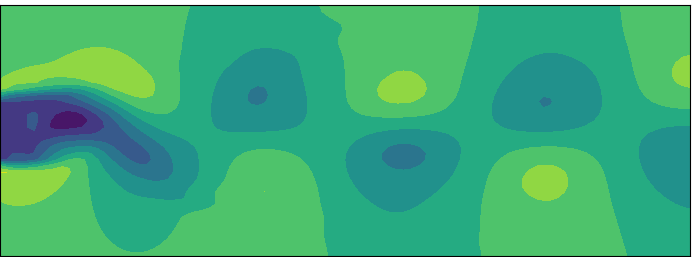}};
            \node[anchor=north, font=\scriptsize] at (3.4,-0.2) {X};
        \end{tikzpicture}
    \end{minipage}

    \caption{(a) Ground truth streamwise velocity components at \(t = 528\). 
    (b) Comparison of the predicted streamwise velocity components for HOSVD (left column) and SVD (right column) across different LSTM architectures. From top to bottom: predictions for LSTM with 1 Dense, 2 Dense, and Time-Distributed architectures.}
    \label{fig:lstm_3d_streamwise}
\end{figure}

% Normal Velocity Figure
\begin{figure}[h!]
    \centering
    \textbf{(a) Ground Truth} \\[5pt]
    \begin{minipage}{0.5\textwidth}
        \centering
        \includegraphics[width=\textwidth]{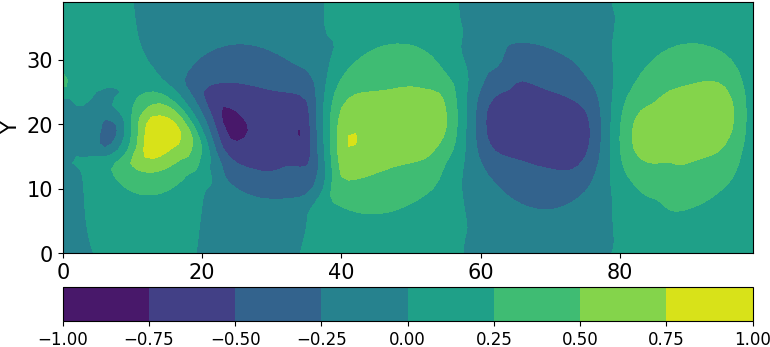}
    \end{minipage}
    \\[15pt]
   % Centered Label for (b) HOSVD and (b) SVD
    \makebox[\textwidth]{\textbf{(b) HOSVD} \hspace{5cm} \textbf{(b) SVD}} \\[5pt]

    % Row 1
    \begin{minipage}{0.48\textwidth}
        \centering
        \begin{tikzpicture}
            \node[anchor=south west, inner sep=0] (image) at (0,0) {\includegraphics[width=0.85\textwidth]{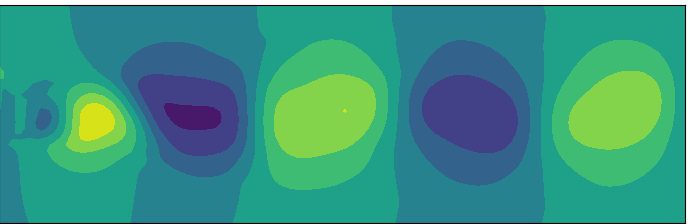}};
            \node[anchor=north, rotate=90, font=\scriptsize] at (-0.6,1) {Y};
            \node[anchor=north, font=\scriptsize] at (3.4,-0.2) {X};
        \end{tikzpicture}
    \end{minipage}
    \hfill
    \begin{minipage}{0.48\textwidth}
        \centering
        \begin{tikzpicture}
            \node[anchor=south west, inner sep=0] (image) at (0,0) {\includegraphics[width=0.85\textwidth]{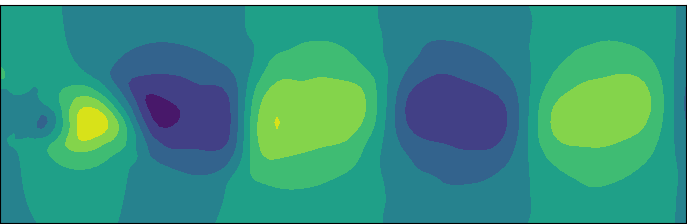}};
            \node[anchor=north, font=\scriptsize] at (3.4,-0.2) {X};
        \end{tikzpicture}
    \end{minipage}
    \\[10pt]

    % Row 2
    \begin{minipage}{0.48\textwidth}
        \centering
        \begin{tikzpicture}
            \node[anchor=south west, inner sep=0] (image) at (0,0) {\includegraphics[width=0.85\textwidth]{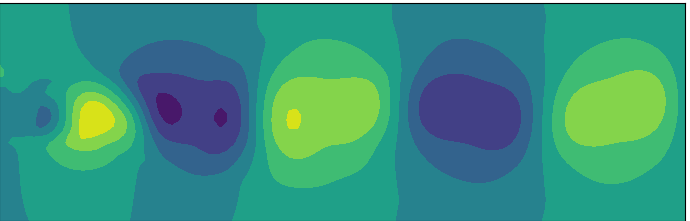}};
            \node[anchor=north, rotate=90, font=\scriptsize] at (-0.6,1) {Y};
            \node[anchor=north, font=\scriptsize] at (3.4,-0.2) {X};
        \end{tikzpicture}
    \end{minipage}
    \hfill
    \begin{minipage}{0.48\textwidth}
        \centering
        \begin{tikzpicture}
            \node[anchor=south west, inner sep=0] (image) at (0,0) {\includegraphics[width=0.85\textwidth]{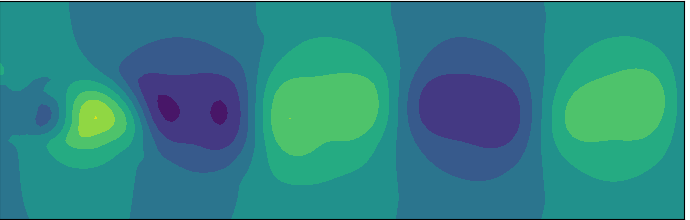}};
            \node[anchor=north, font=\scriptsize] at (3.4,-0.2) {X};
        \end{tikzpicture}
    \end{minipage}
    \\[10pt]

    % Row 3 
    \begin{minipage}{0.48\textwidth}
        \centering
        \begin{tikzpicture}
            \node[anchor=south west, inner sep=0] (image) at (0,0) {\includegraphics[width=0.85\textwidth]{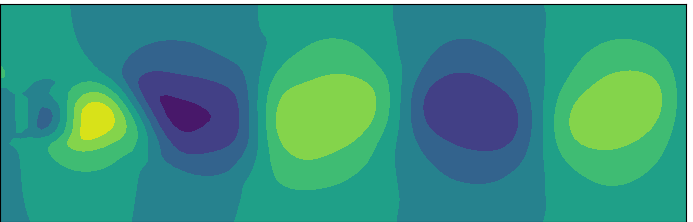}};
            \node[anchor=north, rotate=90, font=\scriptsize] at (-0.6,1) {Y};
            \node[anchor=north, font=\scriptsize] at (3.4,-0.2) {X};
        \end{tikzpicture}
    \end{minipage}
    \hfill
    \begin{minipage}{0.48\textwidth}
        \centering
        \begin{tikzpicture}
            \node[anchor=south west, inner sep=0] (image) at (0,0) {\includegraphics[width=0.85\textwidth]{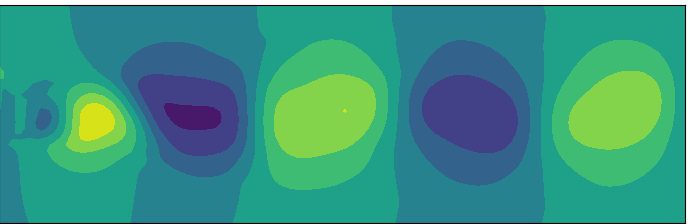}};
            \node[anchor=north, font=\scriptsize] at (3.4,-0.2) {X};
        \end{tikzpicture}
    \end{minipage}

    \caption{(a) Ground truth normal velocity components at \(t = 528\). 
    (b) Comparison of the predicted normal velocity components for HOSVD (left column) and SVD (right column) across different LSTM architectures. From top to bottom: predictions for LSTM with 1 Dense, 2 Dense, and Time-Distributed architectures.}
    \label{fig:lstm_3d_normal}
\end{figure}

% Spanwise Velocity Figure
\begin{figure}[h!]
    \centering
    \textbf{(a) Ground Truth} \\[5pt]
    \begin{minipage}{0.5\textwidth}
        \centering
        \includegraphics[width=\textwidth]{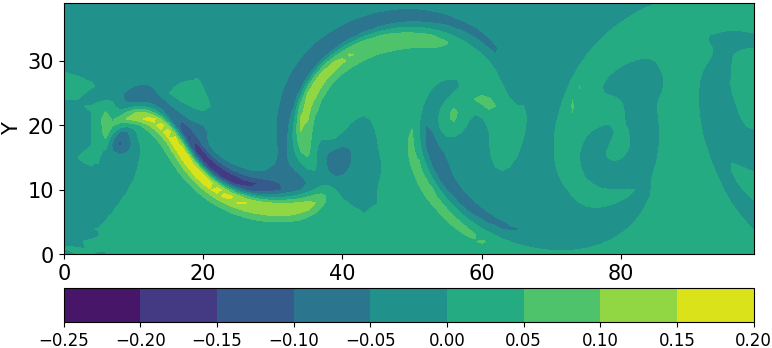}
    \end{minipage}
    \\[15pt]
   % Centered Label for (b) HOSVD and (b) SVD
    \makebox[\textwidth]{\textbf{(b) HOSVD} \hspace{5cm} \textbf{(b) SVD}} \\[5pt]

    % Row 1
    \begin{minipage}{0.48\textwidth}
        \centering
        \begin{tikzpicture}
            \node[anchor=south west, inner sep=0] (image) at (0,0) {\includegraphics[width=0.85\textwidth]{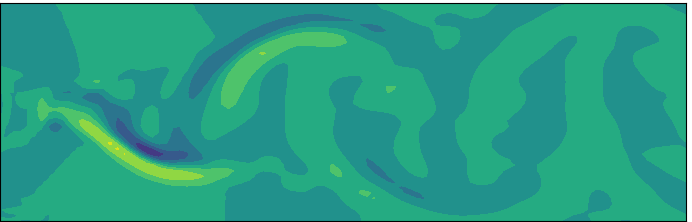}};
            \node[anchor=north, rotate=90, font=\scriptsize] at (-0.6,1) {Y};
            \node[anchor=north, font=\scriptsize] at (3.4,-0.2) {X};
        \end{tikzpicture}
    \end{minipage}
    \hfill
    \begin{minipage}{0.48\textwidth}
        \centering
        \begin{tikzpicture}
            \node[anchor=south west, inner sep=0] (image) at (0,0) {\includegraphics[width=0.85\textwidth]{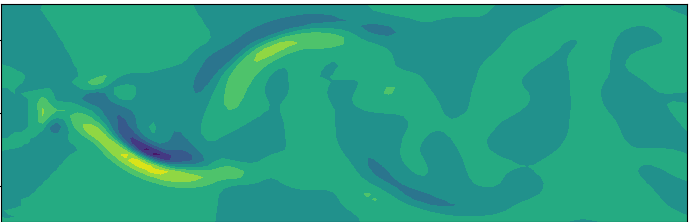}};
            \node[anchor=north, font=\scriptsize] at (3.4,-0.2) {X};
        \end{tikzpicture}
    \end{minipage}
    \\[10pt]

    % Row 2
    \begin{minipage}{0.48\textwidth}
        \centering
        \begin{tikzpicture}
            \node[anchor=south west, inner sep=0] (image) at (0,0) {\includegraphics[width=0.85\textwidth]{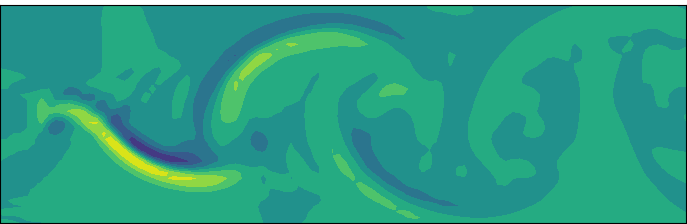}};
            \node[anchor=north, rotate=90, font=\scriptsize] at (-0.6,1) {Y};
            \node[anchor=north, font=\scriptsize] at (3.4,-0.2) {X};
        \end{tikzpicture}
    \end{minipage}
    \hfill
    \begin{minipage}{0.48\textwidth}
        \centering
        \begin{tikzpicture}
            \node[anchor=south west, inner sep=0] (image) at (0,0) {\includegraphics[width=0.85\textwidth]{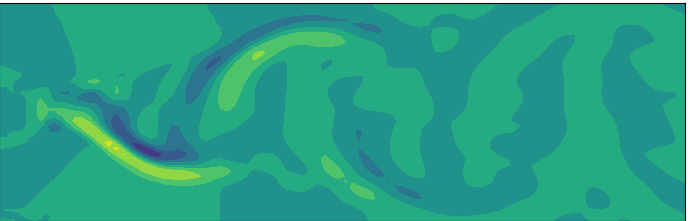}};
            \node[anchor=north, font=\scriptsize] at (3.4,-0.2) {X};
        \end{tikzpicture}
    \end{minipage}
    \\[10pt]

    % Row 3 
    \begin{minipage}{0.48\textwidth}
        \centering
        \begin{tikzpicture}
            \node[anchor=south west, inner sep=0] (image) at (0,0) {\includegraphics[width=0.85\textwidth]{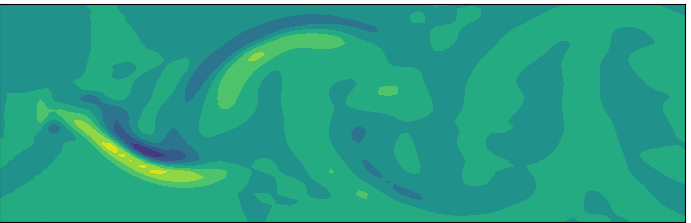}};
            \node[anchor=north, rotate=90, font=\scriptsize] at (-0.6,1) {Y};
            \node[anchor=north, font=\scriptsize] at (3.4,-0.2) {X};
        \end{tikzpicture}
    \end{minipage}
    \hfill
    \begin{minipage}{0.48\textwidth}
        \centering
        \begin{tikzpicture}
            \node[anchor=south west, inner sep=0] (image) at (0,0) {\includegraphics[width=0.84\textwidth]{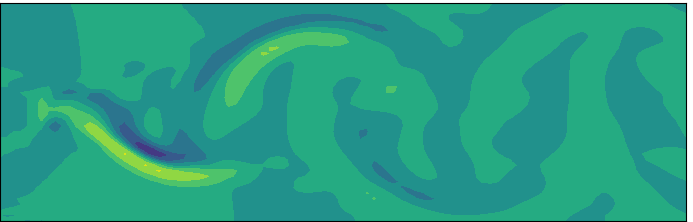}};
            \node[anchor=north, font=\scriptsize] at (3.4,-0.2) {X};
        \end{tikzpicture}
    \end{minipage}

    \caption{(a) Ground truth spanwise velocity components at \(t = 528\). 
    (b) Comparison of the predicted spanwise velocity components for HOSVD (left column) and SVD (right column) across different LSTM architectures. From top to bottom: predictions for LSTM with 1 Dense, 2 Dense, and Time-Distributed architectures.}
    \label{fig:lstm_3d_spanwise}
\end{figure}

Figure~\ref{fig:lstm_3d_streamwise} compares the predicted streamwise velocity components for the 3D cylinder flow, showing that all models effectively capture the main flow structures. HOSVD provides smoother, more coherent predictions, while SVD remains accurate it shows slight inconsistencies. For the normal velocity component, HOSVD and SVD consistently reproduce flow structures with minimal artifacts. 
The spanwise velocity component presents the greatest challenge due to the dominance of zero and near-zero low-velocity regions. SVD and HOSVD struggle to predict the flow features accurately. The 1 dense SVD-based model performs the worst, failing to adequately capture spanwise velocity structures, leading to significant deviations and loss of critical flow features. However, the LSTM 2 dense and the time-distributed SVD model show improved stability, capturing the overall flow dynamics better than the 1 dense architectures. The HOSVD-based LSTM 2 dense architecture performs the best in this scenario, underscoring the advantages of HOSVD in providing a more faithful representation of the flow field. The spanwise velocity results highlight the complexities of accurately capturing the intricate flow physics in regions of near-zero velocities.

The RRMSE values for the 3D cylinder case are summarized in Tables~\ref{tab:rrmse_results_3d_streamwise}, \ref{tab:rrmse_results_3d_normal}, and \ref{tab:rrmse_results_3d_spanwise}. Across all LSTM architectures, HOSVD consistently outperforms SVD, demonstrating its superior ability to capture the underlying flow dynamics with reduced error. Notably, for the spanwise components where HOSVD maintains a significant advantage over SVD by over 8\%. Error values are particularly higher for the spanwise velocity component, where the predominance of very low and near zero velocities in most of the domain leads to extremely high relative errors. In such a case, RRMSE is not a good way to evaluate the model; rather, the focus should be on the UQ curves presented below. While differences between LSTM architectures remain small, the 2 dense model achieves the best performance for the HOSVD approach, followed by the time-distributed architecture. Despite the advantages of HOSVD, achieving high accuracy in this component remains a challenge due to the inherent complexity of three-dimensional flow dynamics.

\begin{table}[h!]
\centering
\begin{tabular}{|c|c|c|}
\hline
\textbf{Architecture} & \textbf{HOSVD (\%)} & \textbf{SVD (\%)} \\ \hline
LSTM 1 Dense          & 3.1                  & 4               \\ \hline
LSTM 2 Dense          & 1.5                  & 3.1               \\ \hline
LSTM Time-Distributed  & 1.6                  & 2.3              \\ \hline
\end{tabular}
\caption{RRMSE values for HOSVD and SVD across LSTM architectures for the streamwise velocity component of the 3D cylinder flow.}
\label{tab:rrmse_results_3d_streamwise}
\end{table}

\begin{table}[h!]
\centering
\begin{tabular}{|c|c|c|}
\hline
\textbf{Architecture} & \textbf{HOSVD (\%)} & \textbf{SVD (\%)} \\ \hline
LSTM 1 Dense          & 11                & 16.3              \\ \hline
LSTM 2 Dense          & 6.8                 & 15.4              \\ \hline
LSTM Time-Distributed  & 7.2                & 10.3              \\ \hline
\end{tabular}
\caption{RRMSE values for HOSVD and SVD across LSTM architectures for the normal velocity component of the 3D cylinder flow.}
\label{tab:rrmse_results_3d_normal}
\end{table}

\begin{table}[h!]
\centering
\begin{tabular}{|c|c|c|}
\hline
\textbf{Architecture} & \textbf{HOSVD (\%)} & \textbf{SVD (\%)} \\ \hline
LSTM 1 Dense          & 42.8                 & 54.1              \\ \hline
LSTM 2 Dense          & 39.4                 & 51             \\ \hline
LSTM Time-Distributed  & 39.6                 & 47.2            \\ \hline
\end{tabular}
\caption{RRMSE values for HOSVD and SVD across LSTM architectures for the spanwise velocity component of the 3D cylinder flow.}
\label{tab:rrmse_results_3d_spanwise}
\end{table}

% Streamwise Velocity
\begin{figure}[h!]
    \centering
    
   \makebox[\textwidth]{\textbf{(a) HOSVD} \hspace{5cm} \textbf{(b) SVD}} \\[5pt]
    % Row 1
    \begin{minipage}{0.48\textwidth}
        \centering
        \begin{tikzpicture}
            \node[anchor=south west, inner sep=0] (image) at (0,0) {\includegraphics[width=1.9\textwidth, height=0.6\textwidth, keepaspectratio]{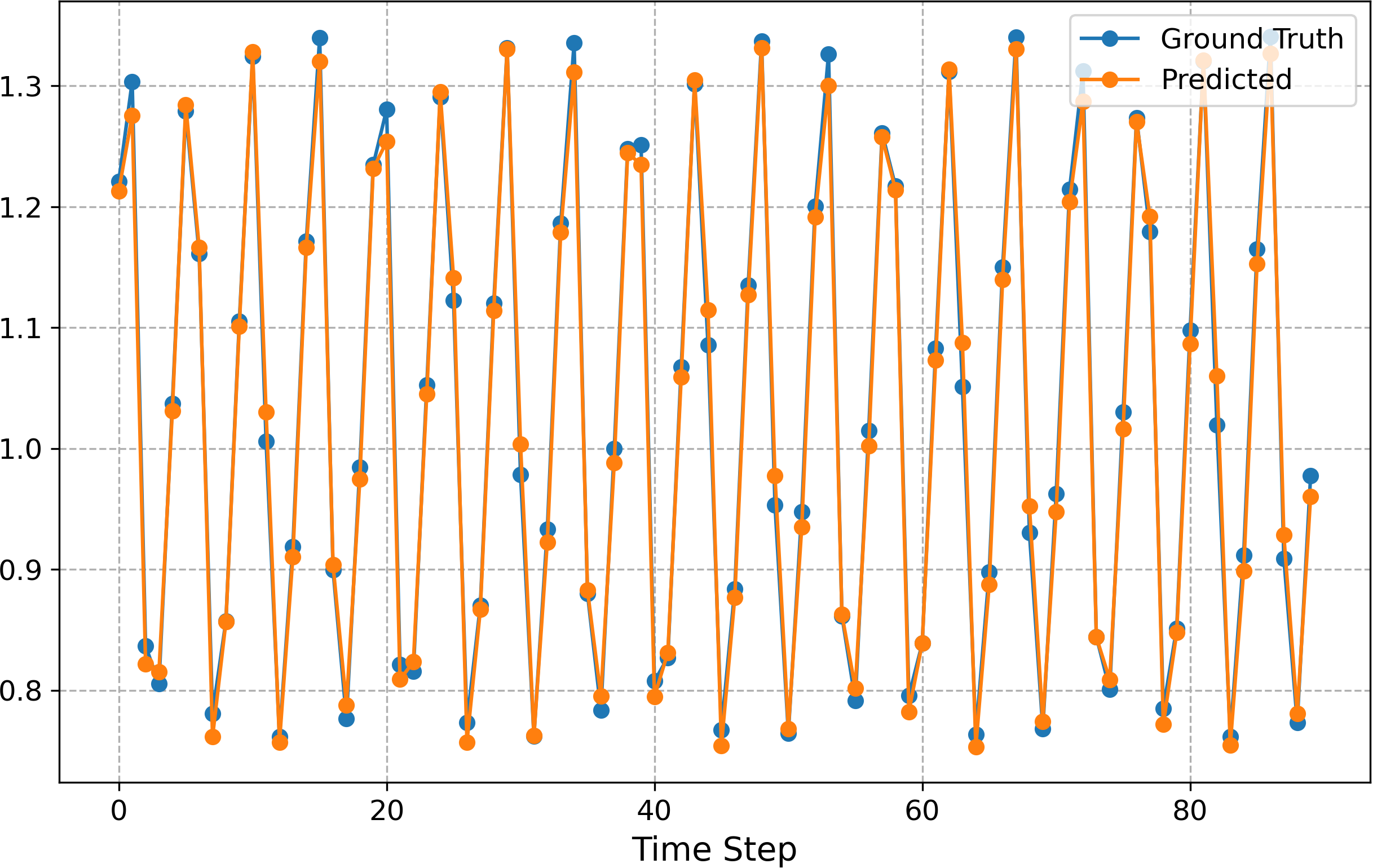}};
            \node[anchor=north, rotate=90, font=\footnotesize\bfseries] at (-0.5,2.5) {u};
        \end{tikzpicture}
    \end{minipage}
    \hfill
    \begin{minipage}{0.48\textwidth}
        \centering
        \includegraphics[width=1.9\textwidth, height=0.6\textwidth, keepaspectratio]{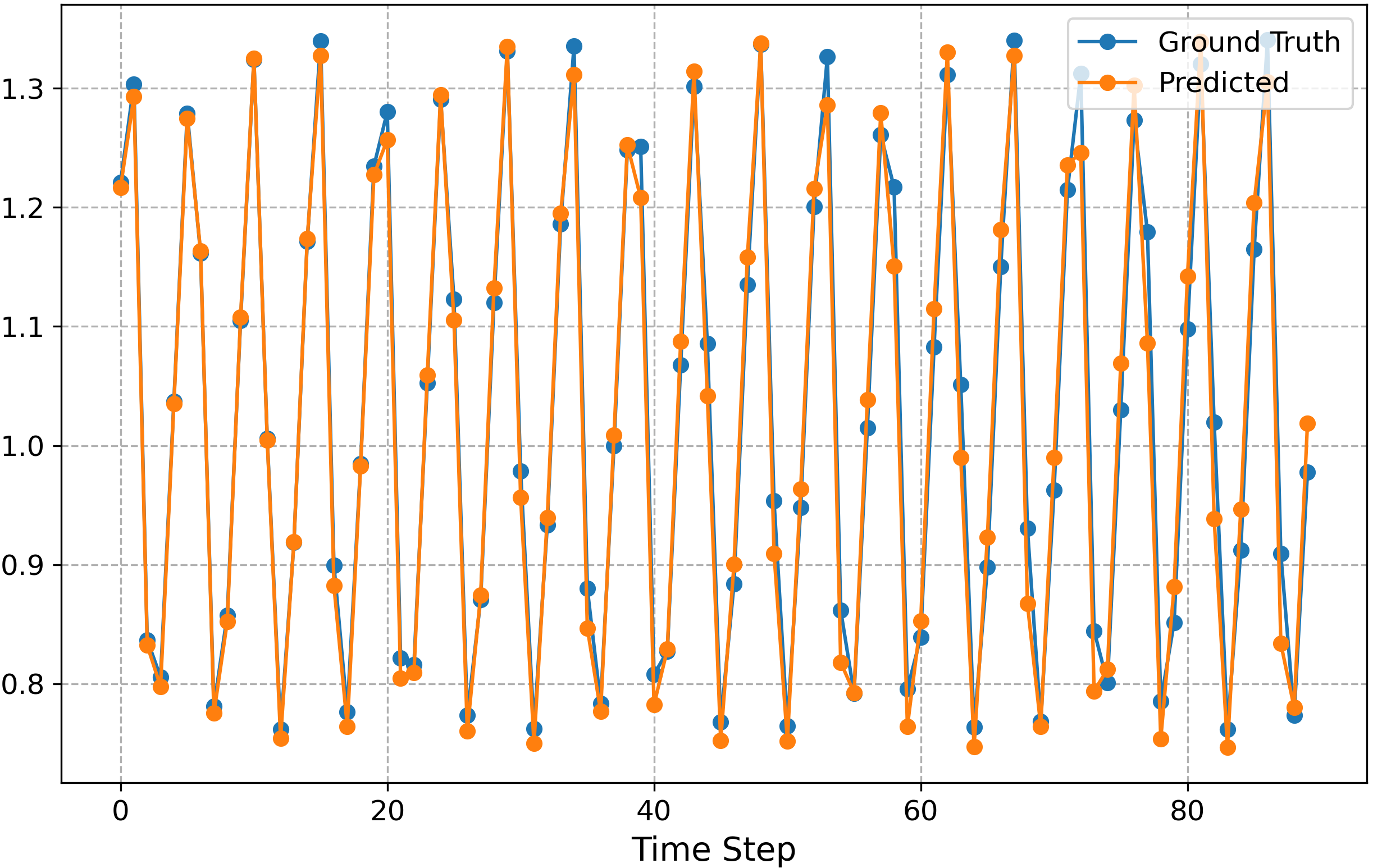}
    \end{minipage}
    \\[15pt]
    % Row 2
    \begin{minipage}{0.48\textwidth}
        \centering
        \begin{tikzpicture}
            \node[anchor=south west, inner sep=0] (image) at (0,0) {\includegraphics[width=1.9\textwidth, height=0.6\textwidth, keepaspectratio]{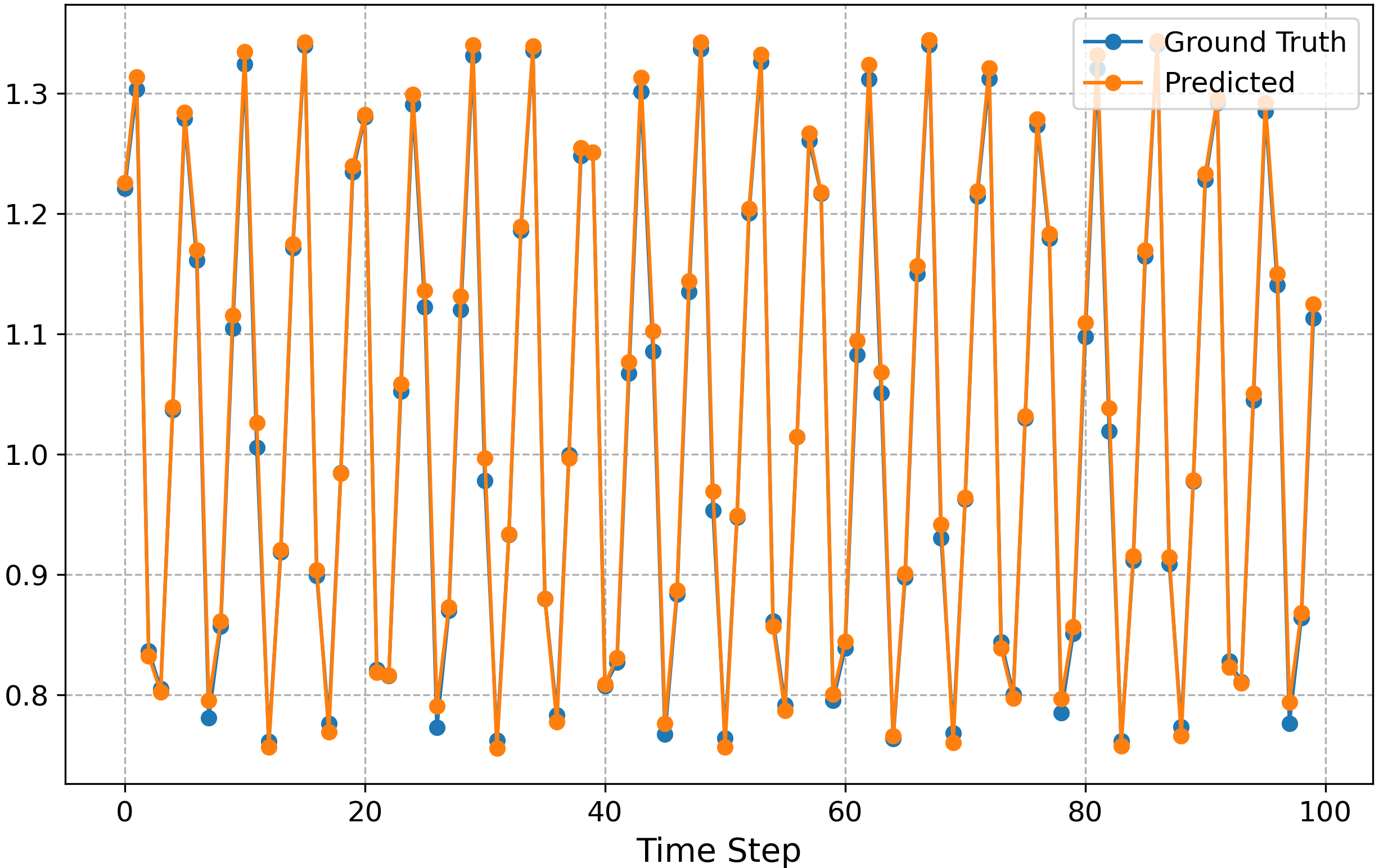}};
            \node[anchor=north, rotate=90, font=\footnotesize\bfseries] at (-0.5,2.5) {u};
        \end{tikzpicture}
    \end{minipage}
    \hfill
    \begin{minipage}{0.48\textwidth}
        \centering
        \includegraphics[width=1.9\textwidth, height=0.6\textwidth, keepaspectratio]{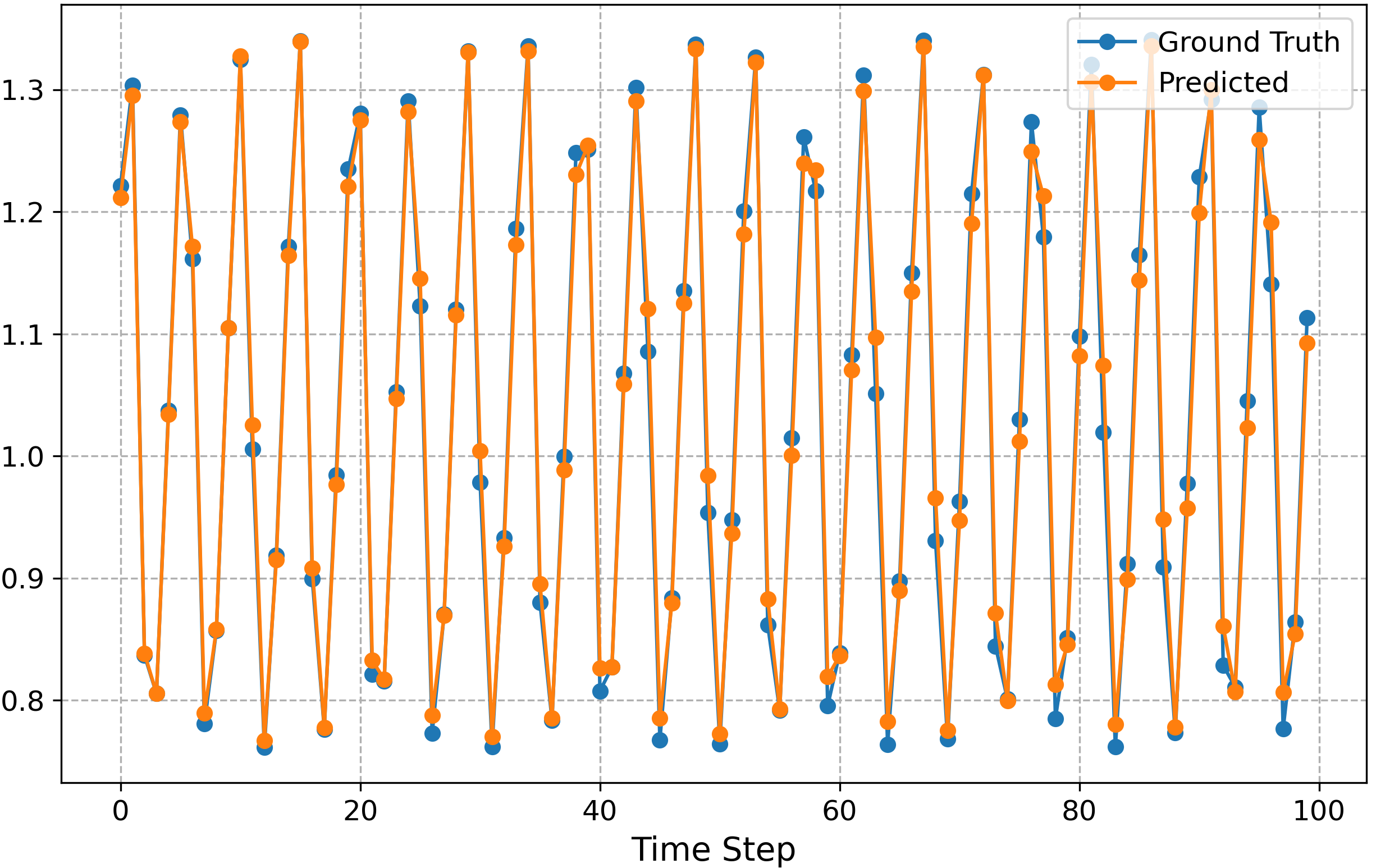}
    \end{minipage}
    \\[15pt]
    % Row 3
    \begin{minipage}{0.48\textwidth}
        \centering
        \begin{tikzpicture}
            \node[anchor=south west, inner sep=0] (image) at (0,0) {\includegraphics[width=1.9\textwidth, height=0.6\textwidth, keepaspectratio]{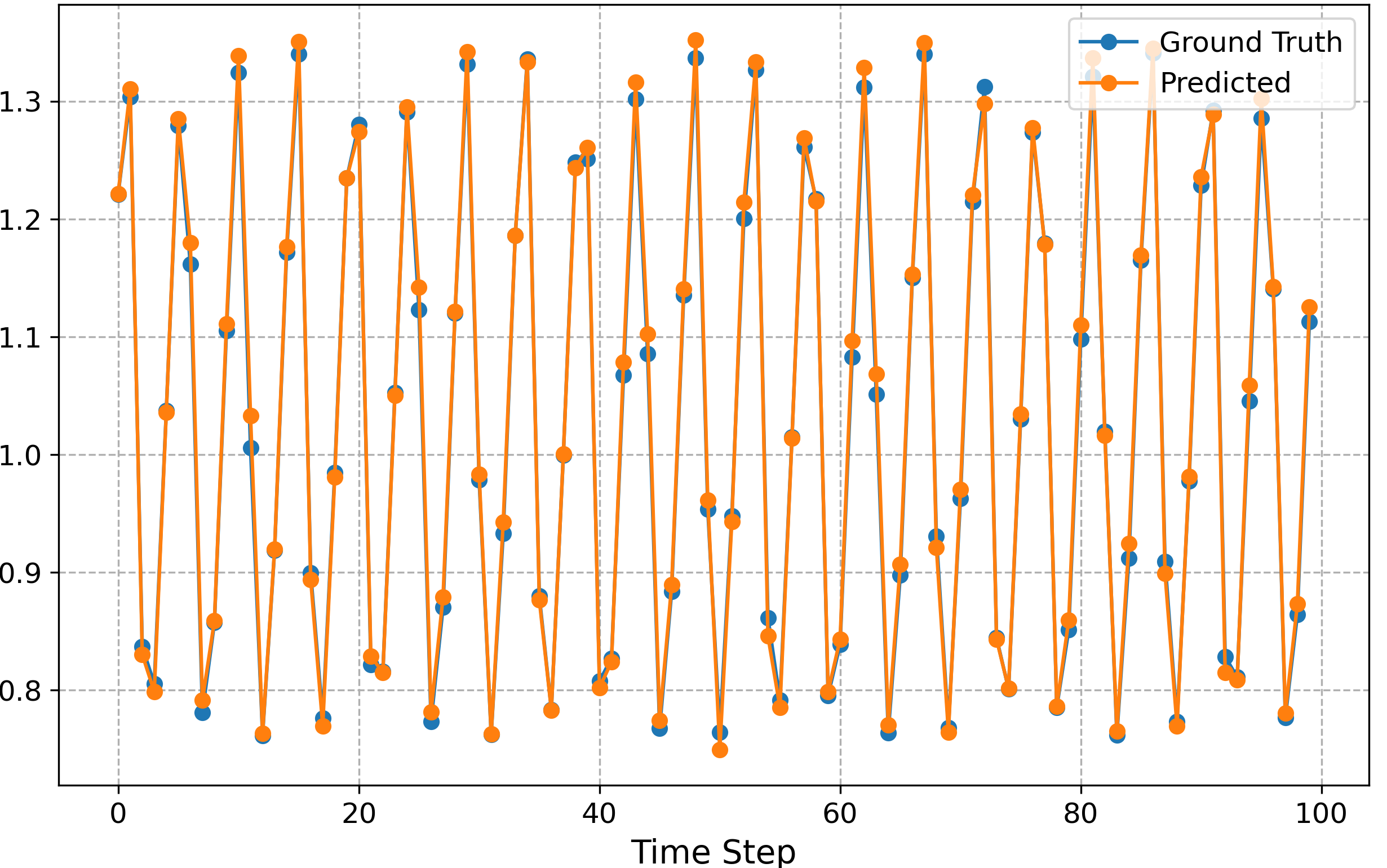}};
            \node[anchor=north, rotate=90, font=\footnotesize\bfseries] at (-0.5,2.5) {u};
        \end{tikzpicture}
    \end{minipage}
    \hfill
    \begin{minipage}{0.48\textwidth}
        \centering
        \includegraphics[width=1.9\textwidth, height=0.6\textwidth, keepaspectratio]{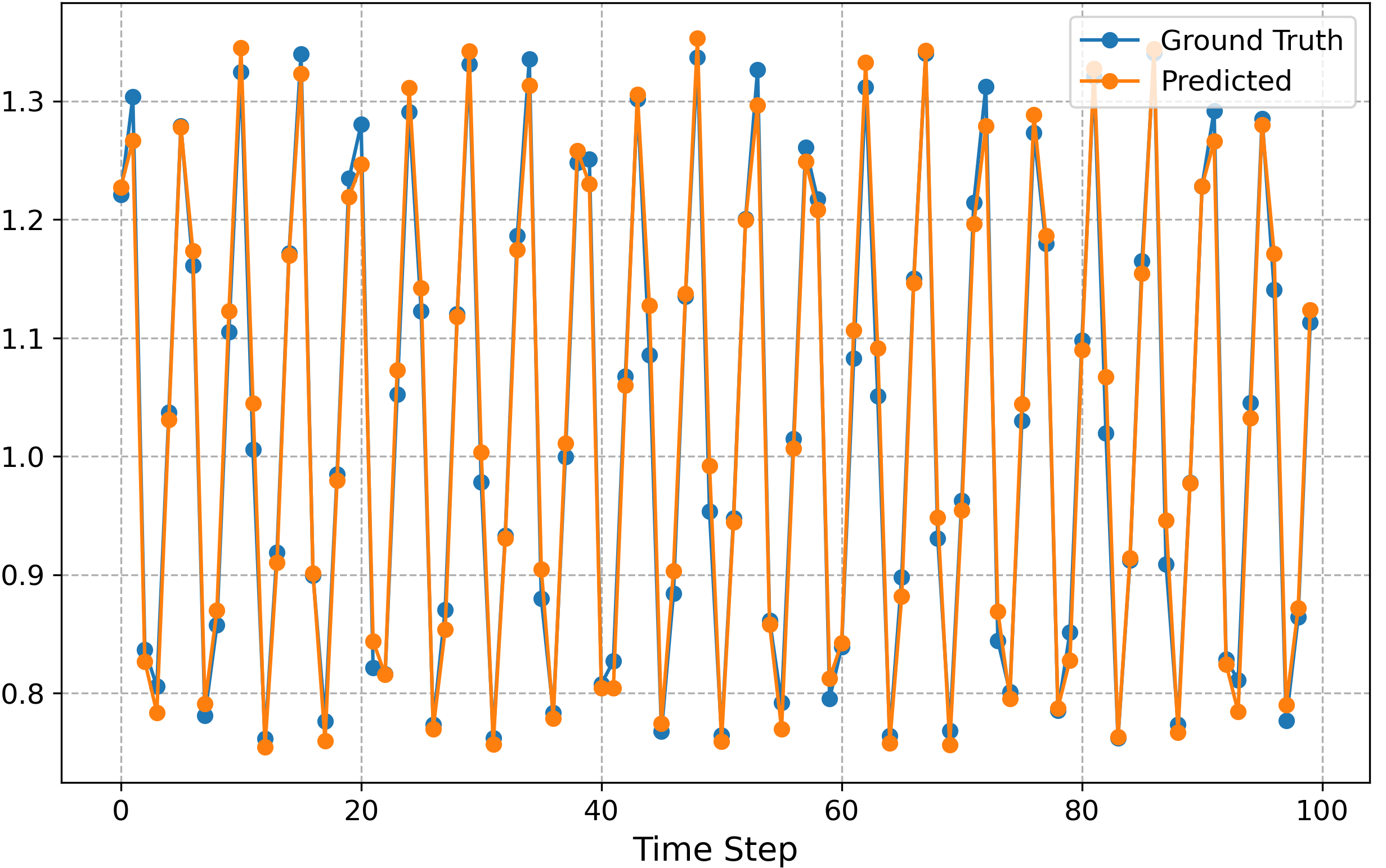}
    \end{minipage}
    \caption{ Comparison of the predicted temporal evolution of streamwise velocity components for (a) HOSVD (left column) and (b) SVD (right column) across different LSTM architectures for the 3D cylinder flow. From top to bottom: LSTM 1 Dense, LSTM 2 Dense, and LSTM Time-Distributed.}
    \label{fig:streamwise_evolution_3d}
\end{figure}

% Normal Velocity
\begin{figure}[h!]
    \centering
    
   \makebox[\textwidth]{\textbf{(a) HOSVD} \hspace{5cm} \textbf{(b) SVD}} \\[5pt]
    % Row 1
    \begin{minipage}{0.48\textwidth}
        \centering
        \begin{tikzpicture}
            \node[anchor=south west, inner sep=0] (image) at (0,0) {\includegraphics[width=1.9\textwidth, height=0.6\textwidth, keepaspectratio]{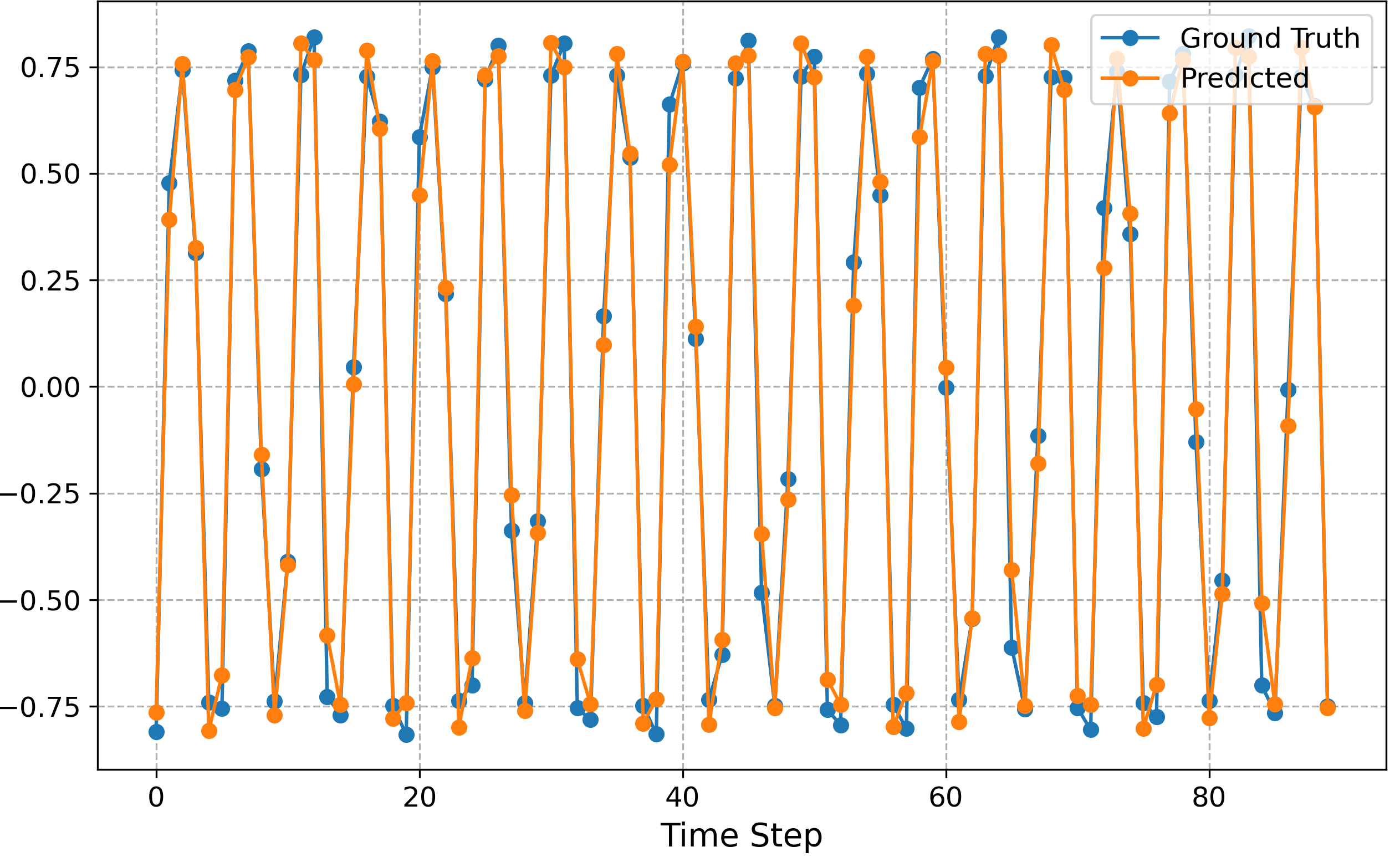}};
            \node[anchor=north, rotate=90, font=\footnotesize\bfseries] at (-0.5,2.5) {v};
        \end{tikzpicture}
    \end{minipage}
    \hfill
    \begin{minipage}{0.48\textwidth}
        \centering
        \includegraphics[width=1.9\textwidth, height=0.6\textwidth, keepaspectratio]{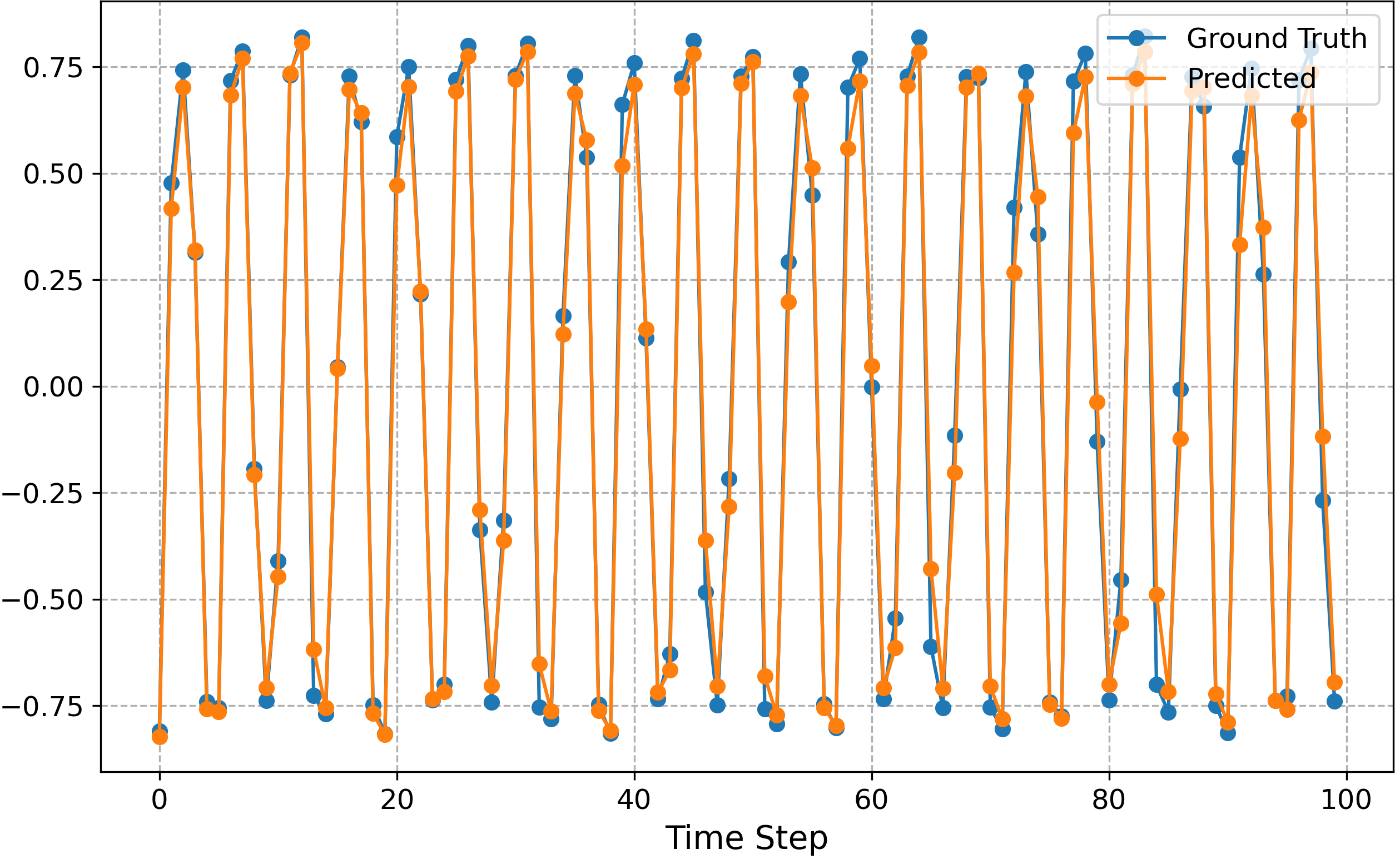}
    \end{minipage}
    \\[15pt]
    % Row 2
    \begin{minipage}{0.48\textwidth}
        \centering
        \begin{tikzpicture}
            \node[anchor=south west, inner sep=0] (image) at (0,0) {\includegraphics[width=1.9\textwidth, height=0.6\textwidth, keepaspectratio]{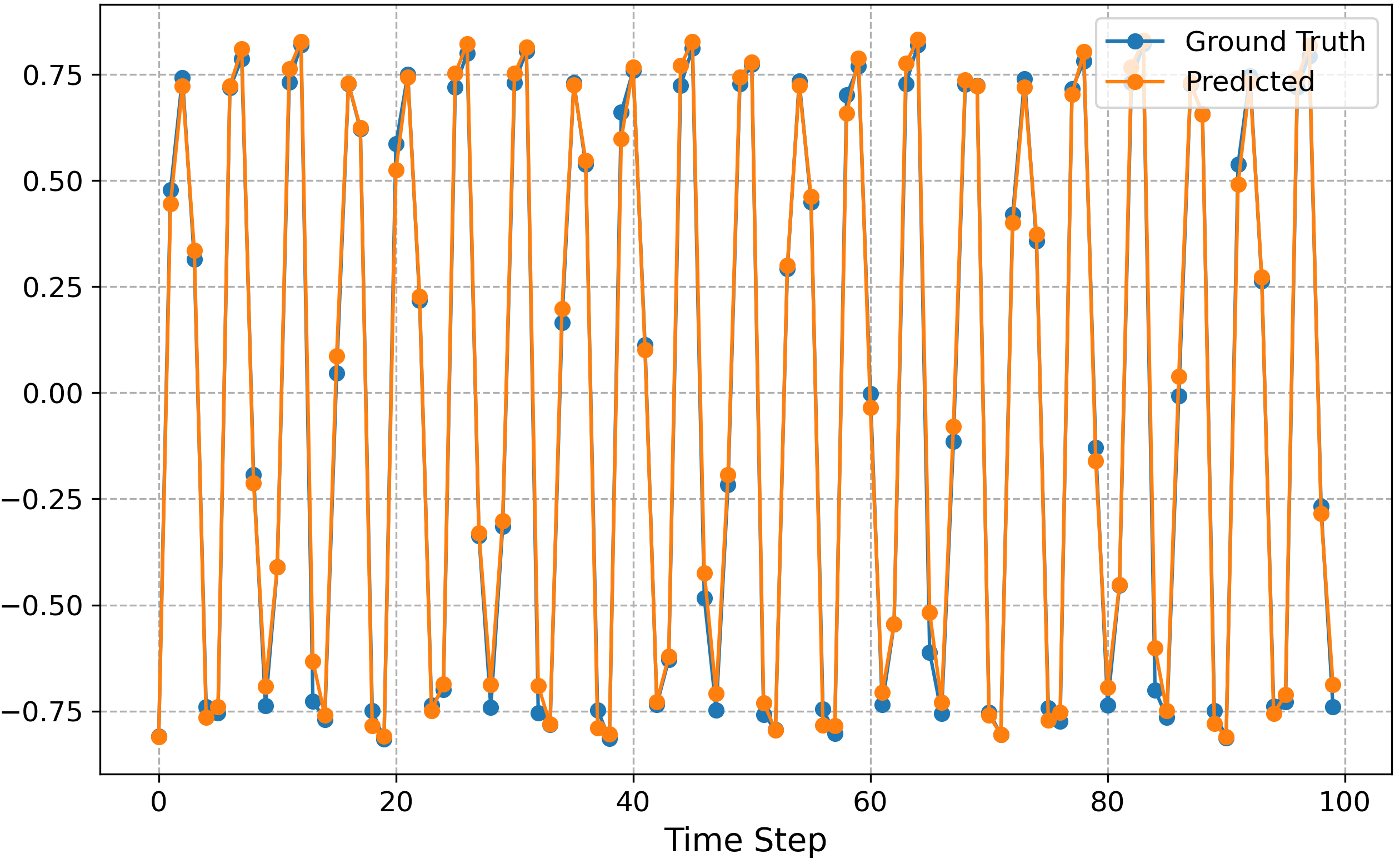}};
            \node[anchor=north, rotate=90, font=\footnotesize\bfseries] at (-0.5,2.5) {v};
        \end{tikzpicture}
    \end{minipage}
    \hfill
    \begin{minipage}{0.48\textwidth}
        \centering
        \includegraphics[width=1.9\textwidth, height=0.6\textwidth, keepaspectratio]{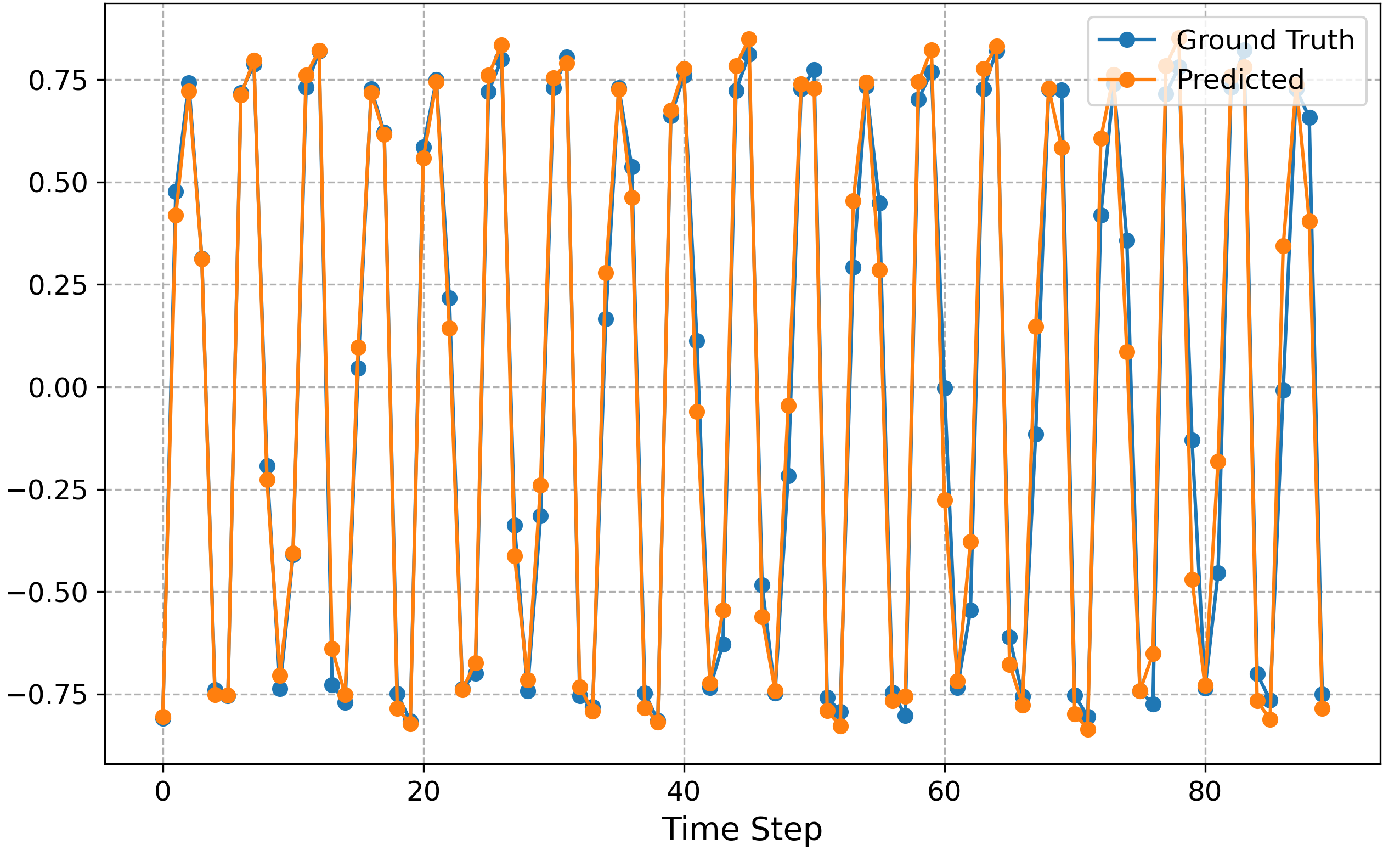}
    \end{minipage}
    \\[15pt]
    % Row 3
    \begin{minipage}{0.48\textwidth}
        \centering
        \begin{tikzpicture}
            \node[anchor=south west, inner sep=0] (image) at (0,0) {\includegraphics[width=1.9\textwidth, height=0.6\textwidth, keepaspectratio]{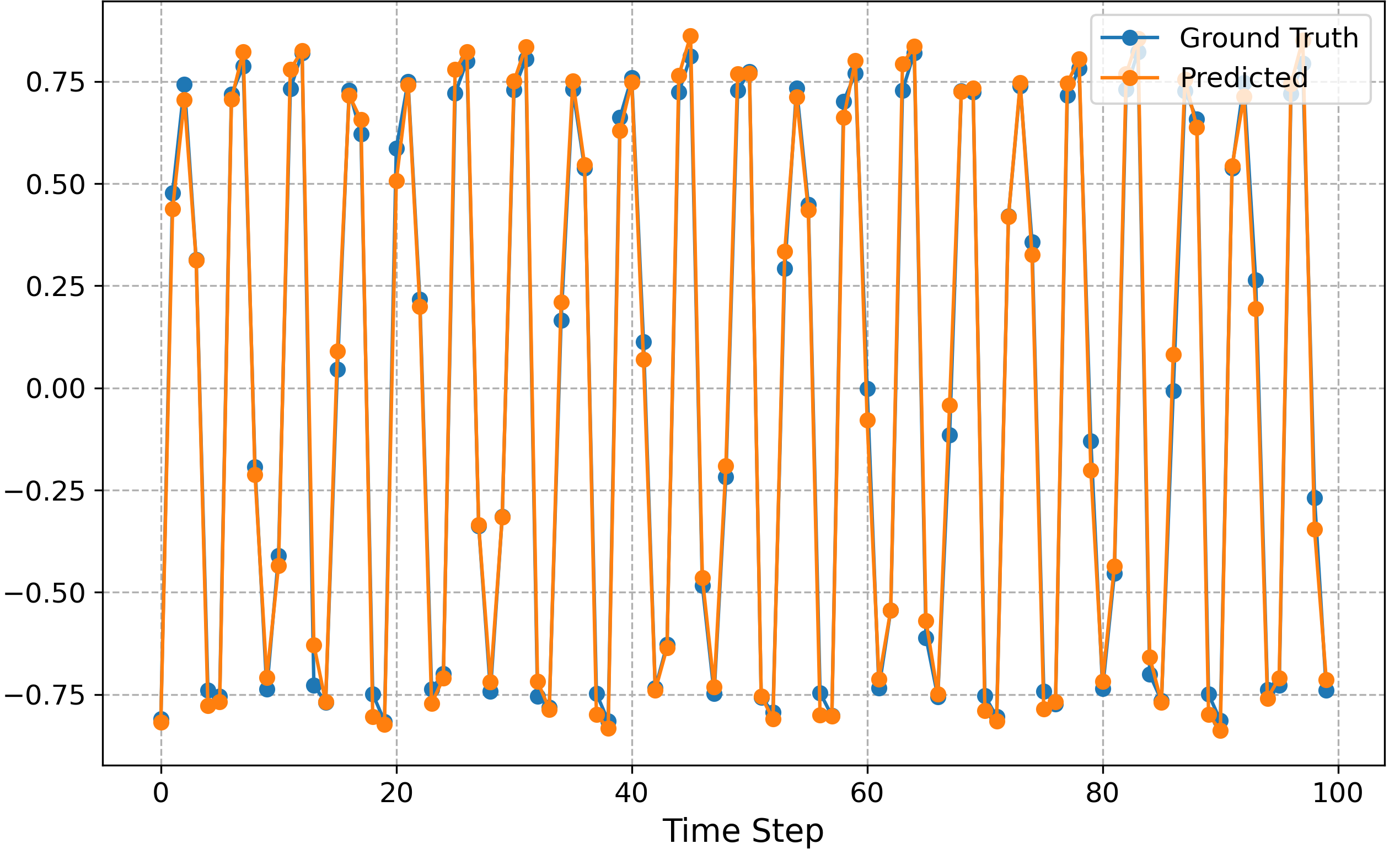}};
            \node[anchor=north, rotate=90, font=\footnotesize\bfseries] at (-0.5,2.5) {v};
        \end{tikzpicture}
    \end{minipage}
    \hfill
    \begin{minipage}{0.48\textwidth}
        \centering
        \includegraphics[width=1.9\textwidth, height=0.6\textwidth, keepaspectratio]{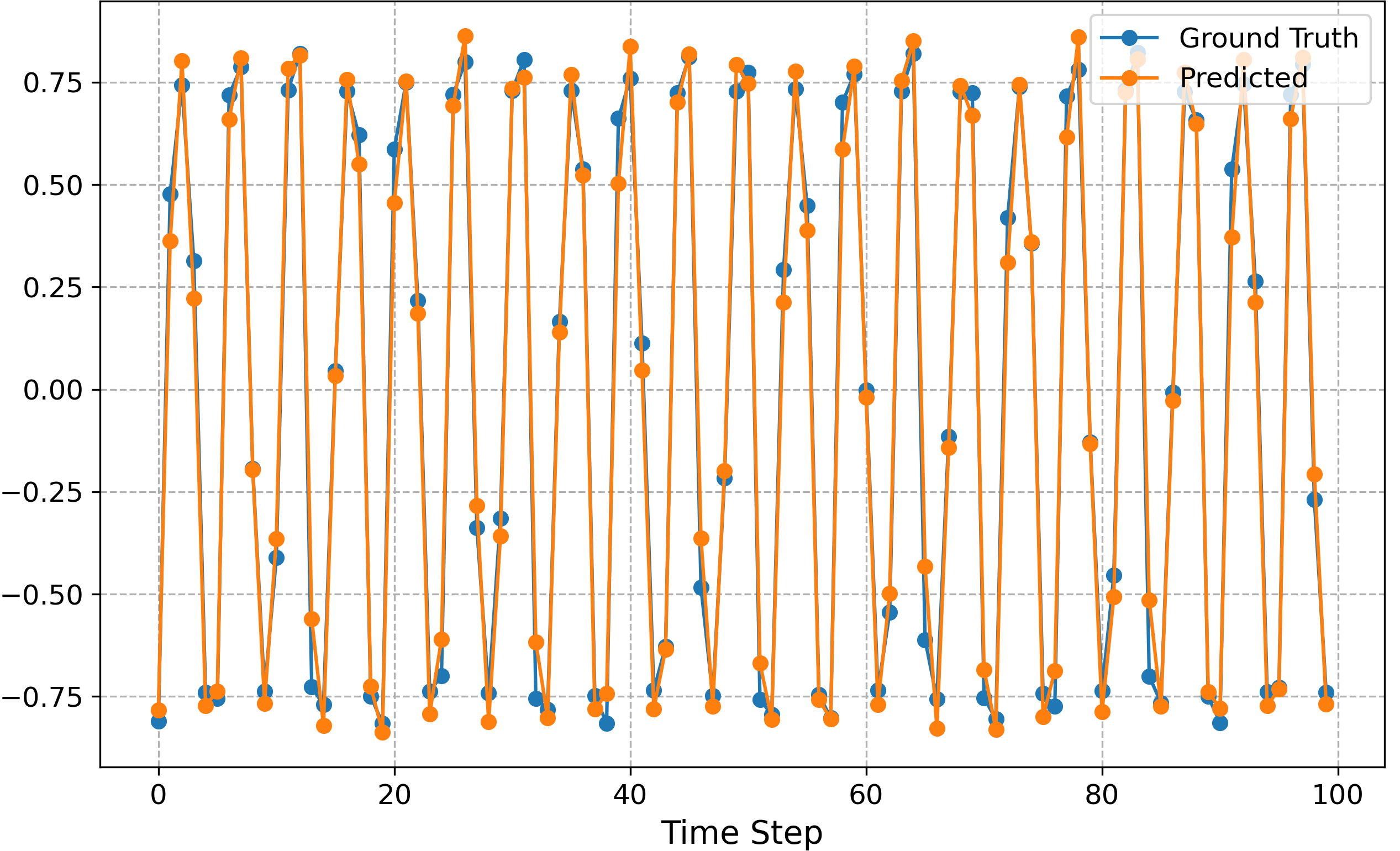}
    \end{minipage}
    \caption{ Comparison of the predicted temporal evolution of normal velocity components for (a) HOSVD (left column) and (b) SVD (right column) across different LSTM architectures for the 3D cylinder flow.From top to bottom: LSTM 1 Dense, LSTM 2 Dense, and LSTM Time-Distributed.}
    \label{fig:normal_velocity_evolution_3d}
\end{figure}

  % Spanwise Velocity
\begin{figure}[h!]
    \centering
    
    \makebox[\textwidth]{\textbf{(a) HOSVD} \hspace{5cm} \textbf{(b) SVD}} \\[5pt]

    % Row 1
    \begin{minipage}{0.48\textwidth}
        \centering
        \begin{tikzpicture}
            \node[anchor=south west, inner sep=0] (image) at (0,0) 
            {\includegraphics[width=1.9\textwidth, height=0.6\textwidth, keepaspectratio]{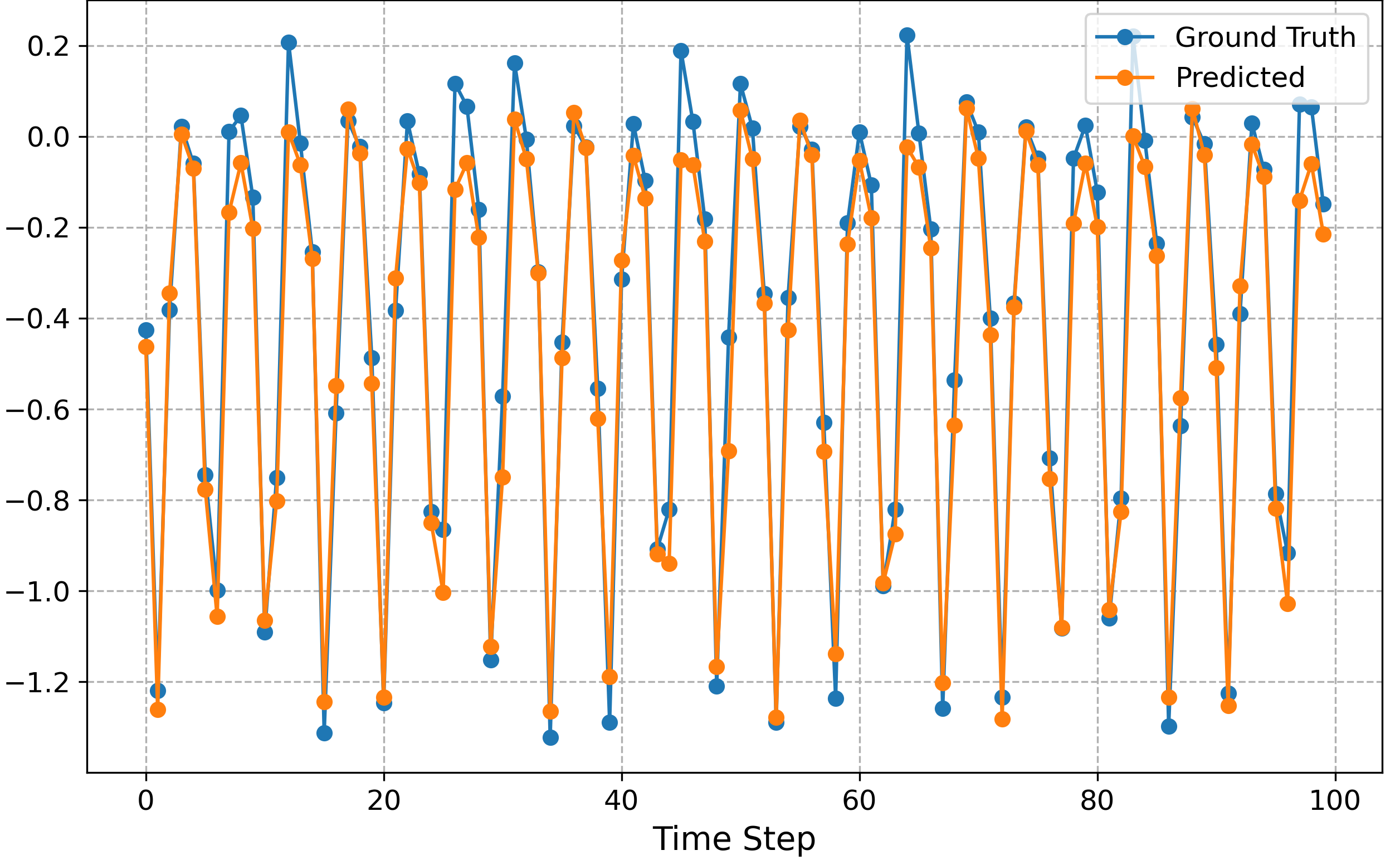}};
            \node[anchor=north, rotate=90, font=\footnotesize\bfseries] at (-0.57,2.5) {\(\mathbf{w} \times 10^{-6}\)};
        \end{tikzpicture}
    \end{minipage}
    \hfill
    \begin{minipage}{0.48\textwidth}
        \centering
        \includegraphics[width=1.9\textwidth, height=0.6\textwidth, keepaspectratio]{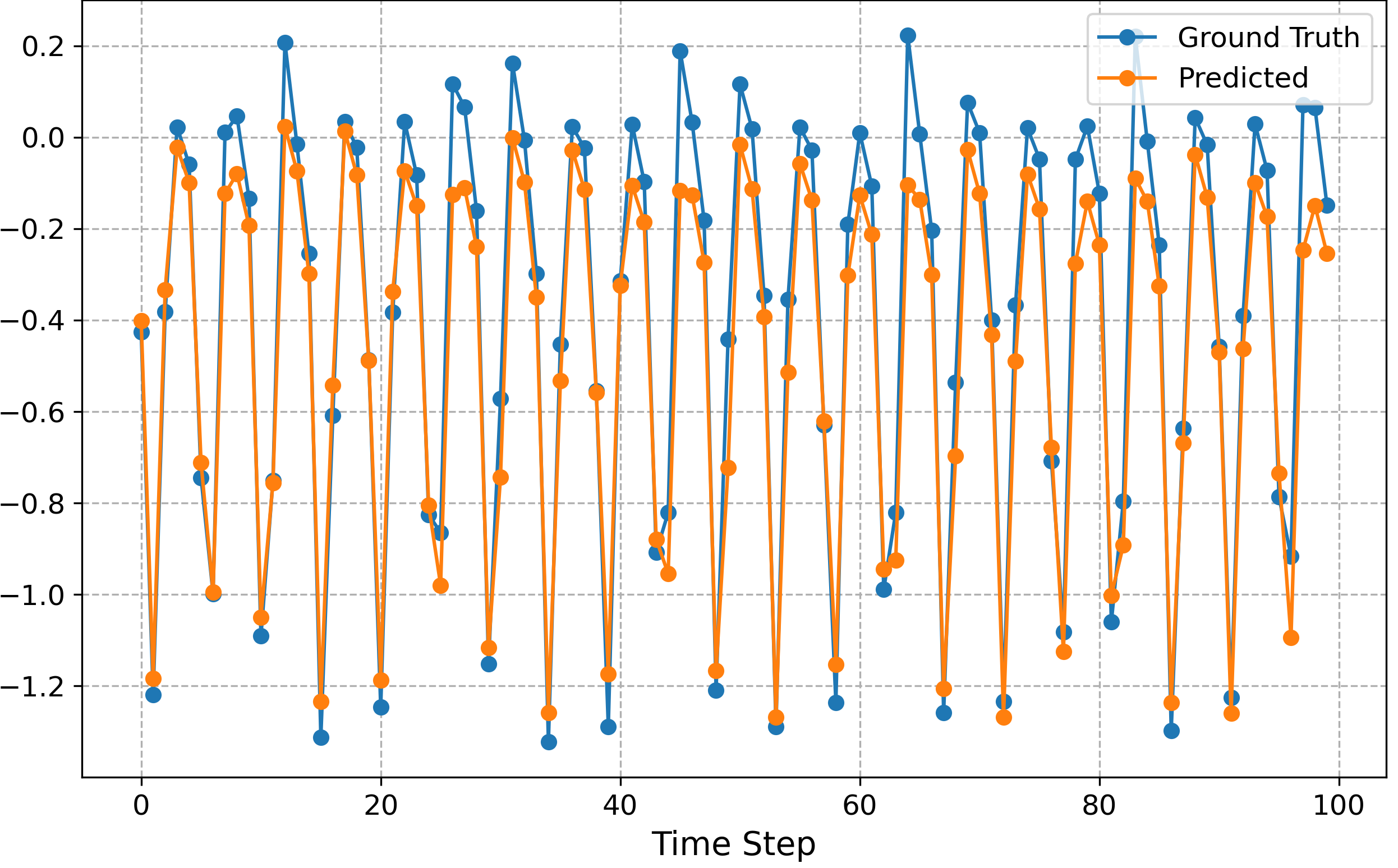}
    \end{minipage}
    \\[15pt]

    % Row 2
    \begin{minipage}{0.48\textwidth}
        \centering
        \begin{tikzpicture}
            \node[anchor=south west, inner sep=0] (image) at (0,0) 
            {\includegraphics[width=1.9\textwidth, height=0.6\textwidth, keepaspectratio]{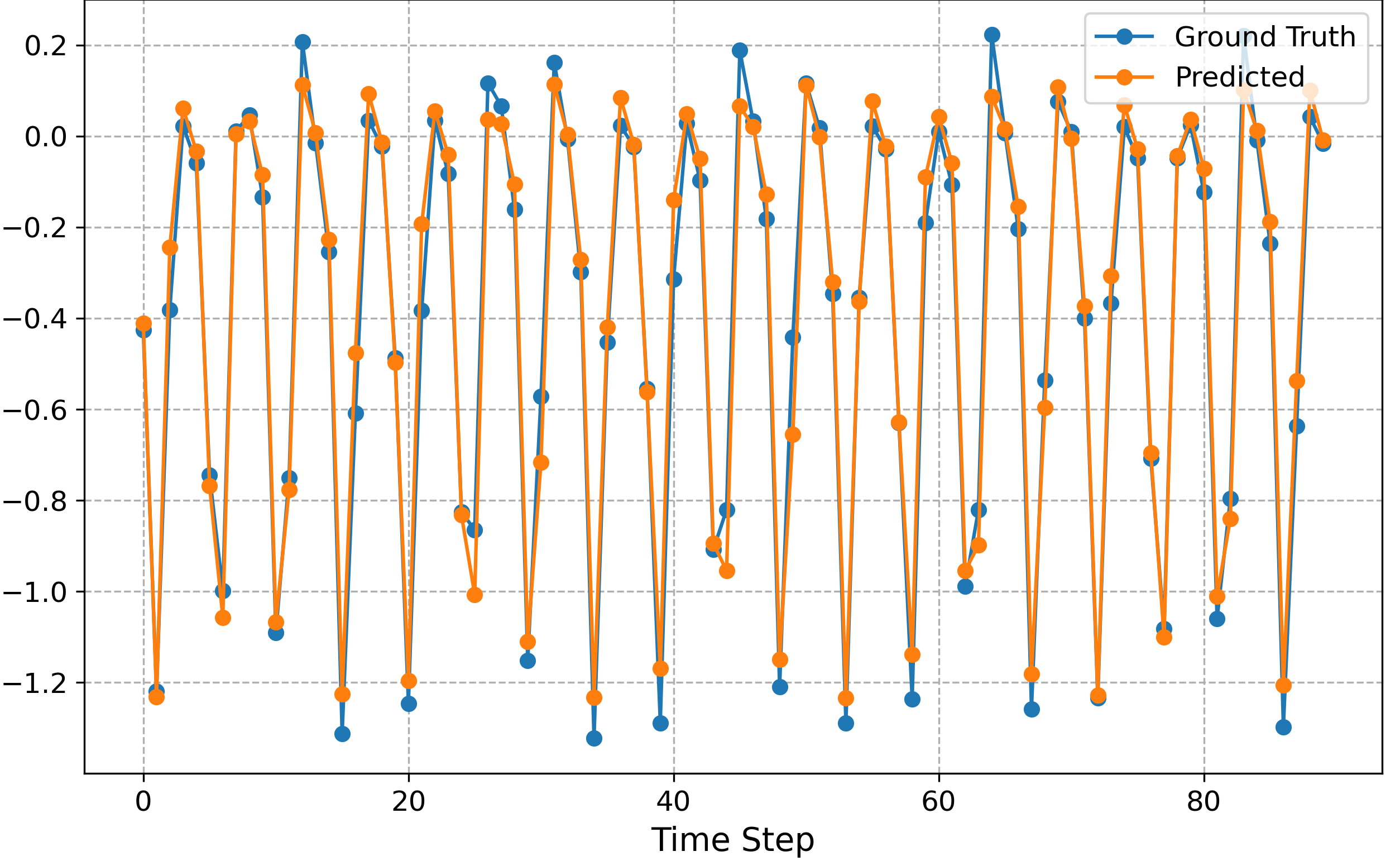}};
            \node[anchor=north, rotate=90, font=\footnotesize\bfseries] at (-0.57,2.5) {\(\mathbf{w} \times 10^{-6}\)};
        \end{tikzpicture}
    \end{minipage}
    \hfill
    \begin{minipage}{0.48\textwidth}
        \centering
        \includegraphics[width=1.9\textwidth, height=0.6\textwidth, keepaspectratio]{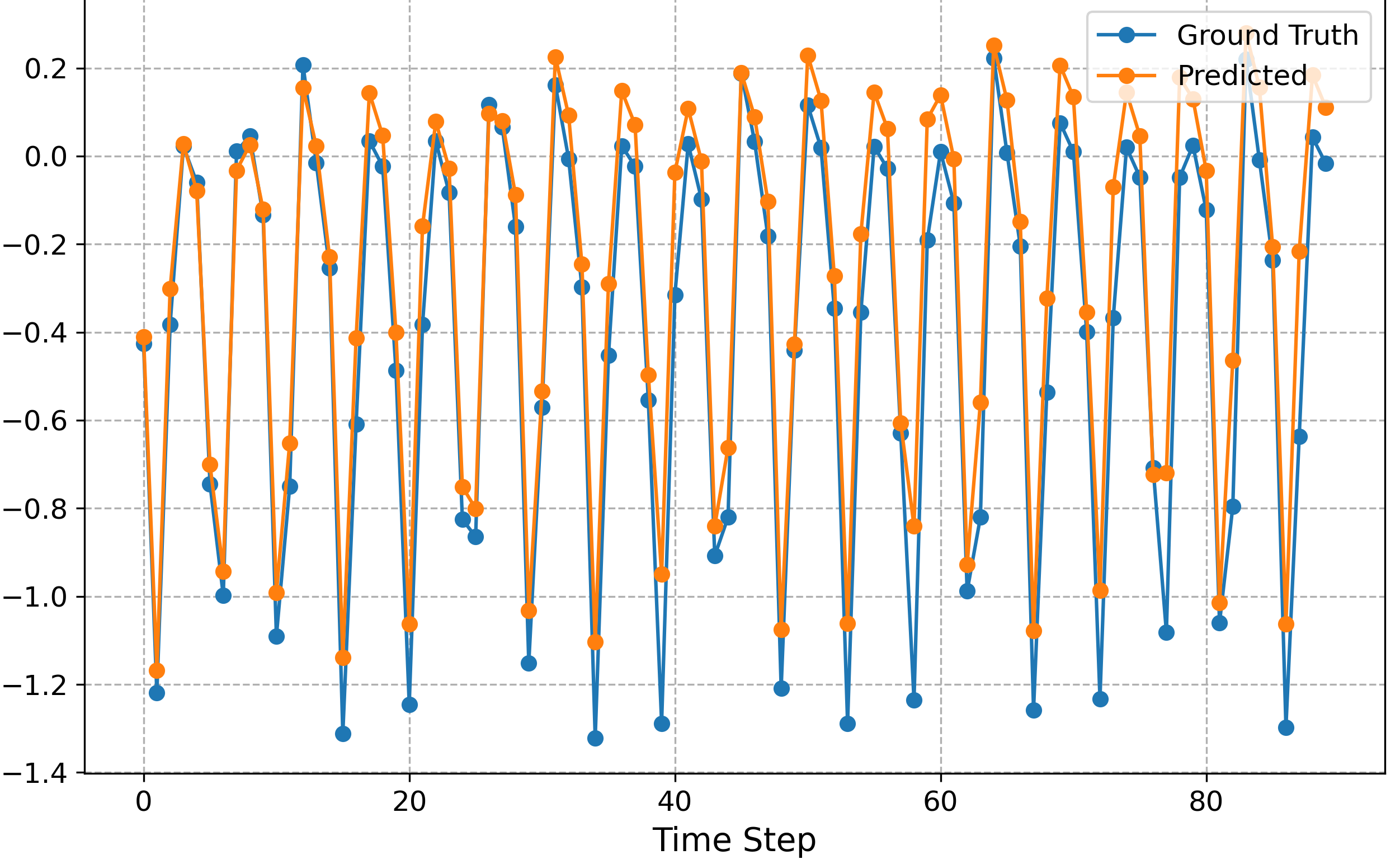}
    \end{minipage}
    \\[15pt]

    % Row 3
    \begin{minipage}{0.48\textwidth}
        \centering
        \begin{tikzpicture}
            \node[anchor=south west, inner sep=0] (image) at (0,0) 
            {\includegraphics[width=1.9\textwidth, height=0.6\textwidth, keepaspectratio]{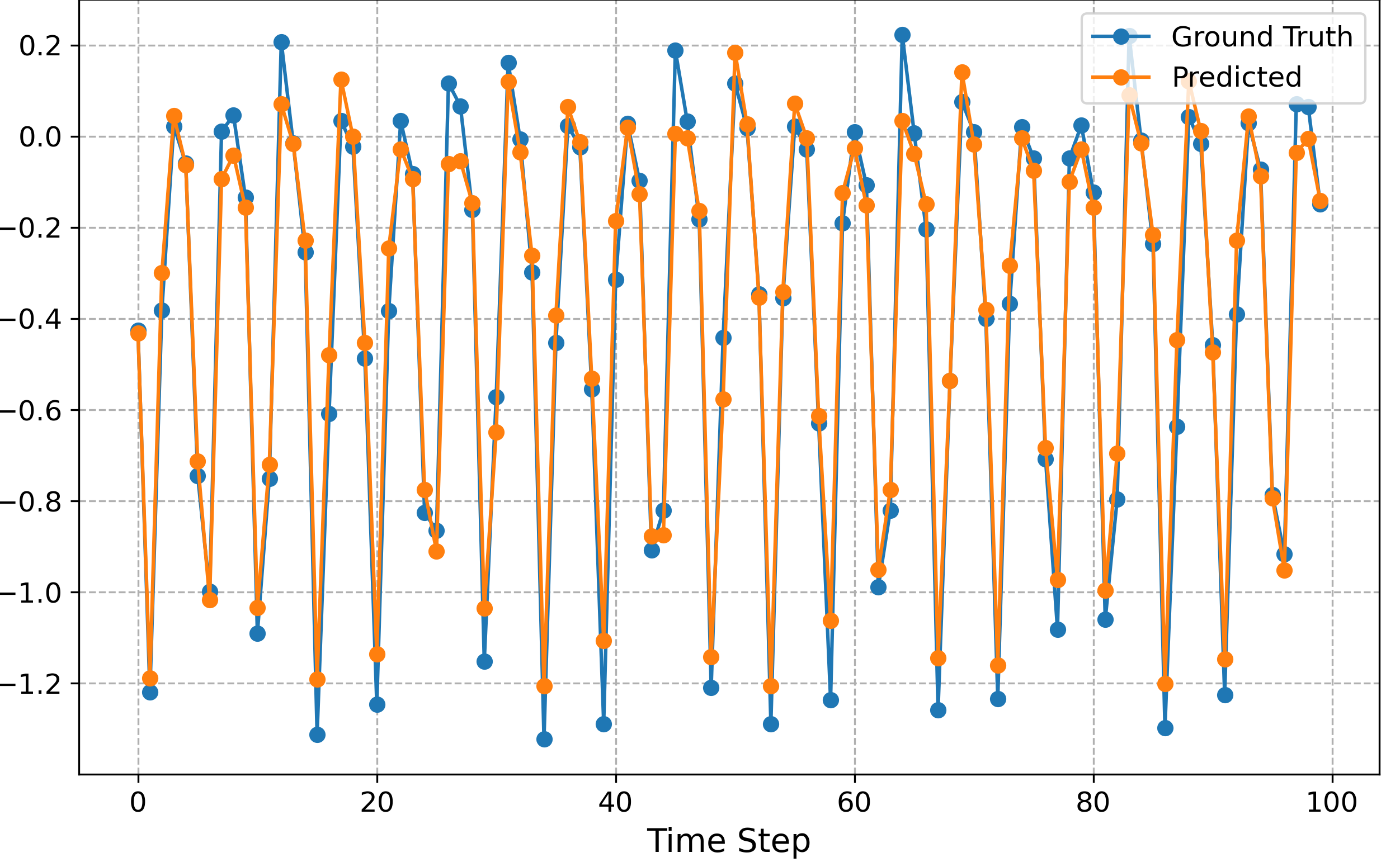}};
            \node[anchor=north, rotate=90, font=\footnotesize\bfseries] at (-0.57,2.5) {\(\mathbf{w} \times 10^{-6}\)};
        \end{tikzpicture}
    \end{minipage}
    \hfill
    \begin{minipage}{0.48\textwidth}
        \centering
        \includegraphics[width=1.9\textwidth, height=0.6\textwidth, keepaspectratio]{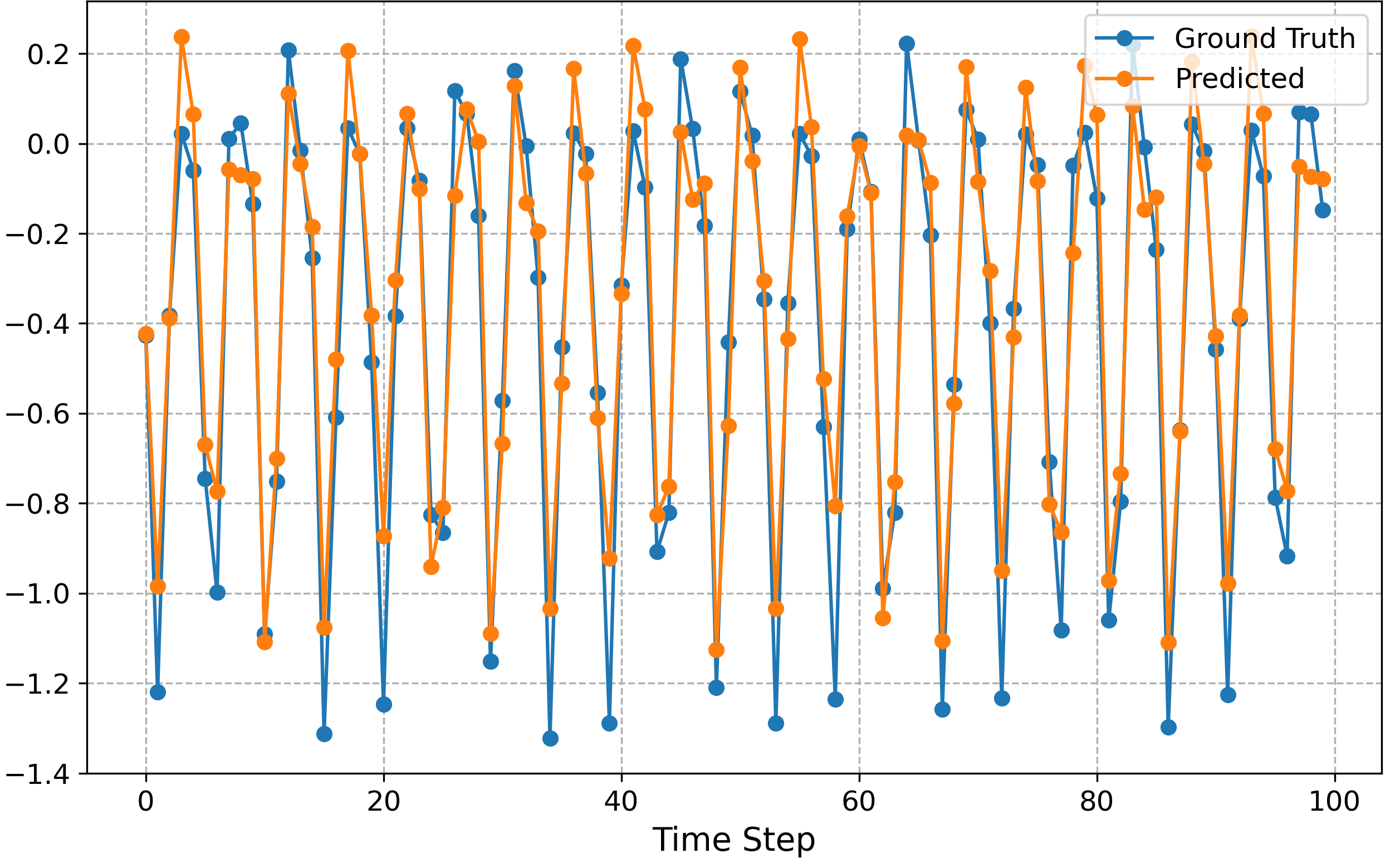}
    \end{minipage}

    \caption{ Comparison of the predicted temporal evolution of spanwise velocity components for (a) HOSVD (left column) and (b) SVD (right column) across different LSTM architectures for the 3D cylinder flow. From top to bottom: LSTM 1 Dense, LSTM 2 Dense, and LSTM Time-Distributed.}
    \label{fig:spanwise_velocity_evolution_3d}
\end{figure}

Figure~\ref{fig:streamwise_evolution_3d} captures the temporal evolution of predicted streamwise velocity components for both HOSVD (left column) and SVD (right column) across LSTM architectures for 3D cylinder flow. HOSVD-based models consistently align more closely with the ground truth, effectively replicating the periodic patterns of the velocity components across time steps. Although minor deviations occur at certain time steps, particularly for the LSTM 1 dense architecture near peaks and troughs, these are minimal and do not significantly affect the overall alignment. In contrast, SVD-based models show less consistency, with deviations from the ground truth, particularly near the peaks of the velocity component. However, both models perform well in predicting the temporal evolution in the spanwise direction. The same can be said for the plots of normal velocity, where both HOSVD and SVD-based models exhibit a consistent match with the ground truth across the time steps.

The spanwise velocity component presents an interesting case, as shown in Figure~\ref{fig:spanwise_velocity_evolution_3d}. The point has been selected at a region comprising near zero velocities to compare the performance between HOSVD and SVD in very low-velocity values. The HOSVD-based models excel in capturing the oscillatory behavior of the velocity components, maintaining consistent alignment with ground truth across all LSTM architectures. However, it does not accurately predict the peak and trough values of this velocity component but does approach the peak values more efficiently than the SVD counterparts. SVD-based models struggle in this scenario. While the general trend is followed, the predictions do not reach the maximum and minimum velocity values observed in the ground truth. HOSVD performs better in this regard.

\begin{figure}[h!]
    \centering
   \makebox[\textwidth]{\textbf{(a) HOSVD} \hspace{5cm} \textbf{(b) SVD}} \\[5pt]
    \begin{minipage}{0.48\textwidth}
        \centering
        \includegraphics[width=1.9\textwidth, height=0.57\textwidth, keepaspectratio]{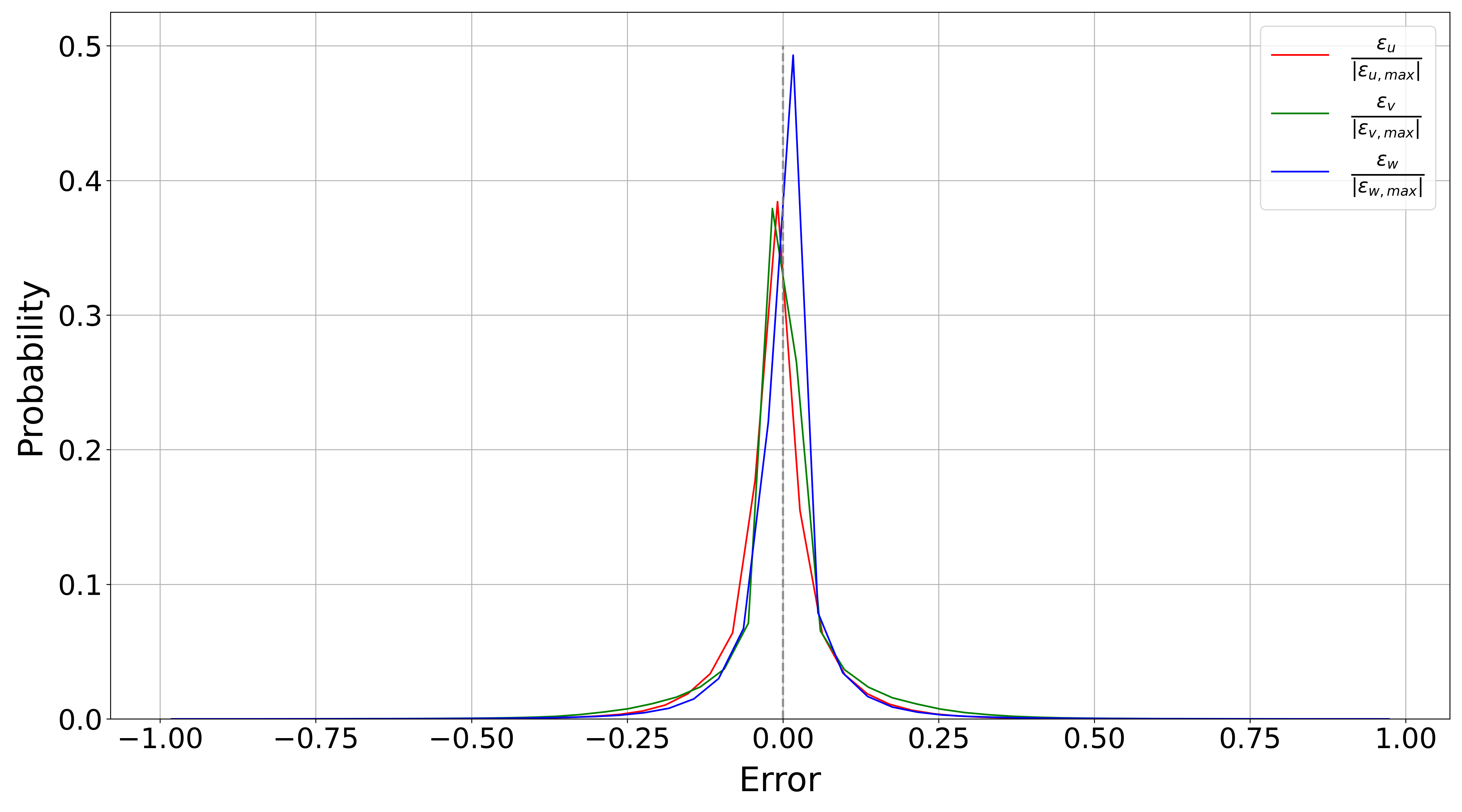}
    \end{minipage}
    \hfill
    \begin{minipage}{0.48\textwidth}
        \centering
        \includegraphics[width=1\textwidth, height=0.57\textwidth]{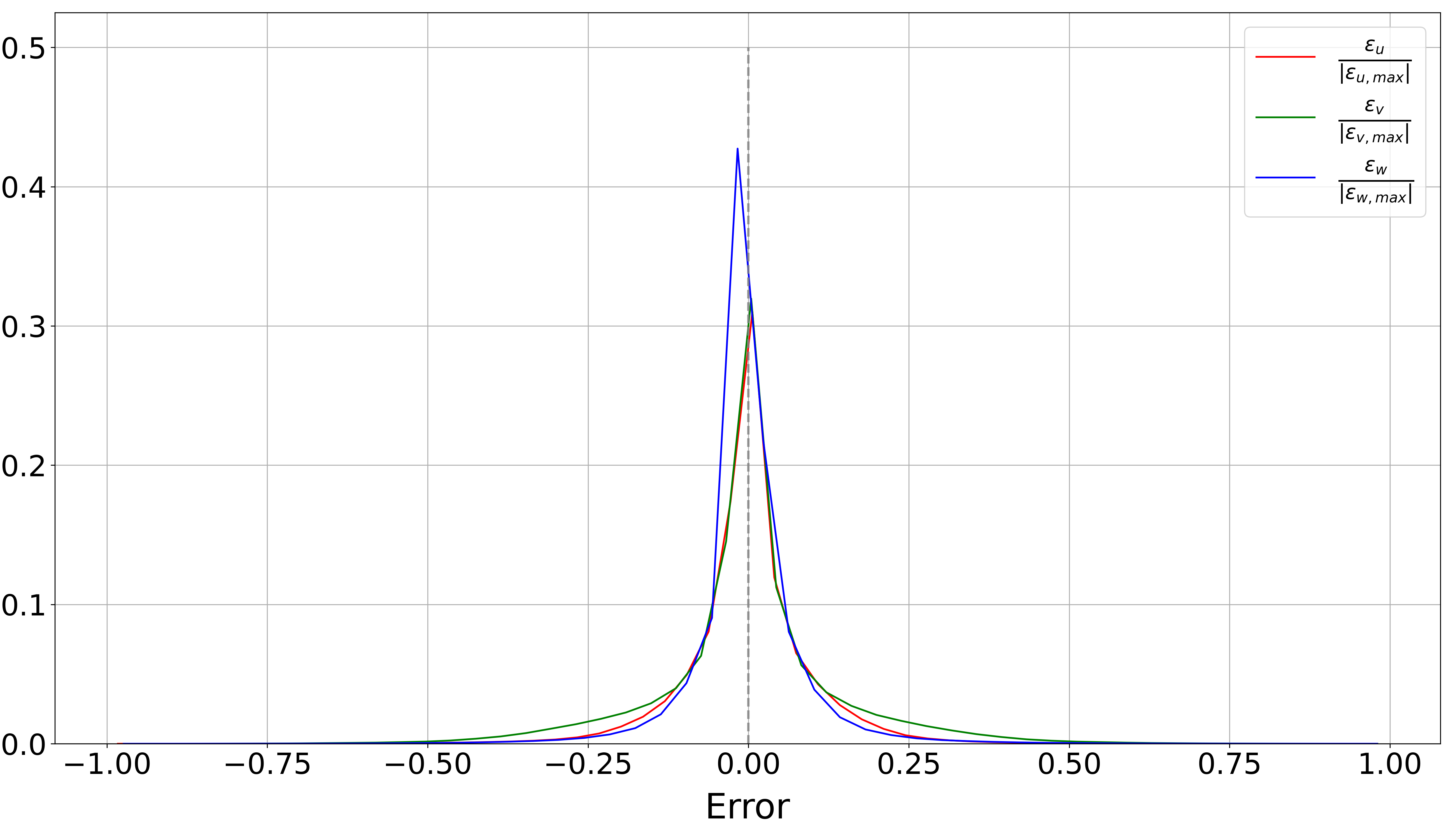}
    \end{minipage}
    \\[10pt]
    \begin{minipage}{0.48\textwidth}
        \centering
        \includegraphics[width=1.9\textwidth, height=0.57\textwidth, keepaspectratio]{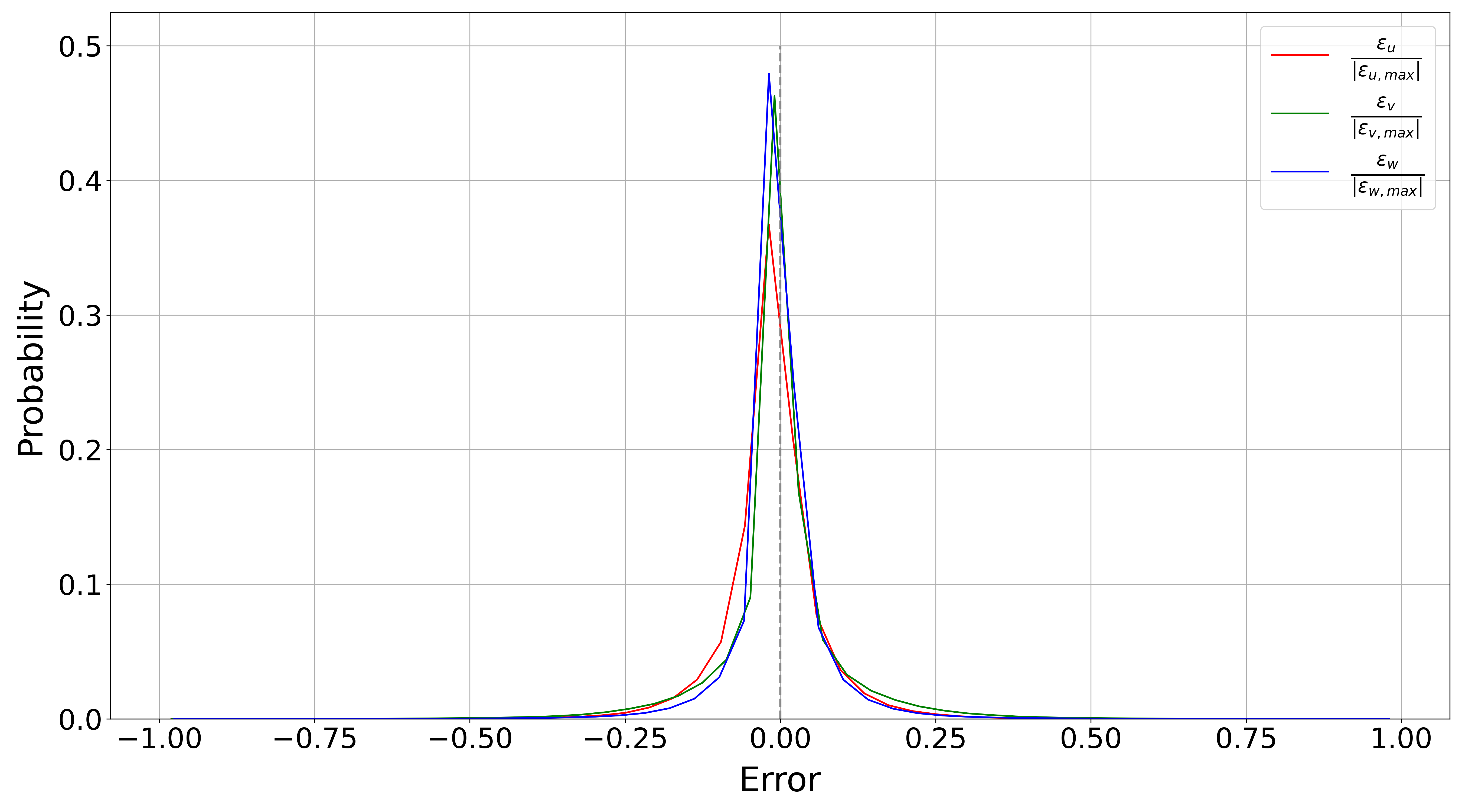}
    \end{minipage}
    \hfill
    \begin{minipage}{0.48\textwidth}
        \centering
        \includegraphics[width=1.9\textwidth, height=0.57\textwidth, keepaspectratio]{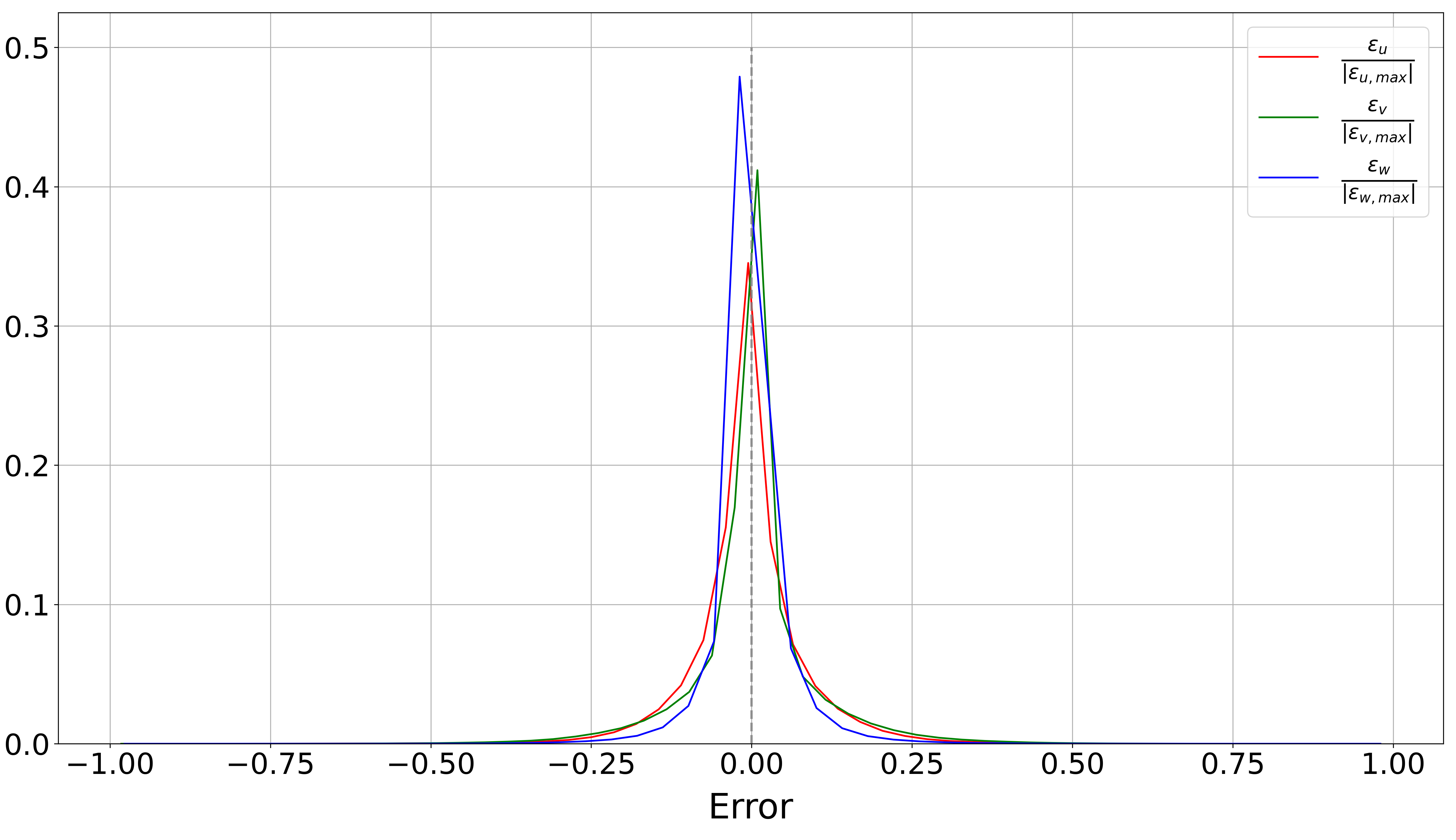}
    \end{minipage}
    \\[10pt]
    \begin{minipage}{0.48\textwidth}
        \centering
        \includegraphics[width=1.9\textwidth, height=0.57\textwidth, keepaspectratio]{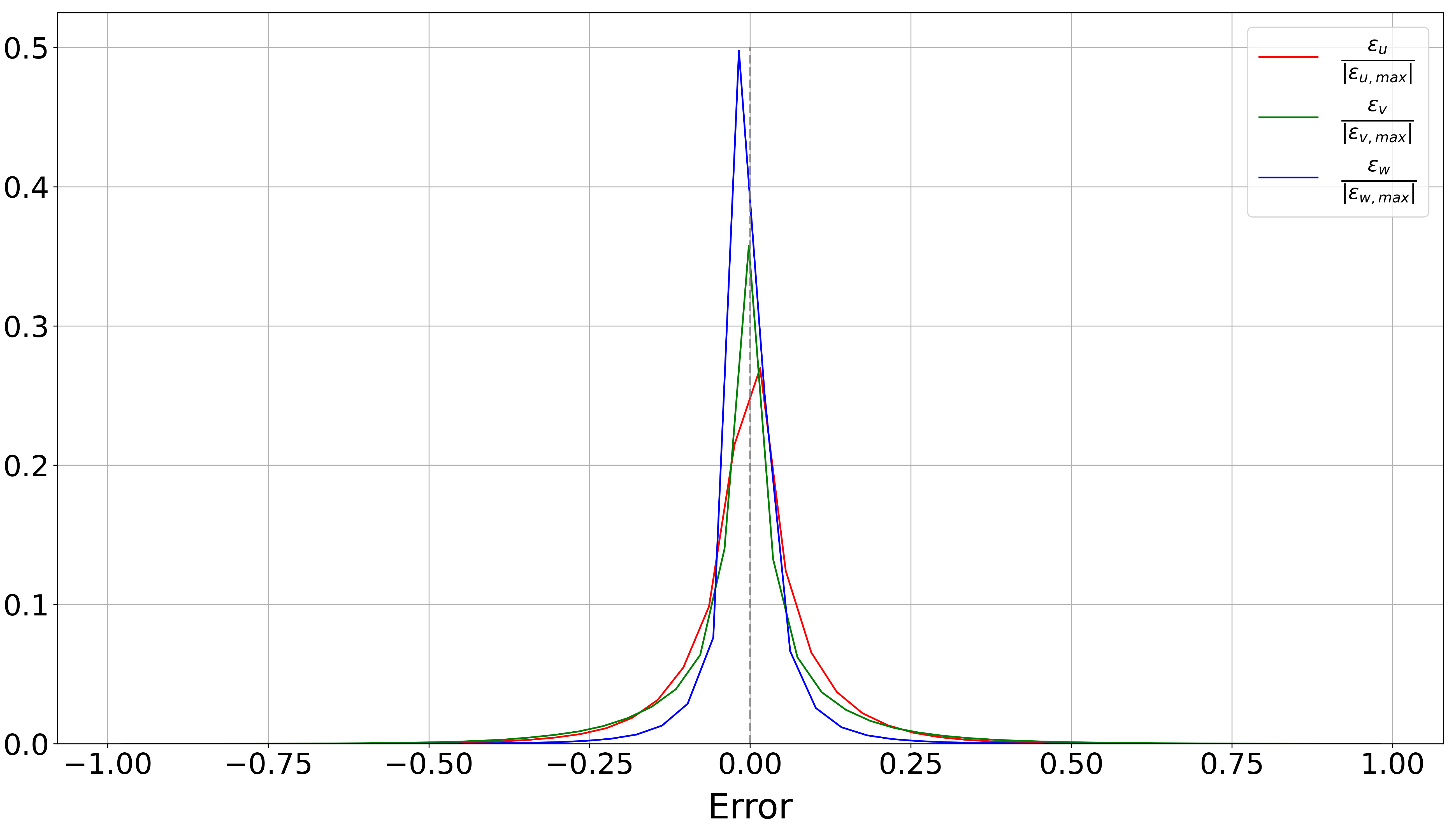}
    \end{minipage}
    \hfill
    \begin{minipage}{0.48\textwidth}
        \centering
        \includegraphics[width=1.9\textwidth, height=0.57\textwidth, keepaspectratio]{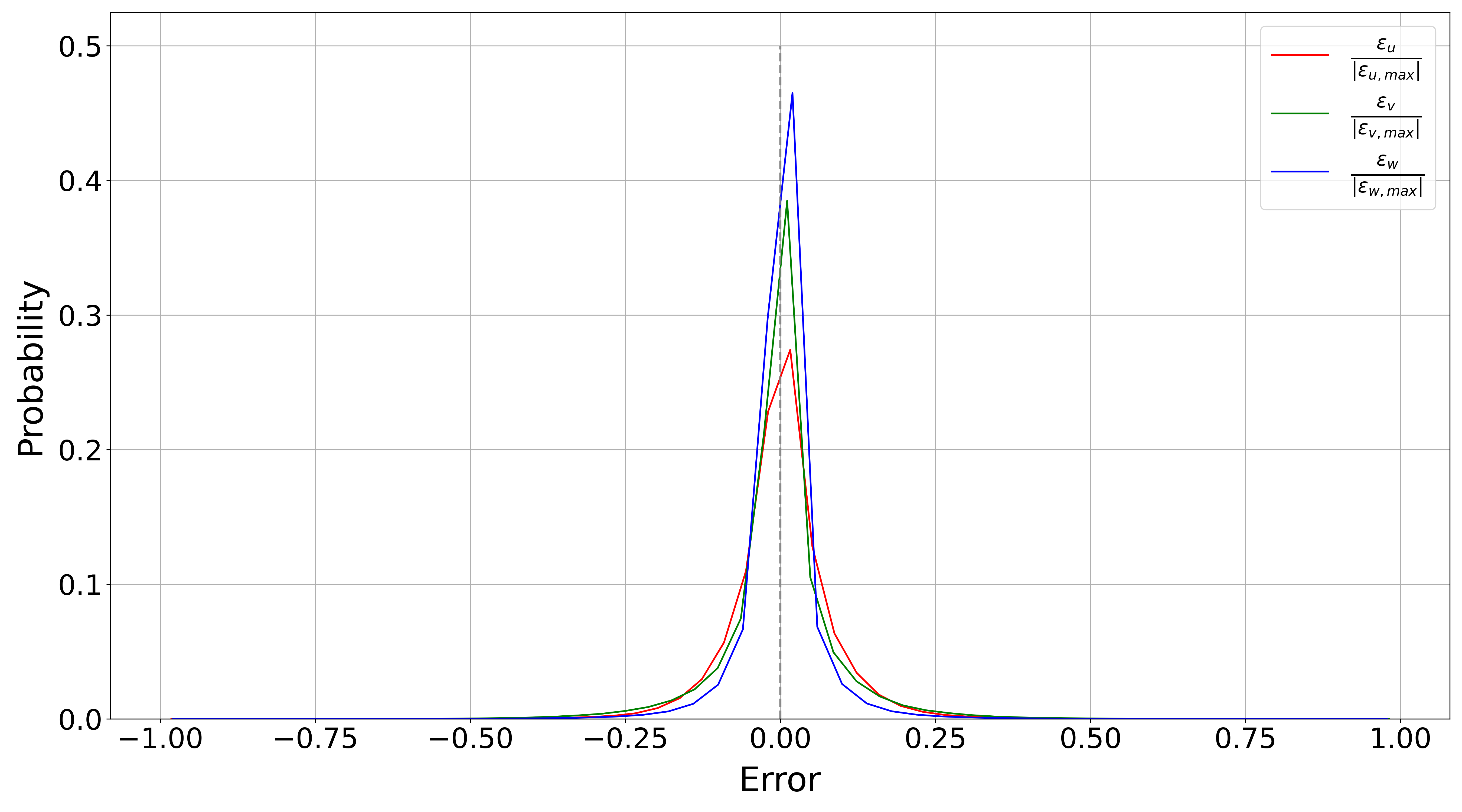}
    \end{minipage}
    \caption{Uncertainty quantification (UQ) results for (a) HOSVD (left column) and (b) SVD (right column) across LSTM architectures. From top to bottom: predictions for LSTM with 1 Dense, 2 Dense, and Time-Distributed architectures for the 3D cylinder flow. Histograms have been constructed using 50 bins.}
    \label{fig:uq_results_3d}
\end{figure}

Building on the above observations, the analysis extends to uncertainty quantification (UQ), providing deeper insights into predictive reliability. For HOSVD (left column), the error distributions are tightly centered around zero for all architectures, with a narrow spread. This indicates a high level of reliability and consistency in the predictions, regardless of the velocity component or the LSTM architecture used. The SVD-based models (right column) demonstrate broader and less consistent error distributions across all LSTM architectures. Although general error trends remain centered around zero, increased spread indicates higher variability in the predictions. The error distributions for SVD tend to show slight asymmetry, particularly in the streamwise and normal components, suggesting that SVD struggles more with capturing complex flow structures compared to HOSVD. Even among the SVD models, the LSTM 2 dense architecture shows the best results for UQ, with sharper peaks and relatively narrower distributions compared to the corresponding architectures, indicating better performance.

\subsubsection{Case: Turbulent Flow Past a Circular Cylinder }

This final data set poses several challenges typical of experimental data. Turbulent flows, inherently chaotic and multi-scale, are difficult to predict accurately, especially with added noise from measurement devices. The presence of noise can obscure critical flow structures and increase the complexity of prediction models. A total of 4000 snapshots are available, of which the last 1000 have been utilized for training and validation purposes using the 80-20 split.

\begin{figure}[h!]
    \centering
    \begin{subfigure}[b]{0.9\textwidth}  % Increased width
        \centering
        \includegraphics[width=\textwidth]{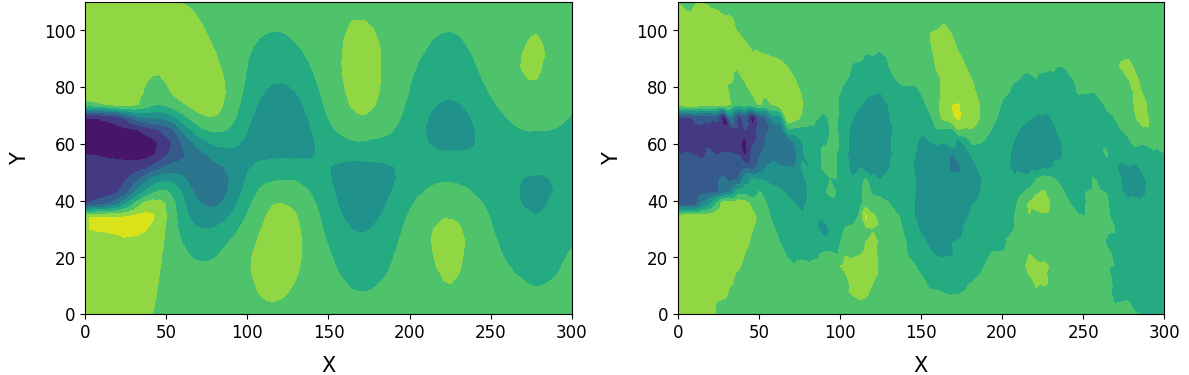}
    \end{subfigure}
    
    \vspace{0.5cm}  % Adds some space between the images
    
    \begin{subfigure}[b]{0.9\textwidth}  % Increased width
        \centering
        \includegraphics[width=\textwidth]{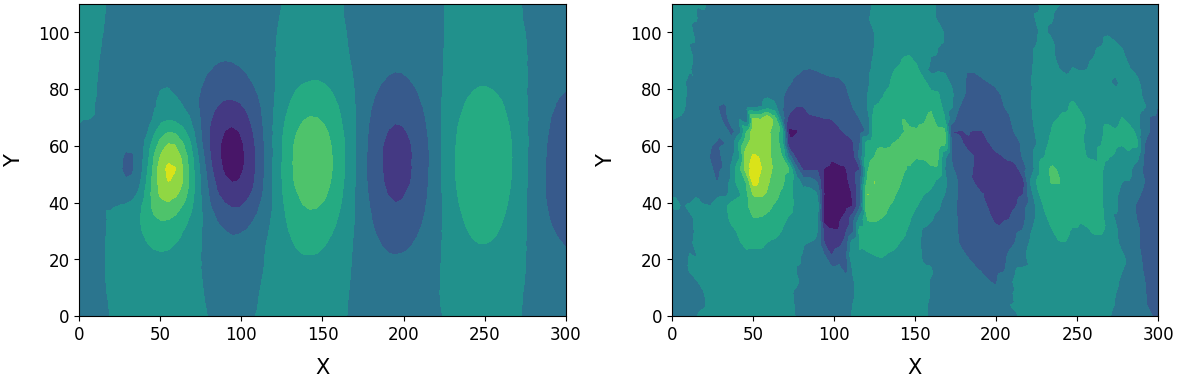}
    \end{subfigure}
    
    \caption{Velocity components of the experimental wake flow. Top: Streamwise velocity (reconstructed vs. ground truth). Bottom: Normal velocity (reconstructed vs. ground truth).}
    \label{fig:wakeflow}
\end{figure}

Given the presence of noise in the experimental data set, the ground truth data have been cleaned and reconstructed to enhance their fidelity and have only been used for the comparison of the results. The training of both the HOSVD and SVD-based models has been performed using the noisy dataset. Figure \ref{fig:wakeflow}  illustrates the original noisy velocity components of the experimental wake flow at \( t = 3828 \) and the cleaned and reconstructed components using SVD. Since only 5 modes have been used, it is only possible to predict the evolution of the coherent flow structures. The cleaning process captures the complexity of the wake while effectively eliminating additional noise, offering a more accurate representation of the flow dynamics.  

\begin{figure}[h!]
    \centering
    \textbf{(a) Ground Truth} \\[5pt]
    \begin{minipage}{0.5\textwidth}
        \centering
        \includegraphics[width=\textwidth]{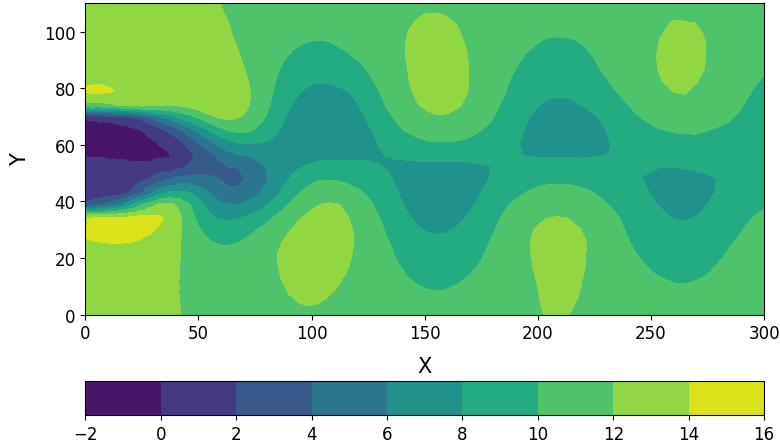}
    \end{minipage}
    \\[15pt]
   % Centered Label for (b) HOSVD and (b) SVD
    \makebox[\textwidth]{\textbf{(b) HOSVD} \hspace{5cm} \textbf{(b) SVD}} \\[5pt]

    % Row 1
    \begin{minipage}{0.48\textwidth}
        \centering
        \begin{tikzpicture}
            \node[anchor=south west, inner sep=0] (image) at (0,0) {\includegraphics[width=0.85\textwidth]{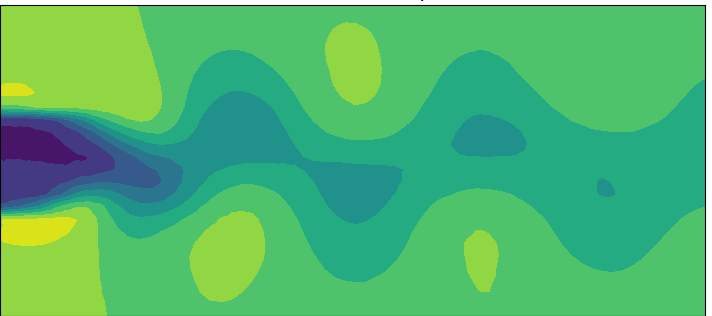}};
            \node[anchor=north, rotate=90, font=\scriptsize] at (-0.6,1.5) {Y};
            \node[anchor=north, font=\scriptsize] at (3.4,-0.2) {X};
        \end{tikzpicture}
    \end{minipage}
    \hfill
    \begin{minipage}{0.48\textwidth}
        \centering
        \begin{tikzpicture}
            \node[anchor=south west, inner sep=0] (image) at (0,0) {\includegraphics[width=0.85\textwidth]{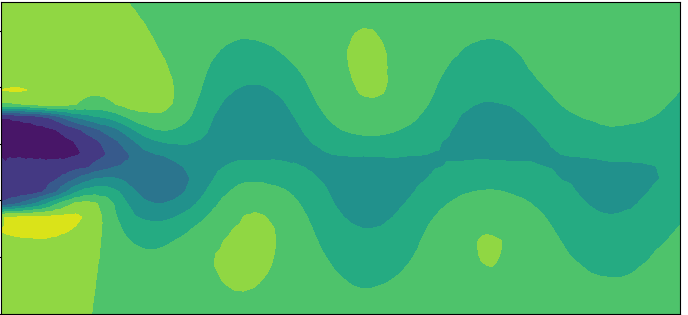}};
            \node[anchor=north, font=\scriptsize] at (3.4,-0.2) {X};
        \end{tikzpicture}
    \end{minipage}
    \\[10pt]

    % Row 2
    \begin{minipage}{0.48\textwidth}
        \centering
        \begin{tikzpicture}
            \node[anchor=south west, inner sep=0] (image) at (0,0) {\includegraphics[width=0.85\textwidth]{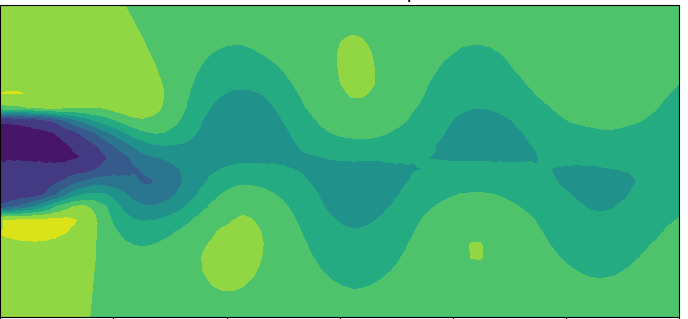}};
            \node[anchor=north, rotate=90, font=\scriptsize] at (-0.6,1.5) {Y};
            \node[anchor=north, font=\scriptsize] at (3.4,-0.2) {X};
        \end{tikzpicture}
    \end{minipage}
    \hfill
    \begin{minipage}{0.48\textwidth}
        \centering
        \begin{tikzpicture}
            \node[anchor=south west, inner sep=0] (image) at (0,0) {\includegraphics[width=0.85\textwidth]{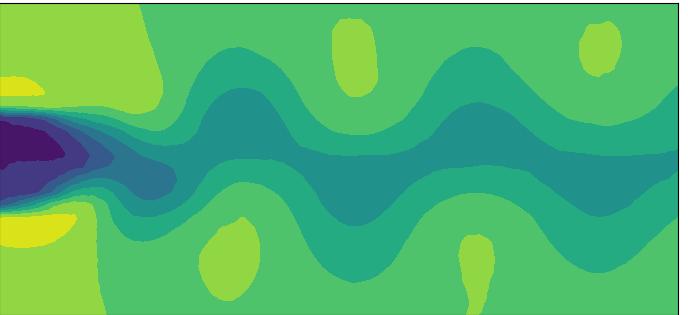}};
            \node[anchor=north, font=\scriptsize] at (3.4,-0.2) {X};
        \end{tikzpicture}
    \end{minipage}
    \\[10pt]

    % Row 3 
    \begin{minipage}{0.48\textwidth}
        \centering
        \begin{tikzpicture}
            \node[anchor=south west, inner sep=0] (image) at (0,0) {\includegraphics[width=0.85\textwidth]{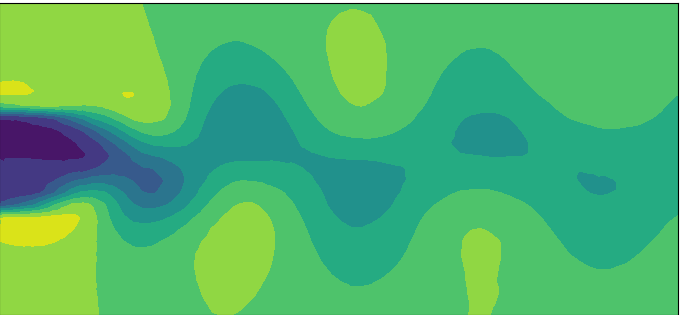}};
            \node[anchor=north, rotate=90, font=\scriptsize] at (-0.6,1.5) {Y};
            \node[anchor=north, font=\scriptsize] at (3.4,-0.2) {X};
        \end{tikzpicture}
    \end{minipage}
    \hfill
    \begin{minipage}{0.48\textwidth}
        \centering
        \begin{tikzpicture}
            \node[anchor=south west, inner sep=0] (image) at (0,0) {\includegraphics[width=0.85\textwidth]{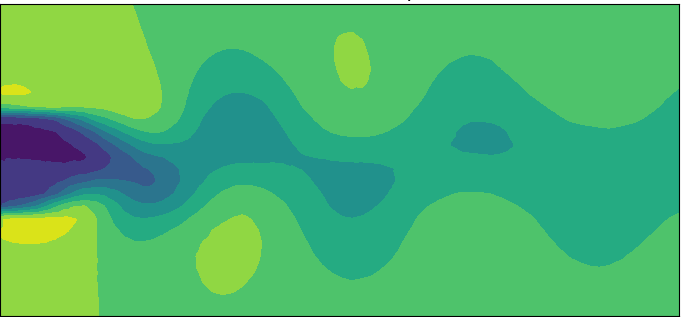}};
            \node[anchor=north, font=\scriptsize] at (3.4,-0.2) {X};
        \end{tikzpicture}
    \end{minipage}

    \caption{(a) Ground truth streamwise velocity components at \(t = 3828\). 
    (b) Comparison of the predicted streamwise velocity components for HOSVD (left column) and SVD (right column) across different LSTM architectures. From top to bottom: predictions for LSTM with 1 Dense, 2 Dense, and Time-Distributed architectures.}
    \label{fig:vki_streamwise_predictions}
\end{figure}

% Normal Velocity Figure
\begin{figure}[h!]
    \centering
    \textbf{(a) Ground Truth} \\[5pt]
    \begin{minipage}{0.5\textwidth}
        \centering
        \includegraphics[width=\textwidth]{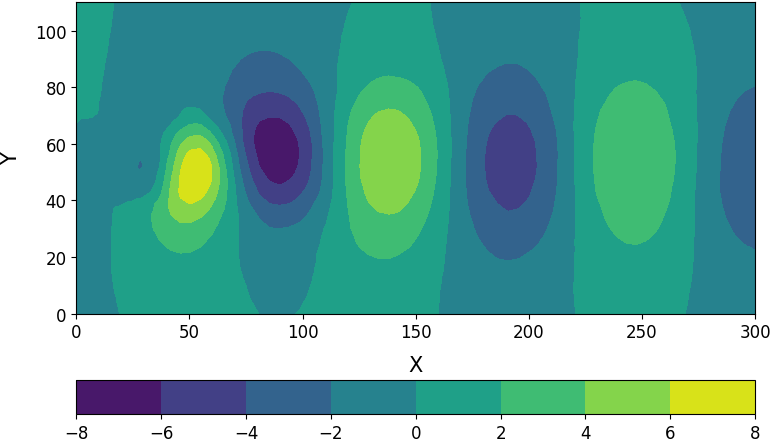}
    \end{minipage}
    \\[15pt]
   % Centered Label for (b) HOSVD and (b) SVD
    \makebox[\textwidth]{\textbf{(b) HOSVD} \hspace{5cm} \textbf{(b) SVD}} \\[5pt]

    % Row 1
    \begin{minipage}{0.48\textwidth}
        \centering
        \begin{tikzpicture}
            \node[anchor=south west, inner sep=0] (image) at (0,0) {\includegraphics[width=0.85\textwidth]{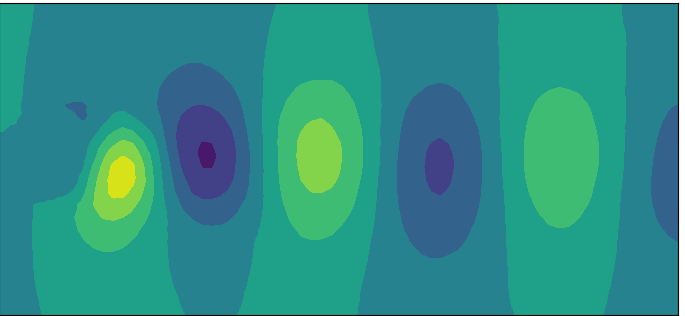}};
            \node[anchor=north, rotate=90, font=\scriptsize] at (-0.6,1.5) {Y};
            \node[anchor=north, font=\scriptsize] at (3.4,-0.2) {X};
        \end{tikzpicture}
    \end{minipage}
    \hfill
    \begin{minipage}{0.48\textwidth}
        \centering
        \begin{tikzpicture}
            \node[anchor=south west, inner sep=0] (image) at (0,0) {\includegraphics[width=0.85\textwidth]{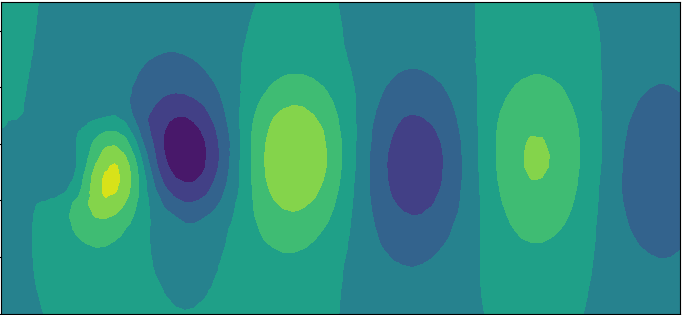}};
            \node[anchor=north, font=\scriptsize] at (3.4,-0.2) {X};
        \end{tikzpicture}
    \end{minipage}
    \\[10pt]

    % Row 2
    \begin{minipage}{0.48\textwidth}
        \centering
        \begin{tikzpicture}
            \node[anchor=south west, inner sep=0] (image) at (0,0) {\includegraphics[width=0.85\textwidth]{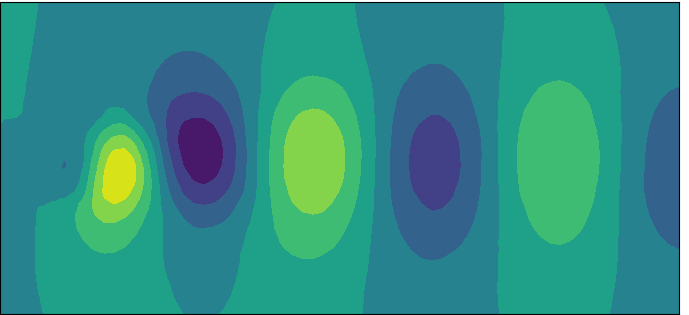}};
            \node[anchor=north, rotate=90, font=\scriptsize] at (-0.6,1.5) {Y};
            \node[anchor=north, font=\scriptsize] at (3.4,-0.2) {X};
        \end{tikzpicture}
    \end{minipage}
    \hfill
    \begin{minipage}{0.48\textwidth}
        \centering
        \begin{tikzpicture}
            \node[anchor=south west, inner sep=0] (image) at (0,0) {\includegraphics[width=0.85\textwidth]{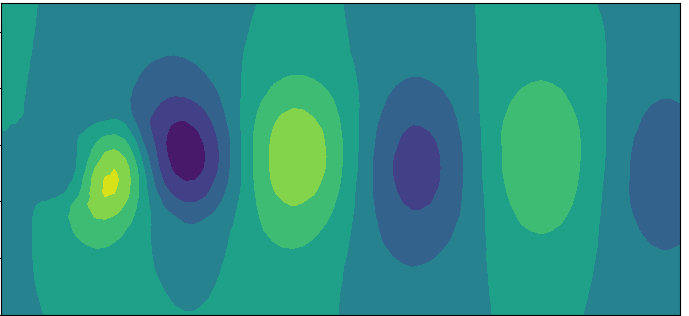}};
            \node[anchor=north, font=\scriptsize] at (3.4,-0.2) {X};
        \end{tikzpicture}
    \end{minipage}
    \\[10pt]

    % Row 3 
    \begin{minipage}{0.48\textwidth}
        \centering
        \begin{tikzpicture}
            \node[anchor=south west, inner sep=0] (image) at (0,0) {\includegraphics[width=0.85\textwidth]{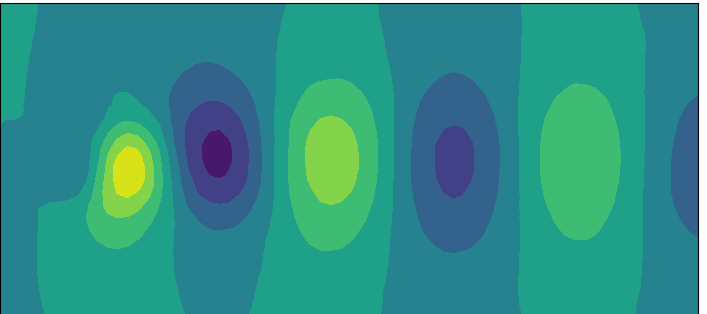}};
            \node[anchor=north, rotate=90, font=\scriptsize] at (-0.6,1.5) {Y};
            \node[anchor=north, font=\scriptsize] at (3.4,-0.2) {X};
        \end{tikzpicture}
    \end{minipage}
    \hfill
    \begin{minipage}{0.48\textwidth}
        \centering
        \begin{tikzpicture}
            \node[anchor=south west, inner sep=0] (image) at (0,0) {\includegraphics[width=0.85\textwidth]{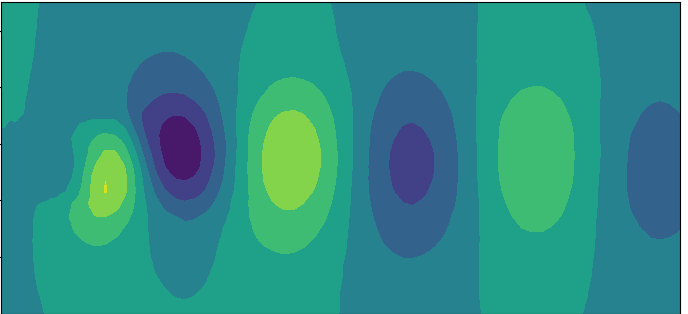}};
            \node[anchor=north, font=\scriptsize] at (3.4,-0.2) {X};
        \end{tikzpicture}
    \end{minipage}

    \caption{(a) Ground truth normal velocity components at \(t = 3828\). 
    (b) Comparison of the predicted normal velocity components for HOSVD (left column) and SVD (right column) across different LSTM architectures. From top to bottom: predictions for LSTM with 1 Dense, 2 Dense, and Time-Distributed architectures.}
    \label{fig:vki_normal_predictions}
\end{figure}

Figure~\ref{fig:vki_streamwise_predictions} and Figure~\ref{fig:vki_normal_predictions} present the predicted streamwise and normal velocity components for HOSVD and SVD  across the three LSTM architectures. For the streamwise velocity, HOSVD performs well, effectively capturing the primary flow structures across all architectures. The SVD-based LSTM models, for certain snapshots, provide an averaged flow scenario just like the cylinder 2D case, except for the time-distributed architecture. However, the predicted normal velocity snapshots align very closely with the cleaned ground truth across the architectures based on the two decomposition techniques.

\begin{table}[h!]
\centering
\begin{tabular}{|c|c|c|}
\hline
\textbf{Architecture} & \textbf{HOSVD (\%)} & \textbf{SVD (\%)} \\ \hline
LSTM 1 Dense          & 4.7                 & 8.6              \\ \hline
LSTM 2 Dense          & 3.1                 & 7.2              \\ \hline
LSTM Time-Distributed  & 3.9                  & 5.4               \\ \hline
\end{tabular}
\caption{RRMSE values for HOSVD and SVD across LSTM architectures for the experimental wake flow (streamwise velocity).}
\label{tab:rrmse_results_vki_streamwise}
\end{table}

\begin{table}[h!]
\centering
\begin{tabular}{|c|c|c|}
\hline
\textbf{Architecture} & \textbf{HOSVD (\%)} & \textbf{SVD (\%)} \\ \hline
LSTM 1 Dense          & 33.6                 & 42.7              \\ \hline
LSTM 2 Dense          & 29.4                 & 37.2              \\ \hline
LSTM Time-Distributed  & 29.3                 & 34.5              \\ \hline
\end{tabular}
\caption{RRMSE values for HOSVD and SVD across LSTM architectures for the experimental wake flow (normal velocity).}
\label{tab:rrmse_results_vki_normal}
\end{table}

For the streamwise velocity component (Table~\ref{tab:rrmse_results_vki_streamwise}), HOSVD achieves lower RRMSE values across all LSTM architectures compared to SVD. The LSTM 2 Dense architecture demonstrates the best performance, achieving the lowest RRMSE of 3.1\%, compared to 7.2\% for SVD. The LSTM 1 Dense and LSTM Time-Distributed architectures also show significant improvement with HOSVD, achieving RRMSE values of 4.7\% and 3.9\%, respectively, compared to 8.6\% and 5.4\% for SVD.

For the normal velocity component (Table~\ref{tab:rrmse_results_vki_normal}), the trend remains consistent, though the overall errors are higher compared to the streamwise case. HOSVD achieves lower RRMSE values across all architectures, with the LSTM Time-Distributed architecture achieving the best performance at 29.3\% compared to SVD's 34.5\%. The LSTM 2 Dense and LSTM 1 Dense architectures also maintain a significant advantage, with RRMSE values of 29.4\% and 33.6\%, respectively, compared to 37.2\% and 42.7\% for SVD.

% Streamwise Velocity
\begin{figure}[h!]
    \centering
    
   \makebox[\textwidth]{\textbf{(a) HOSVD} \hspace{5cm} \textbf{(b) SVD}} \\[5pt]

    % Row 1
    \begin{minipage}{0.48\textwidth}
        \centering
        \begin{tikzpicture}
            \node[anchor=south west, inner sep=0] (image) at (0,0) 
            {\includegraphics[width=1.9\textwidth, height=0.6\textwidth, keepaspectratio]{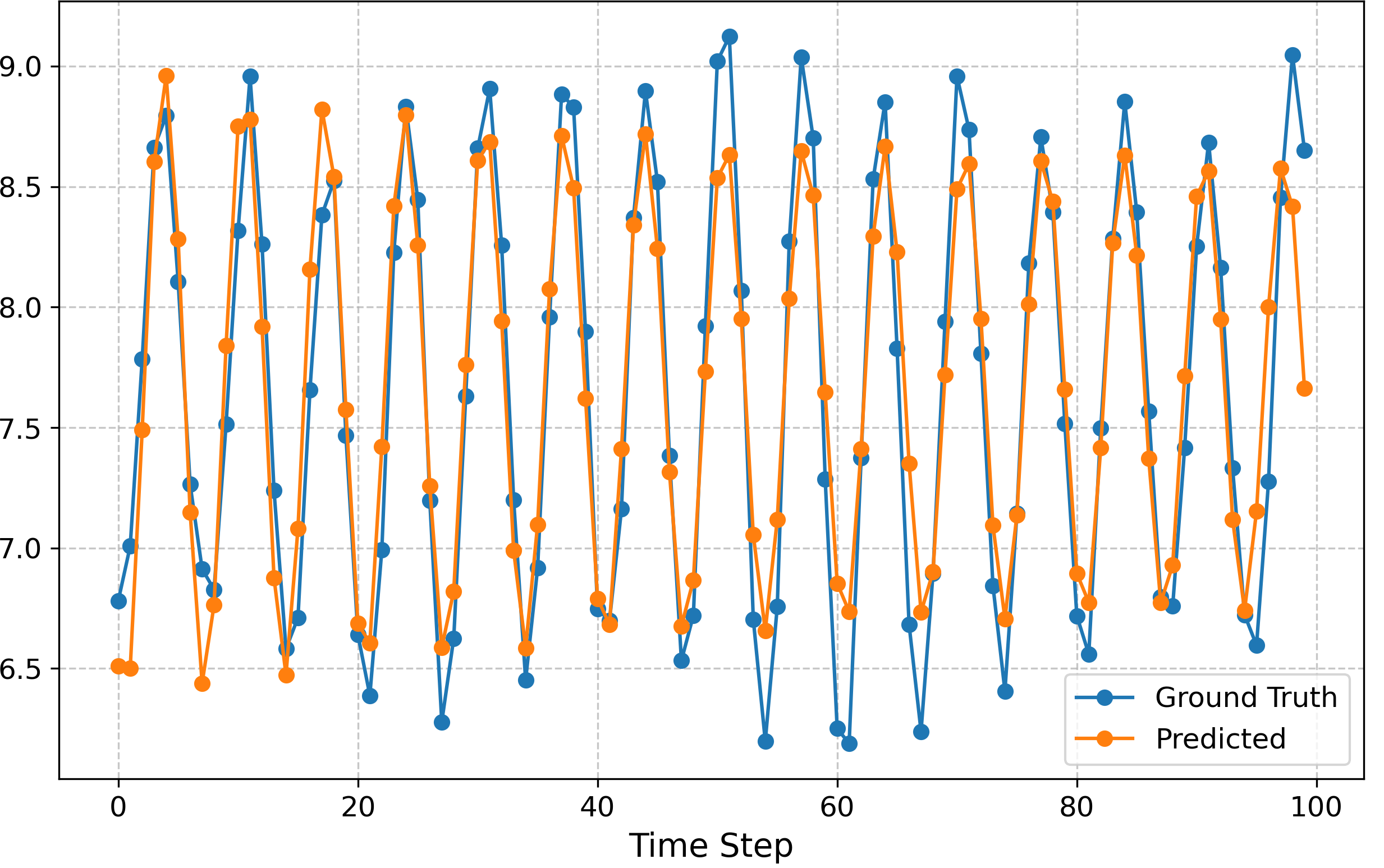}};
            \node[anchor=north, rotate=90, font=\footnotesize\bfseries] at (-0.5,2.5) {u};
        \end{tikzpicture}
    \end{minipage}
    \hfill
    \begin{minipage}{0.48\textwidth}
        \centering
        \includegraphics[width=1.9\textwidth, height=0.6\textwidth, keepaspectratio]{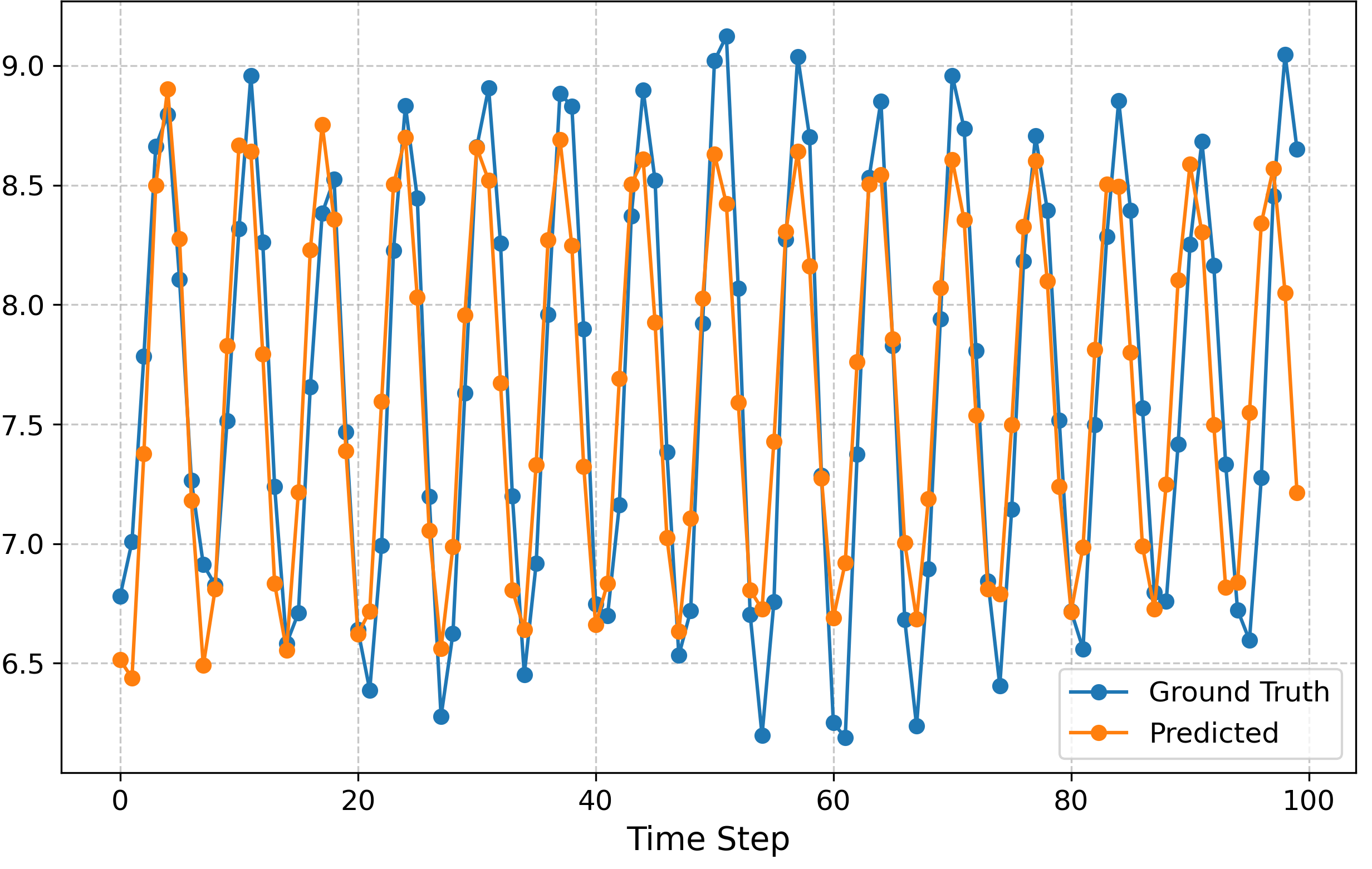}
    \end{minipage}
    \\[15pt]

    % Row 2
    \begin{minipage}{0.48\textwidth}
        \centering
        \begin{tikzpicture}
            \node[anchor=south west, inner sep=0] (image) at (0,0) 
            {\includegraphics[width=1.9\textwidth, height=0.6\textwidth, keepaspectratio]{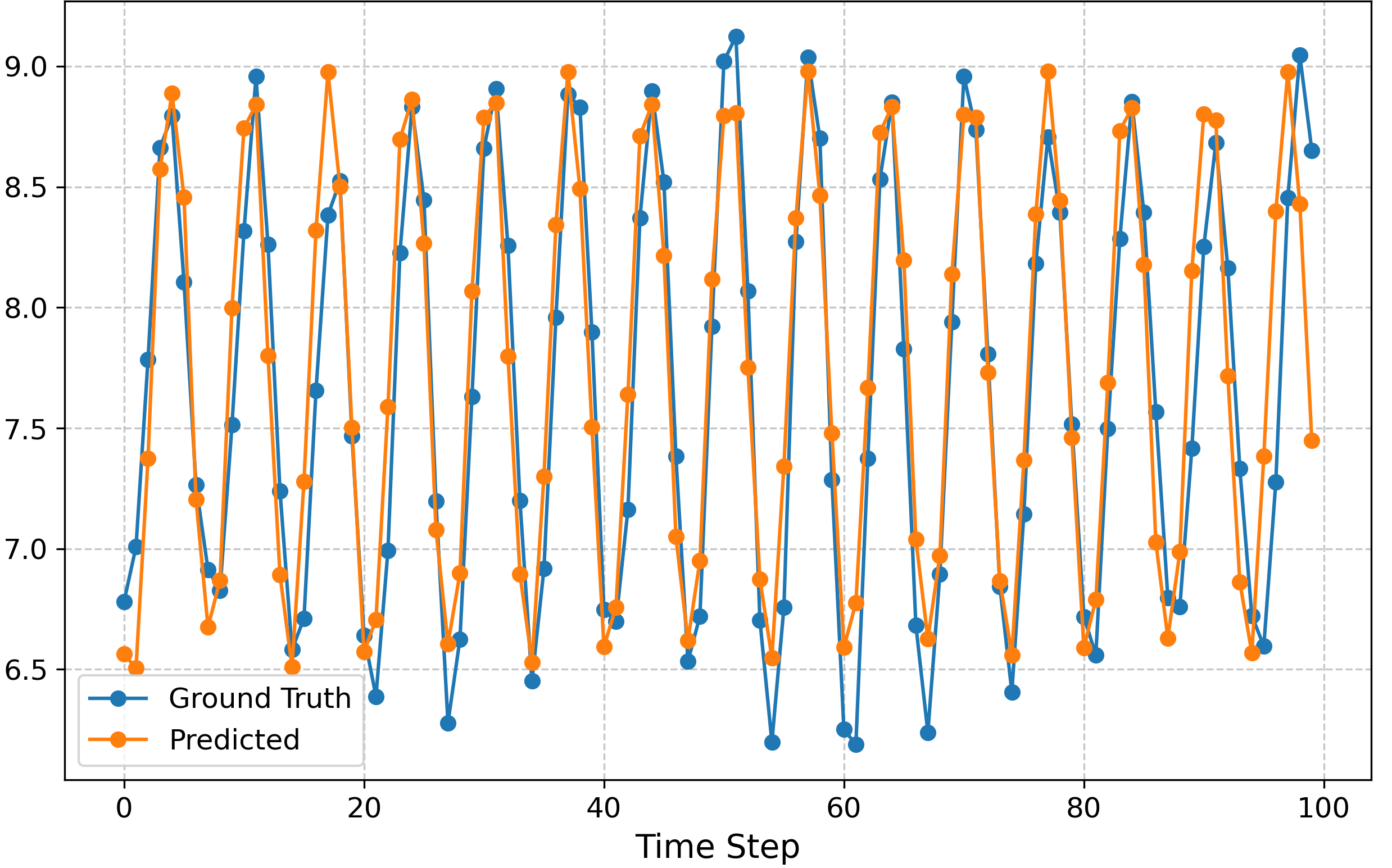}};
            \node[anchor=north, rotate=90, font=\footnotesize\bfseries] at (-0.5,2.5) {u};
        \end{tikzpicture}
    \end{minipage}
    \hfill
    \begin{minipage}{0.48\textwidth}
        \centering
        \includegraphics[width=1.9\textwidth, height=0.6\textwidth, keepaspectratio]{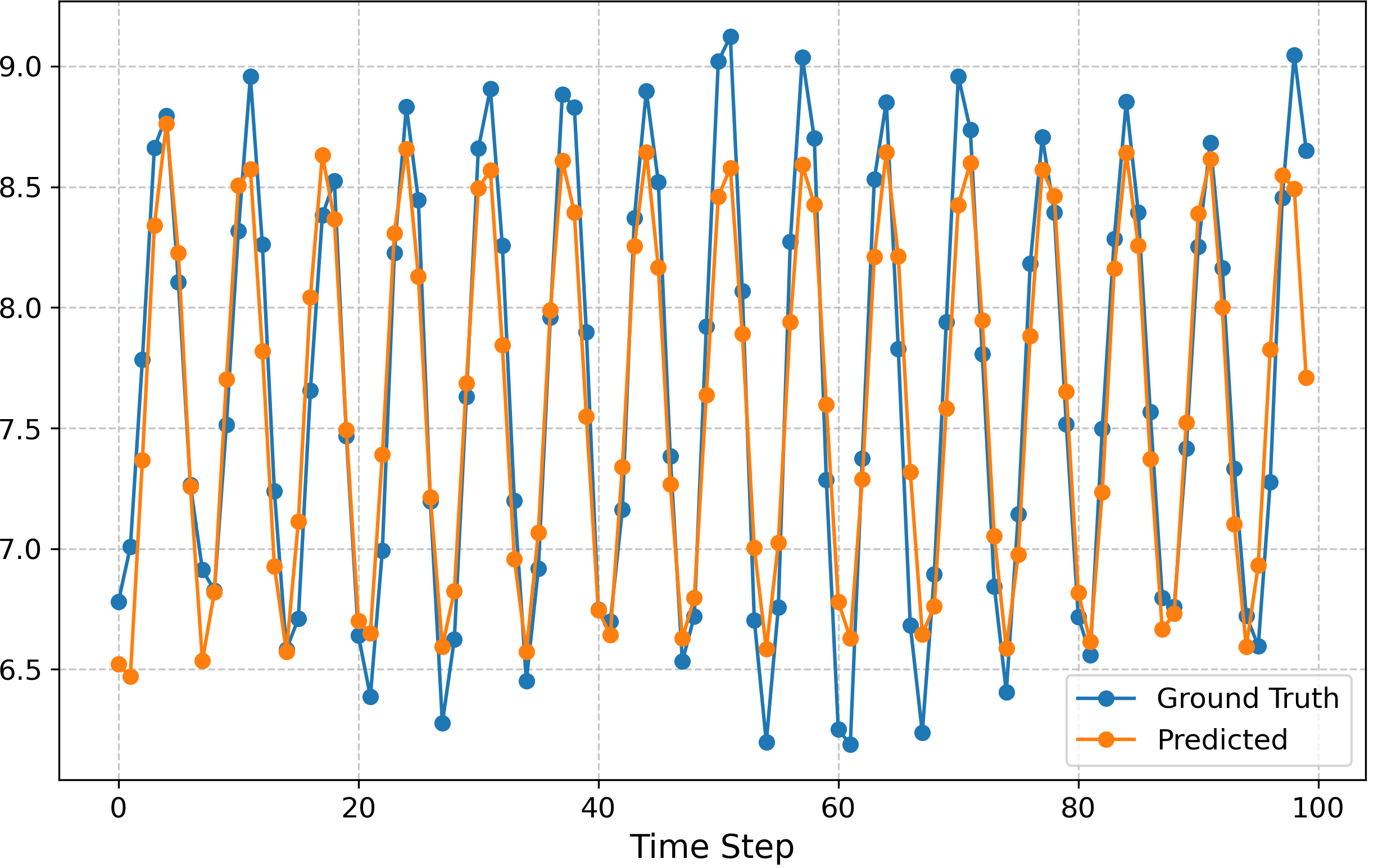}
    \end{minipage}
    \\[15pt]

    % Row 3
    \begin{minipage}{0.48\textwidth}
        \centering
        \begin{tikzpicture}
            \node[anchor=south west, inner sep=0] (image) at (0,0) 
            {\includegraphics[width=1.9\textwidth, height=0.6\textwidth, keepaspectratio]{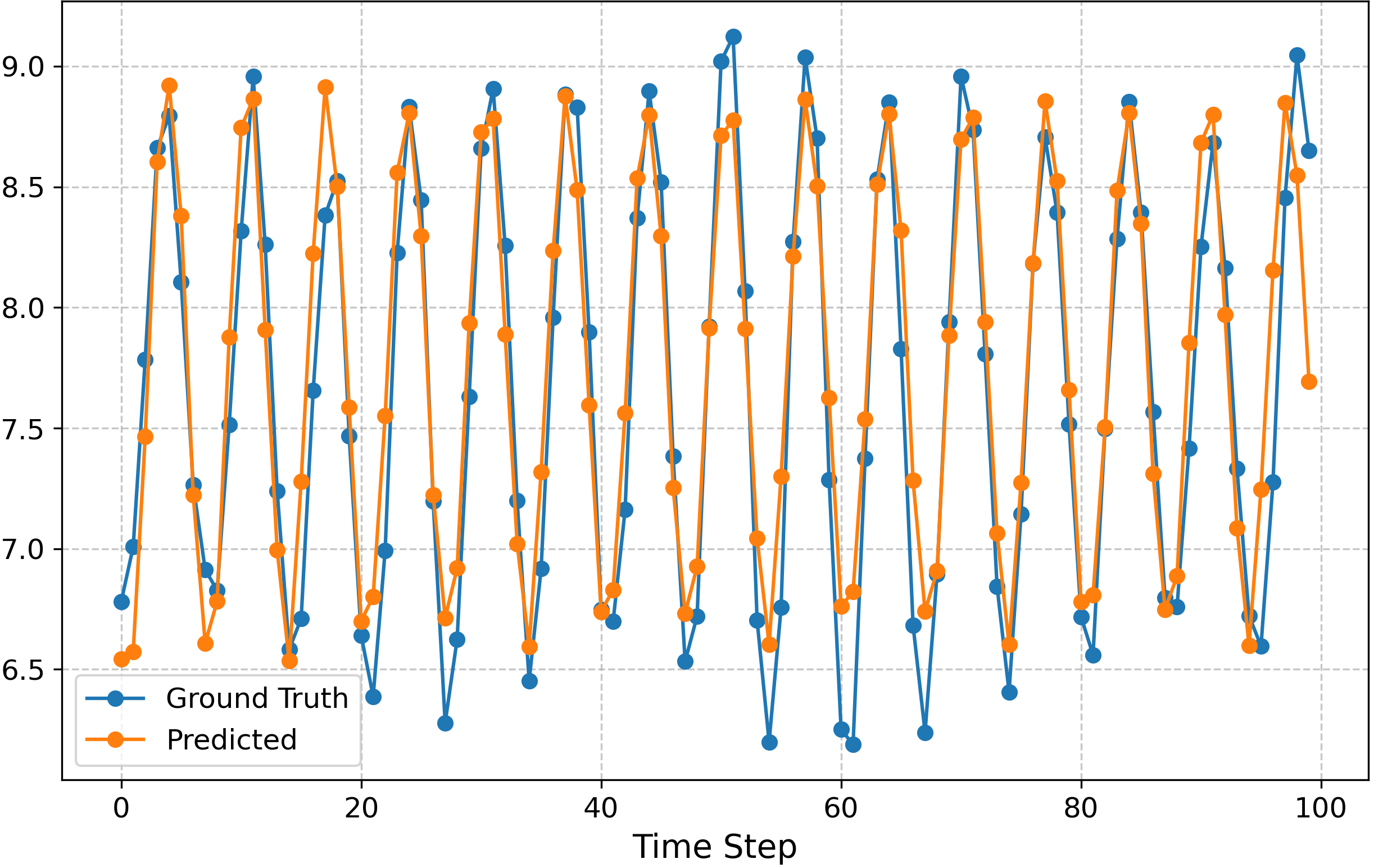}};
            \node[anchor=north, rotate=90, font=\footnotesize\bfseries] at (-0.5,2.5) {u};
        \end{tikzpicture}
    \end{minipage}
    \hfill
    \begin{minipage}{0.48\textwidth}
        \centering
        \includegraphics[width=1.9\textwidth, height=0.6\textwidth, keepaspectratio]{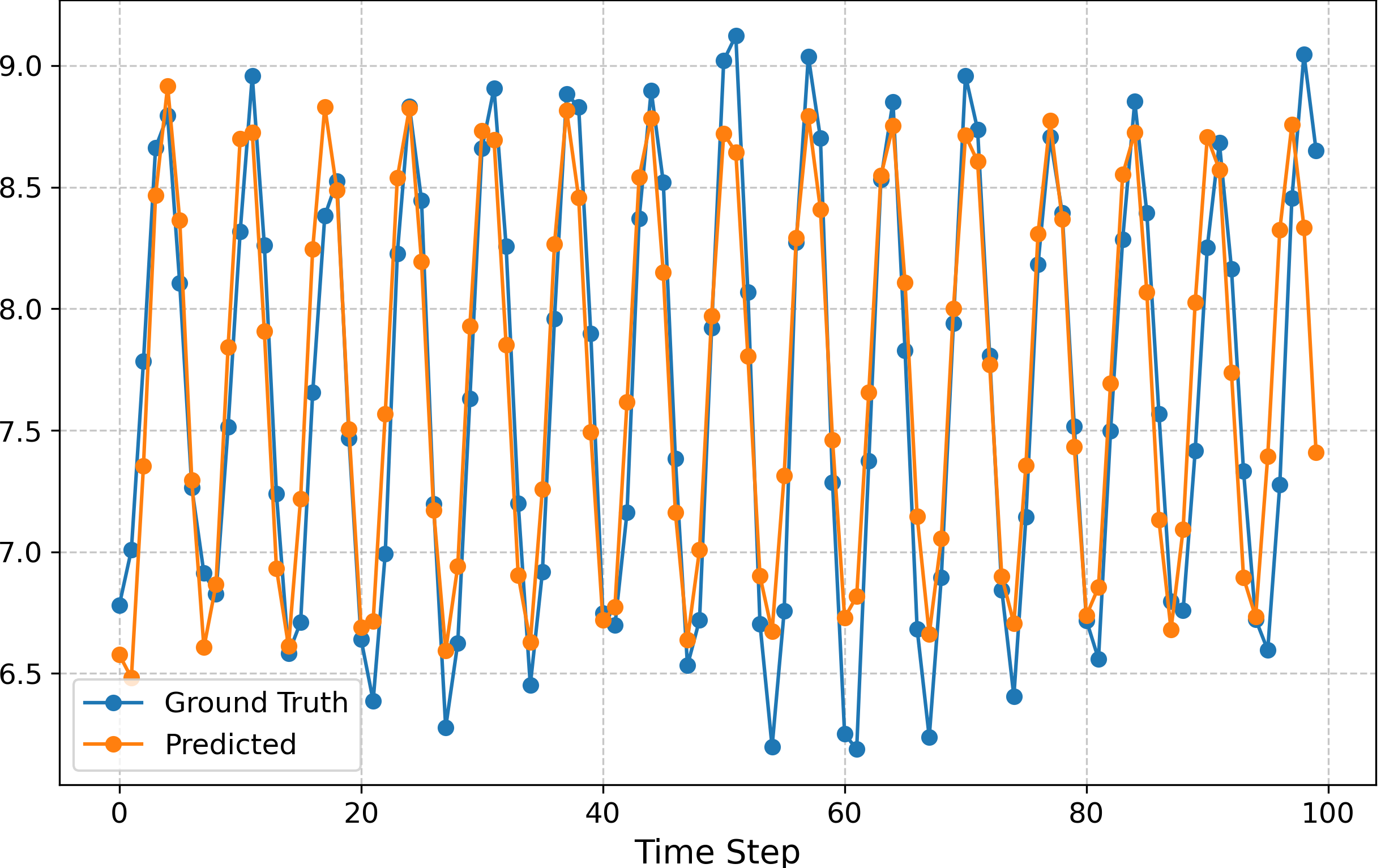}
    \end{minipage}

    \caption{ Comparison of the predicted temporal evolution of streamwise velocity components for (a) HOSVD (left column) and (b) SVD (right column) across different LSTM architectures for the experimental wake flow. From top to bottom: LSTM 1 Dense, LSTM 2 dense, and LSTM Time-Distributed.}
    \label{fig:streamwise_evolution_vki}
\end{figure}

% Normal Velocity
\begin{figure}[h!]
    \centering
    
    \makebox[\textwidth]{\textbf{(a) HOSVD} \hspace{5cm} \textbf{(b) SVD}} \\[5pt]

    % Row 1
    \begin{minipage}{0.48\textwidth}
        \centering
        \begin{tikzpicture}
            \node[anchor=south west, inner sep=0] (image) at (0,0) 
            {\includegraphics[width=1.9\textwidth, height=0.6\textwidth, keepaspectratio]{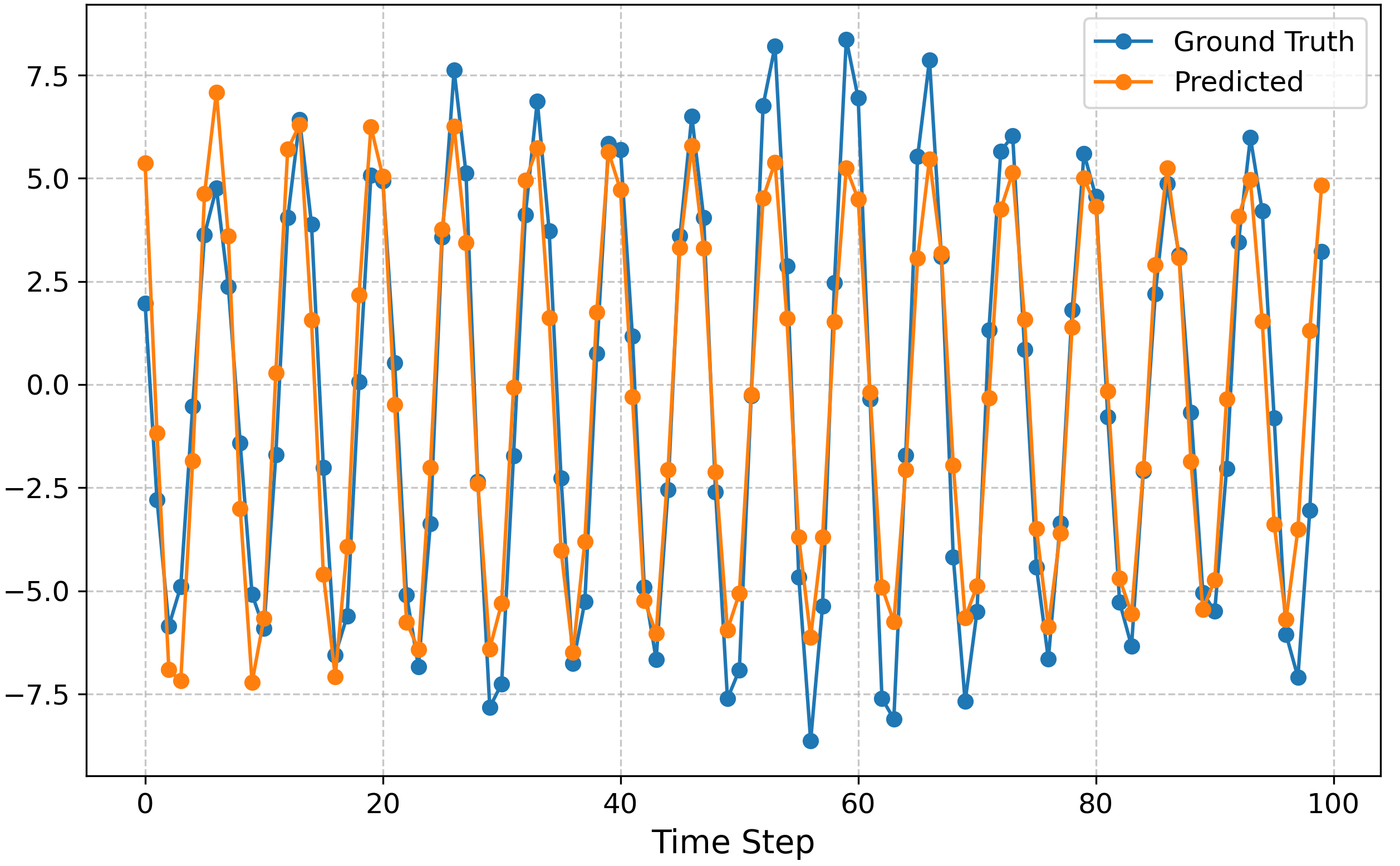}};
            \node[anchor=north, rotate=90, font=\footnotesize\bfseries] at (-0.5,2.5) {v};
        \end{tikzpicture}
    \end{minipage}
    \hfill
    \begin{minipage}{0.48\textwidth}
        \centering
        \includegraphics[width=1.9\textwidth, height=0.6\textwidth, keepaspectratio]{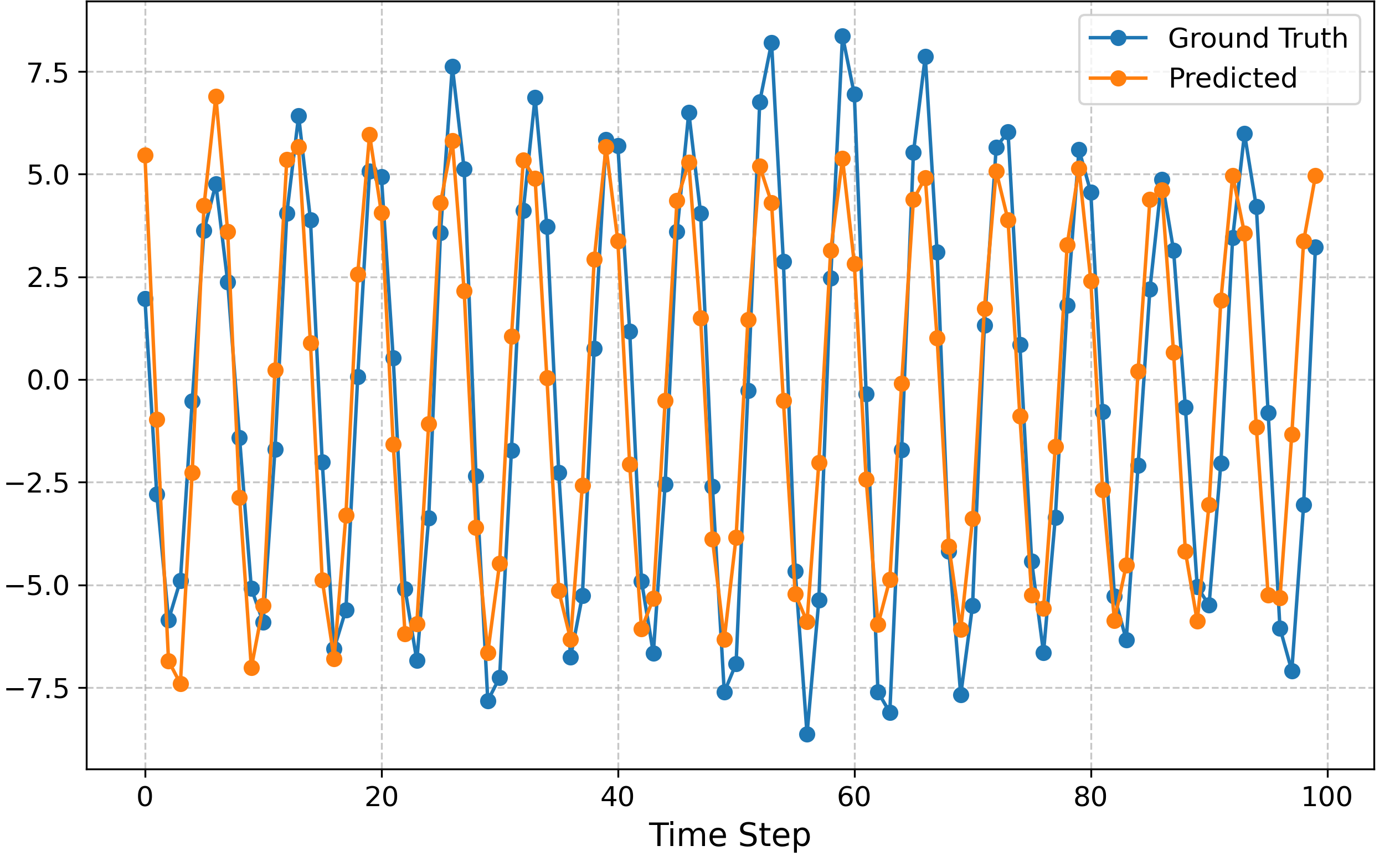}
    \end{minipage}
    \\[15pt]

    % Row 2
    \begin{minipage}{0.48\textwidth}
        \centering
        \begin{tikzpicture}
            \node[anchor=south west, inner sep=0] (image) at (0,0) 
            {\includegraphics[width=1.9\textwidth, height=0.6\textwidth, keepaspectratio]{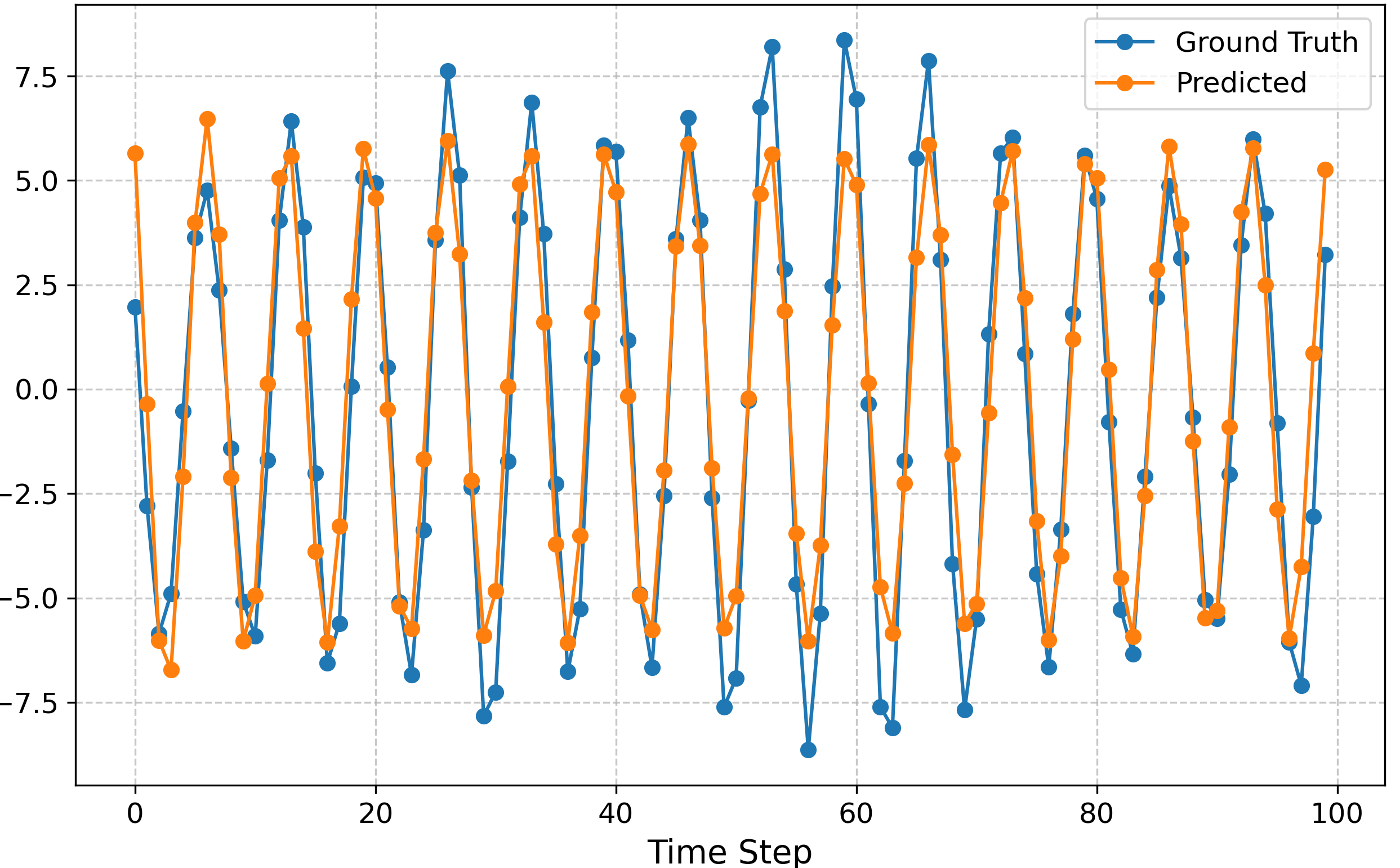}};
            \node[anchor=north, rotate=90, font=\footnotesize\bfseries] at (-0.5,2.5) {v};
        \end{tikzpicture}
    \end{minipage}
    \hfill
    \begin{minipage}{0.48\textwidth}
        \centering
        \includegraphics[width=1.9\textwidth, height=0.6\textwidth, keepaspectratio]{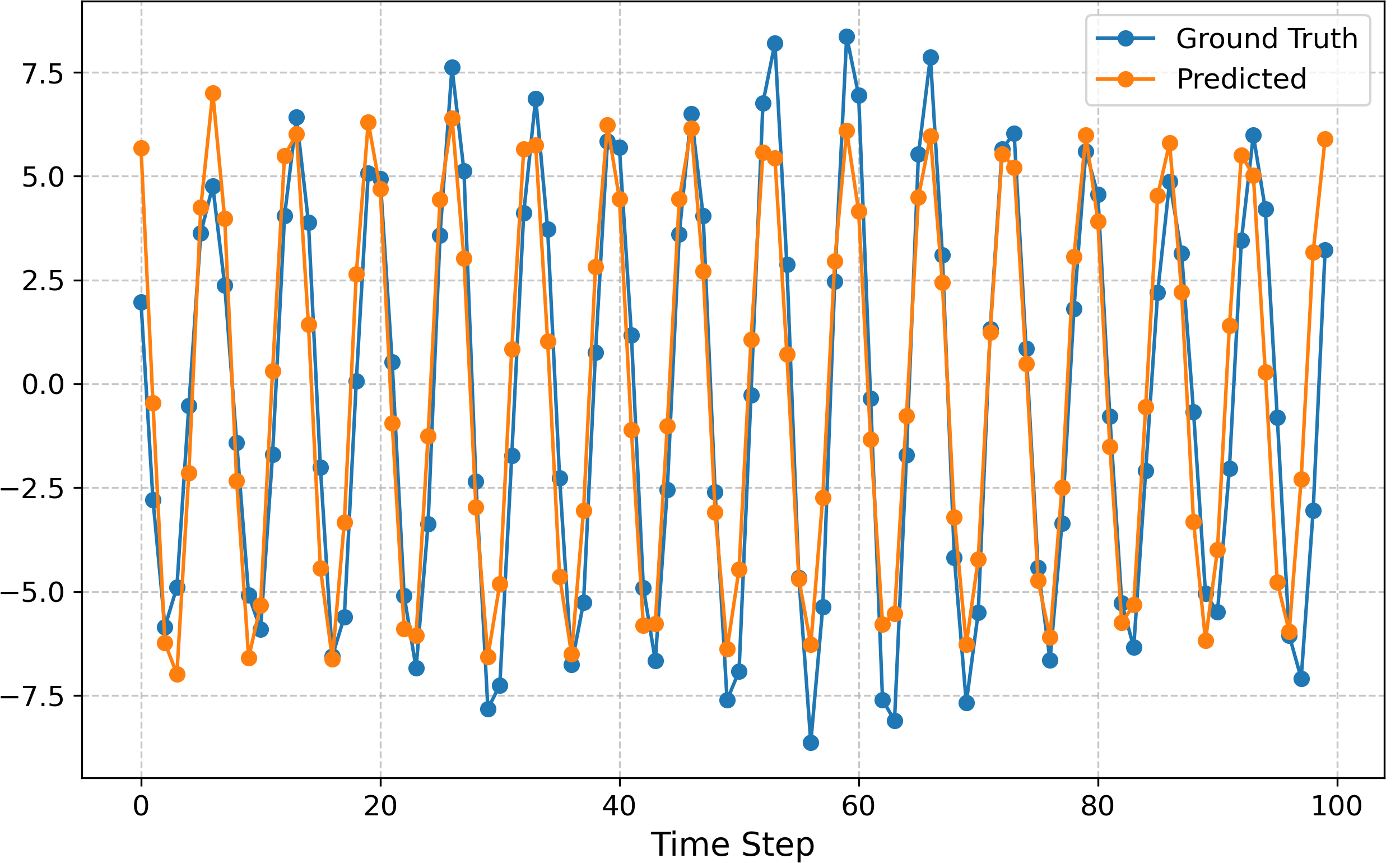}
    \end{minipage}
    \\[15pt]

    % Row 3
    \begin{minipage}{0.48\textwidth}
        \centering
        \begin{tikzpicture}
            \node[anchor=south west, inner sep=0] (image) at (0,0) 
            {\includegraphics[width=1.9\textwidth, height=0.6\textwidth, keepaspectratio]{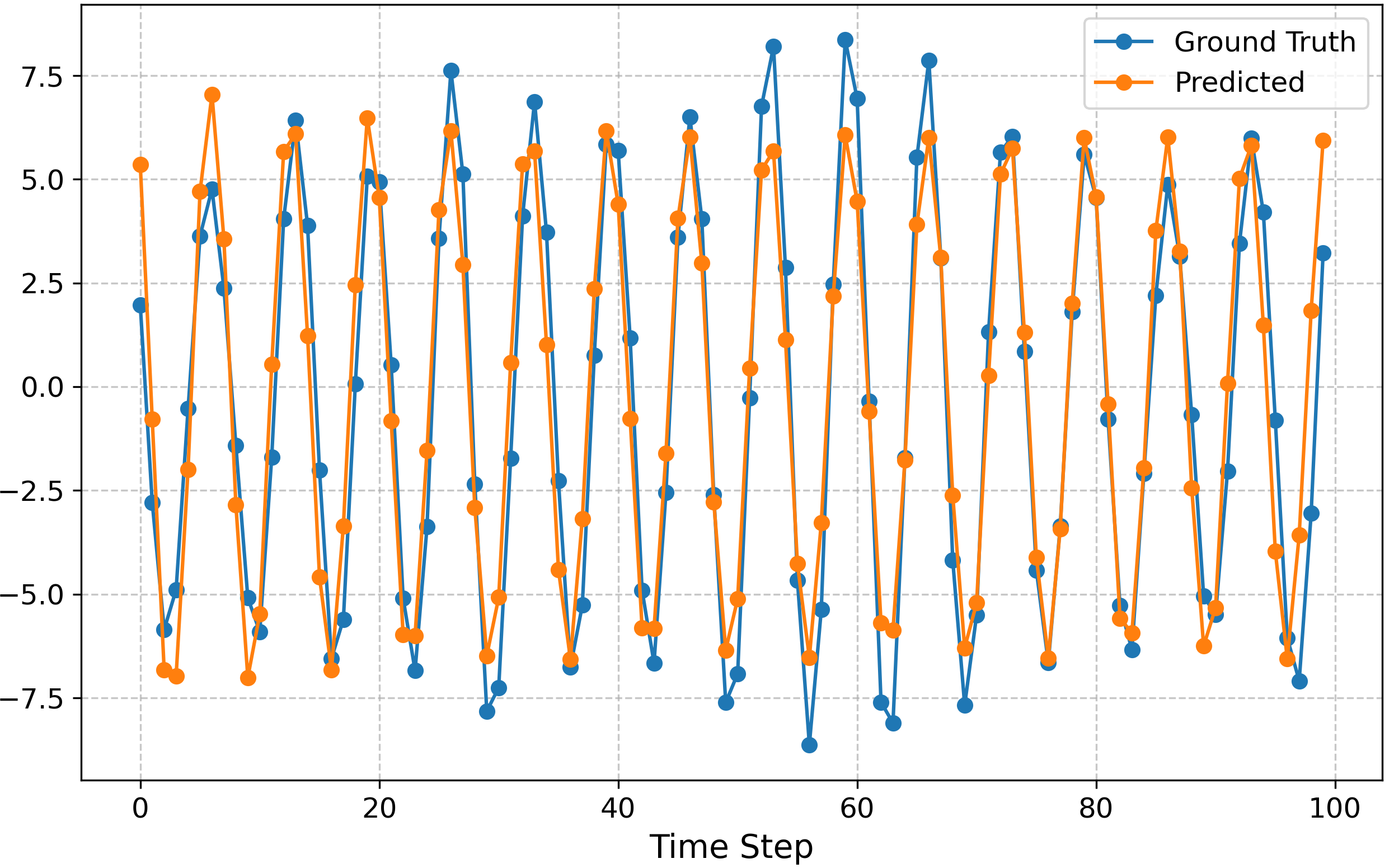}};
            \node[anchor=north, rotate=90, font=\footnotesize\bfseries] at (-0.5,2.5) {v};
        \end{tikzpicture}
    \end{minipage}
    \hfill
    \begin{minipage}{0.48\textwidth}
        \centering
        \includegraphics[width=1.9\textwidth, height=0.6\textwidth, keepaspectratio]{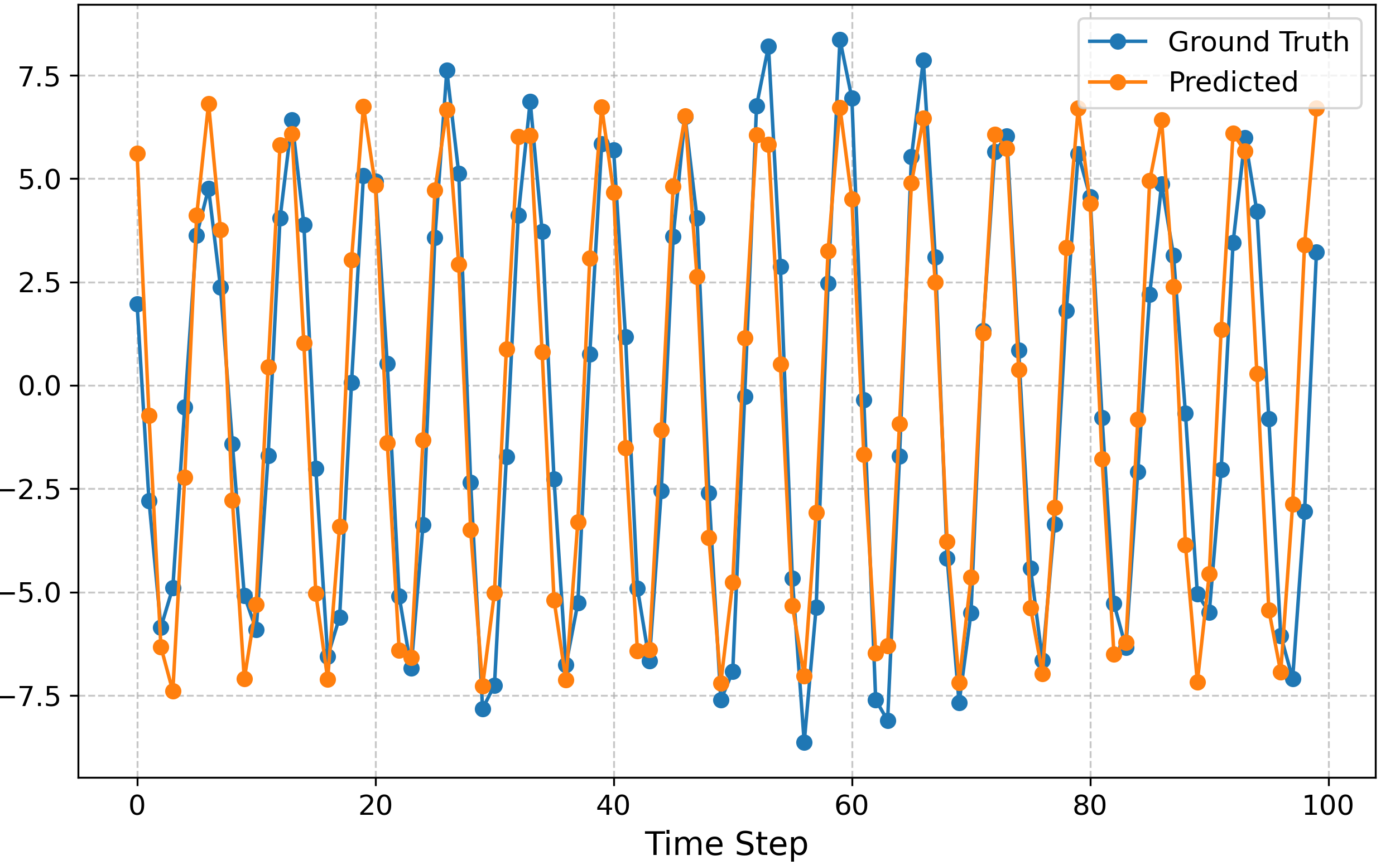}
    \end{minipage}

    \caption{ Comparison of the predicted temporal evolution of normal velocity components for (a) HOSVD (left column) and (b) SVD (right column) across different LSTM architectures for the experimental wake flow. From top to bottom: LSTM 1 Dense, LSTM 2 dense, and LSTM Time-Distributed.}
    \label{fig:normal_velocity_evolution_vki}
\end{figure}

The streamwise and normal velocity components, depicted in Figure~\ref{fig:streamwise_evolution_vki} and ~\ref{fig:normal_velocity_evolution_vki}, reveal that HOSVD and SVD maintain considerable agreement with the ground truth. Both LSTM 2 dense and LSTM time-distributed architectures demonstrate exceptional performance, closely replicating the periodic nature of the streamwise oscillations while minimizing deviations across the time steps. These results highlight the robustness of decomposition techniques in capturing complex temporal dynamics. In comparison, the LSTM 1 dense architecture, while still effective, displays minor discrepancies, particularly at the peaks of the oscillations and over increasing time steps. The HOSVD-based LSTM 2 Dense and LSTM Time-Distributed architectures exhibit a superior ability to capture finer-scale variations at the smaller crests and troughs compared to the SVD-based models, specifically at the streamwise temporal evolution plots.

\begin{figure}[h!]
    \centering
    \makebox[\textwidth]{\textbf{(a) HOSVD} \hspace{5cm} \textbf{(b) SVD}} \\[5pt]
    \begin{minipage}{0.48\textwidth}
        \centering
        \includegraphics[width=1.9\textwidth, height=0.57\textwidth, keepaspectratio]{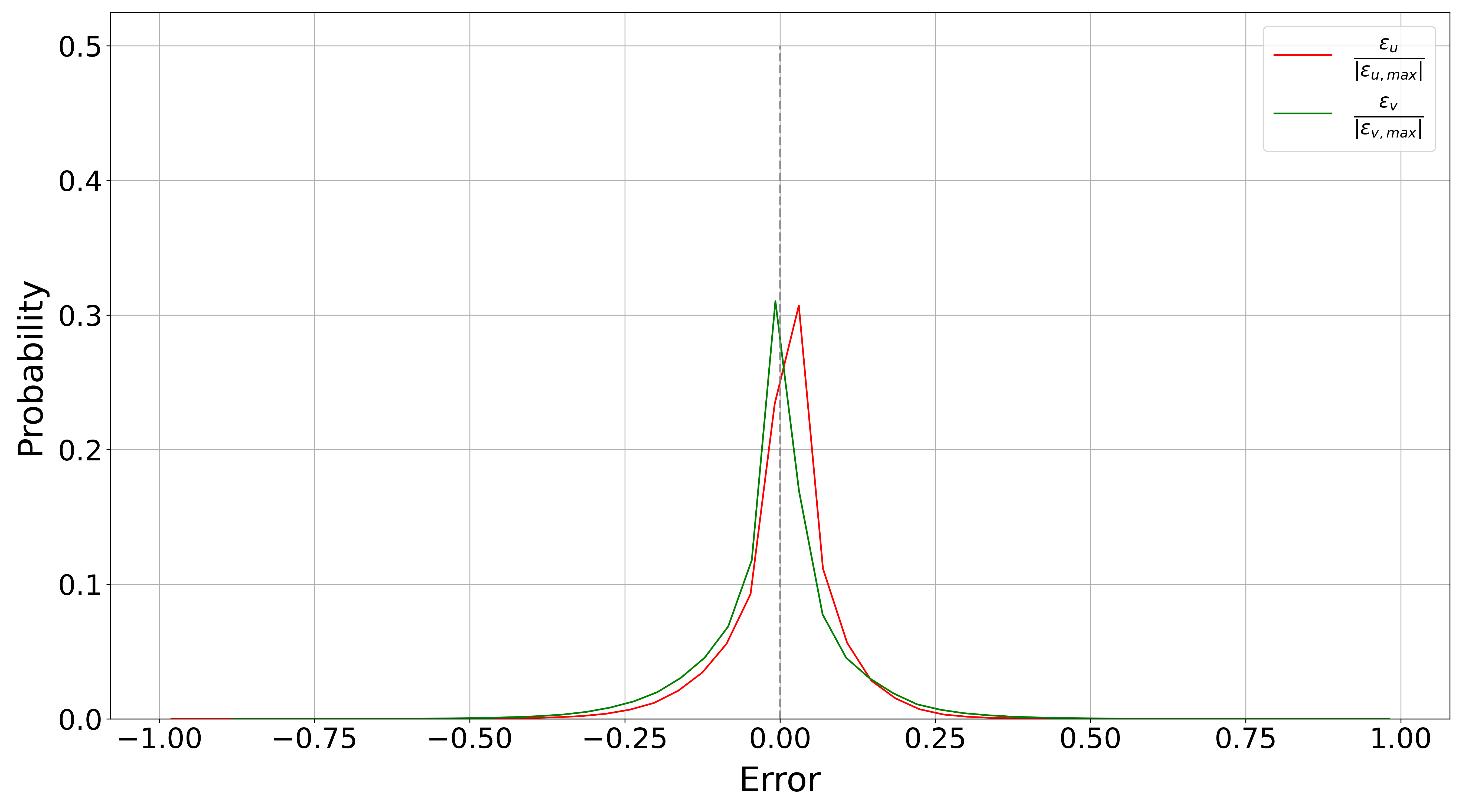}
    \end{minipage}
    \hfill
    \begin{minipage}{0.48\textwidth}
        \centering
        \includegraphics[width=1\textwidth, height=0.57\textwidth]{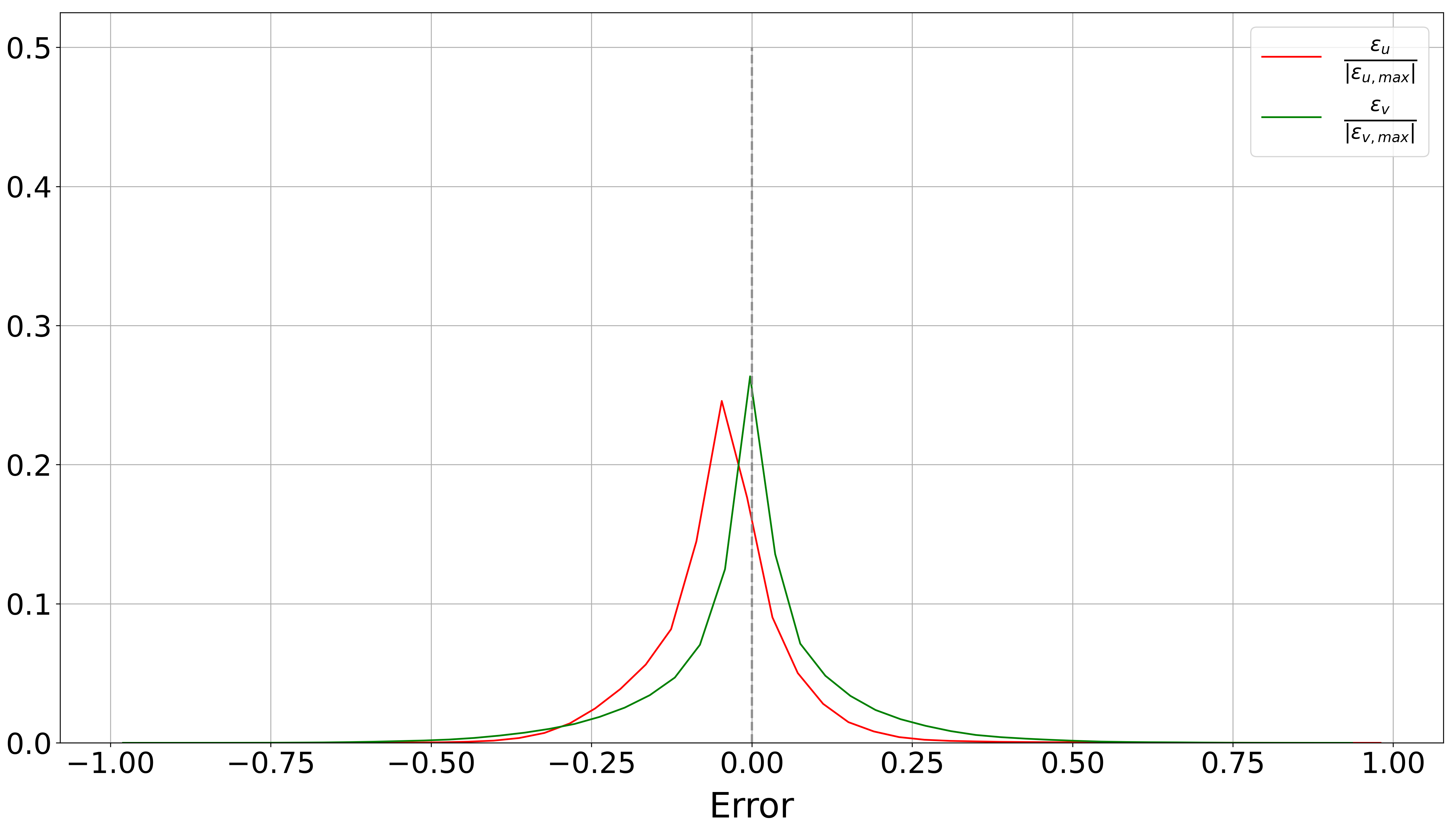}
    \end{minipage}
    \\[10pt]
    \begin{minipage}{0.48\textwidth}
        \centering
        \includegraphics[width=1\textwidth, height=0.57\textwidth]{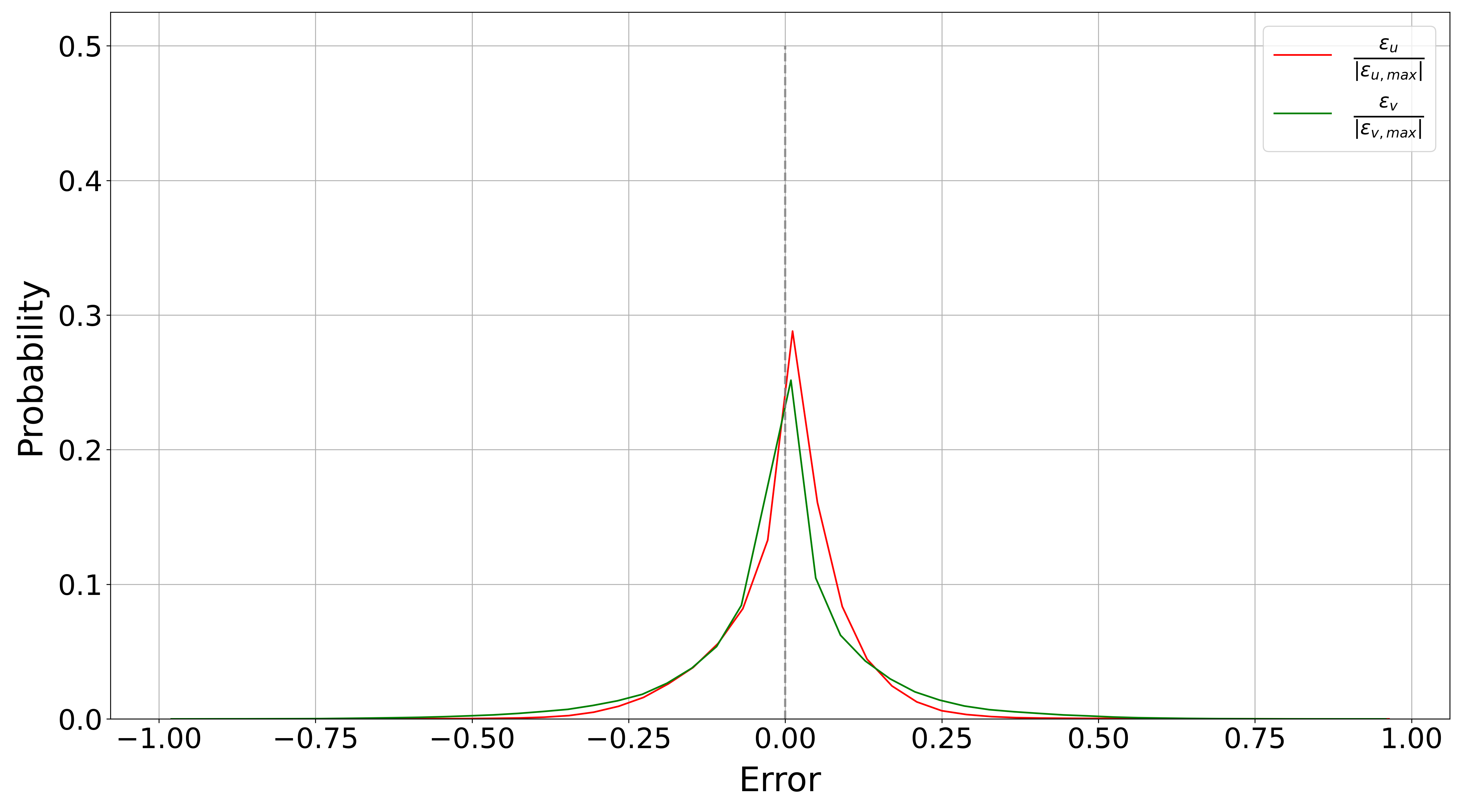}
    \end{minipage}
    \hfill
    \begin{minipage}{0.48\textwidth}
        \centering
        \includegraphics[width=1\textwidth, height=0.57\textwidth]{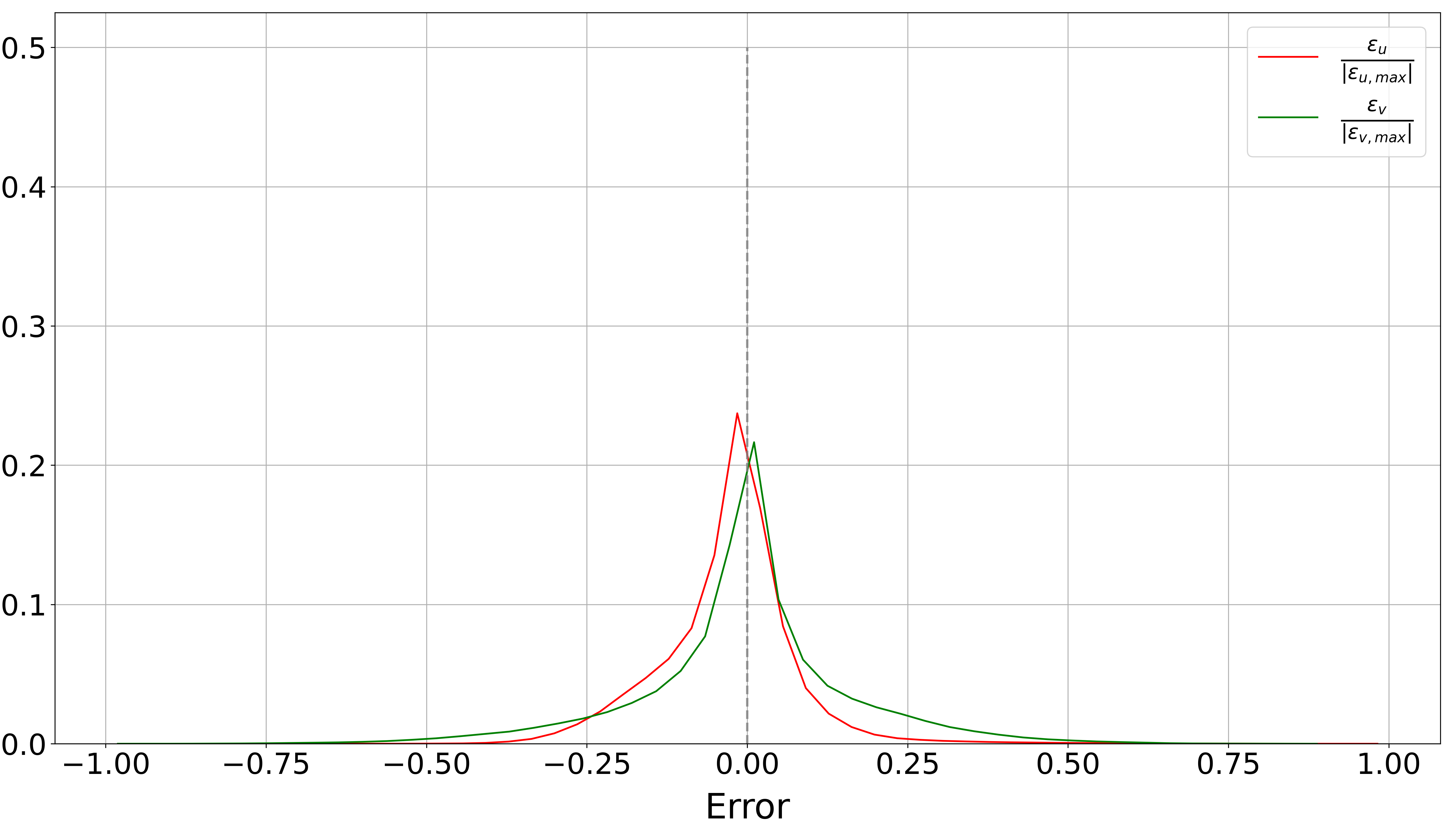}
    \end{minipage}
    \\[10pt]
    \begin{minipage}{0.48\textwidth}
        \centering
        \includegraphics[width=1\textwidth, height=0.57\textwidth]{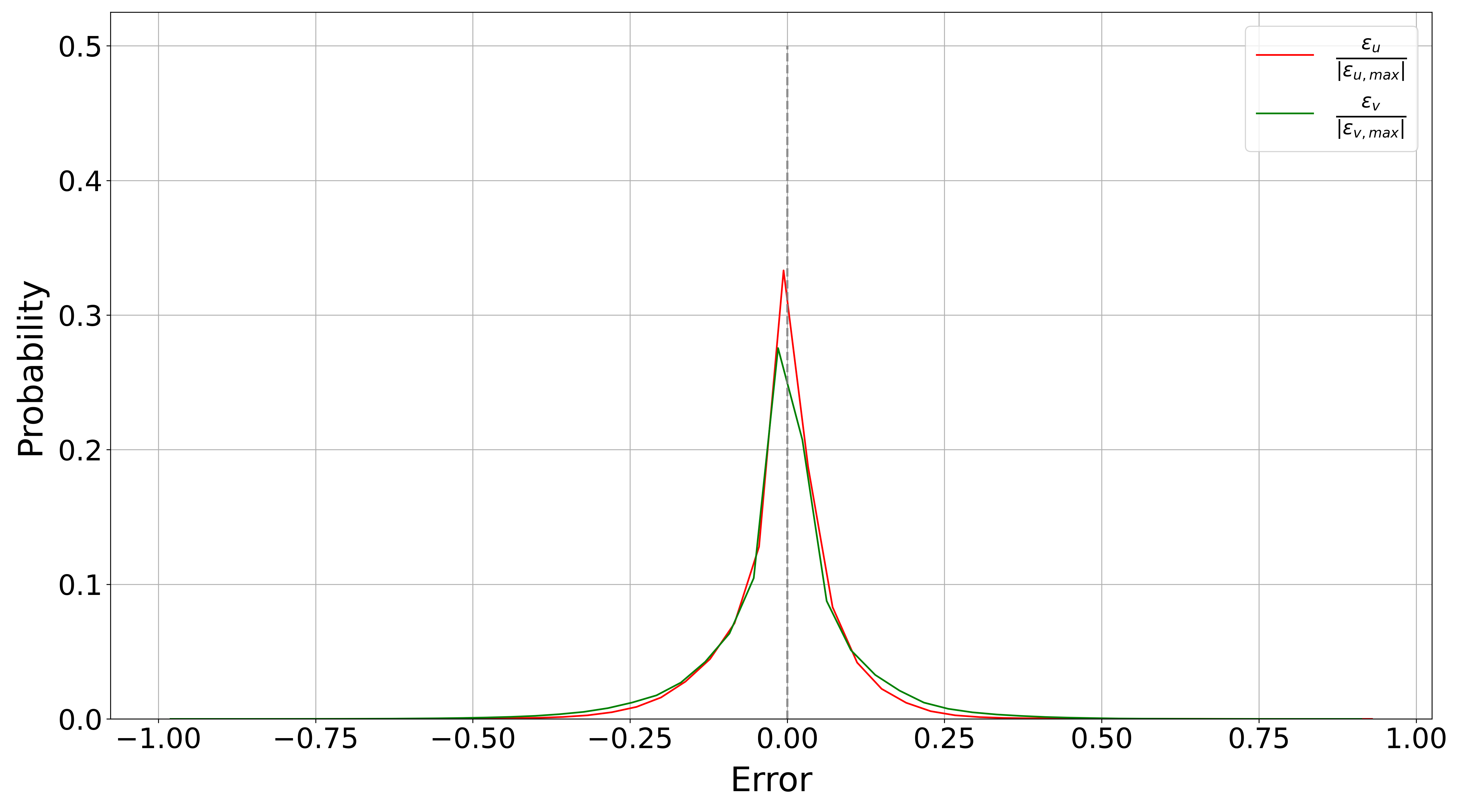}
    \end{minipage}
    \hfill
    \begin{minipage}{0.48\textwidth}
        \centering
        \includegraphics[width=1\textwidth, height=0.57\textwidth]{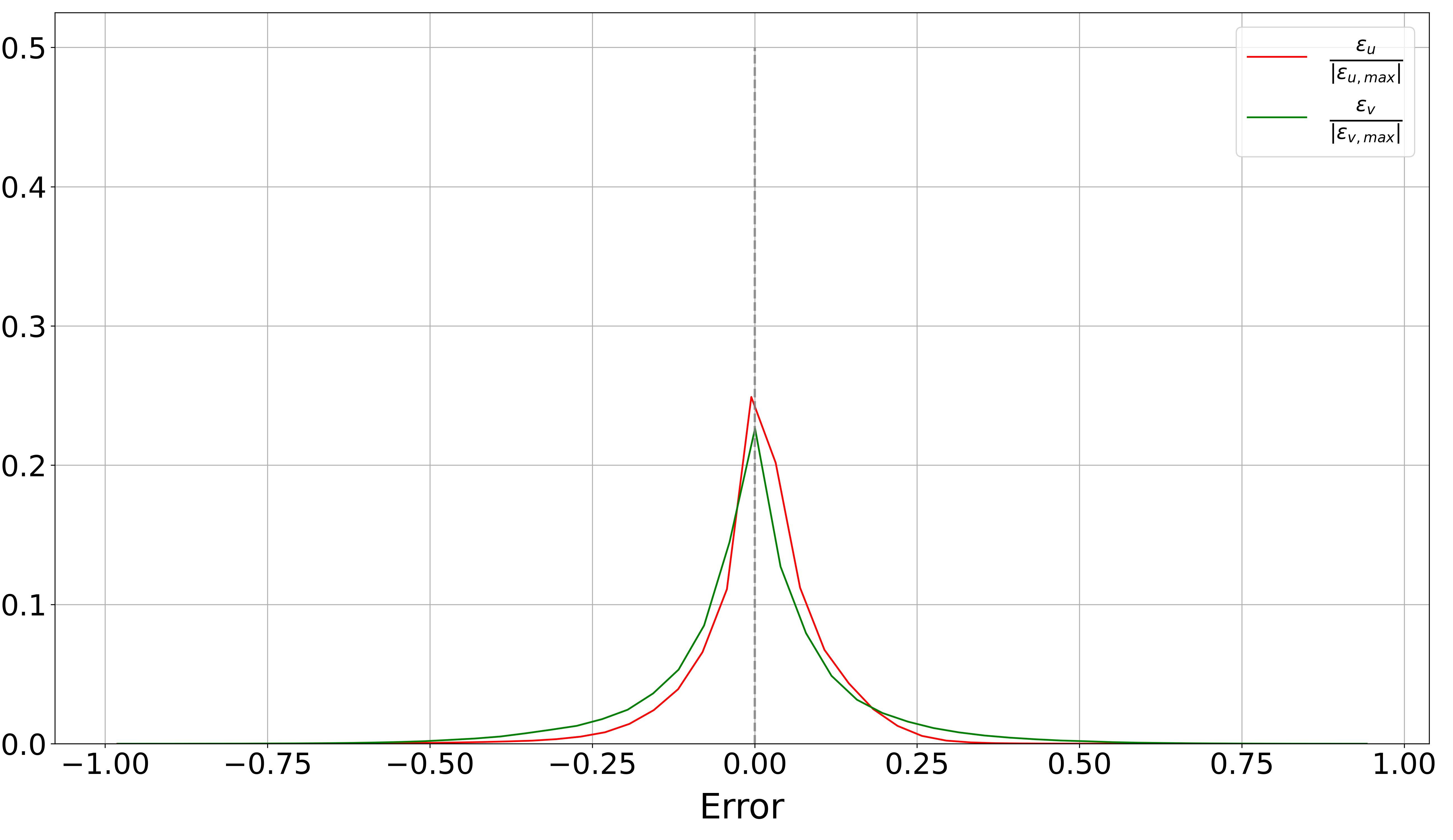}
    \end{minipage}
    \caption{Uncertainty quantification (UQ) results for (a) HOSVD (left column) and (b) SVD (right column) across LSTM architectures. From top to bottom: predictions for LSTM with 1 Dense, 2 Dense, and Time-Distributed architectures for the experimental cylinder flow. Histograms have been constructed using 50 bins.}
    \label{fig:uq_results_vki}
\end{figure}
The UQ results for the experimental wake flow case, shown in Figure~\ref{fig:uq_results_vki}, provide a compelling perspective on predictive reliability in this challenging noise-influenced turbulent data set. For the HOSVD-based models, the error distributions are consistently narrow and sharply centered around zero for all LSTM architectures. This indicates high reliability and minimal variability in the predictions. The sharp peaks in the UQ plots underscore the robustness of HOSVD in capturing the underlying flow features with a high degree of confidence. Among the architectures, the LSTM 2 dense and time-distributed show the best performance, with the narrowest error distribution. The LSTM 1 dense architecture also demonstrates strong performance, but the streamwise velocity is not centered at the zero error line.

In contrast, SVD-based models display broader error distributions across all LSTM architectures, indicating higher variability and reduced reliability in the predictions. The peaks are less pronounced compared to HOSVD, and the tails of the distributions extend further, suggesting that SVD struggles to capture the intricate flow features. The streamwise velocity component is not centered at zero. Notably, the results for the SVD-based LSTM time-distributed architecture are relatively better, with a more pronounced peak and reduced variability compared to the other SVD architectures. However, even in this case, SVD falls short of the precision exhibited by HOSVD.

To further assess the robustness of HOSVD,  noisy modes were introduced in the training set. Simulations were carried out using 20 modes and 50 modes. While the RRMSE values for the streamwise velocity component follow similar trends as obtained with 5 modes, owing to its stronger flow features, the normal velocity component warrants closer attention. This component comprises very low-velocity regions, and even a small amount of noise will have a considerable impact. The RRMSE values reported in Table~\ref{tab:rrmse_noisy_modes} demonstrate that HOSVD exhibits superior performance over SVD, where the error is reduced by 22\% for 20 modes and 18.1\% for 50 modes. This underscores HOSVD’s greater effectiveness in capturing flow dynamics under noisy conditions in complex flow fields.
\begin{table}[h!]
\centering
\begin{tabular}{|c|c|c|c|c|}
\hline
\textbf{Modes} 
& \multicolumn{2}{c|}{\textbf{HOSVD (RRMSE \%)}} 
& \multicolumn{2}{c|}{\textbf{SVD (RRMSE \%)}} \\ \cline{2-5}
& \textbf{Streamwise} & \textbf{Normal} & \textbf{Streamwise} & \textbf{Normal} \\ \hline
20 & 6.9 & 42 & 8.7 & 64 \\ \hline
50 & 9.3 & 55.9 & 11 & 74 \\ \hline
\end{tabular}
\caption{Comparison of RRMSE values (\%) for HOSVD and SVD with added noisy modes in the training data.}
\label{tab:rrmse_noisy_modes}
\end{table}

The results obtained sufficiently demonstrate the superiority of HOSVD over SVD in modeling flow dynamics across all test cases. For the laminar datasets, the LSTM 1 dense architecture, when paired with HOSVD, was sufficient to achieve accurate predictions, effectively capturing simpler dynamics. However, the addition of one or more dense layers in the LSTM architecture displayed an improvement across various metrics, specifically along the spanwise and normal directions. The LSTM 2 dense and LSTM time-distributed configurations enhanced the predictive accuracy across all the datasets. The RRMSE metrics further reinforced these observations, with HOSVD achieving lower error values across all LSTM architectures. In particular, the LSTM 2 dense architecture achieved the lowest RRMSE for HOSVD, underscoring its ability to balance complexity and prediction accuracy in turbulent flow cases. They help to preserve uncorrelated dynamics connected to small flow scales typically found in transient and turbulent flows.

The comparison between the HOSVD-based LSTM 2 dense and the time-distributed model reveals minimal differences in metrics and snapshot predictions across all cases. However, the time-distributed architecture performs better in the spanwise predictions for the 3D cylinder and for both the components in the experimental and 2D cylinder cases for the SVD-based models. SVD-based models decompose the original dataset, but they do not explicitly model temporal dependencies like HOSVD. The time-distributed model helps reconstruct more effectively by applying transformations uniformly to each output across each time step, thereby improving forecasting accuracy. Comparisons were also drawn with another architecture developed with an additional dense layer (Appendix A1) without substantial improvements. This suggests that increasing the depth of the network beyond certain layers does not provide any significant advantage in terms of accuracy and performance. Adding additional layers should be avoided to minimize computational costs unless they provide a significant performance improvement.

\section{Conclusion}

The developed models have demonstrated how integrating HOSVD with LSTM architectures enables the development of efficient and reliable hybrid ROMs for fluid flow prediction. By combining advanced dimensionality reduction techniques with sequential deep learning, this work addresses the challenges of capturing complex spatio-temporal dynamics across a range of laminar and turbulent flow cases, including 2D cylinder, 3D cylinder, and experimental wake flows.

The results presented above highlight the superiority of HOSVD over SVD in both predictive accuracy and robustness. HOSVD effectively preserves the multidimensional structure of the data, allowing for more accurate reconstruction of key flow features while minimizing errors. This is particularly evident in turbulent data sets, where SVD struggled to model chaotic and nonlinear dynamics, as shown by broader error distributions and higher RRMSE values. Across all data sets, HOSVD consistently achieved narrower UQ distributions, indicating greater reliability and confidence in the predictions. In the current implementation of HOSVD, all spatial modes are retained, and only the temporal dimension is reduced. While this preserves the full spatial variability, it limits the denoising and compression advantages that HOSVD offers. In future work, mode selection across all tensor dimensions will be explored, taking advantage of HOSVD's ability to retain different numbers of modes per direction.

The study also investigated the interplay between network depth, complexity, and predictive performance, focusing on determining the optimal complexity required for effective optimization. Simpler LSTM architectures with a single dense layer demonstrated sufficient capability to model the flows, effectively capturing periodic patterns while maintaining computational efficiency. Although these simple architectures produced reasonable predictions, incorporating an additional dense layer improved the network's ability to represent the inherent non-linearities and chaotic dynamics of such data sets. The additional dense layer expands the network's representational capacity, enabling it to better capture intricate flow patterns. These findings emphasize the necessity of tailoring the network depth to the complexity of the flow regime, striking a balance between computational simplicity and model expressiveness to achieve optimal performance.

The integration of HOSVD and LSTM architectures provides a flexible and scalable framework for modeling fluid dynamics, enabling applications beyond the test cases presented in this study. While the current architectures have performed well for the cylinder flow cases, future studies can be conducted to further improve these models and test them on more complicated data sets, such as transitory flow regimes. For more intricate and higher-dimensional flow regimes, such as those involving multiphase systems or significantly higher Reynolds numbers, it may be necessary to employ more complex and deeper architectures.

Future research can build on these findings by extending the proposed methodologies to study complex fluid flow phenomena and other multi-physics problems, exploring the role of additional hyperparameter tuning, and integrating more advanced LSTM variants such as attention-based mechanisms. As the need for accurate and efficient ROMs continues to grow, the methods developed in this work offer a robust pathway to advancing predictive modeling capabilities in fluid dynamics, with implications for both scientific research and industrial applications.

%--- Section ---%
\section*{Conflicts of Interest} 
The authors declare that they have no known financial or personal interests that could have influenced the work reported in this paper.

%--- Section ---%

\section*{Code Availability}
The code developed for this study is available at:  
\noindent\url{https://modelflows.github.io/modelflowsapp/deeplearning/}.
%--- Section ---%
\section*{Acknowledgments}
The authors acknowledge the MODELAIR project that has received funding from the European Union’s Horizon Europe research and innovation programme under the Marie Sklodowska-Curie grant agreement No. 101072559. S.L.C. acknowledges the ENCODING project that has received funding from the European Union’s Horizon Europe research and innovation programme under the Marie Sklodowska-Curie grant agreement No. 101072779. The results of this publication reflect only the author's view and do not necessarily reflect those of the European Union. The European Union can not be held responsible for them. The authors acknowledge the grant PLEC2022-009235 funded by MCIN/AEI/ 10.13039/501100011033 and by the European Union “NextGenerationEU”/PRTR and the grant PID2023-147790OB-I00 funded by MCIU/AEI/10.13039 /501100011033 /FEDER, UE. The authors gratefully acknowledge the Universidad Politécnica de Madrid (www.upm.es) for providing computing resources on the Magerit Supercomputer.

%-------------------------------------------
% References
%-------------------------------------------

% Print bibliography
%\printbibliography
%\input{paper.bbl}
%\bibliography{jsbgf}

%-------------------------------------------
% Appendix
%-------------------------------------------
% Activate the appendix in the doc
% from here on sections are numerated with capital letters 
%\appendix

% Change equation numbering format to be sequential within sections in the appendix
\renewcommand\theequation{\Alph{section}\arabic{equation}} % Redefine equation numbering format
\counterwithin*{equation}{section} % Number equations within sections
\renewcommand\thefigure{\Alph{section}\arabic{figure}} % Redefine equation numbering format
\counterwithin*{figure}{section} % Number equations within sections
\renewcommand\thetable{\Alph{section}\arabic{table}} % Redefine equation numbering format
\counterwithin*{table}{section} % Number equations within sections

\begin{appendices}

%--- Section ---%
\section{Appendix}
This appendix is organized into three sections. The first section presents the results of HOSVD with three dense layers, while the second section focuses on the choice of modes for the experimental dataset. The third section provides a detailed description of the important steps involved in building a DL model. 
\subsection{HOSVD with 3 dense layers}

\begin{figure}[h!]
    \centering
      \begin{minipage}{0.48\textwidth}
        \centering
        \begin{tikzpicture}
            \node[anchor=south west, inner sep=0] (image) at (0,0) {\includegraphics[width=0.85\textwidth]{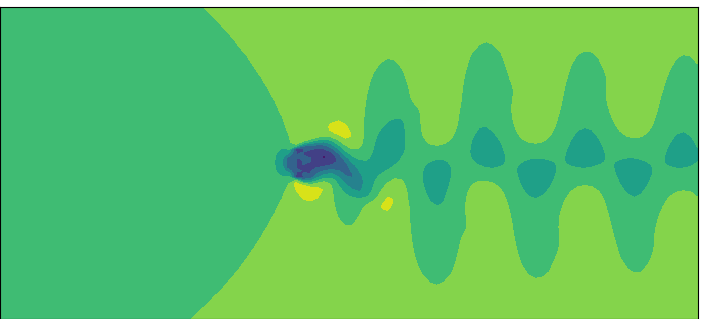}};
            \node[anchor=north, rotate=90, font=\scriptsize] at (-0.6,1.5) {Y};
            \node[anchor=north, font=\scriptsize] at (3.4,-0.2) {X};
        \end{tikzpicture}
    \end{minipage}
    \hfill
    \begin{minipage}{0.48\textwidth}
        \centering
        \begin{tikzpicture}
            \node[anchor=south west, inner sep=0] (image) at (0,0) {\includegraphics[width=0.85\textwidth]{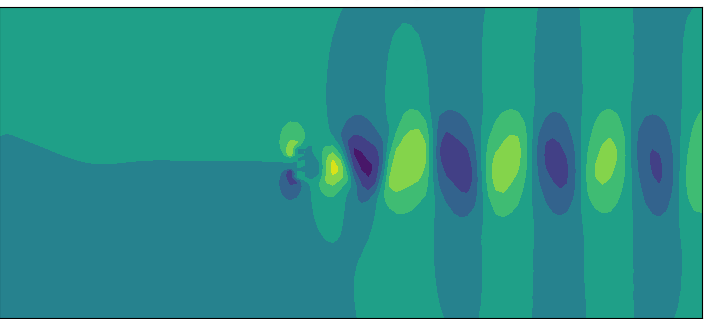}};
            \node[anchor=north, font=\scriptsize] at (3.4,-0.2) {X};
        \end{tikzpicture}
    \end{minipage}
    \\[10pt]
    \caption{Predicted snapshots of LSTM 3 Dense architecture. From left to right: streamwise velocity(left), and normal velocity (right) components.}
    \label{fig:svd_snapshots_mean}
\end{figure}

Figure~\ref{fig:svd_snapshots_mean} shows the predicted snapshots of the HOSVD-based model for the LSTM architecture with 3 dense layers, and the hyper-tuned parameters are shown in Table \ref{tab:3dense_cylinder}.

\begin{table}[h!]
\centering
\begin{tabular}{|c|c|c|c|}
\hline
\multicolumn{4}{|c|}{\textbf{2D Cylinder}} \\ \hline
\textbf{Architecture} & \textbf{Learning Rate} & \textbf{Batch Size} & \textbf{Sequence Length} \\ \hline
LSTM 3 Dense          & 0.001                  & 28                  & 10                       \\ \hline
\end{tabular}
\caption{Tuned hyperparameters for 2D cylinder Flow.}
\label{tab:3dense_cylinder}
\end{table}

\begin{figure}[h!]
    \centering
    \begin{minipage}{0.48\textwidth}
        \centering
        \begin{tikzpicture}
            \node[anchor=south west, inner sep=0] (image) at (0,0) {\includegraphics[width=1\textwidth, height=0.6\textwidth]{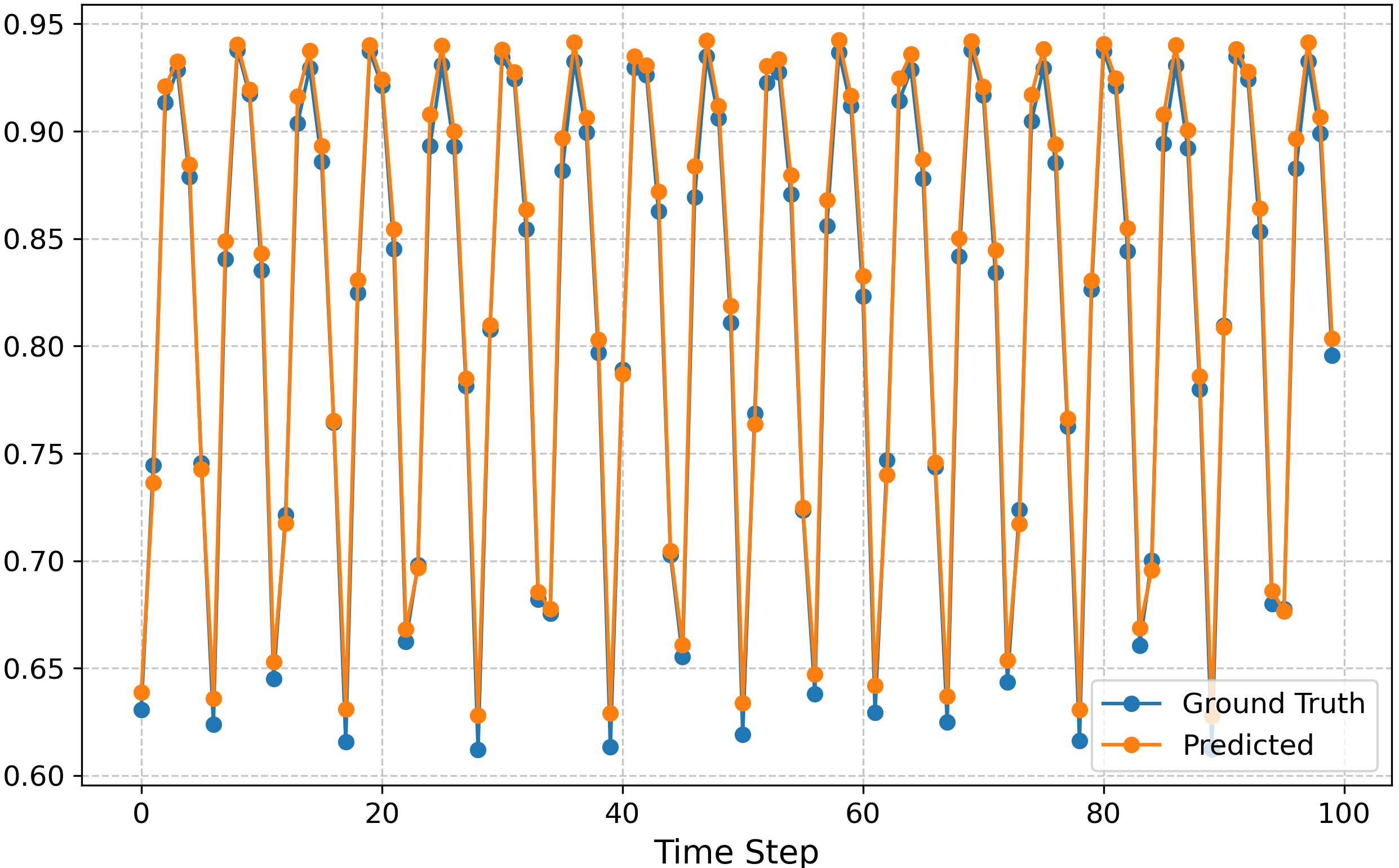}};
            \node[rotate=90, font=\scriptsize] at (-0.3,2.5) {u}; % Adjust position if needed
        \end{tikzpicture}
    \end{minipage}
    \hfill
    \begin{minipage}{0.48\textwidth}
        \centering
        \begin{tikzpicture}
            \node[anchor=south west, inner sep=0] (image) at (0,0) {\includegraphics[width=1\textwidth, height=0.6\textwidth]{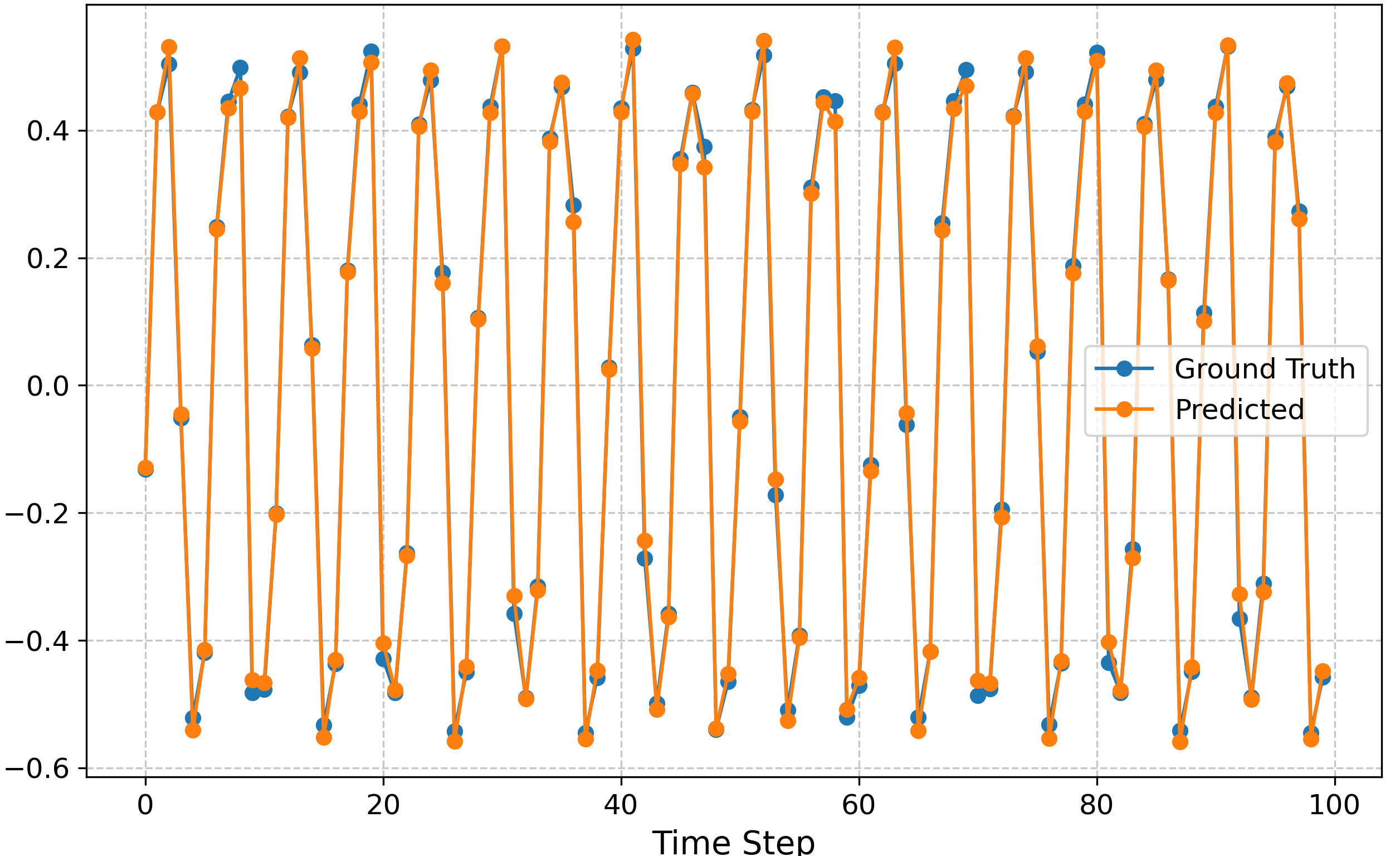}};
            \node[rotate=90, font=\scriptsize] at (-0.3,2.5) {v}; % Adjust position if needed
        \end{tikzpicture}
    \end{minipage}
    \caption{Temporal evolution for HOSVD with LSTM 3 Dense architecture. From left to right: Streamwise velocity and Normal velocity components.}
    \label{fig:temporal_evolution_3d}
\end{figure}

Figure~\ref{fig:temporal_evolution_3d} illustrates the temporal evolution of streamwise and normal velocity components for the same configuration.

\begin{figure}[h!]
    \centering
    \includegraphics[width=0.8\textwidth]{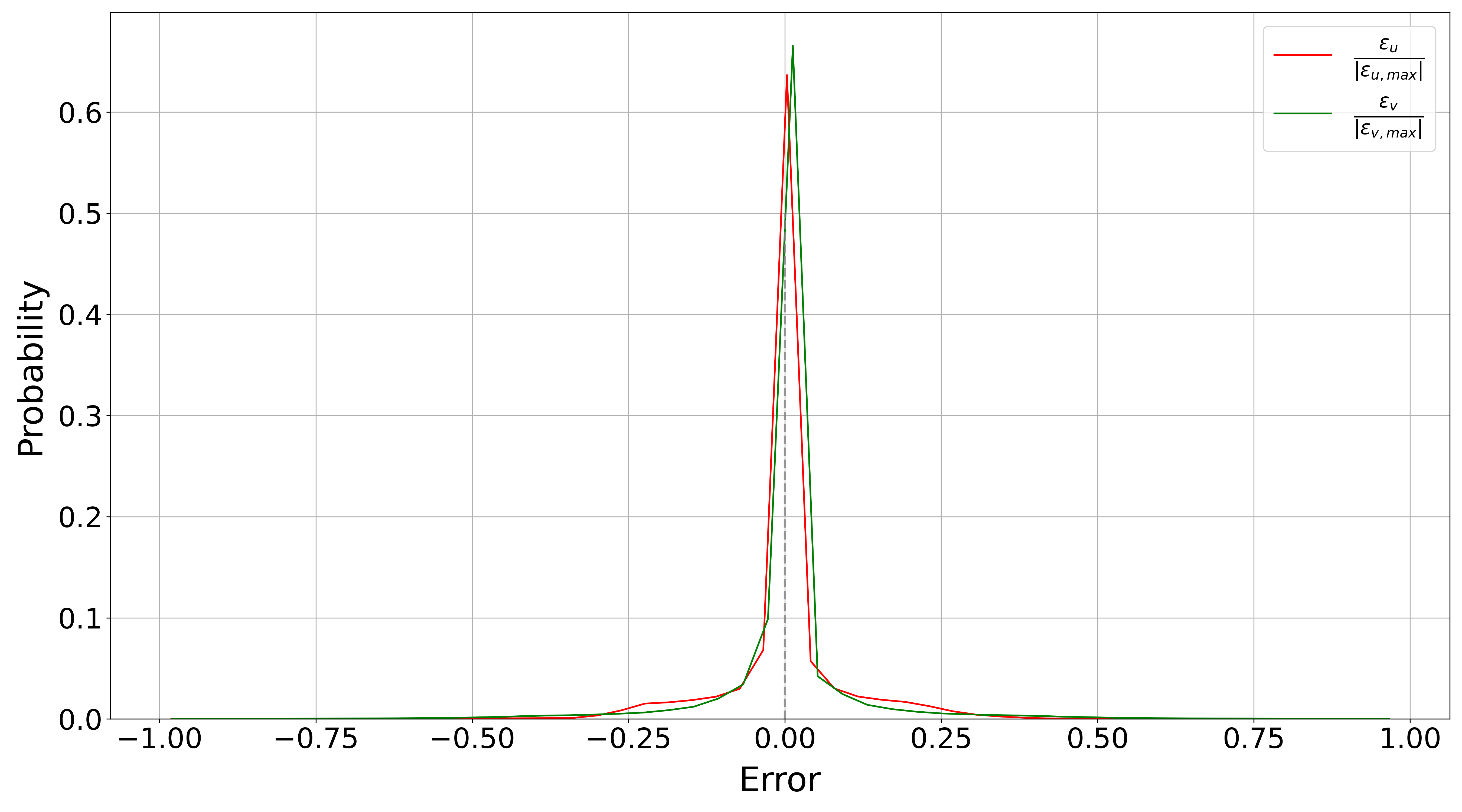}
    \caption{Uncertainty quantification (UQ) results for HOSVD with  LSTM 3 Dense architecture.}
    \label{fig:uq_3d}
\end{figure}

Figure~\ref{fig:uq_3d} presents the UQ results for the configuration.

The average RRMSE for this configuration is 0.37\% (streamwise Component) and 4.7\% (Normal Component).

\subsection{Analysis of the POD Modes}

\begin{figure}[h!]
    \centering
    \includegraphics[width=0.85\textwidth]{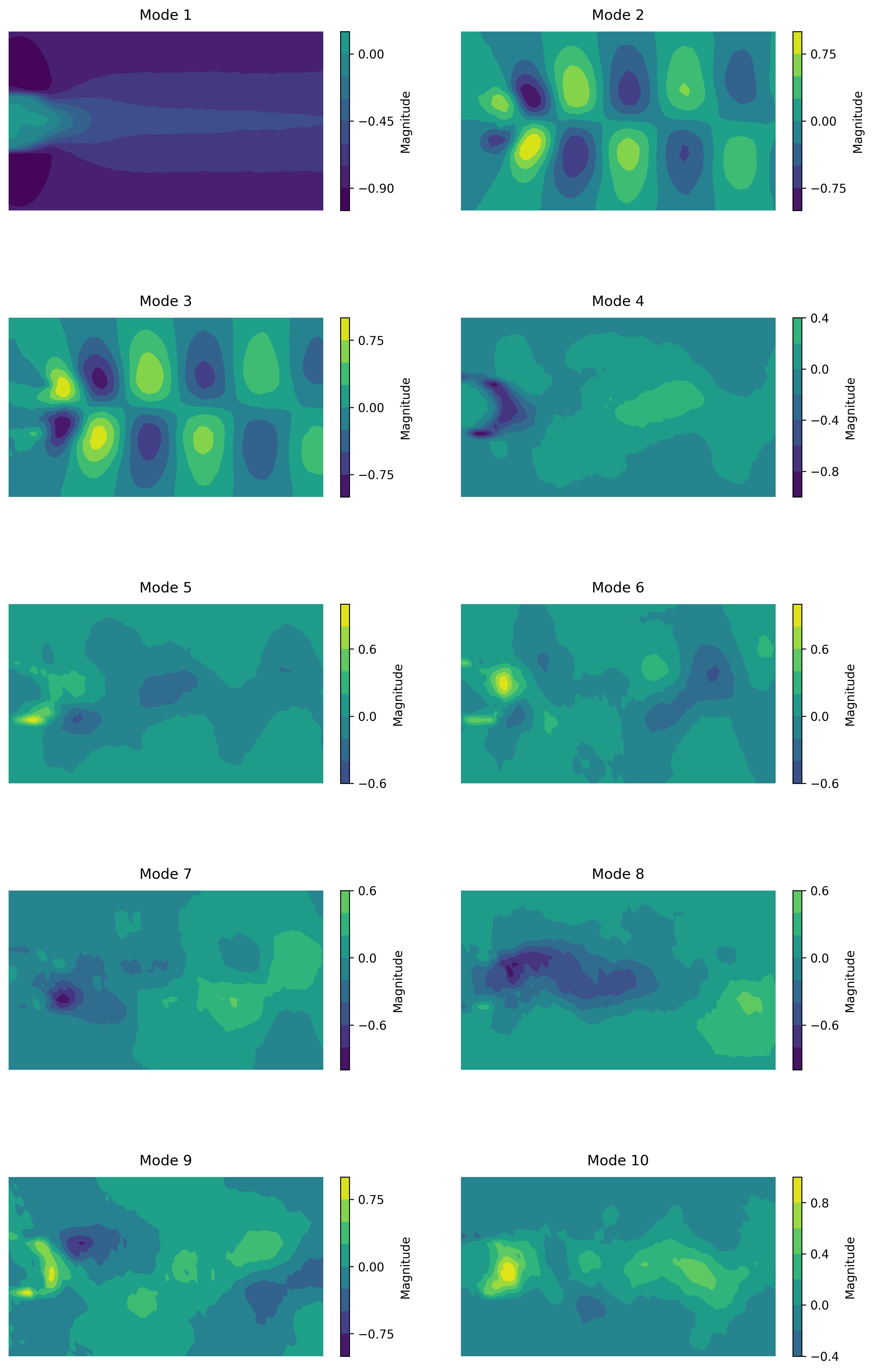} % Adjust the path if needed
    \caption{First ten spatial POD modes from the turbulent experimental cylinder dataset. }
    \label{fig:POD_spatial_modes}
\end{figure}

Figure \ref{fig:POD_spatial_modes} presents the first ten normalized spatial POD modes from the turbulent experimental cylinder case, highlighting the different flow regions. The spatial POD modes from 1 to 5 capture the dominant large-scale structures in the flow, including the wake and the near-field structures, while the remaining modes predominantly capture noise or uncorrelated events. These higher-order modes contain small-scale flow structures that are difficult to predict accurately.

Also, the number of snapshots retained in the snapshot matrix significantly influences the noise content in the POD modes. Increasing the number of snapshots leads to more noise accumulation in modes 3, 4, and 5, making it harder for the method to extract clean, coherent structures. This accumulation suggests that for large datasets, applying HOSVD instead of SVD may provide better noise filtering and improved mode separation.

The evolution of the first ten POD mode coefficients from the turbulent experimental cylinder dataset has been plotted below in Figure \ref{fig:POD_modes}. 500 snapshots were used to create the snapshot matrix. The trend followed by the temporal coefficients is in good agreement with the previous description. In modes 2 and 3, it is possible to identify a periodic behavior. The complexity of these temporal patterns increases for the remaining modes. For modes higher than 6, the noise component becomes very strong. 

\begin{figure}[h!]
    \centering
    \includegraphics[width=0.85\textwidth]{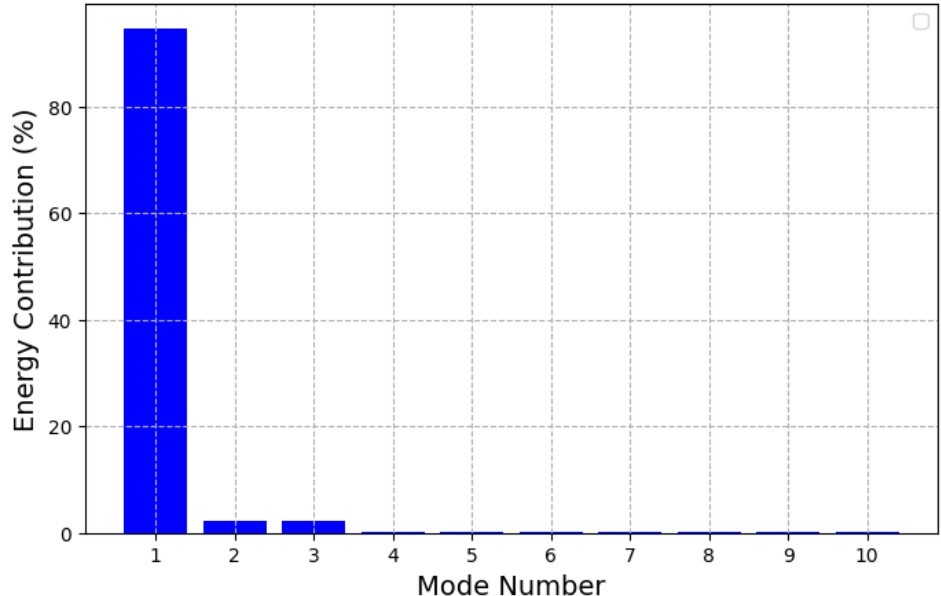} % Adjust the path if needed
    \caption{Energy contained in the first ten POD modes from the turbulent experimental cylinder dataset.}
    \label{fig:Energy_modes}
\end{figure}

\begin{figure}[h!]
    \centering
    \includegraphics[width=0.85\textwidth]{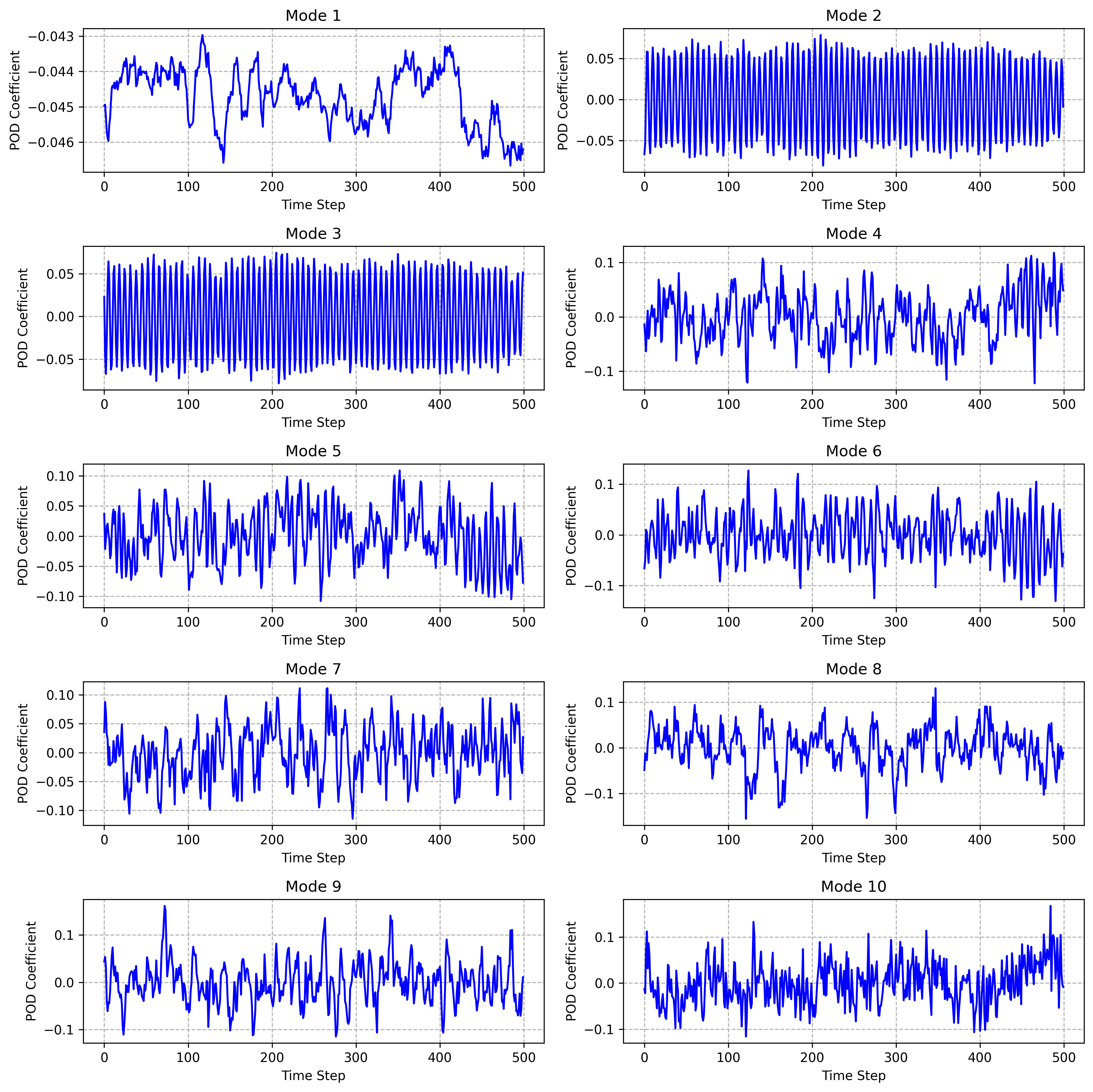} % Adjust the path if needed
    \caption{Time evolution of the first ten POD mode coefficients from the turbulent experimental cylinder dataset.. }
    \label{fig:POD_modes}
\end{figure}

\subsection{Key Considerations for Developing Deep Learning Models}

While some information has already been provided earlier in the paper, this section provides a detailed discussion of the key considerations for developing a deep learning (DL) model.
\subsubsection{Data Management}
Beyond dividing data into training, validation, and test sets, other critical factors must also be considered. The other main points of focus while designing these sets should be the following:
\begin{itemize}
    \item \textbf{Consistency Across Dev and Test Sets:} It should be ensured that the dev and test sets come from the same data distribution to ensure the generalization of the model to unseen data. Misaligned distributions can lead to wasted efforts in optimizing dev performance without achieving comparable test accuracy \cite{coursera_deeplearning}.
    
    \item \textbf{Handling Imbalanced data sets:} When dealing with imbalanced data sets, techniques such as stratified sampling or data augmentation can ensure that all classes are adequately represented in each set, preserving the integrity of the evaluation process \cite{he2013imbalanced}.
    \item \textbf{Cross-Validation for Limited Data:} In cases where data is scarce, cross-validation can be employed to maximize the use of available data while ensuring robust evaluation \cite{coursera_deeplearning}. By rotating the dev and test sets across folds, the model's performance can be assessed comprehensively.
\end{itemize}
\subsubsection{Normalization and Vectorization}
Normalization is a crucial preprocessing step in neural network optimization, especially for deep learning applications.  Normalization plays a critical role in mitigating the vanishing and exploding gradient problems in deep networks. In very deep architectures, gradients can shrink or grow exponentially as they propagate through layers, leading to inefficient or unstable training. Normalization helps stabilize the gradient flow by ensuring that the input values are within a range suitable for neural network computations, thus reducing the risk of extreme gradient behavior \cite{coursera_deeplearning}.

\begin{figure}[h!]
    \centering
      \begin{minipage}{0.48\textwidth}
        \centering
        \begin{tikzpicture}
            \node[anchor=south west, inner sep=0] (image) at (0,0) {\includegraphics[width=0.85\textwidth]{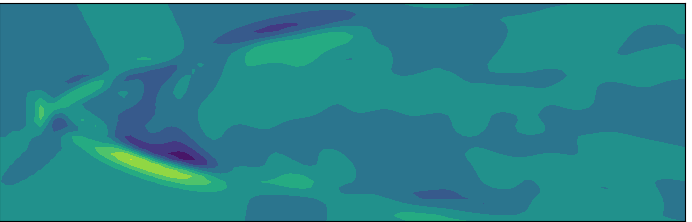}};
            \node[anchor=north, rotate=90, font=\scriptsize] at (-0.6,1) {Y};
            \node[anchor=north, font=\scriptsize] at (3.4,-0.2) {X};
        \end{tikzpicture}
    \end{minipage}
    \hfill
    \begin{minipage}{0.48\textwidth}
        \centering
        \begin{tikzpicture}
            \node[anchor=south west, inner sep=0] (image) at (0,0) {\includegraphics[width=0.85\textwidth]{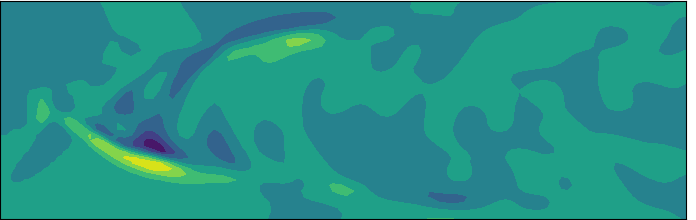}};
            \node[anchor=north, font=\scriptsize] at (3.4,-0.2) {X};
        \end{tikzpicture}
    \end{minipage}
    \\[10pt]
    \caption{Predicted snapshots of spanwise velocity from the 3D cylinder case. From left to right: SVD-based LSTM 1 Dense (left) and LSTM 2 Dense (right) architectures.}
    \label{fig:svd_snapshots}
\end{figure}

The choice of an appropriate normalization technique is critical for ensuring accurate model performance. For comparison, Z-score and min-max normalization were implemented and examined across each case. The results indicate that min-max normalization performed significantly worse for both the laminar and turbulent datasets. As shown in Figure~\ref{fig:svd_snapshots}, the spanwise predictions generated by the SVD-based LSTM 1 and 2 dense models for the 3D cylinder case fail to capture the flow features and dynamics, with RRMSE values exceeding 70\%. These findings underscore the importance of selecting an effective normalization technique to enhance predictive accuracy.

Vectorization is another core technique implemented in deep learning to enhance computational efficiency. Instead of processing each data point individually, vectorization uses matrix operations to handle multiple training examples simultaneously, reducing computations between layers \cite{coursera_deeplearning}. By representing inputs, weights, and activations as matrices, forward and backward propagation can be executed more efficiently, reducing the computational overhead for large data sets.

\subsubsection{Activation Functions and Initialization}

Deep learning models are dependent on well-structured architectures and proper initialization to achieve effective learning. While the design of the neural network provides the framework, initialization plays a pivotal role in ensuring the stability and efficiency of the optimization process. This subsection discusses the core components of neural networks, focusing on the importance of activation functions and initialization techniques.

\paragraph{Activation Functions:}
Activation functions are essential for introducing nonlinearity into a neural network, enabling it to model complex relationships \cite{geron2022hands}. Without them, a network would only compute linear transformations, regardless of its depth, limiting its ability to capture non-linear patterns in data. The most common activation functions are:
\begin{itemize}

    \item \textbf{Sigmoid Function:} The sigmoid activation function produces values between 0 and 1. However, it suffers from the vanishing gradient problem, especially in deeper networks, as gradients become exceedingly small for large positive or negative input values, thereby slowing down training.

    \item \textbf{Tanh Function:} The hyperbolic tangent (\textit{tanh}) function produces outputs between -1 and 1, centering activations around zero. This property often results in faster convergence during training compared to the sigmoid function. Nonetheless, \textit{tanh} also experiences the problem of vanishing gradients in very deep networks, which limits its effectiveness.

    \item \textbf{ReLU (Rectified Linear Unit):} The ReLU activation function, defined as \( \max(0, z) \), has become the most widely used activation function for hidden layers due to its computational efficiency and simplicity. However, it is prone to the "dead neuron" problem, where neurons stop updating weights because gradients become zero for negative inputs.

   \item \textbf{Leaky ReLU:} To address the limitations of ReLU, the Leaky ReLU function introduces a small gradient for negative inputs. This variation mitigates the dead neuron issue, allowing the network to learn more robustly.
\end{itemize}

As mentioned earlier, the choice of activation function often depends on the task and the network architecture. ReLU is typically the default for hidden layers due to its efficiency, while sigmoid or tanh may be employed in specific cases, such as binary classification tasks or when centering activations around zero is beneficial.

\begin{figure}[h!]
    \centering
    \begin{minipage}[t]{0.45\textwidth}
        \centering
        \includegraphics[width=1\textwidth, height=0.7\textheight, keepaspectratio]{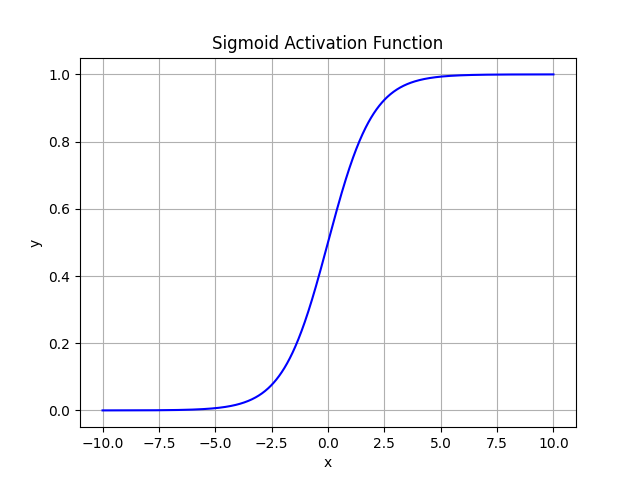}
    \end{minipage}
    \hfill
    \begin{minipage}[t]{0.45\textwidth}
        \centering
        \includegraphics[width=1\textwidth, height=0.7\textheight, keepaspectratio]{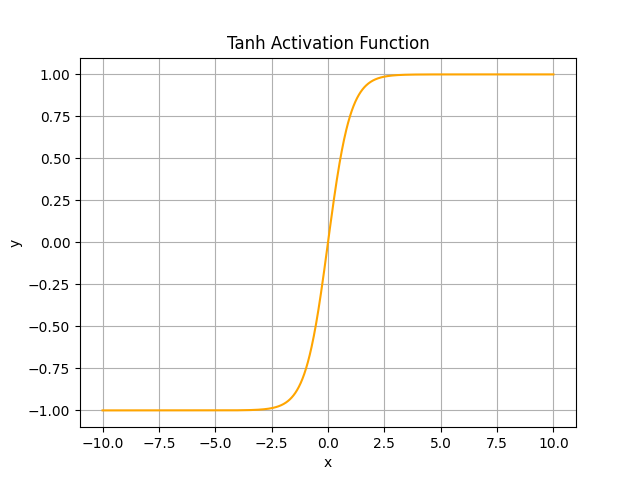}
    \end{minipage}
    \\[15pt]
    \begin{minipage}[t]{0.45\textwidth}
        \centering
        \includegraphics[width=1\textwidth, height=0.7\textheight, keepaspectratio]{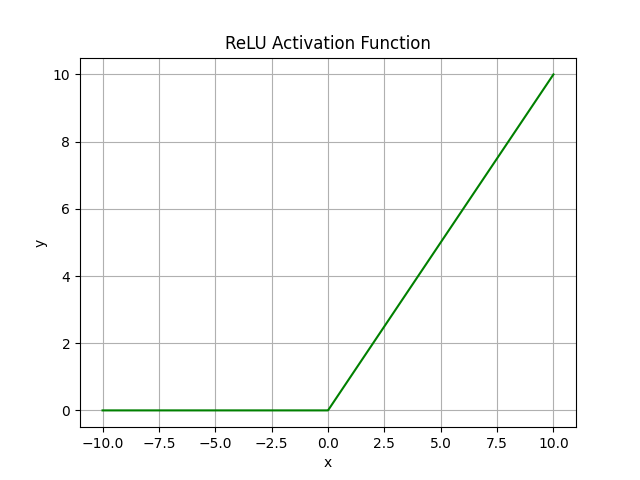}
    \end{minipage}
    \hfill
    \begin{minipage}[t]{0.45\textwidth}
        \centering
        \includegraphics[width=1\textwidth, height=0.7\textheight, keepaspectratio]{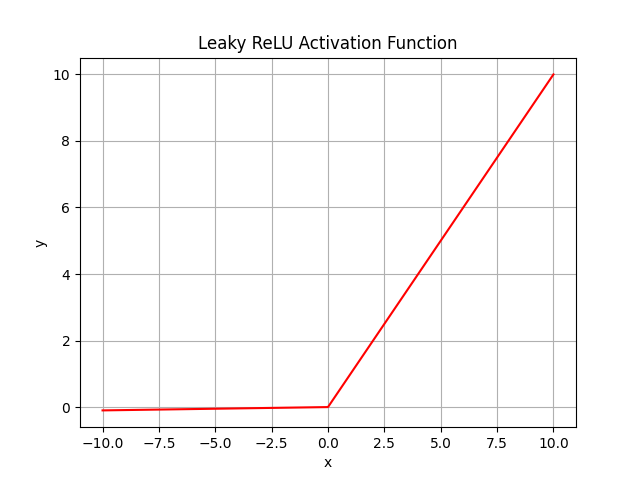}
    \end{minipage}
    \caption{Visualization of activation functions: Top row (left to right): Sigmoid and Tanh. Bottom row (left to right): ReLU and Leaky ReLU.}
    \label{fig:activation_functions}
\end{figure}

\paragraph{Initialization:}
Initialization is critical to ensure that a neural network starts training on the right track. Proper initialization prevents problems such as vanishing or explosion gradients, which can severely hamper learning in deep networks. Key aspects of initialization include \cite{coursera_deeplearning}:
\begin{itemize}
    \item \textbf{Random Initialization}: The weights are initialized randomly to break the symmetry between the neurons, ensuring that different nodes learn different features. Without random initialization, all neurons in a layer would compute identical outputs and gradients, rendering the network ineffective.
    \item \textbf{Scaling Weights}: The weight values must be scaled appropriately based on the activation function used. For example:
    \begin{itemize}
        \item For sigmoid and tanh functions, smaller weights (e.g., scaled by \( 0.01 \)) help prevent activations from saturating, which can lead to vanishing gradients.
        \item For ReLU activation functions, Xavier or He initialization is commonly used to maintain stable gradient magnitudes throughout the network.
    \end{itemize}
    \item \textbf{Bias Initialization}: Bias terms can typically be initialized to zero without causing symmetry issues as long as the weights are initialized randomly.
\end{itemize}

By combining appropriate activation functions and careful initialization techniques, neural networks can effectively learn from data while avoiding common pitfalls, such as vanishing gradients or dead neurons. In this methodology, weight initialization has been implemented using the default strategy provided by TensorFlow/Keras. Specifically, for most layers, Keras applies the \textit{Glorot Uniform} (Xavier Uniform) initializer for the weights and initializes biases to zero. This method ensures that the variance of activations remains stable across layers during training.

\subsubsection{Bias and Variance Tradeoff}

The performance of a deep learning model is often dictated by its ability to balance bias and variance. Bias refers to errors introduced due to overly simplistic models, leading to underfitting, where the model struggles to capture the patterns in the training data. Conversely, variance refers to errors caused by excessive sensitivity to fluctuations in training data, leading to overfitting, where the model performs well in the training set but poorly on unseen data \cite{geman1992neural}.

Diagnosing these issues involves evaluating the training and development (dev) set errors. The most common types are listed below \cite{coursera_deeplearning}:
\begin{itemize}
    \item \textbf{High training error and comparable dev error:} Indicates high bias or underfitting. This occurs when the model is not complex enough to capture the data patterns effectively.
    \item \textbf{Low training error but high dev error:} Suggests high variance or overfitting. The model fits the training data too closely but fails to generalize to new data.
    \item \textbf{High training error with even higher dev error:} Reflects both high bias and high variance, where the model is neither learning the training data nor generalizing well.
\end{itemize}
Strategies to address these issues include increasing model complexity (e.g., adding layers or units) to mitigate high bias. Applying regularization or gathering more data can help to reduce variance. 
\begin{figure}[h]
    \centering
    \includegraphics[width=1\textwidth]{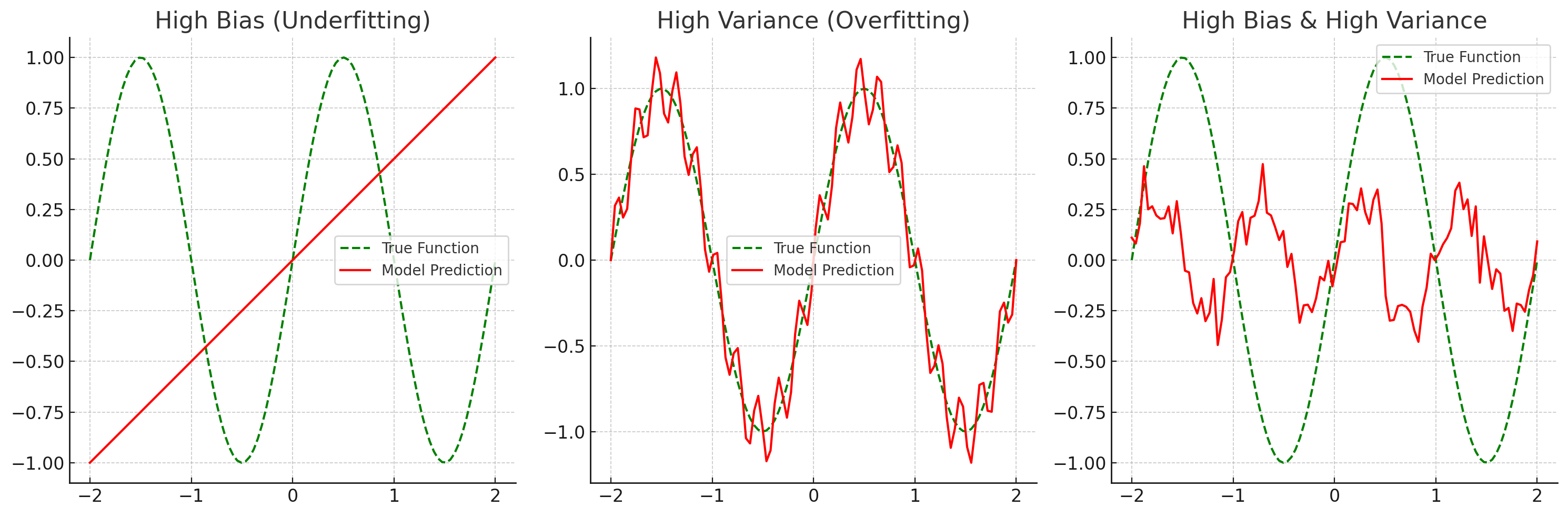}
    \caption{Illustration of different model errors from left to right: high bias (underfitting), high variance (overfitting), and a combination of both.}
    \label{fig:model_errors}
\end{figure}

\subsubsection{Hyperparameter Tuning}

The key hyperparameters, ranked by their influence on model training and generalization, are listed in Table \ref{tab:hyperparameters} \cite{coursera_deeplearning}. These parameters can be tuned using various approaches, with some of the most commonly applied methods being:
\begin{itemize}
     \item \textbf{Grid Search:} This is a systematic and exhaustive search where a predefined set of hyperparameter values is tested across all possible combinations, however, it is computationally expensive. 
    \item \textbf{Random Search:} This technique samples values randomly from a predefined range, enabling efficient exploration of the parameter space. Unlike a grid search, it does not evaluate all possible combinations, making it particularly useful when only a subset of hyperparameters significantly impacts model performance.
    \item \textbf{Coarse-to-Fine Adjustment:} This begins with a broad search throughout the parameter space to identify promising regions, followed by finer searches within those regions to refine the values.
   
    \item \textbf{Bayesian Optimization:} In this work, Bayesian optimization was employed for hyperparameter tuning. The optimization is performed using the Bayesian optimization class from Keras Tuner. Bayesian optimization leverages a Gaussian Process (GP) as a surrogate model to approximate the objective function and an acquisition function to guide sampling towards promising regions \cite{brochu2010tutorial}. This approach efficiently balances exploration and exploitation, reducing the number of function evaluations needed to find the optimal solution.
\end{itemize}

\begin{table}[h!]
\centering
\begin{tabular}{|p{4cm}|p{11cm}|}
\hline
\textbf{Hyperparameter} & \textbf{Description} \\ \hline
\textbf{Learning Rate} & Determines the step size during gradient descent. An optimal learning rate is crucial; a value that is too large can cause divergence, while a value that is too small leads to slow convergence. \\ \hline
\textbf{Batch Size} & Defines the number of training samples processed before updating model weights. Smaller batch sizes provide noisier but more frequent updates, whereas larger batch sizes yield smoother gradients. \\ \hline
\textbf{Number of Layers and Units per Layer} & Specifies the network's depth and capacity to learn complex patterns. Deeper networks can capture intricate relationships but require proper regularization to avoid overfitting. \\ \hline
\textbf{Regularization Strength (\(\lambda\))} & Controls the penalty applied to the loss function, helping prevent overfitting by simplifying the model's complexity. \\ \hline
\end{tabular}
\caption{Key hyperparameters and their impact on model performance.}
\label{tab:hyperparameters}
\end{table}

Efficient hyperparameter tuning is crucial for avoiding underfitting or overfitting. By prioritizing impactful parameters, such as the learning rate and batch size, and using advanced techniques like Bayesian optimization, the model's ability to generalize to unseen data is enhanced.

\subsubsection{Debugging Deep Learning Models}

Debugging plays a pivotal role in ensuring the reliability and robustness of deep learning models. Unlike traditional software debugging, addressing issues in neural networks often requires identifying subtle problems related to data, architecture, hyperparameters, and optimization strategies. This section outlines key aspects of debugging deep learning models, supported by systematic techniques and tools to enhance model performance and reliability.

\paragraph{Key Aspects of Debugging:}
Debugging neural networks involves a wide range of activities, from ensuring data consistency to diagnosing architectural inefficiencies. The following key aspects are critical for this process \cite{coursera_deeplearning}:

\paragraph{Data and Preprocessing:}
Data quality and consistency are fundamental to the success of deep learning models. Debugging in this context focuses on the following:
\begin{itemize}
    \item \textbf{Input Data Inspection:} Ensuring that input data are properly normalized and scaled while avoiding distortions that could mislead the model.
    \item \textbf{Consistency Across Splits:} Verifying that training, development, and test sets share the same distribution. Mismatched distributions can lead to biased evaluations and poor generalization.
\end{itemize}

\paragraph{Monitoring Loss and Metrics:}
Loss curves and evaluation metrics provide valuable insights into model performance during training. This includes:
\begin{itemize}
    \item \textbf{Loss Curves:} Monitoring training and development loss curves helps identify potential issues such as diverging or plateauing trends, often caused by inappropriate learning rates or initialization strategies.
    \item \textbf{Evaluation Metrics:} Comparing performance in the dev and test sets allows the detection of overfitting or underfitting, ensuring the model generalizes well to unseen data.
\end{itemize}

\paragraph{Gradient Checking:}
Gradient checking is a numerical technique for validating the accuracy of backpropagation implementations. By comparing analytical gradients (computed during backpropagation) with numerical approximations, the method ensures that gradient computation is error-free, especially when using custom layers or loss functions.

\paragraph{Hyperparameters:}
Hyperparameters play a crucial role in determining the stability and efficiency of model training. The following strategies are recommended:
\begin{itemize}
    \item Experimentation with various learning rates, batch sizes, and optimizers can help identify the best configurations.
    \item Monitoring training instability, often caused by inappropriate hyperparameters such as excessively high learning rates, can also help with debugging.
\end{itemize}

\paragraph{Regularization and Generalization:}
Debugging regularization techniques such as L2 and dropout ensure that the model maintains a balance between bias and variance. Inspecting layer outputs and gradients, intermediate activations can help diagnose problems such as vanishing or exploding gradients, which can affect training.

Moreover, modern deep learning frameworks offer robust tools to help in debugging. For example, TensorBoard is a visualization tool that provides information on loss curves, metrics, and weight distributions over time, enabling informed adjustments during training. Debugging deep learning models is a multifaceted process that involves addressing challenges across data, architecture, and optimization. By leveraging systematic techniques, practical tools, and detailed analysis of loss and metrics, researchers can ensure that the models achieve robust and reliable performance.

\end{appendices}

\end{document}